\documentclass[11pt]{article}
 \usepackage{amsmath}
 \usepackage{graphicx}
 \usepackage[merge,numbers,compress]{natbib}
 \usepackage[T1]{fontenc}
 \usepackage{booktabs}
 \usepackage{xcolor} 
 \usepackage{xspace}
 \usepackage{dcolumn}
 \usepackage{placeins}
 \usepackage{amssymb}
 \usepackage[colorlinks=true,citecolor=blue!50!black,linkcolor=black]{hyperref}
 \usepackage{caption}
 \usepackage
 [subrefformat=parens,position=top,skip=-15pt,margin=15pt,justification=justified,singlelinecheck=false]
 {subcaption}
 \usepackage{array,multirow}

\newcommand{\Pj}{\ensuremath{\text{j}}\xspace}
\newcommand{\Pp}{\ensuremath{\text{p}}\xspace}
\newcommand{\Pe}{\ensuremath{\text{e}}\xspace}
\newcommand{\Pb}{\ensuremath{\text{b}}\xspace}

\newcommand{\Pt}{\ensuremath{\text{t}}\xspace}
\newcommand{\Pu}{\ensuremath{\text{u}}\xspace}
\newcommand{\Pd}{\ensuremath{\text{d}}\xspace}

\newcommand{\Pg}{\ensuremath{\text{g}}\xspace}

\newcommand{\PW}{\ensuremath{\text{W}}\xspace}
\newcommand{\PZ}{\ensuremath{\text{Z}}\xspace}

\newcommand{\mjj}{\ensuremath{m_{\Pj\Pj}}}
\newcommand{\dejj}{\ensuremath{\Delta \eta_{\Pj\Pj}}}
\newcommand{\dyjj}{\ensuremath{\Delta y_{\Pj\Pj}}}
\newcommand{\ssWW}{\ensuremath{\PW^\pm \PW^\pm \Pj\Pj}}
\newcommand{\osWW}{\ensuremath{\PW^+ \PW^- \Pj\Pj}}
\newcommand{\WZjj}{\ensuremath{\PW^\pm \PZ \Pj\Pj}}
\newcommand{\ZZjj}{\ensuremath{\PZ\PZ\Pj\Pj}}

\newcommand{\Wgamma}{\ensuremath{\PW^\pm \gamma \Pj\Pj}}
\newcommand{\Zgamma}{\ensuremath{\PZ \gamma \Pj\Pj}}
\newcommand{\ggWW}{\ensuremath{\gamma \gamma \rightarrow \PW^+\PW^-}}

\newcommand{\GeV}{\ensuremath{\,\text{GeV}}\xspace}
\newcommand{\TeV}{\ensuremath{\,\text{TeV}}\xspace}

\newcommand{\fb}{{\ensuremath\unskip\,\text{fb}}\xspace}

\newcommand{\pt}{\ensuremath{p_\text{T}}\xspace}

\newcommand{\rT}{{\mathrm{T}}}
\newcommand{\alphas}{\ensuremath{\alpha_\text{s}}\xspace}

\newcommand{\Sherpa}{{\sc Sherpa}\xspace}
\newcommand{\PHANTOM}{{\sc Phantom}\xspace}
\newcommand{\POWHEG}{{\sc POWHEG}\xspace}
\newcommand{\HERWIG}{{\sc HERWIG}\xspace}
\newcommand{\madgraph}{{\sc\small MadGraph5\_aMC@NLO}\xspace}
\newcommand{\madgraphbis}{{\sc\small MG5\_aMC}\xspace}

\usepackage{color}

\begin{document}

\title{\hfill ~\\[-30mm]
\phantom{h} \hfill\mbox{\small FR-PHENO-2021-05, TIF-UNIMI-2021-2, VBSCAN-PUB-02-21}
\\[1cm]
\vspace{13mm}   \textbf{Vector-Boson Scattering at the LHC: \\ unravelling the Electroweak sector}}

\date{}
\author{
Roberto Covarelli$^{1\,}$\footnote{E-mail:  \texttt{roberto.covarelli@unito.it}},
Mathieu Pellen$^{2\,}$\footnote{E-mail:  \texttt{mathieu.pellen@physik.uni-freiburg.de}},
Marco Zaro$^{3\,}$\footnote{E-mail:  \texttt{marco.zaro@mi.infn.it}}
\\[9mm]
{\small\it $^1$ Physics Department of University of Torino and INFN, Via Pietro Giuria 1,} \\ %
{\small\it Torino, I-10125, Italy}\\[3mm]
{\small\it $^2$ Albert-Ludwigs-Universit\"at Freiburg, Physikalisches Institut,} \\ %
{\small\it Hermann-Herder-Stra\ss e 3, D-79104 Freiburg, Germany}\\[3mm]
{\small\it $^3$ TIF lab, Dipartimento di Fisica, Universit\`a degli Studi di Milano,} \\ %
{\small\it and INFN, Sezione di Milano, Via Celoria 16, 20133 Milano, Italy}\\[3mm]
    }
\maketitle

%
%
%
%
%
%
%
%
%

\begin{abstract}

Vector-boson scattering (VBS) processes probe the innermost structure of electroweak interactions in the Standard Model, and provide a unique
sensitivity for new physics phenomena affecting the gauge sector. In this
review, we report on the salient aspects of this class of processes, both from the theory and experimental point of view. We start by
discussing recent achievements relevant for their theoretical description, some of which have set important milestones in improving
the precision and accuracy of the corresponding simulations. We continue by covering the development of experimental 
techniques aimed at detecting these rare processes and improving the signal sensitivity over large backgrounds.
We then summarise the details of the most relevant VBS signatures and review the related measurements available to 
date, along with their comparison with Standard-Model predictions. We conclude by discussing the perspective at the upcoming
Large Hadron Collider runs and at future hadron facilities.

\end{abstract}


\newpage

\tableofcontents

\newpage

\section{Introduction}
\label{sec:intro}

The Standard Model (SM) of fundamental interactions is a theory which explains natural phenomena
at the smallest distances that can be probed by human-built scientific facilities. Despite the
very simple assumptions it is based upon, Lorentz invariance, locality, and gauge symmetries, 
it is able to explain an astonishing wide range of phenomena, from atomic spectroscopy to particle
collisions at the highest possible energy. In a Lagrangian formulation, forces are represented by gauge fields, with
a symmetry group $\mathcal G = \textrm{SU}(2)_L \otimes \textrm{U}(1)_Y \otimes \textrm{SU}(3)_C$~\cite{Glashow:1961tr,Weinberg:1967tq,Salam:1968rm}, 
where the three groups respectively gauge
the weak isospin, hypercharge, and color charge quantum numbers. Matter is represented by spin-$\frac{1}{2}$ fields, which fall under 
different representations of the gauge groups: the quarks, which are triplets for $\textrm{SU}(3)_C$, and the leptons which are
singlets. Quark and lepton fields come in three families, \emph{i.e.}\ three copies
with identical gauge quantum numbers, but with different masses. Each family is organised in doublets. Particles within each doublet differ by their electric charge, so there are
up- and down-type quarks ($\Pu, \Pd$), as well as neutrinos and charged leptons ($\nu, \ell$). Remarkably, the interactions with the $\textrm{SU}(2)_L \otimes \textrm{U}(1)_Y$ fields are
dictated by the fermion helicity: indeed, only left-handed doublets are charged under $\textrm{SU}(2)_L$, which makes gauge interactions in the SM \emph{chiral}.

This seemingly neat and simple structure, however, is not sufficient to explain the origin of mass of gauge bosons, as well as fermions, since an inclusion of mass terms in the Lagrangian would
unavoidably lead to breaking gauge symmetries. This problem
was solved in the 1960's by 
Higgs, Brout, Englert, Guralnik, Hagen, and Kibble~\cite{Higgs:1964ia,Higgs:1964pj,Englert:1964et,Guralnik:1964eu,Higgs:1966ev,Kibble:1967sv}. By adding a new 
scalar field to the theory, it is possible to trigger the so-called spontaneous Electroweak Symmetry Breaking (EWSB), giving mass to the gauge fields and preserving gauge symmetries
at the same time. After spontaneous symmetry breaking, the hypercharge and isospin fields mix and give rise to the mediators of the electromagnetic and weak interactions:
the massless photon, $\gamma$, and the massive weak bosons, $\PW^\pm$ and $\PZ$. Similarly, the Yukawa-type interactions between the scalar field and the fermions give rise to the 
fermionic masses~\cite{Weinberg:1967tq}. The remainder of the scalar field is the so-called Higgs boson, a new particle whose existence is a prediction of the SM. The quest for
this new particle culminated in 2012, almost fifty years after its existence was postulated, with the discovery announced by the ATLAS and CMS experiments at CERN~\cite{Aad:2012tfa,Chatrchyan:2012ufa}. 

After the discovery of the Higgs boson, while the SM is a complete and consistent theory, some phenomena remain unexplained:
the dominance of matter over anti-matter in the universe, the non-natural pattern of fermionic masses, and the evidence for neutrino oscillations are some of these. This is why
extensions of the SM have been hypothesized, which either predict new particles, or deviations of parameters from the SM predictions, or both. Experimental
searches and measurements, such as those carried out by the ATLAS and CMS experiment at the CERN Large Hadron Collider (LHC), scrutinise many different scattering processes in order
to find any sign of physics beyond the SM (BSM). Among the various processes which trigger the attention of theory and experimental experts, vector-boson
scattering (VBS) is certainly one prominent example. Indeed, it probes two key aspects of the SM together: gauge interactions on the one hand, being one
of the few processes with tree-level sensitivity to the quadrilinear (or quartic) gauge couplings; the couplings between the Higgs and gauge bosons on the other hand, which are probed
at energy scales which can sensibly differ from the Higgs mass. Indeed, a typical undergraduate textbook exercise is to show that, in the SM, the three
classes of Feynman diagrams consisting of i) diagrams featuring only trilinear gauge couplings, ii) diagrams featuring the quartic gauge coupling, and iii) diagrams
featuring the Higgs boson violate unitarity, if considered on their own. However, when considering all the three classes together, unitarity is restored.

At the time when this review is written, between the LHC Run-2 and Run-3, experimental collaborations have collected the first evidences for rare VBS processes,
while on the theory side several key advances in their description have been achieved. The scope of the review is therefore to present such advancements both on
the theoretical and experimental side, and to outline the possible improvements that could be attained with the LHC Run-3 data. In particular, 
this work focuses on SM results, while BSM physics in VBS is only briefly reviewed. The reader interested in BSM can find a more extensive discussion in Ref.~\cite{Gallinaro:2020cte}.
In addition, complementary literature on the topic of VBS exists and can be found in Refs.~\cite{Szleper:2014xxa,Green:2016trm,Rauch:2016pai,Tricoli:2020vgn} and references therein.

This review is structured as follows.
The first part is devoted to general aspects of VBS at the LHC and begins with a definition of what is actually meant by VBS in this context. It then turns to the theoretical aspects and experimental techniques use to predict and measure VBS at the LHC.
The second part reviews all the possible VBS signatures at the LHC from both an experimental and a theoretical perspective.
After this part, which is the core of the review, a short section is dedicated to the future of VBS measurements, especially concerning the High-Luminosity phase of the LHC as well as higher-energy regimes.
Finally, the review ends with a summary and concluding remarks in the last section.

\section{General aspects of vector-boson scattering at the LHC}
\label{sec:vbs}

In this section, general aspects of VBS at the LHC are addressed.
First, a clear definition of VBS at the LHC is given.
In particular, emphasis is put on potential differences between theory and experiment in that respect. The concept of \emph{polarised} VBS cross section is then introduced and results related to that topic are reviewed.

After these introductory sections, the theoretical and experimental status of VBS at the LHC is discussed.
For the theory part, all aspects related to Quantum Chromo-Dynamics (QCD) and electroweak (EW) calculations are presented.
The experimental part starts with the description of the ATLAS and CMS detectors, before moving to the main aspects of VBS analyses:
the reconstructions of tagging jets and of the VBS final states.

This section ends with a short review of current limits on anomalous quartic couplings, obtained by the ATLAS and CMS collaborations during the first two runs of the LHC.
In particular, no bounds on concrete models are discussed as these can be found in other reviews such as Ref.~\cite{Gallinaro:2020cte}.

\subsection{Definition of vector-boson scattering}
\label{sec:sob}

From an experimental perspective, a scattering process at the LHC is defined through its measured final state.
It includes particle jets, leptons, photons or missing energy from neutrinos, which define the signature of the process.
From a theoretical point of view, a scattering process is defined through its external particles as well as its strong and electroweak couplings in perturbative theory.
While these two definitions partially overlap, they also induce some ambiguities when comparing measurements with theory predictions.

\begin{figure*}[t]
\centering
          \includegraphics[width=0.30\linewidth]{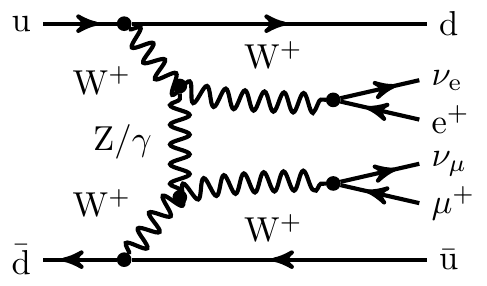}
          \includegraphics[width=0.30\linewidth]{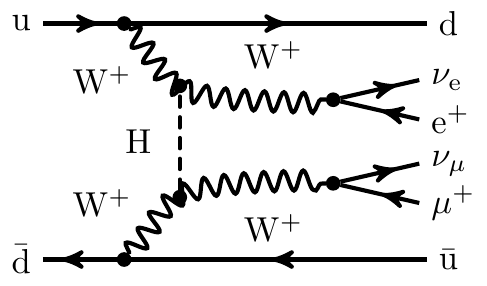}
          \includegraphics[width=0.30\linewidth]{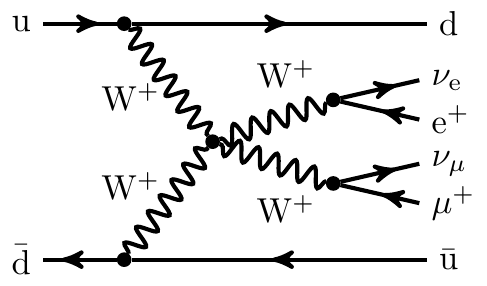}\\[2ex]
          \raisebox{.5ex}{\includegraphics[width=0.35\linewidth]{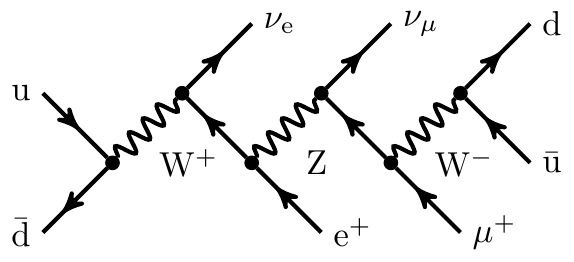}}
          \raisebox{-1.8ex}{\includegraphics[width=0.32\linewidth]{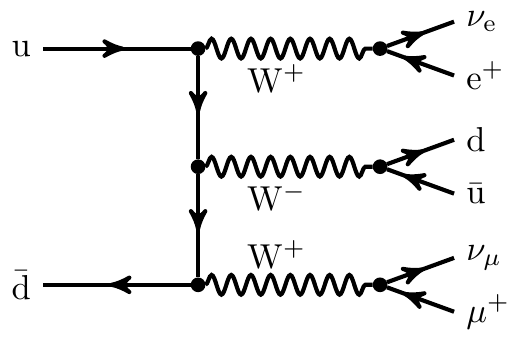}}
\caption{Typical Feynman diagrams for VBS contributions (top) as well as non-VBS contributions (bottom) 
contributing to the process ${\rm p}{\rm p}\to\mu^+\nu_\mu{\rm e}^+\nu_{\rm e}{\rm j}{\rm j}$.
These figures are taken from Ref.~\cite{Biedermann:2017bss}.}
\label{fig:diag}
\end{figure*}

The measurement of VBS at the LHC is exemplary in this respect.
The typical picture of VBS consists of two gauge bosons radiated off two separate quarks lines to scatter.
Typical Feynman diagrams are shown in the top row of Fig.~\ref{fig:diag}.
In case of heavy vector bosons (all except $\gamma$), the VBS process is thus defined at Born level at the order $\mathcal{O}\left( \alpha^6\right)$, upon including the decay products of the heavy gauge bosons.
It has quarks in the initial state and at least two quarks and up to four leptons in the final state.
This implies that three possible VBS signatures at the LHC are: 2 jets and 4 leptons (\emph{fully leptonic}), 4 jets and 2 leptons (\emph{semi-leptonic}/\emph{semi-hadronic}), or 6 jets (\emph{fully hadronic}).
This definition has the advantage to be clearly gauge invariant and to describe all non-resonant, off-shell, and interference effects.
In particular, it means that many other diagrams beyond the VBS ones such as tri-boson contributions are included (some of them are shown in the bottom row of Fig.~\ref{fig:diag}).
In order to select VBS diagrams only, approximations to the full process like the effective vector-boson \cite{Dawson:1984gx,Duncan:1985vj,Cahn:1983ip,Kuss:1995yv,Accomando:2006hq} or the vector-boson scattering \cite{Figy:2003nv,Ballestrero:2018anz} ones have to be used.

Measuring experimentally a VBS signature necessarily implies also measuring non-VBS contributions. On an event-by-event basis, the quantum-mechanical nature of the process does not allow to distinguish a VBS event from a tri-boson event for example.
Therefore, experimental analyses targeting VBS measurements sometimes include specific cuts in order to suppress undesired contributions in the definition of fiducial regions for the measurements.
In the case of tri-boson contributions, a typical cut would be an invariant-mass cut on the decay products of a $\PW/\PZ$ boson around their mass.
Alternatively, such contributions can also be subtracted from the measurements using Monte Carlo simulations.

In addition to the purely EW contributions at order $\mathcal{O}\left( \alpha^6\right)$, VBS signatures also feature irreducible contributions of orders
$\mathcal{O}\left( \alpha_{\rm s} \alpha^5\right)$ and $\mathcal{O}\left( \alpha_{\rm s}^2 \alpha^4\right)$.
These three contributions are usually referred to as \emph{EW (or VBS) signal}, \emph{interference}, and \emph{QCD background}, respectively.
Again, on an event-by-event basis, an event cannot be unambiguously attributed to any of these contributions.
Nonetheless, experimental collaborations have measured VBS in signal-enriched phase-space regions upon applying specific kinematic cuts.

\begin{figure*}[t]
    \centering
    \includegraphics[width=0.49\linewidth]{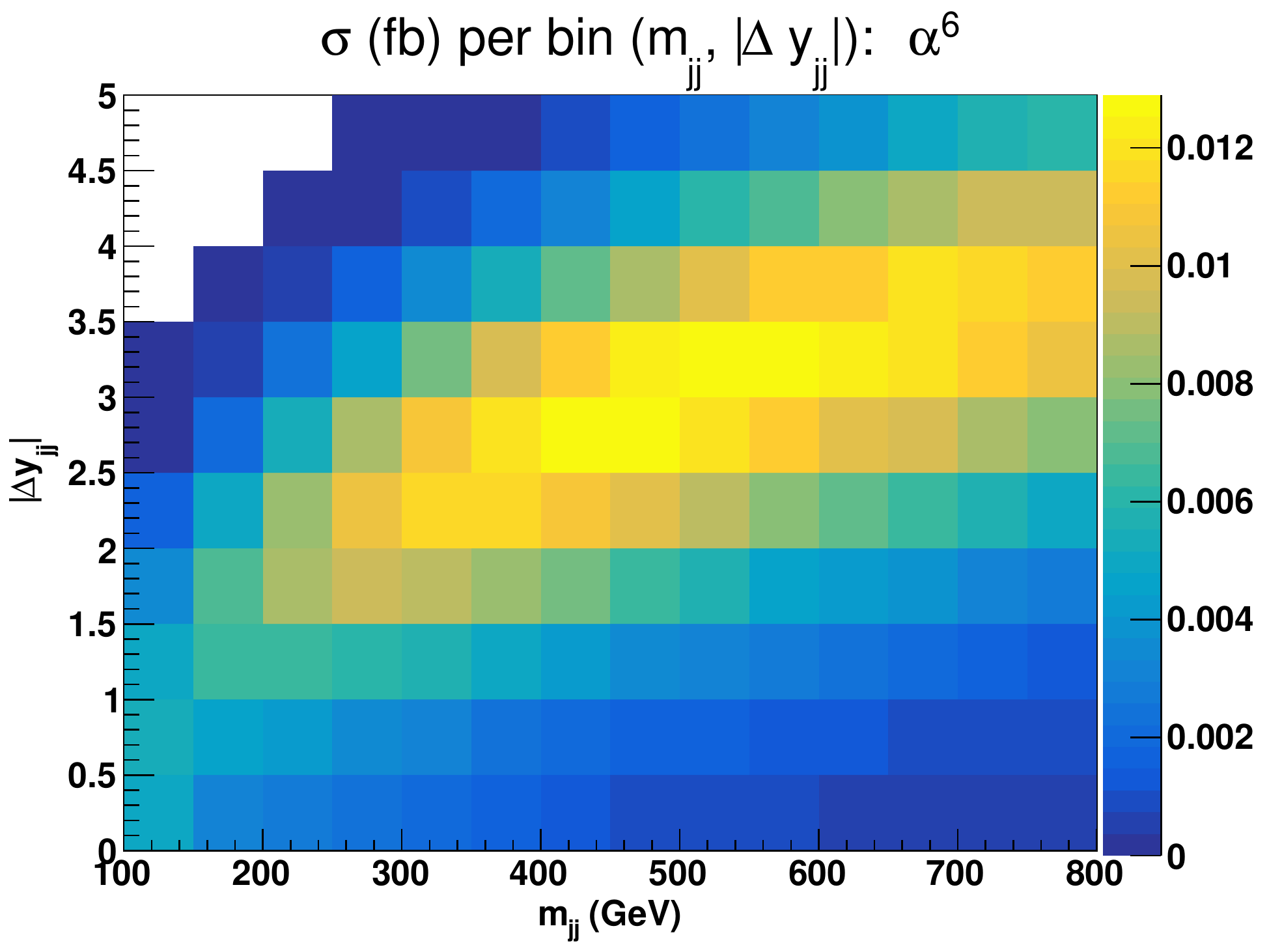}
    \includegraphics[width=0.49\linewidth]{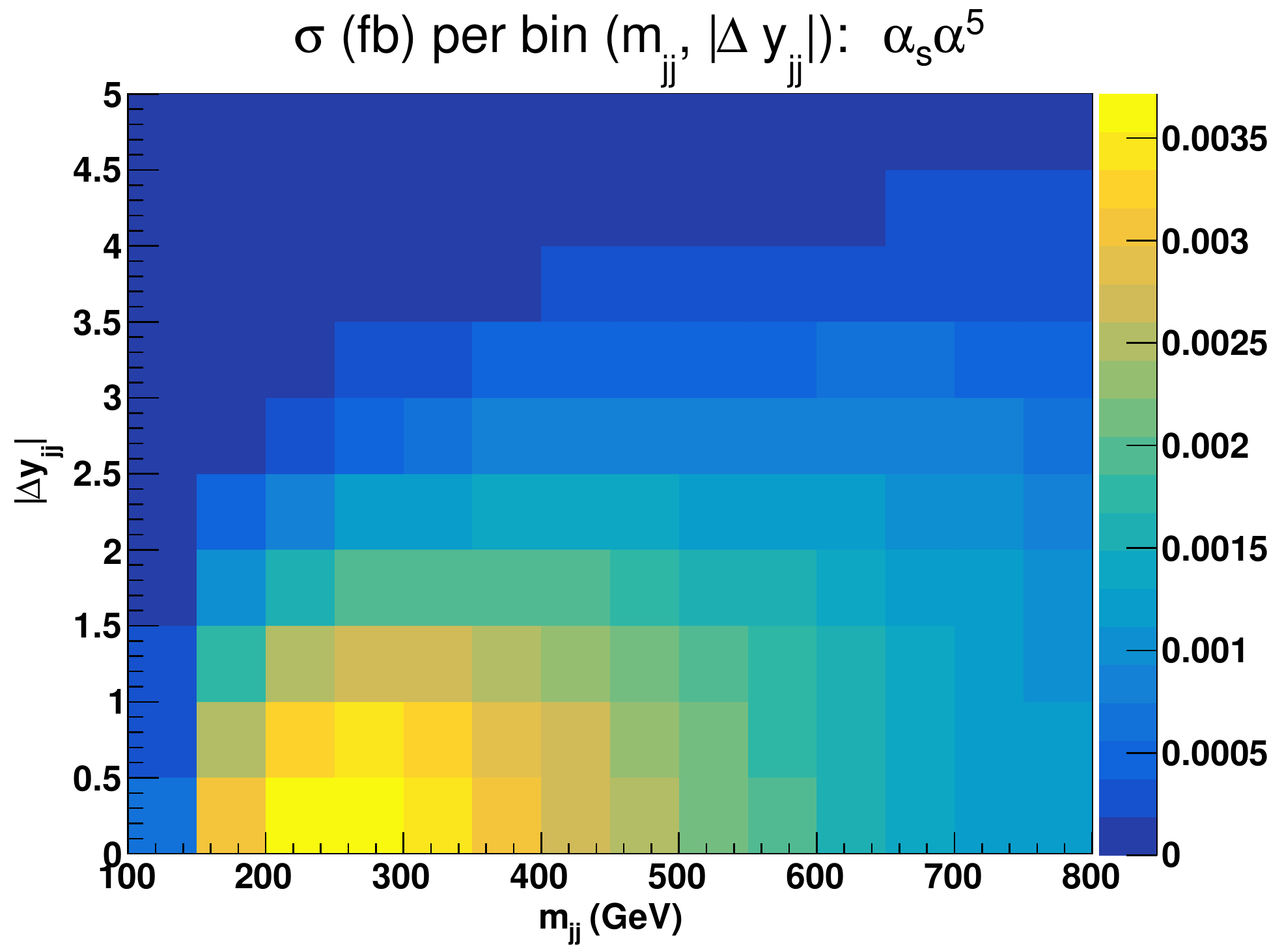} \\
    \includegraphics[width=0.49\linewidth]{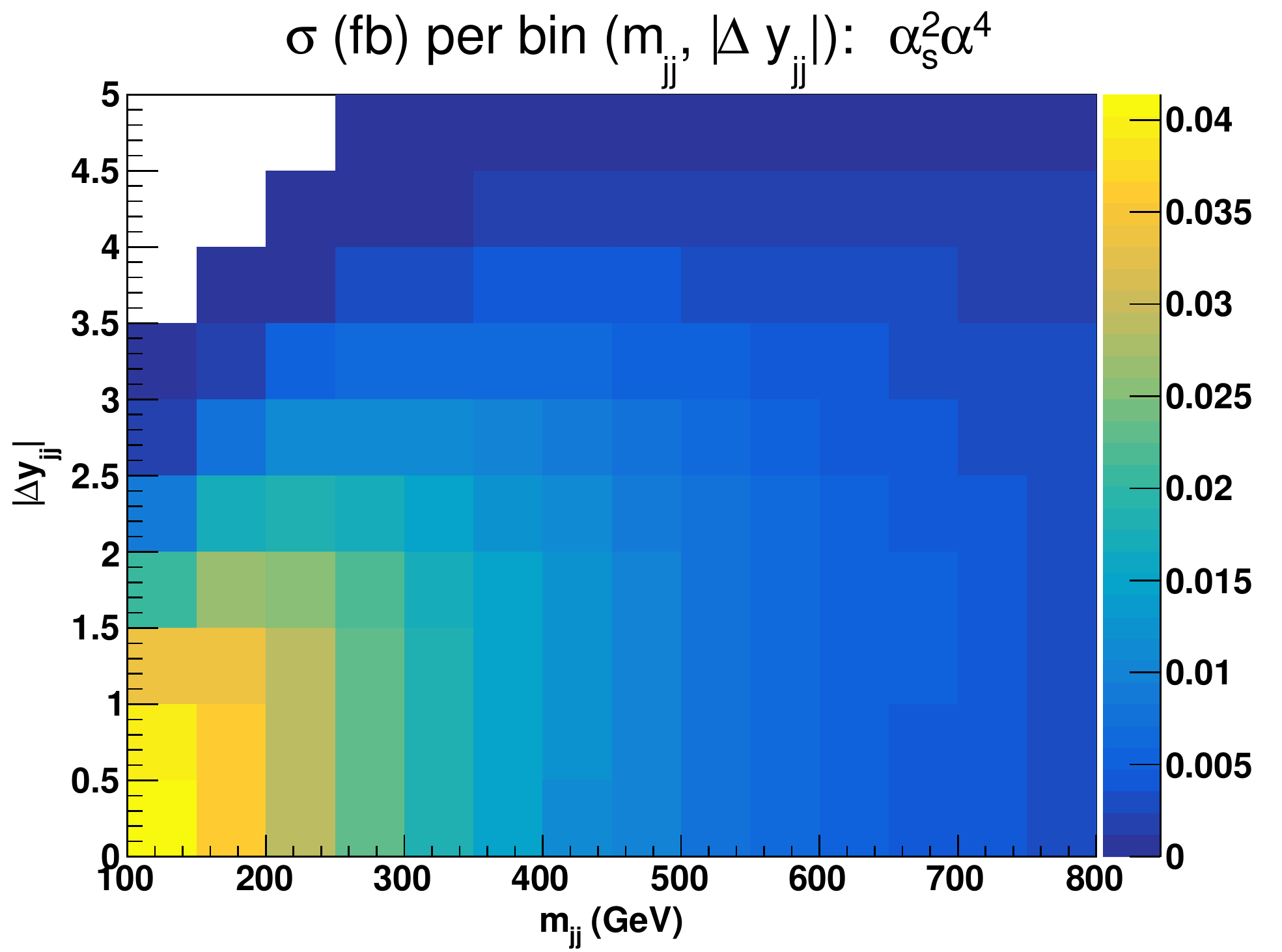}
    \caption{Double-differential distributions in the variables $m_{{\rm j}{\rm j}}$ and $|\Delta y_{{\rm j}{\rm j}}|$ for the three LO contributions of orders $\mathcal{O}(\alpha^6)$ (top left), 
    $\mathcal{O}(\alpha_{\rm s}\alpha^5)$ (top right), and $\mathcal{O}(\alpha_{\rm s}^2 \alpha^4)$ (bottom).
    Typical event selection is applied apart from cuts on $m_{{\rm j}{\rm j}}$ and $|\Delta y_{{\rm j}{\rm j}}|$.
    These figures are taken from Ref.~\cite{Ballestrero:2018anz}.}
    \label{fig:scan}
    \end{figure*}

The justification of such a procedure is that the EW and the QCD contributions behave rather differently.
In particular, given their rather different QCD structures, they tend to be maximal in different phase-space regions.
The EW component does not feature QCD exchanges between the two quark lines while the QCD component does.\cite{Rauch:2016pai}
It implies that the differential cross section as a function of the invariant mass or the rapidity difference of the two scattered quarks (that form what will be called the \emph{tagging jets}) is very different for the two components:\cite{Ballestrero:2018anz}
typically, the EW component is characterized by tagging jets with large invariant masses (\mjj) and rapidity differences (\dyjj), leaving the central part of the scattering free from QCD activity, at least in the fully leptonic signature,
while it is the opposite for the QCD component.
This is particularly well illustrated by Fig.~\ref{fig:scan} which shows two-dimensional differential distributions of the three contributions as a function of the invariant mass and rapidity difference of the two tagging jets.
Therefore, in the same way as for non-VBS contributions, cuts can thus be applied in order to suppress the interference and/or QCD background: the invariant mass of the tagging jets and their rapidity separation are the basic discriminating cuts for VBS measurements at the LHC.

\begin{figure*}[t]
\centering
          \includegraphics[width=0.90\linewidth]{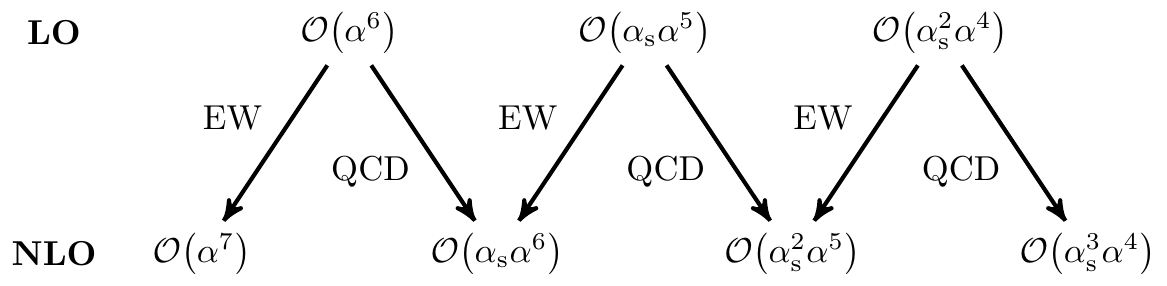}
\caption{All contributing orders at both LO and NLO for the VBS processes at the LHC.
This figure is taken from Ref.~\cite{Biedermann:2017bss}.}
\label{fig:orders}
\end{figure*}

Alternatively, the QCD and interference contributions can again be subtracted using Monte Carlo predictions, the aim still being to isolate VBS contributions from its irreducible background.
While this procedure is already questionable at LO, it is not meaningful at NLO and beyond.
At LO, it is an arbitrary choice to include interference contribution to either the EW signal or to the QCD background as by definition the interference possesses both amplitudes.
At NLO, the number of contributions goes up to 4 as shown in Fig.~\ref{fig:orders}.
In particular, some of these corrections are of mixed type which means that two different Born processes (with the two different amplitudes) received corrections.
Because the two types of corrections are linked by infrared singularities, they cannot be separated in contributions that have only one type of amplitude (EW or QCD).\cite{Biedermann:2017bss}
Hence, it is not possible to define an NLO signal or background without making assumptions.
As a result, a measurement of the VBS signal necessarily relies on such theoretical inputs and their approximations.

In order to rely as less as possible on theoretical inputs, as well as to enable the soundest comparison between theory and data, 
fiducial measurements should be presented without subtracting any process contributing to the VBS signatures (neither irreducible QCD or interference background nor tri-boson contributions).
From this physical measurement, various subtractions can subsequently be applied in order to single out the salient features of VBS.
Such results are naturally subject to approximations and can then be studied as such.
We believe that this is the best way to get most of the VBS physics in a transparent way and hence foster fruitful exchange between the theory and the experimental community.
It is worth emphasising that first steps in this direction have been already taken for example in Refs.~\cite{CMS:ssWWandWZ,CMS:ZZ2} 
by presenting fiducial cross sections of the EW and QCD component separately as well as their sum.

Along these lines, the presentation of several fiducial regions in experimental analyses is also welcome as it enables the study of various physical effects.
Looser selections are typically dominated by the QCD background or tri-boson contributions, while tighter VBS cuts will highlight mostly the characteristics of the vector-boson scattering at high energy.
These physical effects are the multiple sides of the same physical process and such measurements have therefore the power to explore all of them.

\subsection{Polarised vector-boson scattering}
\label{sec:polar}

Among the various features of the SM which are especially relevant for VBS, the possibility to access the different polarisation states of the vector bosons is certainly one
of the most intriguing ones. After EWSB, massive vector bosons
feature three physical polarisation states: two transverse (left and right handed), $\epsilon_\mu^{\rm L/R}$, and one longitudinal,  $\epsilon_\mu^0$. In the SM,
the vector-boson masses and their longitudinal polarisation are generated by the Higgs mechanism, and the presence of the Higgs boson unitarises the scattering
amplitude for longitudinal polarisations. Hence the ability to study different polarisation states is a precious asset in order
to validate the SM, or to spot possible deviations from its predictions. BSM physics may disrupt the unitarisation of longitudinally-polarised vector 
bosons~\cite{Lee:1977yc,Lee:1977eg,Chanowitz:1984ne,Chanowitz:1985hj}, or 
alter the relative impact of different polarisation states. 

The possibility to explore different polarisation states is a feature of all processes involving vector bosons (see \emph{e.g.}\ Ref.~\cite{Stirling:2012zt}).
For example, single vector-boson production, possibly in association with jets, is dominated by left-handed polarisation states~\cite{Bern:2011ie}, while the $\PW$ boson from
the top decay is dominantly longitudinal~\cite{AguilarSaavedra:2006fy}.

The definition of the production of a specific polarisation state is therefore possible. In
the following, we will show that this is true only in an idealised situation, as a realistic environment poses significant difficulties. We will discuss
how these difficulties can be overcome and the necessary conditions for polarised cross sections to be well defined. We will conclude the section by presenting some
results for polarised VBS at hadron colliders in the SM and in some of its extensions.

\subsubsection{Definition of polarised cross section}

\begin{figure*}[t]
\centering
          \includegraphics[width=0.45\linewidth,clip=true,trim={0.2cm 0cm 1.5cm 0.cm}]{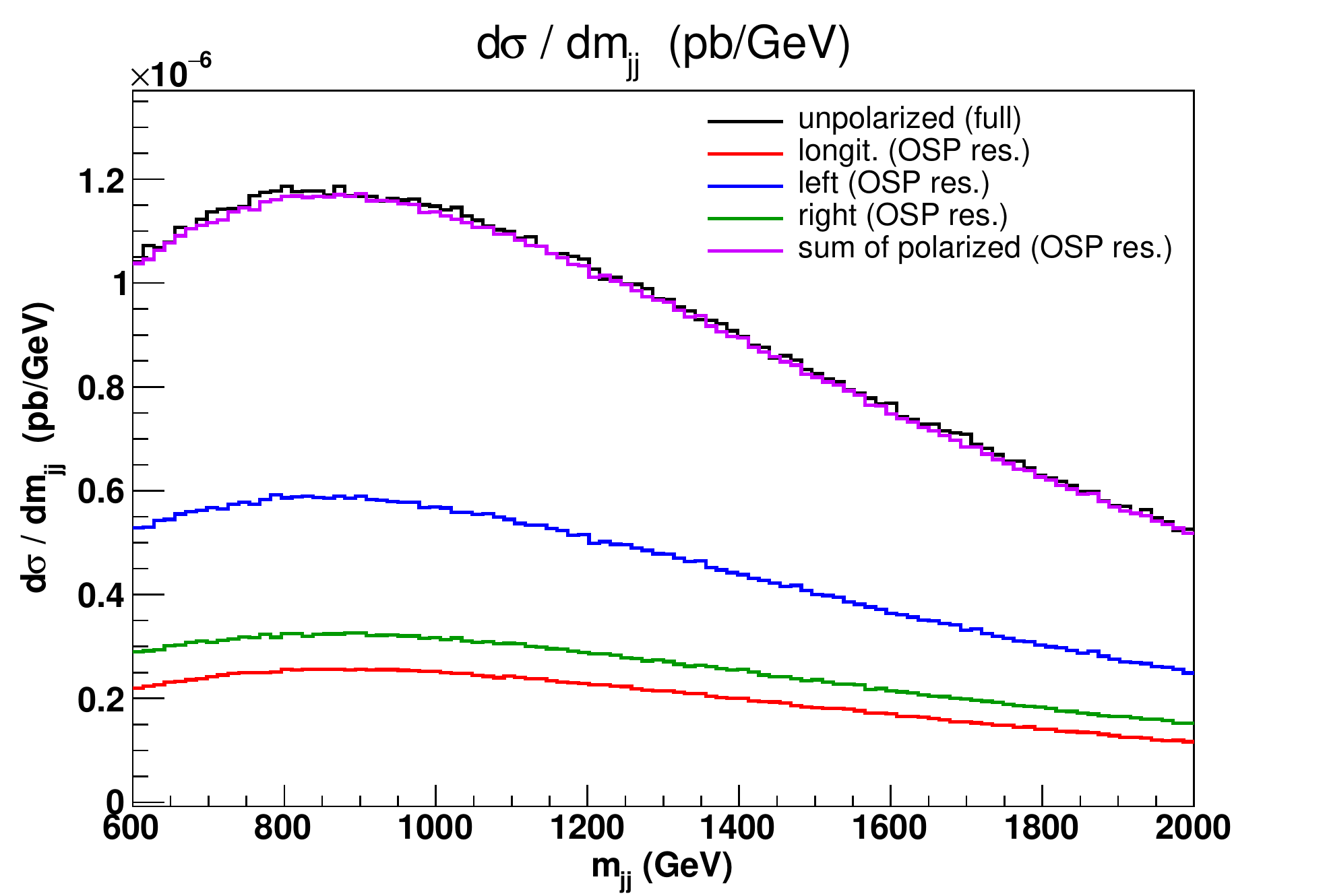}
          \includegraphics[width=0.45\linewidth,clip=true,trim={0.2cm 0cm 1.5cm 0.cm}]{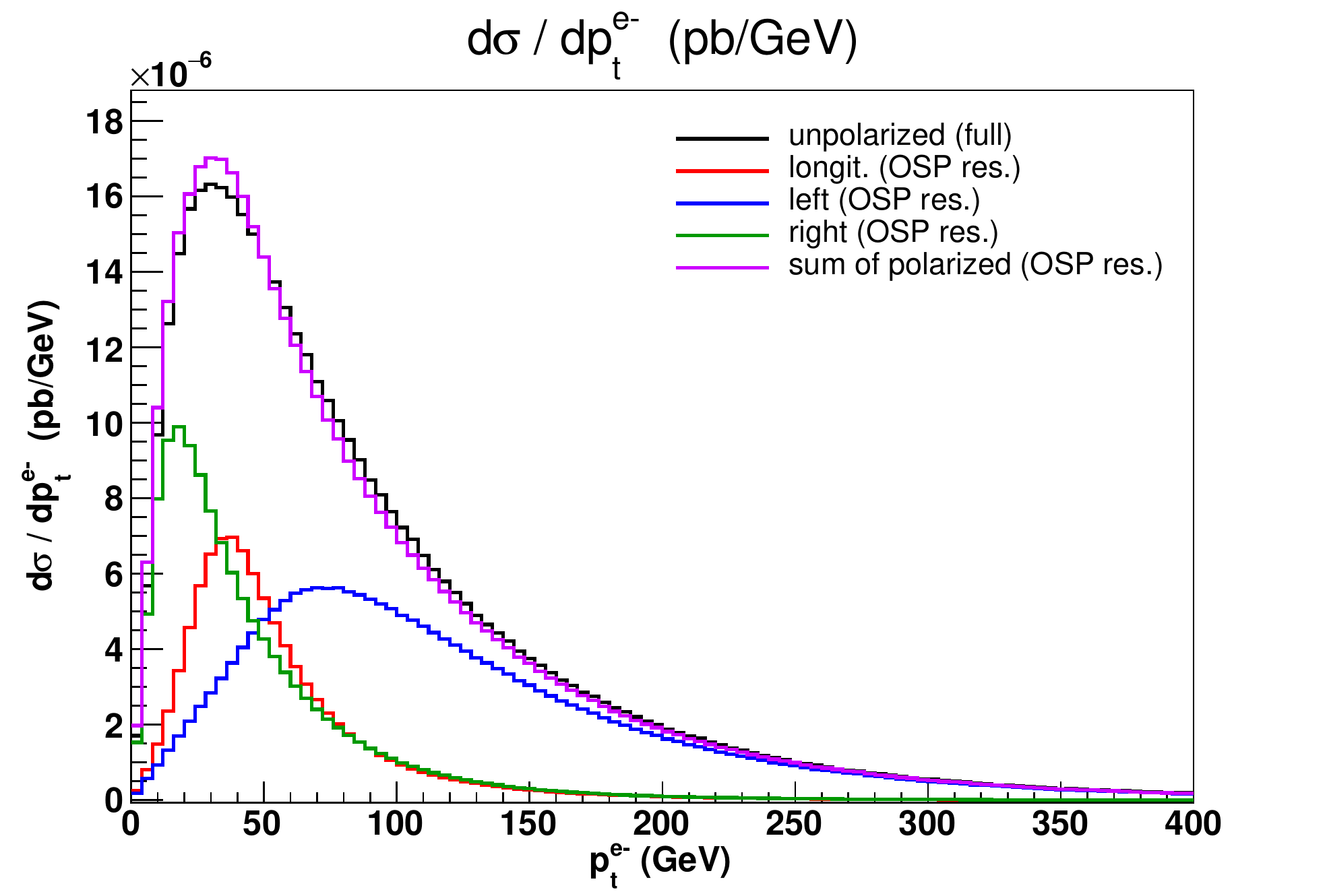} \\
          \includegraphics[width=0.45\linewidth,clip=true,trim={0.2cm 0cm 1.5cm 0.cm}]{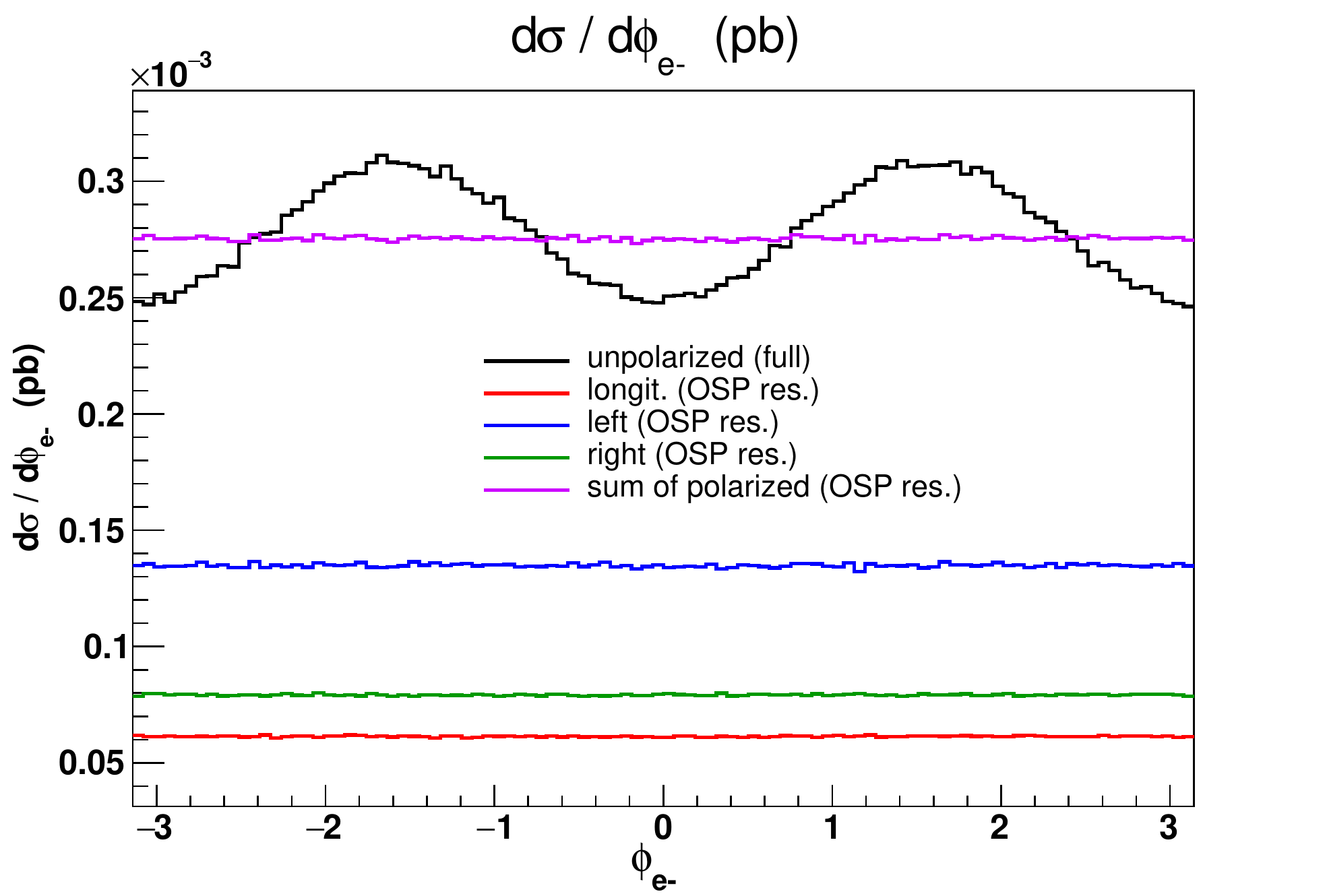}
\caption{Differential singly-polarised cross sections for opposite-sign W scattering at the LHC. Polarisations are defined in the laboratory frame. The unpolarised prediction (black)
    is compared with the incoherent sum (purple) of the polarized
 ones (blue, green and red; these use  on-shell projection and take into account only the resonant diagrams). No cut is imposed on leptonic
variables. These figures are taken from Ref.~\cite{Ballestrero:2017bxn}.}
\label{fig:vbspol1}
\end{figure*}

Several aspects need to be considered when one tries to define the production cross section for a specific polarisation state of a vector boson. In this
discussion, we will follow the discussion of Ref.~\cite{Ballestrero:2017bxn}, which also documents the implementation of the polarised cross 
section for \osWW\ in the \PHANTOM code~\cite{Ballestrero:2007xq}. Extension to the cases of WZ and ZZ production are documented in Ref.~\cite{Ballestrero:2019qoy}.
The first and most immediate aspect is that vector bosons are unstable particles which undergo a decay. Hence any information on their polarisation must be kept through the decay
processes. If we consider the case of a single vector boson (the generalisation to the case of multiple vector bosons is trivial) which is produced from
an initial state $I$ and decays into a final state $F$:
\begin{equation}
 I \to V \to F,
\end{equation}
the corresponding matrix element can be written as:
\begin{equation}
    \mathcal M _{I \to V \to F} = \mathcal M_{I \to V}^\mu \frac{-g_{\mu\nu} + \frac{p_\mu p_\nu}{M_V^2}}{\left(p^2-M_V^2\right)^2 + M_V^2\Gamma_V^2}
                                \mathcal M_{V \to F}^\nu .
                                \label{eq:polme}
\end{equation}
Here, $p$ is the momentum of the intermediate vector boson. The projector $-g_{\mu\nu} + \frac{p_\mu p_\nu}{M_V^2}$ can be expressed as the sum over
the polarisations of the intermediate vector bosons:
\begin{equation}
    -g_{\mu\nu} + \frac{p_\mu p_\nu}{M_V^2} = \sum_\lambda \epsilon^\lambda _{\mu}  {\epsilon^\lambda _{\nu}}^*.
\end{equation}
Now, when the amplitude in Eq.~\eqref{eq:polme} is squared, one obtains
\begin{eqnarray}
\left|\mathcal M _{I \to V \to F}\right|^2 & = &
\mathcal M_{I \to V}^\mu \frac{ \sum_\lambda \epsilon^\lambda _{\mu}  {\epsilon^\lambda _{\nu}}^*} {\left(p^2-M_V^2\right)^2 + M_V^2\Gamma_V^2}
                                 \mathcal M_{V \to F}^\nu \nonumber\\
                                 & & \times \left(\mathcal M_{I \to V}^\mu \frac{ \sum_{\lambda'} \epsilon^{\lambda'} _{\mu}  {\epsilon^{\lambda'} _{\nu}}^*} {\left(p^2-M_V^2\right)^2 + M_V^2\Gamma_V^2}
                                 \mathcal M_{V \to F}^\nu \right)^* \nonumber\\
            &\ne& 
\sum_\lambda
\left[
\mathcal M_{I \to V}^\mu \frac{ \epsilon^\lambda _{\mu}  {\epsilon^\lambda _{\nu}}^*} {\left(p^2-M_V^2\right)^2 + M_V^2\Gamma_V^2}
                                 \mathcal M_{V \to F}^\nu \right. \nonumber\\
                                 & & \left. \times \left(\mathcal M_{I \to V}^\mu \frac{\epsilon^{\lambda} _{\mu}  {\epsilon^{\lambda} _{\nu}}^*} {\left(p^2-M_V^2\right)^2 + M_V^2\Gamma_V^2}
                             \mathcal M_{V \to F}^\nu \right)^* \right]. \label{eq:polinterf} 
\end{eqnarray}
The meaning of Eq.~\eqref{eq:polinterf} is that, since the vector bosons are not external particles, their polarisation states interfere with each other. This, in principle,
jeopardises the definition of a polarised cross section. However, interference terms integrate to zero over the whole range of the decay azimuth angle, 
and this makes it possible, at least in principle, to have a well-defined polarised cross section.
It has to be stressed that, whenever cuts are imposed (as it is the case in any realistic setup) or when
one is interested in observables sensitive to the decay degrees of freedom, particularly to the azimuth angle, the cancellation of interferences is not bound to happen. This can
be observed in Fig.~\ref{fig:vbspol1} for \osWW\ VBS, where singly-polarised cross sections (the positively-charged \PW\ remains unpolarised) and their incoherent sum are compared with the full cross sections: 
the upper left plot shows the dijet invariant mass, which has no dependence on the lepton azimuth angle. The upper right plot shows the lepton transverse momentum, which has an indirect dependence on the azimuth angle.
Finally, the bottom plot shows the lepton azimuth angle. It can be observed that, while for the dijet invariant mass the incoherent sum of the polarisation and the full cross section 
are indistinguishable, small but visible differences appear for the electron transverse momentum, and obvious effects appear for the azimuth angle. Observables 
which display good agreement between the incoherent sum of the different polarisation states and the full result can be employed to extract the polarisation fractions for
the different states. While, as mentioned above, cuts on the leptons can in principle spoil such an agreement, in practice their effect is generally mild on most observables.

There are other aspects which need to be considered in the definition of polarised cross section. First, vector bosons may be produced off-shell, \emph{i.e.}\ far from the resonance peak. 
This issue is typically addressed by using the so-called on-shell projection (OSP), where some momentum transformation is used changing the momenta such that the intermediate vector-boson is on its mass shell.
This is similar to the so-called pole scheme approximation, usually employed in the computation of higher-order corrections (for example in 
Refs.~\cite{Aeppli:1993cb,Aeppli:1993rs,Beenakker:1998gr,Billoni:2013aba,Denner:2000bj,Denner:2016jyo,Denner:2020bcz} and therein).
Second, non-resonant diagrams may exists, \emph{i.e.}\ diagrams for the process $I\to F$ that do not feature an intermediate vector boson in the $s$-channel (\emph{e.g.}\ the left diagram in the second row of Fig.~\ref{fig:diag}), but those are usually assumed not to contribute significantly to the cross section.
The last relevant aspect to be considered is that polarisation vectors are not Lorentz-covariant, hence a given reference frame must be chosen. Typical choices
are the partonic centre-of-mass frame, the laboratory frame or the diboson centre-of-mass frame. In particular the latter has been used in the analysis of 
Ref.~\cite{Aaboud:2019gxl}. The choice of a given frame is mostly dictated by practical reasons, like the experimental capability to reconstruct the frame.
\footnote{For a recent study on different reconstruction techniques, see \emph{e.g.}\ Ref.~\cite{Grossi:2020orx}.}
Studies available to date, see \emph{e.g.}\ Ref.~\cite{Ballestrero:2020qgv}, show that no frame choice has particular advantage over the others.

To summarize, the definition of a polarised cross section relies on the following assumptions: interferences cancel in 
Eq.~\eqref{eq:polinterf} (strictly true only for integrated quantities); non-resonant diagrams are neglected; and an OSP is introduced to reshuffle the external
momenta onto the vector-boson mass shell. In this context, the polarisation fractions $f_{L/R}$ and $f_0$ can be introduced (with polarisation vectors in a frame of choice) 
for a specific kinematic variable $X$. If one considers the case of a $\PW^\pm$ boson decaying into lepton and neutrino, where $\theta$ is the polar angle 
in the $\PW$ rest frame (and the azimuth angle $\phi$ is integrated over), one obtains
\begin{equation}
    \frac{1}{\frac{d\sigma(X)}{dX}} \frac{d\sigma(\theta,X)}{d\cos \theta dX} = 
        \frac{3}{8}\left(1\mp\cos\theta\right)^2 f_{\rm L}(X)+ \frac{3}{8}\left(1\pm\cos\theta\right)^2 f_{\rm R}(X)+ \frac{3}{4}\sin^2\theta f_0(X).
\end{equation}

Using this equation, one could extract the polarisation fractions from data by fitting the angular distributions.
Experimental analysis is in practice more complicated, since selections alter the shape as a function of $\cos\theta$ and angular-dependent acceptance/efficiency factors must be taken into account.

The method discussed above is general, in the sense that it can be applied to any process featuring intermediate vector bosons. While the discussion
has been carried assuming there is a single polarised vector boson, it can be easily extended to the case where more vector bosons appear, such as VBS. Besides the case mentioned
above (single vector boson, and top production), the method of Ref.~\cite{Ballestrero:2017bxn} has been also applied to vector-boson pair production in Refs.~\cite{Baglio:2019nmc,Denner:2020bcz,Denner:2020eck}, 
and in Ref.~\cite{BuarqueFranzosi:2019boy} it has been automatised using \madgraph~\cite{Alwall:2014hca} (henceforth denoted as \madgraphbis) and {\sc MadSpin}~\cite{Artoisenet:2012st}, 
paving the path to the possibility of including NLO QCD corrections in VBS analysis.

\subsubsection{Phenomenological results}

Having introduced the polarisation fractions and their prerequisites to be well defined, we will show some phenomenological results 
which highlight how these fractions can be employed to probe beyond the SM effects.

The first case we consider is discussed in Ref.~\cite{Ballestrero:2017bxn}, and it is the case of a Higgs-less SM, \emph{i.e.}\ corresponds to pushing the Higgs boson mass to infinity.
\begin{figure*}[t]
\centering
          \includegraphics[width=0.7\linewidth,clip=true]{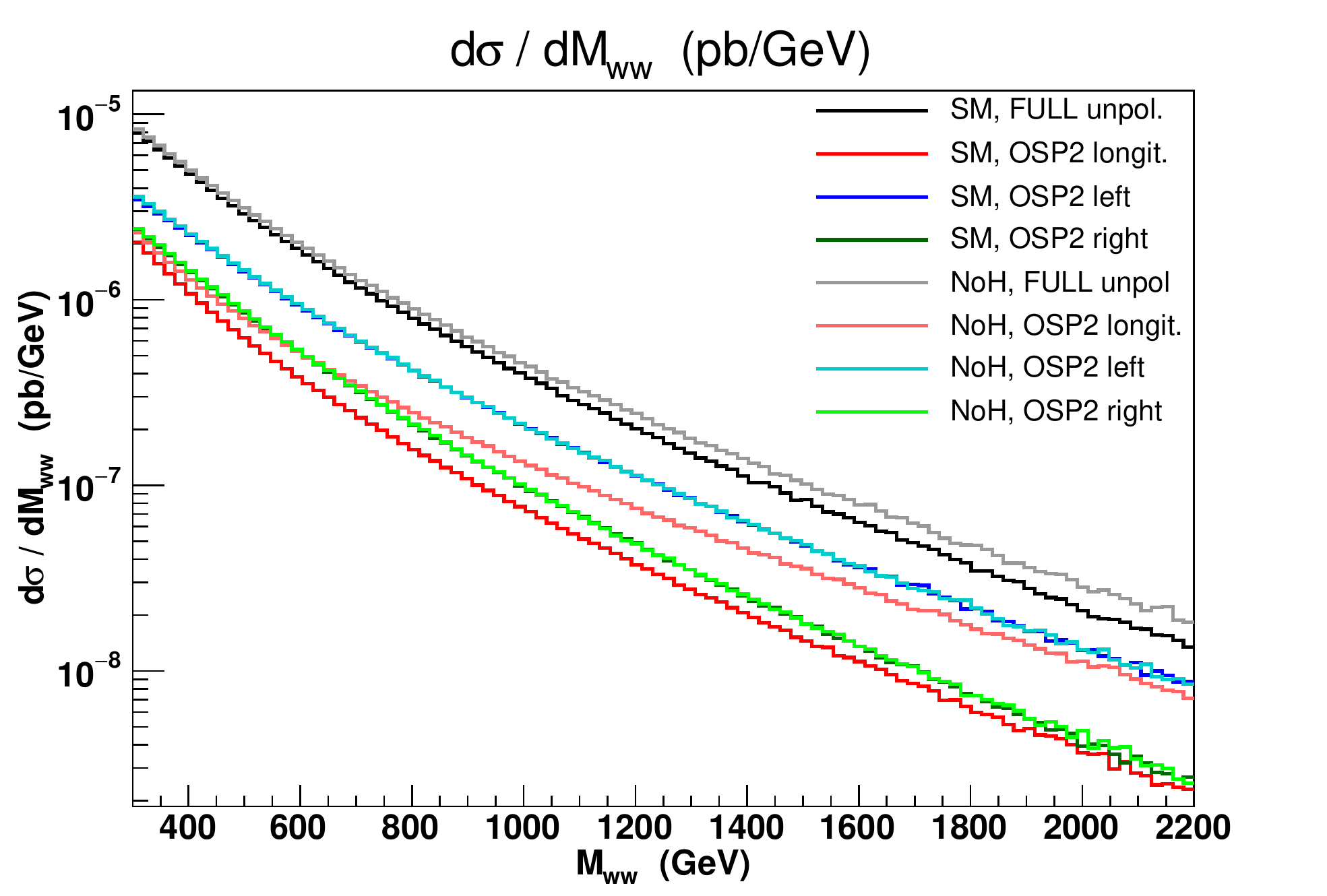}
\caption{Polarisation fractions in the SM and in the Higgs-less SM, in opposite-sign W scattering at the LHC, for the invariant mass of the W 
    pair. Polarisations are defined in the laboratory frame. The (negatively-charged) lepton from
    the polarised W boson is required
    to satisfy the cuts $p_T(\ell) > 20 \GeV$, $|\eta(\ell)| < 2.5$. This figure is taken from Ref.~\cite{Ballestrero:2017bxn}.}
\label{fig:vbspolnoH}
\end{figure*}

In Fig.~\ref{fig:vbspolnoH}, the $\PW\PW$ invariant-mass distribution is shown, for the unpolarised process as well as for the different polarisation states of one of the the negatively-charged vector boson. The SM and the Higgs-less SM (dubbed \emph{NoH} in the figure) are shown. Since the Higgs boson unitarises the scattering of longitudinal
vector bosons, one expects the longitudinally-polarised component in the Higgs-less SM to display a harder behaviour with respect to the SM case, as can be seen
in the plot of Fig.~\ref{fig:vbspolnoH}. While the left and right polarisations display an identical behaviour in the two models, the behaviour of the longitudinal polarisation 
is radically different at high energies. Indeed, in the SM, $21\%$ of events feature a longitudinally-polarised vector boson when a minimum cut, $M_{\PW\PW} > 300\GeV$ is required 
on their invariant mass. The fraction decreases to $15\%$ when the invariant-mass cut is raised to $1\TeV$. In the Higgs-less case, the two fractions become respectively
$27\%$ and $35\%$, with more than a factor-2 effect when the hardest cut is imposed.

\begin{figure*}[t]
\centering
          \includegraphics[width=0.7\linewidth,clip=true]{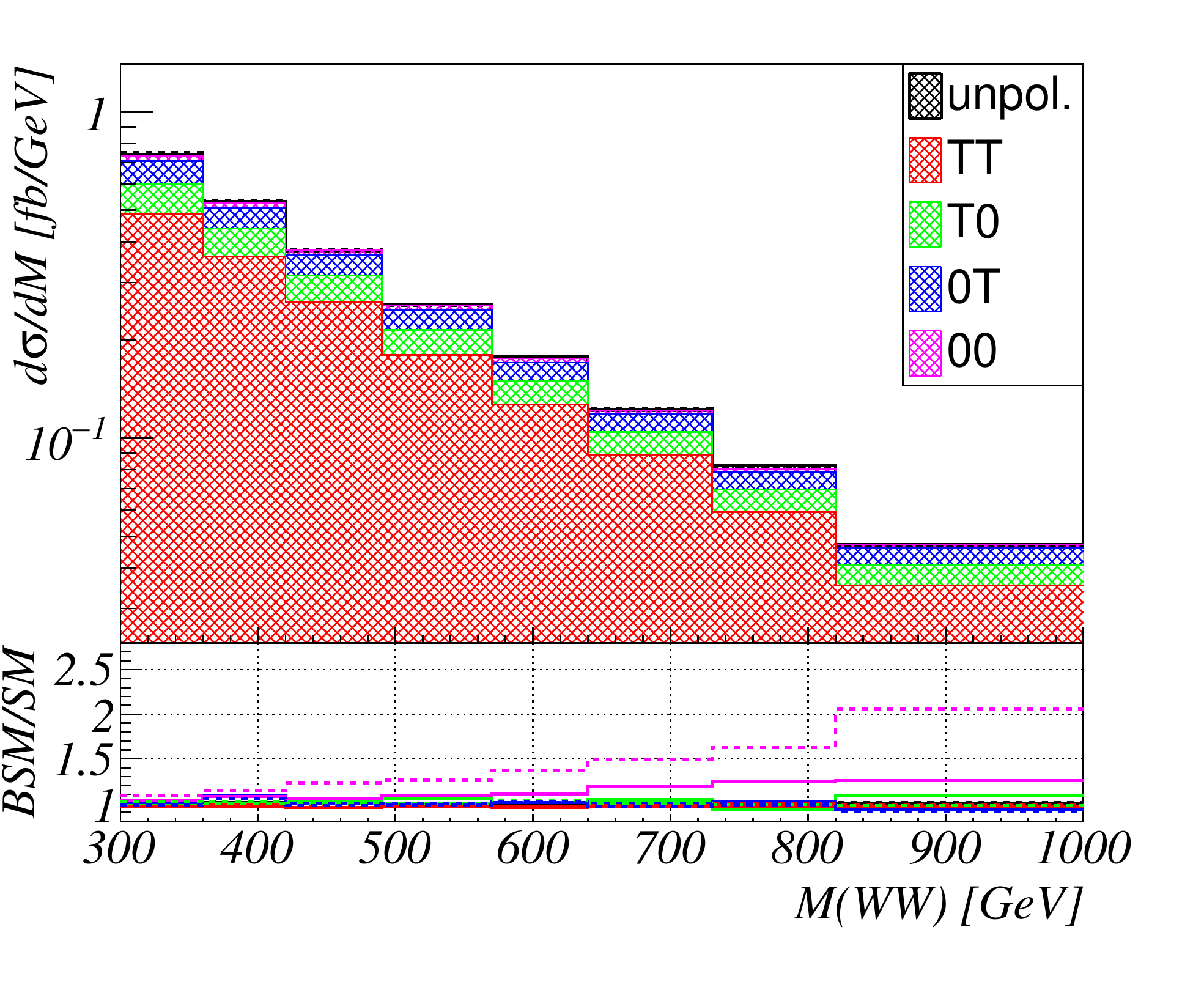}
\caption{Polarisation fractions in the SM in opposite-sign W scattering at the LHC, for the invariant mass of the W pair. Polarisations are defined in the partonic centre-of-mass frame.
The namings 0 and T identify the longitudinal and transverse polarisation, respectively. 
The inset displays the ratios $\mathcal R (M_{WW})$ defined in Eq.~\protect\eqref{eq:vbspolR} for $a=0.8$ (dashed) and $a=0.9$ (solid). 
This figure is taken from Ref.~\cite{BuarqueFranzosi:2019boy}.}
\label{fig:vbspolCH}
\end{figure*}

The second example, from Ref.~\cite{BuarqueFranzosi:2019boy}, compares the SM case with a composite-Higgs 
model~\cite{Kaplan:1983fs,Kaplan:1983sm,Georgi:1984af,Dugan:1984hq,Contino:2003ve,Agashe:2004rs,Contino:2006qr,Agashe:2006at}. In this class of models, or at least in their most recent versions, the interaction of the Higgs
boson and the weak gauge bosons is rescaled with a common factor $a$, and can be described by the following effective Lagrangian~\cite{Bellazzini:2014yua,Panico:2015jxa}
\begin{equation}
  L \supset \left(\frac{m_\PZ^2}{2} \PZ^\mu \PZ_\mu + M_W^2 \PW^\mu \PW_\mu\right)\left(1 + 2 a \frac{h}{v} + \ldots \right)\,.
\end{equation}
The SM case is recovered when the scattering of longitudinal vector bosons is unitarised, which corresponds to $a=1$. Other values different from unity will display a unitary-violating
behaviour. As in the previous case, the process at hand is \osWW\ production and the $\PW\PW$ invariant mass distribution shown in Fig.~\ref{fig:vbspolCH} is examined, 
where the polarisation fractions of both vector bosons are given. The upper plot in the figure
shows the polarisation fractions in the SM ($a=1$), while the lower inset shows how they are affected when one sets $a=0.8$ (dashed) or $a=0.9$ (solid), by plotting the ratio:
\begin{equation}
    \mathcal R(M_{\PW\PW}) 
\equiv 
                         \left. \frac{d \sigma(a)} {d M_{\PW\PW}} \middle/ \frac{d \sigma(a=1)}{d M_{\PW\PW}} \right.\,.
     \label{eq:vbspolR} 
\end{equation}

The most striking behaviour can be observed in the ratio for the longitudinal-longitudinal scattering fraction.
It shows effects of the order of $30\%$ in the largest-mass bin considered for $a=0.9$ and grows up to $100\%$ when $a=0.8$.


\subsection{Theoretical predictions}
\label{sec:theory}
In this section we will review the state of the art of the theoretical predictions for
VBS, and we will discuss the main sources of theoretical uncertainty.

\subsubsection{Effects of QCD origin}
\paragraph{NLO QCD corrections}\mbox{}\\
The anatomy of QCD corrections in VBS processes is quite peculiar, and it is dictated by the underlying structure of
the process. A typical $t$-channel VBS diagram at tree level, such as those in the top row of Fig.~\ref{fig:diag}, features two quark
lines that exchange electroweak bosons. Since no color charge is exchanged between the two quarks, QCD corrections tend to factorise, in 
the sense that they affect one quark line at a time.
While non-factorisable corrections exist, for example in the case of the scattering of identical quarks,
they are suppressed by color considerations and by kinematics. If one neglects non-VBS diagrams, the situation is completely analogous to Higgs production in vector-boson fusion (VBF), where NLO QCD corrections were
first computed in the factorised approximation using the so-called structure-function approach~\cite{Han:1992hr}. Indeed, also for VBS, the first results including NLO QCD
corrections were obtained discarding non-factorisable corrections~\cite{Jager:2006zc,Jager:2006cp,Bozzi:2007ur,Jager:2009xx}. Within this approximation, NLO QCD corrections to VBS are rather mild, and their exact impact depends
on the cuts employed to define the VBS signal, on the choice of renormalisation and factorisation scales and, of course, on the specific process.

\begin{figure*}[t]
\centering
\includegraphics[width=0.45\linewidth,height=0.35\textheight,clip=true,trim={0.cm -2cm 1.cm 0.1cm}]{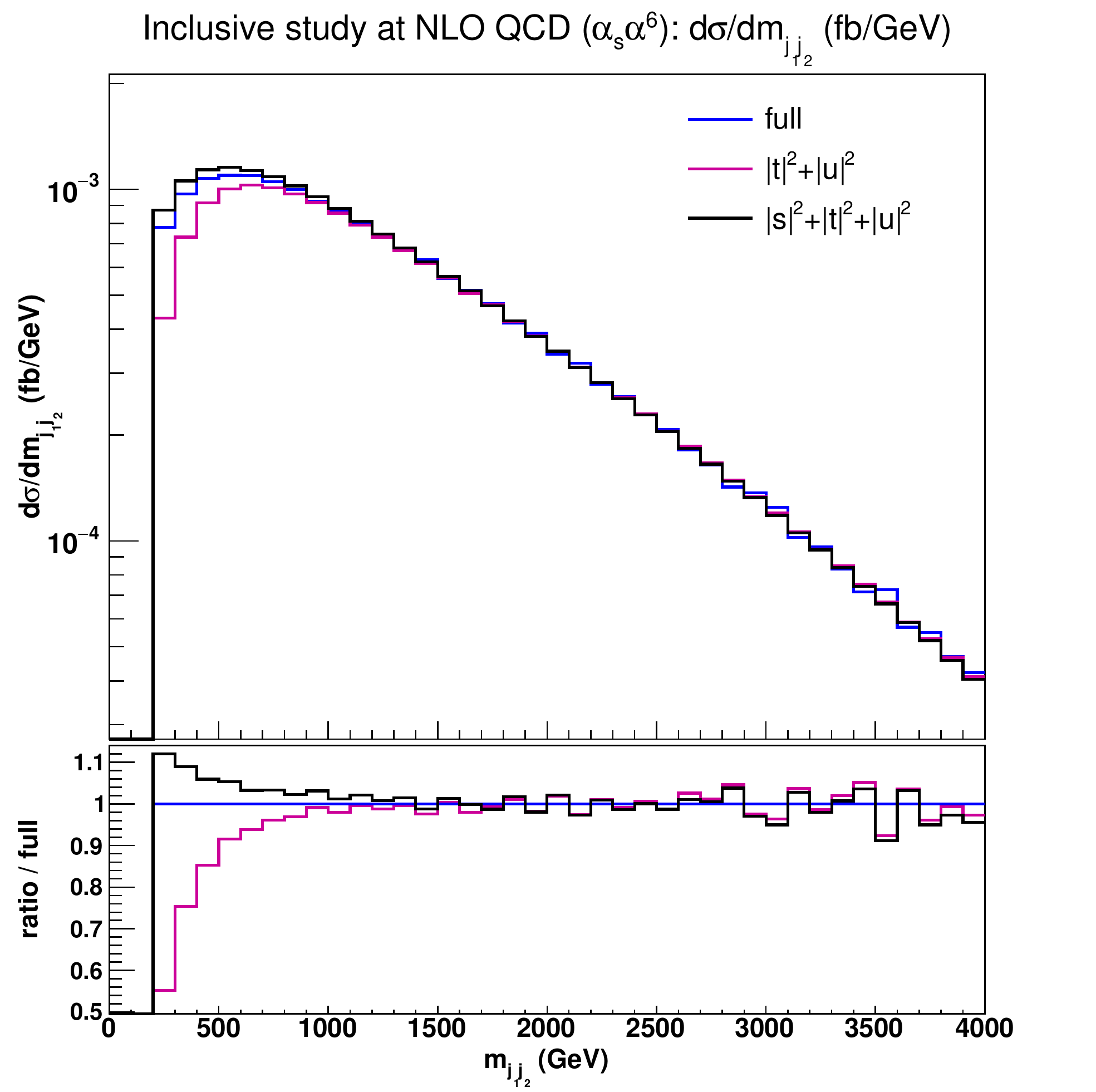}
\includegraphics[width=0.45\linewidth,height=0.35\textheight,clip=true,trim={0.4cm 2.15cm 0.cm 0.7cm}]{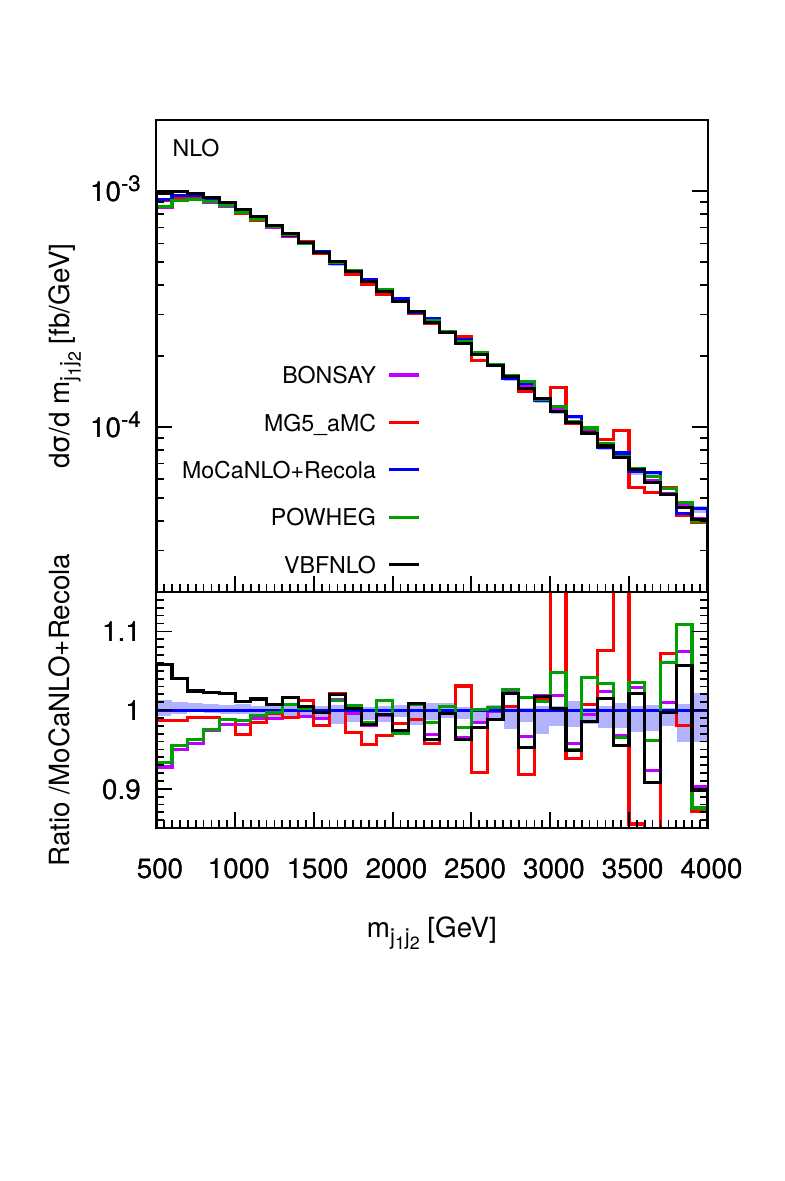}
\caption{Differential distributions in the invariant mass of the two tagging jets at the NLO (fixed order).
Various theoretical predictions with different approximations are here compared in an inclusive set-up and a more exclusive one (as in the experimental analysis).
The best prediction is denoted by either \emph{full} or \textsc{MoCaNLO}+\textsc{Recola}.
These figures are taken from Ref.~\cite{Ballestrero:2018anz}.}
\label{fig:mjjnlo}
\end{figure*}

Going beyond this approximation, \emph{i.e.}\ including non-factorisable corrections, entails a major step in computational complexity.
On the one hand, loops with
many (six or more) external legs and high-rank numerators, due to the presence of vector bosons, have to be evaluated.
On the other hand, non-factorisable corrections are in general not Infra-Red (IR)-finite, and hence possibly dependent on the specific IR regulator.
This could be avoided only by considering all contributions of $\mathcal O (\alpha_s \alpha^6)$, including those which classify as EW corrections to the LO QCD-EW interference of VVjj production, as depicted in Fig.~\ref{fig:orders} (in that figure, the NLO QCD corrections to VBS signal correspond to the $\mathcal O (\alpha_s \alpha^6)$ contribution). Only recently, with advanced techniques having paved
the way to the automation of EW and mixed QCD-EW corrections, non-factorisable contributions have been included in the NLO QCD corrections
to VBS~\cite{Biedermann:2017bss,Denner:2019tmn,Denner:2020zit}. When they are compared to 
the approximation that assumes factorisation, the impact of non-factorisable QCD corrections is found to be small in typical VBS phase spaces~\cite{Ballestrero:2018anz}, exceeding
$5\%$ only in more inclusive phase spaces. This can be seen in Fig.~\ref{fig:mjjnlo} which shows two differential distributions in the invariant mass of the two tagging jets.
The left-hand side shows a comparison of NLO predictions in a inclusive phase-space, 
while the right-hand one is in a more exclusive phase space, typical of experimental analyses.
The full predictions which include non-factorisable corrections 
are denoted by either \emph{full} or \textsc{MoCaNLO}+\textsc{Recola}.
The differences among other predictions are due either to the inclusion or not of non-VBS contributions, or to the details of the definition
of non-factorisable corrections, or both. 

Once NLO QCD corrections are included, theoretical uncertainties estimated through the variation of renormalisation
and factorisation scale are of the order of few per cent on NLO-accurate observables. In the case of \ssWW\ production,
this can be observed in Fig.~\ref{fig:mjjnlo} for the \mjj\ observable, 
where the theory uncertainty is shown as a blue band around the {\textsc{MoCaNLO}+\textsc{Recola}} prediction.
The scale uncertainty is obtained through 7-fold scale variation (see Sec.~\ref{sec:mcsim}).
The inclusion of non-factorisable corrections does not significantly impact the size of the scale uncertainty.

\paragraph{Beyond NLO-QCD: NNLO and parton-shower effects}\mbox{}\\
For what concerns QCD effects beyond NLO, it is a fair statement that NNLO QCD corrections, even in the factorised approximation, are extremely challenging to compute, 
and will likely not be available on a short-term timescale. Indeed, at variance with the case of single- or even double-Higgs production in 
VBF, where corrections up to NNNLO in QCD have been computed within a factorised approach~\cite{Bolzoni:2010xr,Bolzoni:2011cu,Cacciari:2015jma,Dreyer:2016oyx,Dreyer:2018qbw,Dreyer:2018rfu,Dreyer:2020xaj}, VBS processes
present more complex topologies, since the outgoing vector bosons can couple to the quark lines.\footnote{Some recent achievements towards the computation 
    of non-factorisable corrections to Higgs production in VBF are worth to be cited~\cite{Liu:2019tuy,Dreyer:2020urf}, which may be relevant also for VBS.}

\begin{figure*}[t]
\centering
          \includegraphics[width=0.48\linewidth]{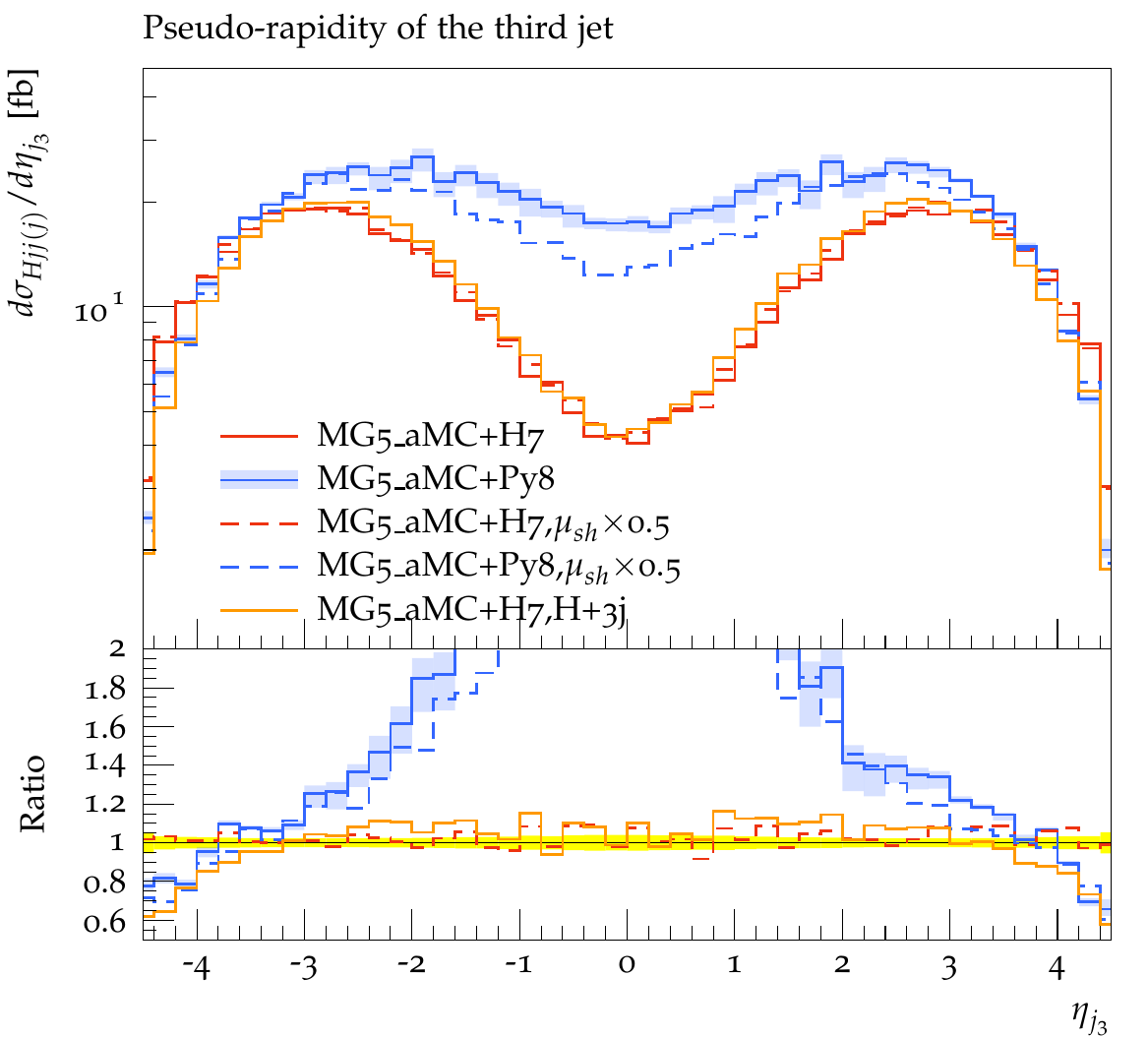}
          \includegraphics[width=0.48\linewidth]{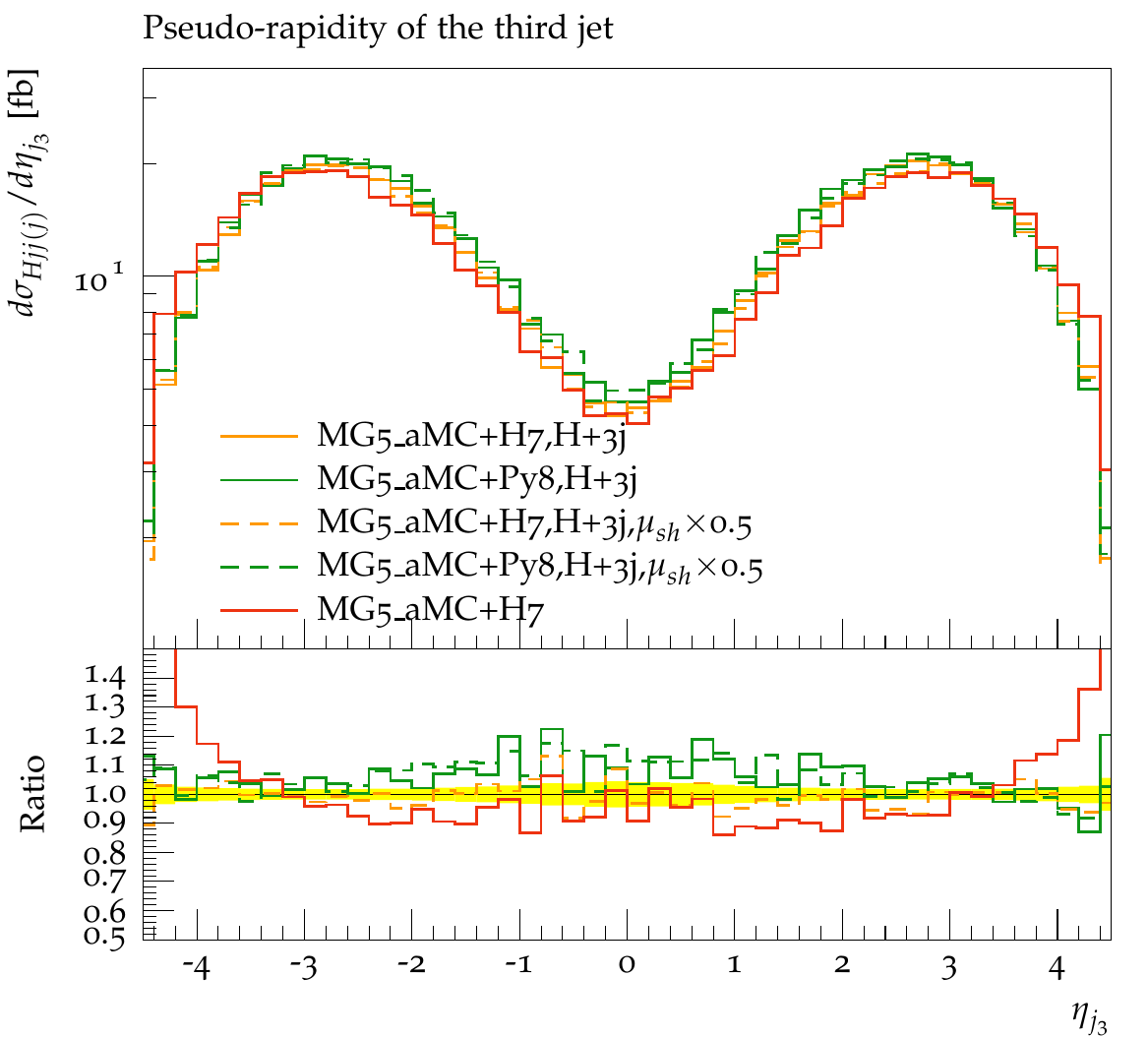}
          \caption{The rapidity of the third jet in Higgs production via VBF, obtained using \madgraphbis + {\sc Pythia8} or {\sc Herwig7}. The left panel shows
          predictions for the production of a Higgs boson plus two jets at NLO+PS matched with {\sc Pythia8} (blue) or {\sc Herwig7} (red), 
          including renormalisation and factorisation scale uncertainties (blue band) as well as variations of the shower starting scale (dashed lines), 
          together with the prediction for Higgs plus three jets at NLO+PS matched with {\sc Herwig7} (orange). The right panel shows
          predictions for the Higgs boson plus three jets at NLO, matched with {\sc Herwig7} (orange) or {\sc Pythia8} (green).
          together with the prediction 
          for Higgs plus two jets at NLO+PS matched with {\sc Herwig7} (red solid). 
          These figures are taken from Ref.~\cite{Jager:2020hkz}.}
\label{fig:VBFj3}
\end{figure*}

However, for almost all processes, NLO QCD corrections have been matched to parton-showers (PS) within different matching 
schemes~\cite{Jager:2011ms, Jager:2013mu, Jager:2013iza, Ballestrero:2018anz, Jager:2018cyo}, and now this kind of processes are within the reach of automatic frameworks. In general, when compared
with NLO predictions at fixed order, NLO+PS results show quite small (10-15\%) effects on the shape and normalisation of distributions. The spread of predictions
obtained with different matching schemes and/or different PS programs is also found to be at the 10\% level, and it has been investigated in detail for \ssWW\ production
in Ref.~\cite{Ballestrero:2018anz}. 

A relevant exception, worth to be discussed, is related to the modelling of the third-jet kinematics. It has been observed that, for 
predictions matched with {\sc Pythia8}~\cite{Sjostrand:2007gs,Sjostrand:2014zea}, the global-recoil scheme leads to a large unphysical enhancement of the third-jet activity in the mid-rapidity region, related
to a wrong assignment of the phase-space boundaries for processes with initial-final color connections. Such an enhancement, absent in predictions obtained with
other parton showers such as {\sc Herwig7}~\cite{Bahr:2008pv,Bellm:2015jjp,Bellm:2019zci}, is observed both with the {\sc Powheg}~\cite{Nason:2004rx, Frixione:2007vw} and
the {\sc MC@NLO}~\cite{Frixione:2002ik}
matching schemes, although it is larger with the latter, owing to the fact that {\sc Powheg} generates the first emission with 
an internal Sudakov factor (and thus shower effects only enter from the second emission on). This effect is discussed in detail for
 same-sign $\PW\PW$ production in Ref.~\cite{Ballestrero:2018anz}, but it is in fact a general issue affecting processes with VBF/VBS-type topologies. Indeed, a similar enhancement has been observed also
 in the measurement of electroweak single-$\PZ$ production~\cite{Sirunyan:2017jej}, and for Higgs production in VBF~\cite{Jager:2020hkz}. 

While the unphysical enhancement disappears when a new recoil scheme, developed for Deep-Inelastic-Scattering processes (dipole recoil~\cite{Cabouat:2017rzi}),
is employed, the way Monte Carlo counterterms are currently implemented \emph{e.g.}\ in \madgraphbis\ prevents the user to employ a recoil scheme different from
the global recoil~\footnote{A new implementation of the Monte Carlo counterterms has been recently presented in Ref.~\cite{Frederix:2020trv}, and future 
developments on allowing a more flexible choice of shower parameters are in progress.}. This does not apply to a {\sc Powheg}-type matching, where the user can instead 
change the shower parameters more freely.   

In Ref.~\cite{Jager:2020hkz}
 it has been shown that, even within the global-recoil scheme, this effect disappears when a NLO-accurate description of the third jet at the matrix-element level is employed, as can be observed in
 Fig.~\ref{fig:VBFj3}. This is a further demonstration that the central-rapidity enhancement observed for predictions matched with {\sc Pythia8} is unphysical and, as such, it should not
 be considered as an uncertainty source for the third-jet description. Given the similarities
 between VBS and VBF from the QCD point of view, these conclusions can be extended from the latter to the former.
They could also be verified explicitly using a NLO prediction for VBS with three jets, which is available at the moment, but should
 not be beyond the reach of modern event generators and matrix-element providers.

\paragraph{PDF uncertainties}\mbox{}\\
Accounting for all sources of uncertainty stemming from QCD requires also the inclusion of those coming from parton distribution functions (PDFs).
At LO, only quarks appear in the initial state of VBS processes,
regardless of the specific final state. Within typical VBS cuts, they mostly feature intermediate values
of the Bjorken $x$'s and scales $Q=\mathcal O (100)\GeV$. 

\begin{figure*}[t]
\centering
          \includegraphics[width=0.58\linewidth]{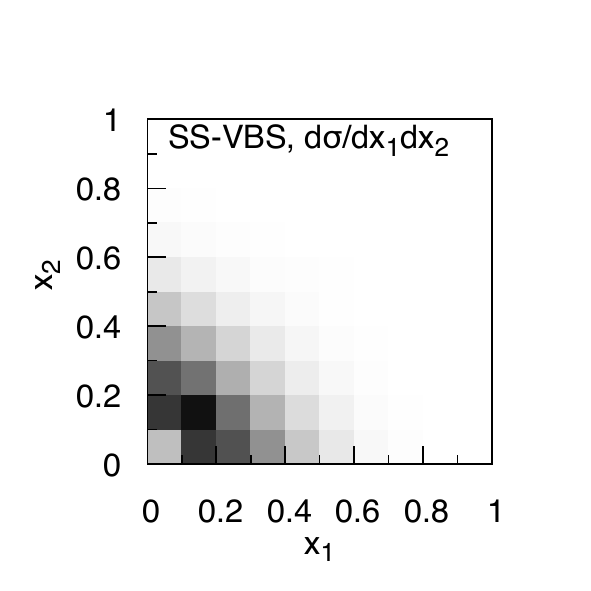}
          \caption{The Bjorken $x$ regions contributing to same-sign $\PW\PW$ production via VBS, at the $13\TeV$ LHC.
          The colour coding is a follow: maximal is for black, minimal is for white.
          The set-up used for this plot is the same as in Ref.~\cite{Ballestrero:2018anz}.}
\label{fig:vbsx12}
\end{figure*}

From Fig.~\ref{fig:vbsx12} one can appreciate that
the bulk of the \ssWW\ cross-section comes from $x\simeq0.2$, a region where quark densities, especially valence ones,
are quite well constrained nowadays, with uncertainties below $5\%$~\cite{Butterworth:2015oua}.
As gluons only enter at NLO, they give a subleading contribution to the
cross section, considering the small size of NLO corrections discussed above. The produced final state,
in particular the charges of the gauge bosons, affect the combination of flavours which can initiate the process: positively charged final states, such as $\PW^+\PW^+\Pj\Pj$ production,
are mostly sensitive to valence quarks, while for neutral or negatively charged ones the contribution of sea quarks becomes more important. Hence, PDF uncertainties are
expected to be quite process-specific. 

\begin{figure*}[t]
\centering
          \includegraphics[width=0.48\linewidth]{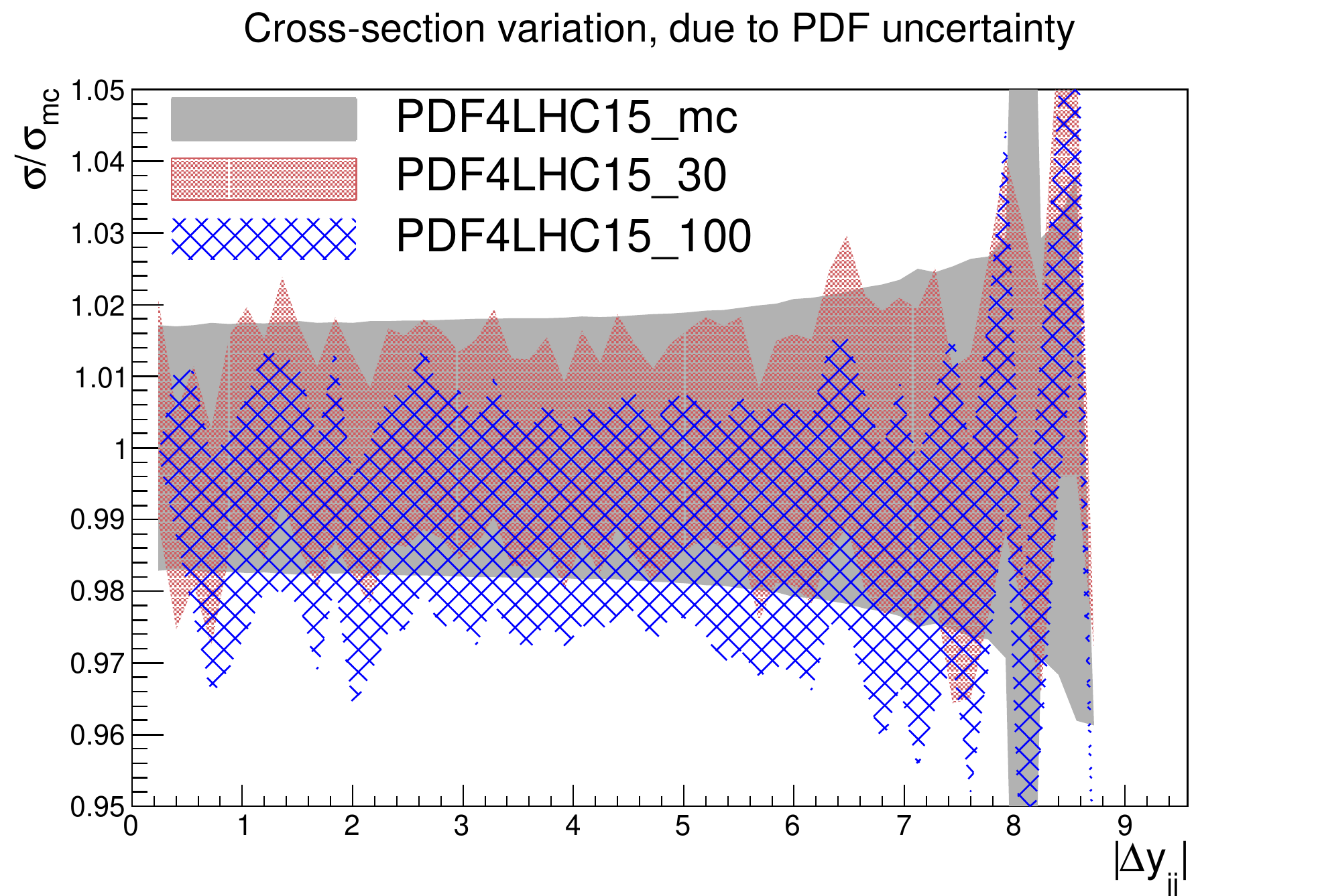}
          \includegraphics[width=0.48\linewidth]{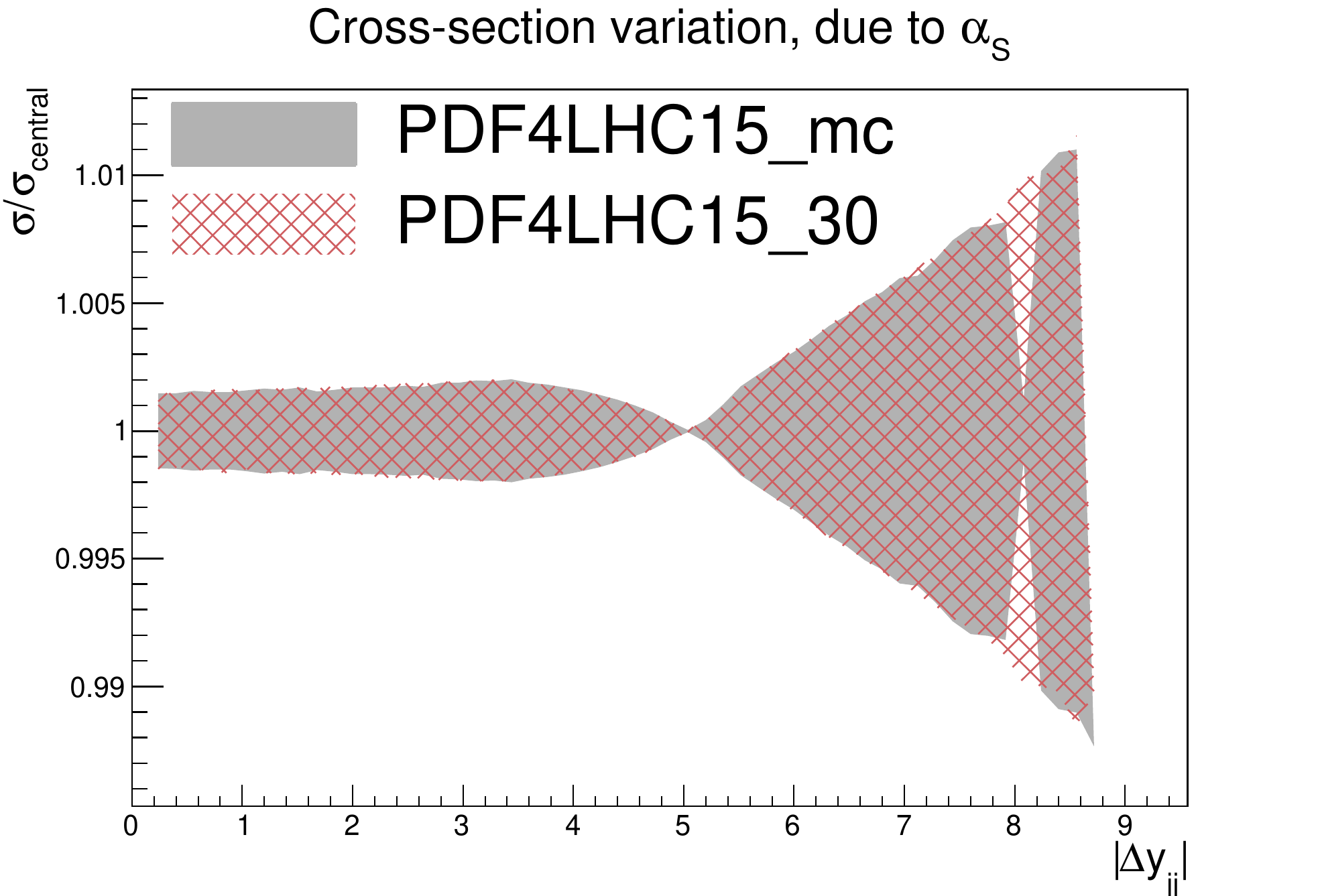}
          \caption{PDF (left) and $\alpha_s$ (right) uncertainty for the rapidity separation of the two tagging jets,
              evaluated with the PDF4LHC15 set. These figures are taken from Ref.~\cite{Bellan:2019xpr}.}
\label{fig:dyjj_pdfunc}
\end{figure*}

If we consider again, as an example, the case of same-sign $\PW$ boson production, and
specifically the rapidity separation of the two jets, one can
appreciate from Fig.~\ref{fig:dyjj_pdfunc} that the PDF uncertainties, evaluated with the PDF4LHC15 set~\cite{Butterworth:2015oua},
are at the level of $\pm 2\%$ for a large part of the range, up to $4\%$ only for extreme separations. For the same observable, 
uncertainties due to $\alpha_s$ are (as expected) totally negligible, below $1\%$ across almost the whole considered range. 

As mentioned above, these numbers are expected to be rather process specific
and are calculated in experimental analyses targeting the corresponding final states. In order to have an idea of how the charge of the final state can
affect their size, one can consider the case of charged-Higgs production via VBF,
for which these studies are available~\cite{Zaro:2010fc,Zaro:2015ika,deFlorian:2016spz}. In this case, the Higgs mass plays the
role of the invariant mass of the vector-boson pair. Estimations in Ref.~\cite{Zaro:2015ika} show that PDF uncertainties never exceed a few percent, 
being smaller for lighter final states and when valence-quark contributions are mostly probed (Fig.~\ref{fig:dyjj_pdfunc}).

\subsubsection{Effects of Electroweak origin}

Given the magnitude of the strong and electroweak couplings, for typical LHC processes NLO EW corrections are generally of the order of NNLO QCD corrections, that is a few percent.
This power-counting argument is usually valid at the level of the total cross sections, but the situation is rather different when considering differential distributions, as EW and QCD corrections exhibit a rather different behaviour and are relevant in different phase-space regions.
In general, they become negative and large (typically several $10\%$) in the high-energy limit because of Sudakov logarithms~\cite{Denner:2019vbn}.

\begin{figure*}[t]
\centering
          \includegraphics[width=0.48\linewidth,height=0.35\textheight,clip=true,trim={2.3cm 0.8cm 0.cm 0.cm}]{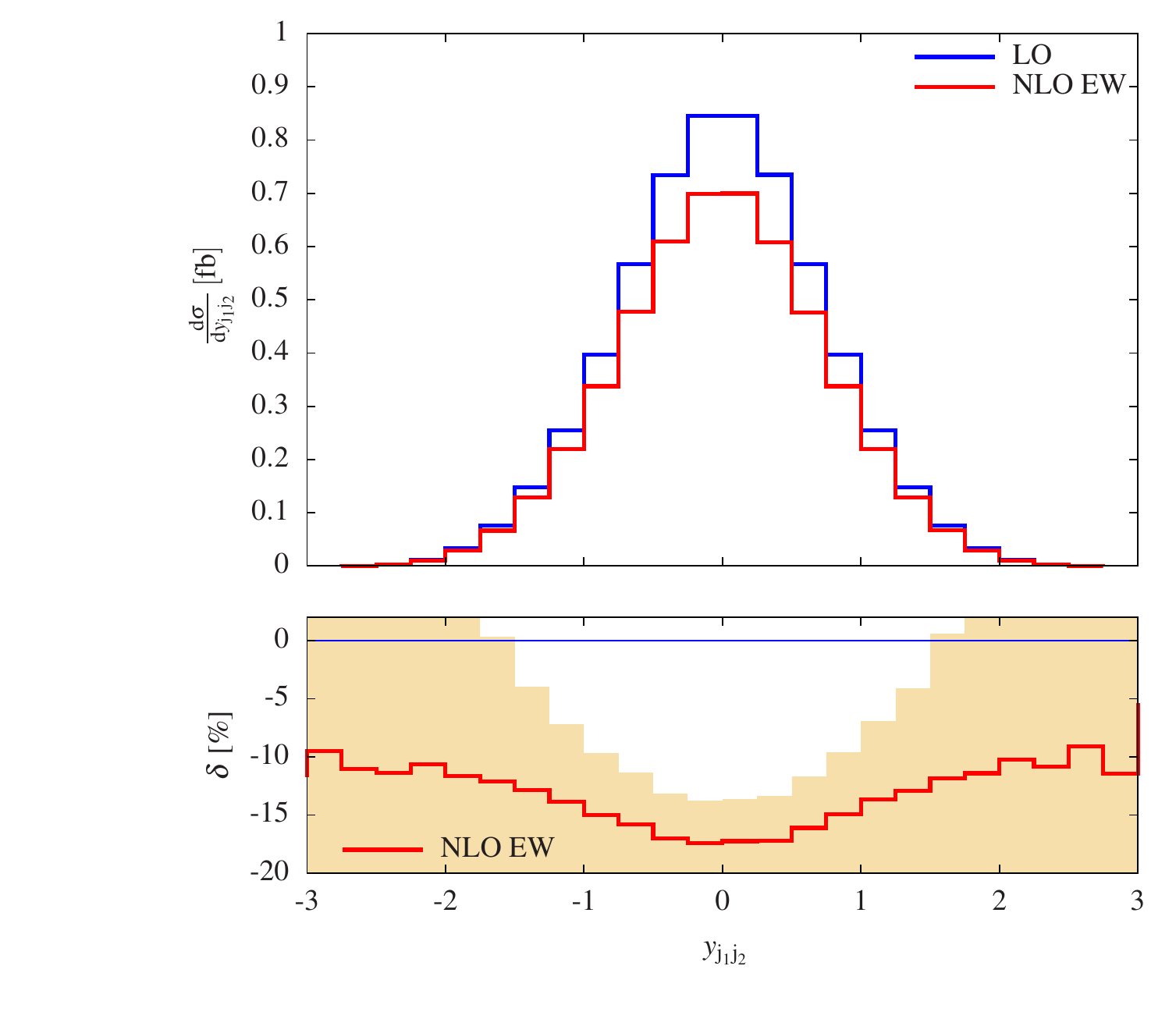}
          \includegraphics[width=0.48\linewidth,height=0.35\textheight,clip=true,trim={2.3cm -0.2cm 0.cm 0.cm}]{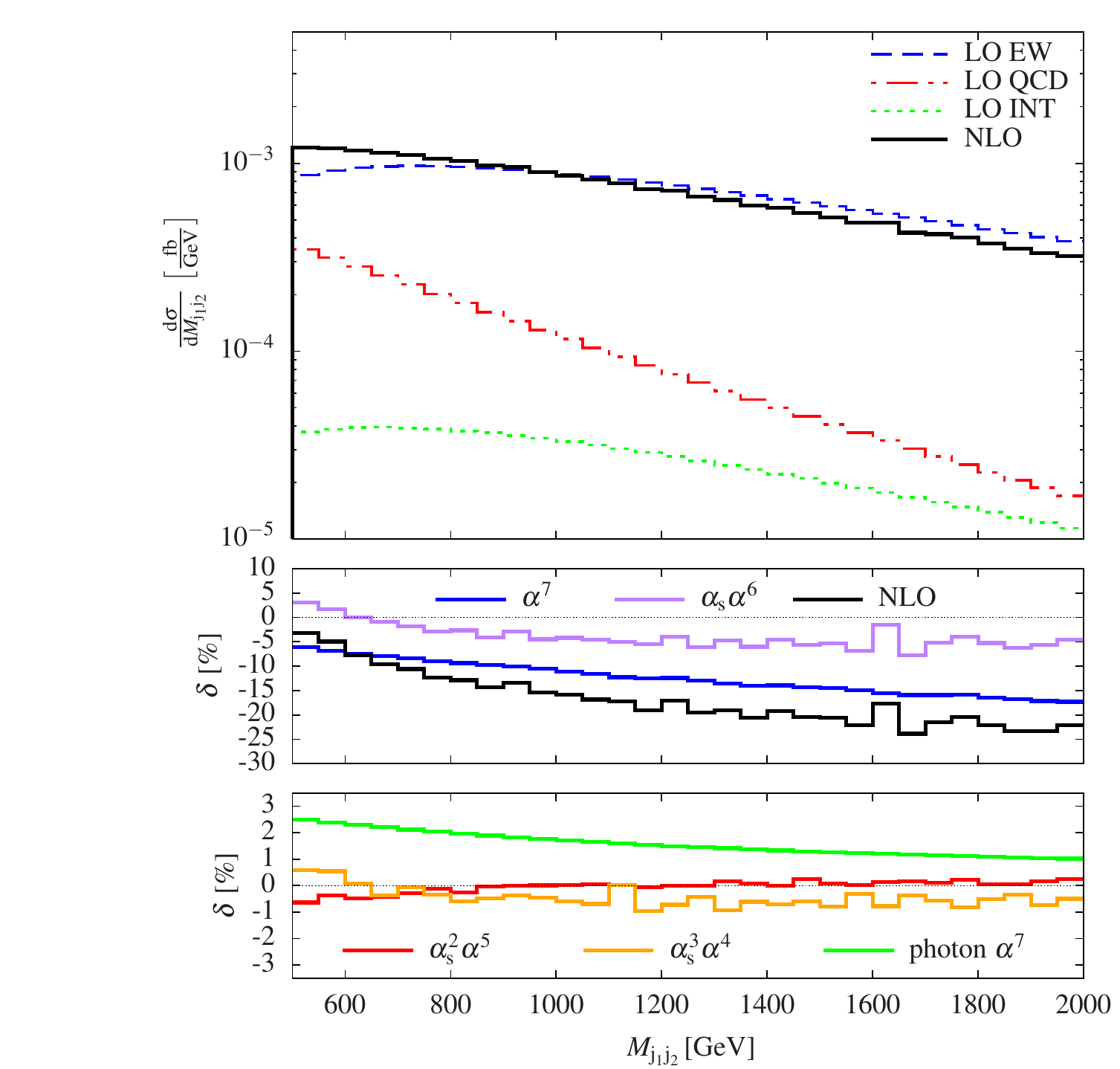}
          \caption{
              NLO EW corrections to same-sign WW scattering for the rapidity distribution of the two tagging jets (left).
              The yellow band in the lower plot represents the expected statistical experimental uncertainty at $3000\fb^{-1}$.
              Full NLO corrections to same-sign WW scattering for the invariant mass of the two tagging jets (right).
              The left figure is taken from Ref.~\cite{Biedermann:2016yds} while the right one is taken from Ref.~\cite{Biedermann:2017bss}.}
\label{fig:vbsew}
\end{figure*}

In the case of VBS, the global picture is quite different, namely the 
NLO EW corrections are large relative to QCD corrections of the same order.
As shown in Ref.~\cite{Biedermann:2016yds}, large EW corrections are an 
intrinsic feature of VBS at the LHC. At the level of the total cross section, they can be of the order of 
$-20\%$ and reach up to $-40\%$ in tails of differential distributions.
Their origin can be attributed to the massive $t$-channel which enhance the typical scale of the process~\cite{Denner:1997kq}, 
as well as the fact that the EW Casimir operators are larger for bosons than fermions~\cite{Denner:2000jv}.
For same-sign WW scattering, where all NLO corrections are known, the EW corrections to the VBS process of order $\mathcal O (\alpha^7)$
are the largest corrections~\cite{Biedermann:2017bss}. Such a pattern has been confirmed for the $\PW\PZ$~\cite{Denner:2019tmn} and $\PZ\PZ$~\cite{Denner:2020zit} signature.
It is also worth mentioning that such EW corrections are largely independent of the charge of the final state as shown in Ref.~\cite{Chiesa:2019ulk} for \ssWW.
Even more, the Leading-Log approximations derived in Refs.~\cite{Biedermann:2016yds,Denner:2019tmn,Denner:2020zit} are rather universal due to the identical $\rm SU(2)_w$ couplings occurring in all scatterings.
In Fig.~\ref{fig:vbsew}, the differential distribution in the rapidity of the two tagging jets is displayed.
In the lower plot, the yellow band represents the expected statistical experimental uncertainty in each bin for the high-luminosity LHC collecting $3000\fb^{-1}$.
Given their magnitude, one can thus expect that high-luminosity measurements will be sensitive to such EW corrections.

Another source of EW effects is the inclusion of photon PDF. 
The determination of the photon PDF has witnessed a complete change of paradigm in 2016, when the LUXqed methodology was introduced~\cite{Manohar:2016nzj,Manohar:2017eqh}, which employs a more robust determination from first principles of the photonic density. Thanks
to these works, the photon density can now be constrained at a level comparable to that of the strong-interacting partons. This was made possible
by relating photon-induced processes with their non-photon induced counterpart, and using this relation to extract the photon density. Before 2016, the only available approaches were
either relying on some a-priori parametrization of the photon density~\cite{Martin:2004dh,Schmidt:2015zda}, or on leaving it completely free to be fit~\cite{Ball:2013hta}.
This led in the first case to an impossible, or very difficult quantification of theoretical uncertainties, and in the second to huge uncertainties, often 
of the order of $100\%$, relative to the impact of the photon density.
The LUXqed methodology is now employed by all major PDF providers, such as NNPDF~\cite{Bertone:2017bme} and MMHT~\cite{Harland-Lang:2019pla} as well as the PDF4LHC working group~\cite{Butterworth:2015oua}.

For example, in same-sign WW scattering \cite{Biedermann:2017bss}, the photon-induced contributions are of the order of $2.7\%$ with NNPDF-3.0 QED \cite{Ball:2013hta} 
while they go down to $1.5\%$ when using the LUXqed\_plus\_PDF4LHC15\_nnlo\_100 set.
Due to charge conservation, LO photon-induced contributions are present for the $\PW\PZ$, $\PZ\PZ$, and $\PW^+\PW^-$ channels as well.
They involve one or two initial-state photons and contribute to the orders $\mathcal (\alpha^6)$ and $\mathcal (\alphas\alpha^5)$.
They amount to about $0.4\%$ with respect of the LO of order $\mathcal (\alpha^6)$ for $\PW\PZ$~\cite{Denner:2019tmn}.

When referring to NLO EW corrections, it was so far implied that only real photon radiations are included.
The radiation of heavy gauge bosons occurs at the same perturbative order, and in principle can also be accounted for.
To date, this effect is relatively unexplored in the context of VBS, while studies exist for other processes~\cite{Frixione:2014qaa,Frixione:2015zaa,Czakon:2017wor}.
In Ref.~\cite{Azzi:2019yne}, the related correction has been estimated to be of the order of few percent for the High-Luminosity LHC at the level of the total cross section.

Finally, for signatures other than same-sign WW scattering, there also exist a photon-to-jet conversion function 
which is necessary to cancel IR divergences associated to photons splitting into a quark-antiquark pair~\cite{Denner:2019zfp}.
While it ensures a proper treatment of this non-perturbative contribution, 
its numerical impact is rather small and has been evaluated to be the order of $0.01\%$ for $\PW\PZ$~\cite{Denner:2019tmn}.


\subsection{Experimental techniques}
\label{sec:experim}
Taking into account the decay of heavy gauge bosons, VBS cross-sections are typically of the order of 
femtobarns in proton-proton (${\rm p}{\rm p}$) collisions at a center-of-mass energy of 13 TeV.
For this reason, the LHC experiments that have sensitivity to these
processes are the ones which benefit from the full amount of integrated luminosity delivered by the accelerator,
ATLAS~\cite{Aad:2008zzm} and CMS~\cite{Chatrchyan:2008aa}.
\footnote{Apart from an exploratory theoretical study for the LHCb experiment \cite{Pellen:2019ywl}, no results are available from other experiments than ATLAS and CMS.}

Both ATLAS and CMS have analysed fully or partially the $13\TeV$ LHC data set delivered between 2016 and 2018, referred
to as \emph{Run-2} of the accelerator, which corresponds to about $135-140\fb^{-1}$ per experiment, depending on the percentage
of high-quality data which can be used to reconstruct a specific final state. In these runs, the mean number
of $\Pp\Pp$ collisions per LHC bunch crossing (\emph{pileup}) varied between 23-27 in 2016 to 35 and more in 2017 and 2018.  
Using such a large data set, the experimental knowledge of VBS has dramatically increased in the recent years. Starting from the pre-Run-2 results
at $7$ and $8\TeV$ where just upper limits on VBS SM cross sections were reported,
both experiments have now claimed evidence or observation for all the main VBS processes.

\subsubsection{The ATLAS and CMS detectors}

\begin{figure*}[t]
\centering
\includegraphics[width=0.75\textwidth]{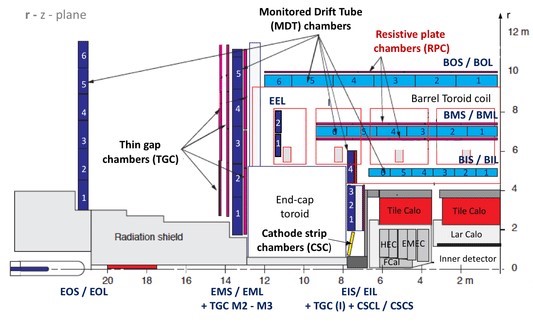} \\
\includegraphics[width=0.70\textwidth]{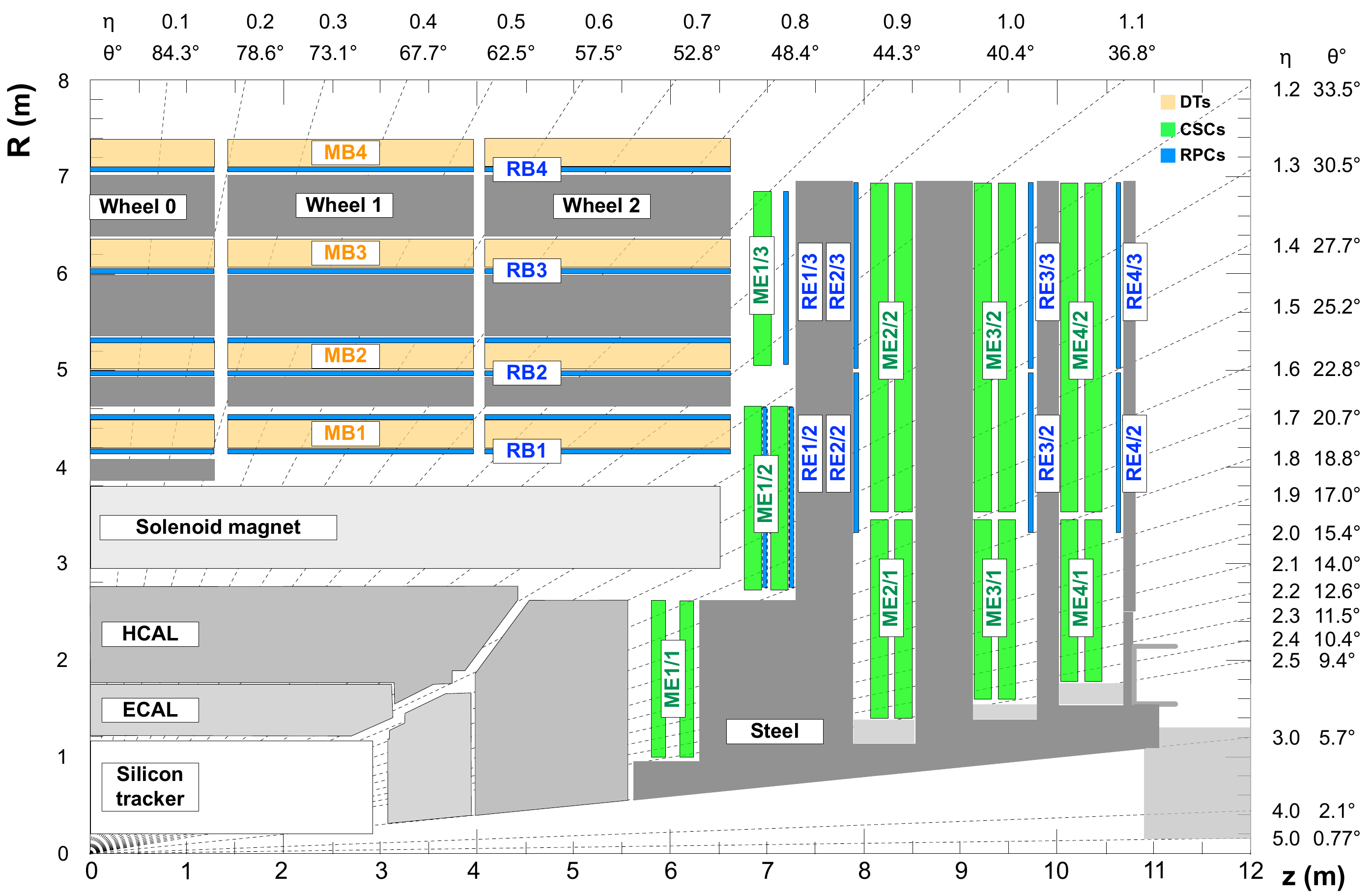}
\caption{Longitudinal views of a quadrant of the ATLAS (top) and CMS (bottom) detectors. Sub-apparati used for
the detection of different particles and their positioning are shown.}
\label{fig:spaccati}
\end{figure*}

ATLAS and CMS are general-purpose detectors with a cylindrical geometry and a nearly hermetic coverage
in $\theta$ and $\phi$\footnote{Both ATLAS and CMS use right-handed coordinate systems with their origin placed
at the nominal interaction point and the $z$-axis running along the beam direction. The $x$- and $y$-axes point to the
centre of the LHC ring and upward, respectively. Cylindrical coordinates $(r,\theta,\phi)$ are used in this
coordinate system. The subscript $T$ refers to quantities measured in the $(x,y)$, or transverse plane, while
the pseudorapidity is defined as $\eta = -\log[\tan(\theta/2)]$.}. Figure~\ref{fig:spaccati} shows longitudinal views of
a quadrant of the ATLAS and CMS detectors.

In both ATLAS and CMS the interaction point is surrounded by tracking detectors. For both the innermost system
consists of a silicon pixel detector, providing precise estimation of track impact parameters and vertices, and
is complemented by outer layers of silicon microstrip detector. In ATLAS tracking information is also provided
by a transition radiation tracker. In both experiments, these inner detectors provide precise measurements
of charged-particle tracks in the pseudorapidity range $|\eta| < 2.5$.

Electromagnetic and hadronic calorimeters cover the region $|\eta| < 3.2$ in ATLAS and $|\eta| < 3.0$ in CMS.
In ATLAS the electromagnetic calorimeter is based on high-granularity, lead/liquid-argon (LAr) sampling
technology, while in CMS it consists of high-resolution lead tungstate crystals.
The ATLAS hadronic calorimeter is comprised of a steel/scintillator-tile sampling detector in the central
region and a copper/LAr detector in the region $1.5 < |\eta| < 3.2$, while CMS uses a brass/scintillator detector.

For VBS signatures, it is of paramount importance to collect electromagnetic and hadronic energies at larger
values of pseudorapidity. To achieve this goal, both ATLAS and CMS are equipped with forward calorimeters,
that have lower granularities but must satisfy stringent radiation hardness requirements. In ATLAS the region of
the detector $3.1 < |\eta| < 4.9$ features a forward calorimeter (FCal), measuring electromagnetic and
hadronic energies in copper/LAr and tungsten/LAr modules. In CMS the forward calorimeter (HF) covers the
region up to $|\eta| \lesssim 5.0$ and consists
of a steel absorber equipped with quartz fibres of two different lengths which distinguish the electromagnetic
and hadronic components.

The magnet arrangement is different in the two experiments. CMS features a large superconducting solenoid
with a 6-m inner diameter providing an axial magnetic field of 3.8~T. In ATLAS, a smaller solenoid providing a
magnetic field of 2~T surrounds the inner tracker, while three large superconducting toroidal magnets are
placed with an eightfold coil symmetry outside the calorimeters.

In both ATLAS and CMS the muon spectrometer comprises trigger and high-precision tracking chambers to measure the
trajectory of muons. Detector technologies include drift tubes, cathode strip chambers in
the forward regions, resistive-plate chambers, and thin-gap chambers. The CMS muon coverage is $|\eta| < 2.4$
while in ATLAS it is $|\eta| < 2.7$ for tracking and $|\eta| < 2.4$ for trigger chambers.

Events of interest are selected in real time using two-tiered trigger systems in both experiments~\cite{Aad:2020wji,Khachatryan:2016bia}.
The first level is composed of specialized hardware processors and uses information from the calorimeters and muon
detectors. The second level (high-level)
consists of farms of processors running a fast, optimized version of the event reconstruction software
that reduce the event rate before data storage. Data are stored in different streams according to which
high-level trigger path(s) find compatibility between an event and a specific particle hypothesis (single electron,
double muon, etc.).

\subsubsection{Tagging-jet reconstruction}
\label{sec:tagjet}

All VBS processes have in common a pair of jets in the final state, originating from the hard scattering process.
These have been precisely defined in Sec.~\ref{sec:sob} and we refer to those jets here as \emph{VBS-tagging jets}.

In ATLAS, jet constituents are topologically-grouped clusters (``topo-clusters'') of electromagnetic and hadronic calorimeter cells~\cite{Aad:2016upy}.
In CMS events are reconstructed using a more detailed particle-flow algorithm~\cite{Sirunyan:2017ulk} that identifies each individual particle with an optimized combination of all subdetector information.
However, at the very high pseudorapidity values covered by the HF only, a hadron or electromagnetic particle-flow candidate is just defined
by its energy release in a $\eta$-$\phi$ HF cell, since information from no other subdetector is available.

In both experiments jets are reconstructed from either particle-flow candidates or topo-clusters using the anti-$k_{\rm T}$ clustering algorithm~\cite{Cacciari:2008gp}, 
as implemented in the FastJet package~\cite{Cacciari:2011ma}, with typically a distance parameter of $0.4$.
This value ensures a good particle containment while reducing the instrumental background, as well as the contamination from pileup. In both ATLAS and CMS identification criteria for jets are very loose and retain
almost all physically meaningful jets. Since these jets include leptons with the surrounding QED activity
(sometimes referred to as \emph{dressed} leptons), analyses with leptonic final states always require a minimum
tagging-jet/lepton $\Delta R$ separation in defining their fiducial regions, usually taken equal to the jet distance parameter.

Pileup effects, which are particularly relevant in the forward regions, affect jet measurement in two ways: 
by adding entire jets that do not originate from the hardest-scattering event (thus requiring \emph{pileup rejection} techniques) 
and by adding particles in the jet that do not belong to signal jets but overlap in space (requiring \emph{pileup subtraction} techniques).
In evaluating both effects, there is an important difference between charged and neutral particles. For charged particles, the hypothesis of being originated in the hardest-scattering \emph {primary vertex} can be evaluated track by track and is based on impact-parameter compatibility\footnote{In both ATLAS and CMS, the hardest-scattering vertex is defined as the primary vertex in the event for which the scalar sum of the $\pt^2$ of the associated tracks is maximum.}.
For neutral particles, this association is not possible and subtraction or rejection must be done on a statistical basis. It has to be noted that outside the tracker coverage ($|\eta| < 2.5$ for both ATLAS and CMS) all particles must be treated as neutral.

ATLAS and CMS apply pileup subtraction as part of the jet energy corrections~\cite{Aaboud:2017jcu,Khachatryan:2016kdb}.
The subtraction has the analytical form suggested in Ref.~\cite{Cacciari:2007fd}
\begin{equation}
p^{corr}_{\rm T} = \pt - \rho \times A_{jet}, 
\end{equation}
where $\rho$ is the estimated average pileup $\pt$ density in specific regions of the detector and
$A_{jet}$ is the jet area. Depending on the experiment and specific analysis, this subtraction can be performed on the original jet, or in combination with charged-hadron subtraction based on vertex compatibility. 
ATLAS~\cite{Aad:2015ina,Aaboud:2017pou} and CMS~\cite{Sirunyan:2020foa} have dedicated pileup rejection methods. Within
the tracking volume, a jet-vertex combined compatibility is computed from the charged tracks found inside a jet,
while in the forward regions different jet shapes are employed to build discriminating variables that isolate signal
jets from pileup jets. Not all VBS analyses apply selections based on these very recent methods. 

Techniques enriching
the selected sample with quark-initiated jets over the more abundant gluon-initiated background were studied both in
ATLAS~\cite{ATLAS:2016wzt} and CMS~\cite{CMS:2013kfa} and could be useful in VBS searches. Their limitations, however,
reside in the limited resolution and granularity of the subdetectors covering the very forward regions. In ATLAS,
performances outside the tracking volume are not
even reported, while in CMS they were found to be poor in the forward region.


Requirements on the transverse momentum ($\pt$) of tagging jets, applied after jet energy corrections, vary among the different analyses and can be symmetric or not between the two.
Pseudorapidity requirements need to be as loose as possible because of the particular spatial distribution of tagging jets: typically all jets with $|\eta| < 4.5$ (4.7) are selected in ATLAS (CMS).
Variables describing kinematics of the jet pair are the most discriminating between VBS and various sources of
background.
As mentioned previously, typical selections include a minimum rapidity or pseudorapidity difference (\dyjj\ or \dejj, usually
taken as unsigned quantities), a minimum invariant mass of the jet pair (\mjj) and, in some cases, selection on
more complex quantities, like the Zeppenfeld variables that we shall define in the following.

A common choice in case of more than two reconstructed jets in an event is to retain the event if fulfilling the selection, and choose the two jets with the largest $\pt$ or energy as the VBS-tagging jets.
Both choices have non-trivial implications on the analyses.
First of all, a third-jet veto is in principle a powerful handle for background rejection, since the VBS topology implies a rapidity gap between the tagging jets with very little hadronic activity. 
However, besides possible pileup contributions occurring in the gap, a long-standing theory problem was the observation of large differences in the third-jet kinematics when comparing predictions obtained with some commonly-used parton-shower programs~\cite{Ballestrero:2018anz}. The origin of these differences has recently been understood (see Sec.~\ref{sec:theory}), thus in principle making it possible to
include veto techniques in future analyses without being hampered by large theoretical systematic uncertainties.
Second, the definition of the tagging jets can be particularly relevant for phase-space regions including heavy-gauge boson resonances decaying into quarks.
Such an example is given in Sec.~\ref{sec:zz} when discussing the case of $\PZ\PZ$ scattering. Finally, it has to
be noted that, as detailed in the next section, in presence of jets originating from heavy gauge-boson decays in the selection,
the choice of the tagging jets is not obvious and varies between different analyses.

\subsubsection{Vector-boson reconstruction}

The analysis techniques to reconstruct and select vector bosons in VBS processes depend strongly on the final state under investigation.

High-energy photons ($\gamma$) are reconstructed in electromagnetic calorimeters with very high efficiencies.
On the other hand, when one or both vector bosons are heavy ($\PW$ or $\PZ$) the reconstruction is based on their decay products and three classes of analyses can be distinguished:

\paragraph{1) Fully leptonic channels:}

Heavy gauge bosons decay into the $\PW^\pm \rightarrow \ell^\pm \nu_\ell$ and $\PZ \rightarrow \ell^+ \ell^-$ final states, where $\ell$ denotes either an electron or a muon.
Even though the branching fractions are approximately only $20\%$ and $6.7\%$, respectively, these final states are the cleanest and can satisfactorily cover the phase spaces of all VBS processes.
For this reason, all evidences or observations of SM VBS processes so far relies on fully leptonic (or lepton $+\,\gamma$) channels. 

The decays of $\PW$ and $\PZ$ bosons into $\tau$ leptons are not considered in existing analyses because, while having identical branching fractions as electrons and muons, 
they are much more challenging to reconstruct due to the presence of the missing neutrinos in the $\tau$ secondary decays.
Nevertheless, the events where the $\tau$ decay leptonically can enter the selected samples in the fully leptonic channels.
This contamination is much smaller in size than the $\tau$ leptonic branching fraction, since secondary leptons from $\tau$ decays have smaller transverse momenta and/or fail invariant mass requirements.
However, all analyses do (or should in principle) state if this small contribution is considered or not in their definition of fiducial analysis volumes.

In the $\PW^\pm \rightarrow \ell^\pm \nu_\ell$ case, due to the presence of neutrinos, the process cannot be fully reconstructed.
As opposed to the $\PZ\to\ell^+\ell^-$ case, it implies that non-resonant contributions also enter the selected sample, and therefore must also be included in the simulation.

\paragraph{2) Semi-leptonic channels:}

One heavy gauge bosons is reconstructed from the $\PW^\pm \rightarrow \ell^\pm \nu_\ell$ or $\PZ \rightarrow \ell^+ \ell^-$ final state, 
and the other one from the $\PW^\pm \rightarrow q{\bar q}'$ or $\PZ \rightarrow q{\bar q}$ final state.

These final states exhibit larger cross sections because of the higher branching ratios.
However, performing a standard reconstruction of the jets from $\PW$ or $\PZ$ hadronic decays, as described in Sec.~\ref{sec:tagjet}, 
results in samples overwhelmingly dominated by the production of single-bosons in association with jets, in which sensitivity to SM VBS is negligible compared to fully leptonic channels.

However, special reconstruction techniques apply in case of \emph{boosted} vector bosons, \emph{i.e.}\ when their Lorentz $\gamma$-factor is large~\cite{jetsatLHC}.
In particular if the aperture angle of the quark-antiquark pair is $\Delta R \simeq 2/\gamma \simeq 2M(q{\bar q})/\pt(q{\bar q}) < 0.8$, that is for $\pt(q{\bar q}) \gtrsim 220$ GeV, 
the hadronic decay products of the gauge boson do not cluster into two separated $R =0.4$ jets but are instead merged into a larger-area jet.
In this case, hadronic $\PW$ and $\PZ$ decays are identified by anti-$k_{\rm T}$ jets with $R =0.8$ which contain all the products of the decay.
As opposed to standard jets, these \emph{merged} jets have two important characteristics that help distinguishing them from regular-jet background: 
after removing soft QCD radiation, the invariant mass of all jet constituents peaks at the $\PW$ or $\PZ$ mass; 
and the inner structure of the jet is such that two \emph{subjets} with smaller radii can be identified.

There are two main characteristics in analyses employing boosted vector bosons: first, they usually address mixed final
states because jet-mass resolutions are such that $\PW$ and $\PZ$ cannot be easily separated. Therefore these final
states are indicated by V (indicating generically a vector boson, either $\PW$ or $\PZ$). Secondly, the requirement
$\pt({\rm V}) \gtrsim 220$ GeV implies that only small parts of the SM VBS phase space are accessible, which compensates
for the higher branching fraction. On the other hand, BSM effects in EFT approaches produce cross sections with
larger components in the boosted phase space as stated in Sec.~\ref{sec:bsm}.
Hence, the most stringent limits on the Wilson coefficients of EFT operators are obtained from semi-leptonic channels.

\paragraph{3) Fully hadronic channels:}

Because of the dominant multijet background, these analyses can only be
performed for final states with two boosted gauge bosons, VV. While potentially they could have even better
sensitivities than semi-leptonic channels on EFT operators, there are no public LHC analyses to date employing these
final states.

\paragraph{Lepton and missing energy reconstruction}

A brief review of charged-lepton reconstruction follows, which is common to many VBS analyses.
Photon and merged-jet reconstruction and selection are specific to some analyses and will be discussed in Sec. 3. 

In ATLAS and CMS, muons are reconstructed by combining information from the inner tracking system with the signals
in the muon chambers and finding matches between reconstructed tracks in the two detectors. In most analyses, muons
must satisfy identification criteria which are called
\emph{medium} in ATLAS~\cite{Aad:2016jkr} and \emph{tight} in CMS~\cite{Sirunyan:2018fpa} but in both
cases correspond to efficiencies exceeding 90\% after fiducial selections. A minimum number of
hits in the related subdetectors is required, which rejects fake matchings, as well as kaon and pion decays in flight.
Tight primary-vertex compatibility criteria select only \emph{prompt} muons, rejecting those
originating from long-lived particle decays. CMS utilizes the transverse and longitudinal track impact
parameter $d_{xy}$ and $d_{z}$ as selection variables for primary-vertex track compatibility, while ATLAS uses the
significance of the impact parameter in the transverse plane $|d_{xy}/\sigma_{d_{xy}}|$ and the
vertex-track distance computed from the longitudinal impact parameter, $d_{z} \sin \theta$, with tighter
requirements. Isolation requirements, further reducing the $B$- and $D$-meson decay background, are in general
loose for both experiments.

Similarly, in both ATLAS and CMS, electron reconstruction combines information from inner-detector tracks
and electromagnetic energy clusters. Primary vertex compatibility of the electron track is evaluated in a similar way
as for muons, with equal or tighter requirements. However, identification criteria, again defined as
\emph{medium} in ATLAS~\cite{Aaboud:2019ynx,Aad:2019tso} and \emph{tight} in CMS~\cite{Khachatryan:2015hwa}, 
are more complex in order to cope with potentially large backgrounds of misidentified photons and jets. These criteria
involve many aspects of the reconstruction, including: angular and energy-momentum matching between track and cluster,
electromagnetic shower shape variables, energy ratios between the central cluster cell and the surrounding ones,
and maximum energy released in hadronic calorimeters in close-by regions of the detector. In ATLAS, these criteria
are combined using a likelihood-ratio method, while in CMS selections are either applied sequentially or combined
in a Boosted-Decision Tree (BDT)\footnote{A decision tree is an algorithm which takes a set of input features and splits input data recursively based on those.
Boosting is a machine-learning method which combines several decision trees to make a stronger signal-background classifier.
BDT algorithms are coded in commonly used programs like TMVA~\cite{Speckmayer:2010zz} or Keras (\texttt{https://keras.io}).}. Electron isolation considers separately the energy/momentum reconstructed
around the electron direction in trackers, electromagnetic calorimeters, and hadronic calorimeters, taking into
account possible \emph{bremsstrahlung} effects in the traversed detectors: isolation requirements are part of the
CMS identification criteria, while they are applied separately in ATLAS. Total efficiencies are lower than for muons,
ranging in about 80-85\% for typical electrons from $\PW$ or $\PZ$ decays.

The only analyses significantly departing from the above choices are those having $4\ell$ in the final state,
because the simultaneous presence of four charged leptons removes many types of backgrounds and looser selections
can be used to recover efficiency.

Another important ingredient in the case of final states involving one or more $\PW^\pm$ boson(s) is the
reconstruction of the missing transverse momentum $p^{\mathrm{miss}}_{\rT}$, that can be identified as the (total) transverse momentum of
the undetected high-energy neutrino(s). In both ATLAS and CMS this is defined as the opposite of the vector sum of all
reconstructed particle momenta, so its precise determination depends on all energy/momentum corrections applied
to visible particles and in particular to jets~\cite{Aaboud:2018tkc,Sirunyan:2019kia}.

\subsubsection{Monte Carlo simulation}
\label{sec:mcsim}

Monte Carlo (MC) simulation of VBS signals and backgrounds which cannot be estimated from data-driven techniques (\emph{e.g.}\ QCD background, which has inherently the
same signature) is an essential ingredient of experimental analyses.

Regarding simulations at NLO QCD, matched to parton shower or merged with higher parton multiplicities, the most used generator tools for physics events are
\madgraphbis ~\cite{Alwall:2014hca}, version 2.3 (v2.3) and above, \POWHEG~\cite{Nason:2004rx,Frixione:2007vw,Alioli:2010xd} v2, and \Sherpa~\cite{Bothmann:2019yzt} v2.1 and above. 

Generation parameters can vary in different processes and experimental analyses,
but there are some common choices. In general, the central renormalization and factorization scales are set automatically
by \madgraphbis to the central $m_{\rm T}^2$ scale after $k_{\rm T}$-clustering of 
the event, while in \POWHEG\ and \Sherpa\ the default choice is 
process-dependent (for diboson and VBS processes a common choice is to use the
diboson invariant mass). Uncertainties from renormalization and factorization scales are mostly derived from the \emph{7-fold scale variation scheme}, where both are
varied independently by a factor of two up and down, but avoiding the cases
where the two vary in opposite directions (that is, differ by a factor four).

Quite peculiarly, in CMS, the standard sets of parton distribution functions (PDFs) used are different for the simulation of the 2016 detector conditions (NNPDF3.0 NLO) and for the 2017-18 conditions (NNPDF3.1 NNLO).
In ATLAS, the the NNPDF3.0 NNLO PDF set is used in most cases. 
The estimation of PDF uncertainties follow prescriptions from the NNPDF collaboration~\cite{Ball:2014uwa}.

While \Sherpa\ has an internal parton shower (PS) and underlying-event 
program, other Monte Carlo generators need an external tool providing PS,
which are usually {\sc Pythia8}~\cite{Sjostrand:2007gs} or {\sc HERWIG}~\cite{Bellm:2015jjp}. Underlying-event tuning is slightly different in the two experiments and tunes have also
been updated in some cases during the course of Run-2~\cite{Khachatryan:2015pea,Sirunyan:2019dfx}. 

Detector simulation is obtained through the {\sc GEANT4} software~\cite{GEANT,GEANT2} in both experiments.


\subsection{Impact on Beyond-the-Standard-Model theories}
\label{sec:bsm}
VBS studies can constrain the existence of Beyond-the-Standard-Model (BSM) physics in several ways, which are reviewed in detail in Ref.~\cite{Gallinaro:2020cte}.
Therefore, in the present review we simply restrict ourselves to outline the main results obtained at the LHC.
Searches for new physics in VBS channels can be divided into those based on an explicit (and possibly simplified) new physics model and
general model-independent searches, usually parameterized as Effective Field Theories (EFTs)~\cite{Degrande:2012wf}.

EFT constraints obtained in experimenal analyses, in ATLAS and CMS, use the parameterization of Refs.~\cite{Eboli:2006wa,Almeida:2020ylr}, where dimension-8 operators are considered. Unlike at dimension-6,
where quartic and trilinear gauge couplings are intrinsically related, at dimension-8 one can assume
the presence of anomalous quartic gauge couplings (aQGC) and no anomalous triple gauge couplings (aTGC).
There are 18 independent bosonic dimension-8 operators relevant for 2-to-2 scattering processes involving Higgs or gauge bosons at tree level, and conserving parity and charge conjugation.
They can be classified as \emph{scalar}, \emph{mixed}, and \emph{transverse} according to the number of gauge-boson strength fields contained in the operator (0, 2, and 4, respectively).
Several CMS Run-2 analyses constrain physics from non-zero dimension-8 operators, while ATLAS Run-2 measurements mostly focus on SM
VBS observations and do not provide explicit constraints on BSM physics: ATLAS results using the same model are
however available in $7$ and $8\TeV$ analyses.

All these operators have the common feature that non-zero Wilson coefficients lead to modifications of the high-energy tail of differential distributions of the scattering process.
Therefore, in the experimental analyses, events are first selected in VBS enhanced phase-space regions;
second, in the selected sample, a distribution sensitive to this modification is used to set constraints on the couplings.
At the LHC, such distributions include the invariant mass of the diboson system (or approximations based on reconstruction of the missing neutrino flight directions), or the transverse momentum of either scattered gauge boson.

\begin{figure*}[htbp]
\centering
\includegraphics[width=0.65\textwidth]{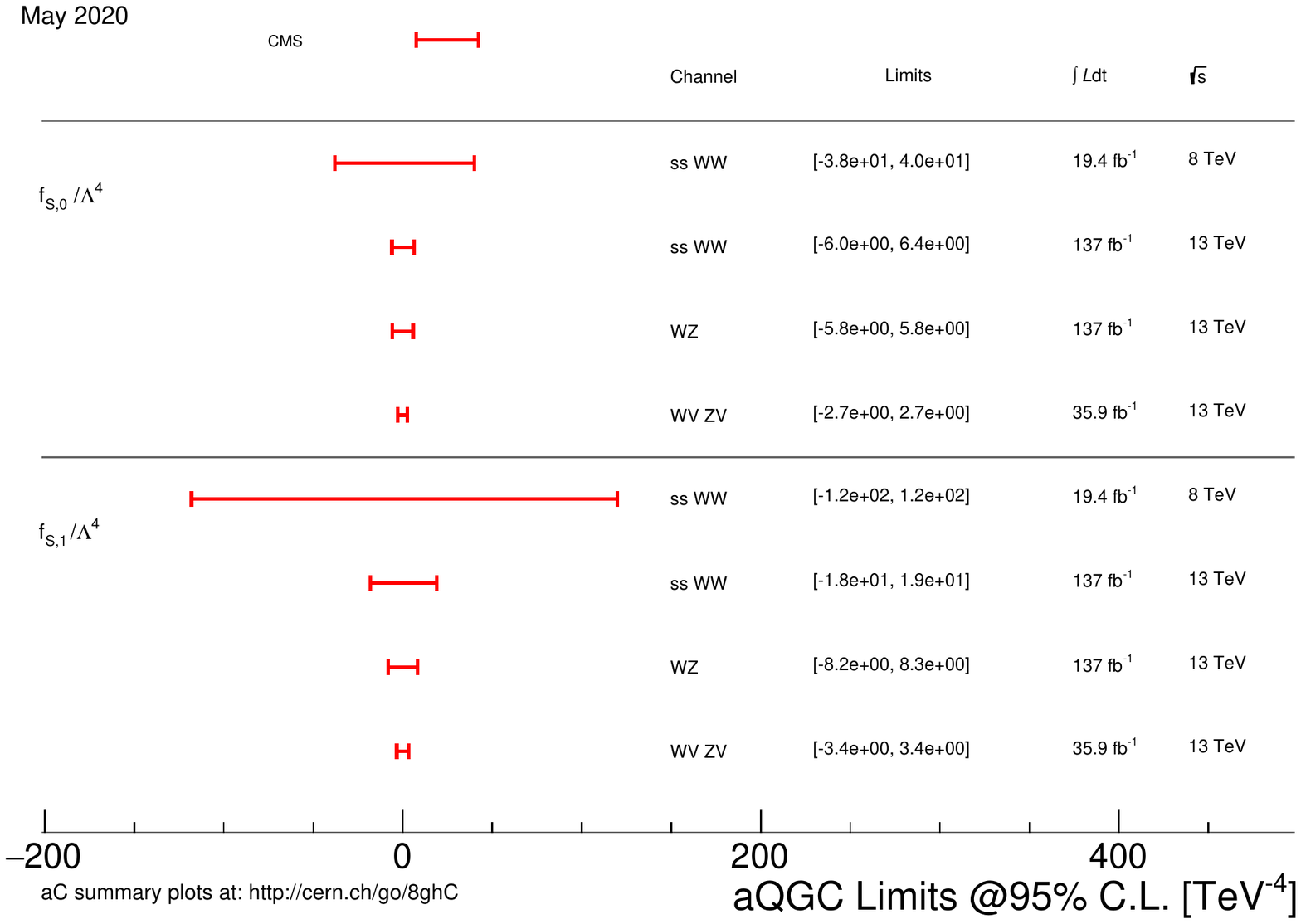}  \\
\caption{Current constraints on scalar dimension-8 operator couplings from various ATLAS (blue) and CMS (red) analyses at $7$, $8$, and $13\TeV$, with corresponding integrated luminosities. The figure is taken from Ref.~\cite{aQGCcomp}.}
\label{fig:aqgc}
\end{figure*}

\begin{figure*}[htbp]
\centering
\includegraphics[width=0.75\textwidth]{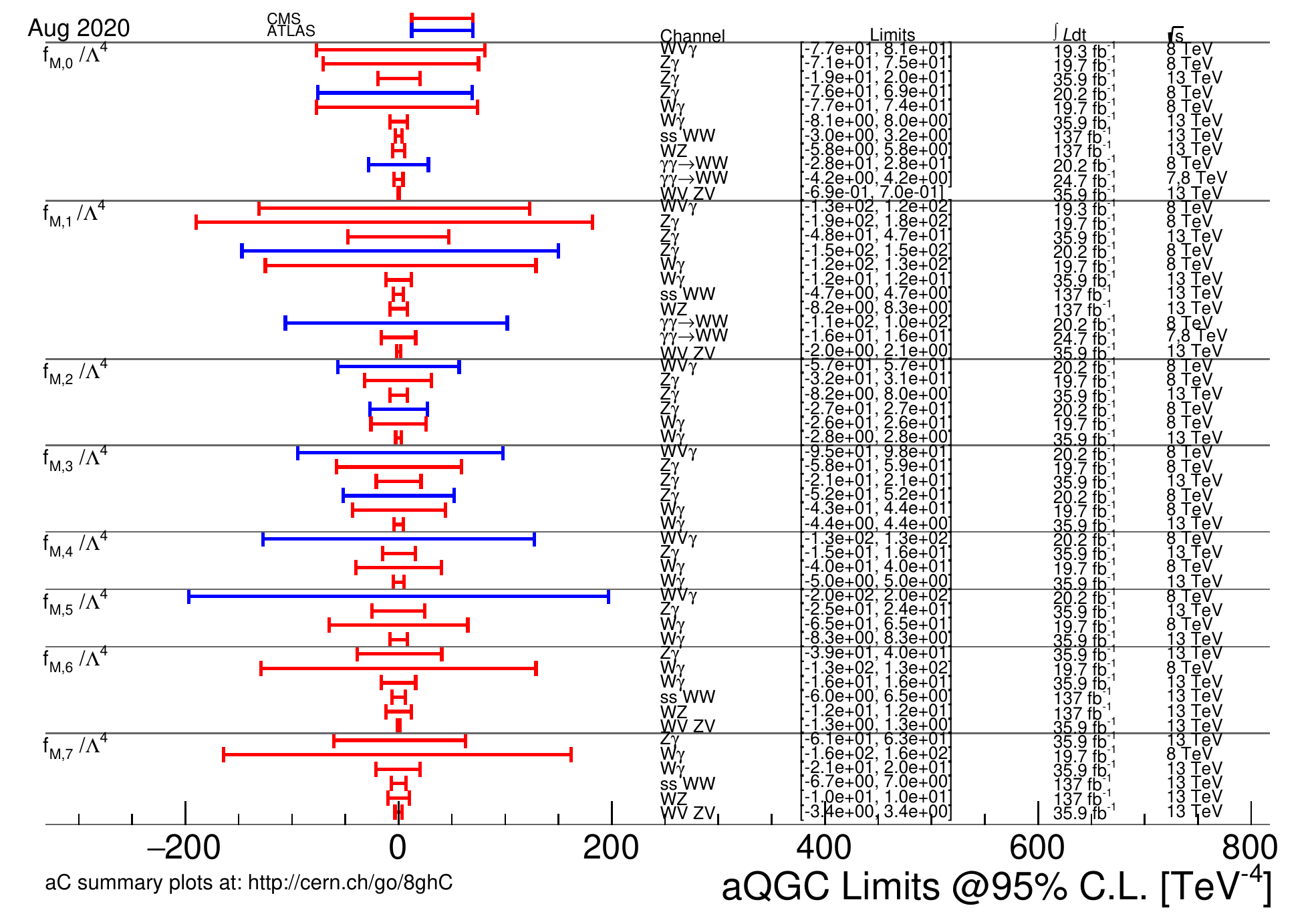}   \\
\caption{Current constraints on mixed dimension-8 operator couplings from various ATLAS (blue) and CMS (red) analyses at $7$, $8$, and $13\TeV$, with corresponding integrated luminosities.
The operator M,6 is reported but recently found to be redundant and hence not independent from other mixed 
operators~\cite{Perez:2018kav}. The figure is taken from Ref.~\cite{aQGCcomp}.}
\label{fig:aqgc2}
\end{figure*}

\begin{figure*}[thbp]
\centering
\includegraphics[width=0.75\textwidth]{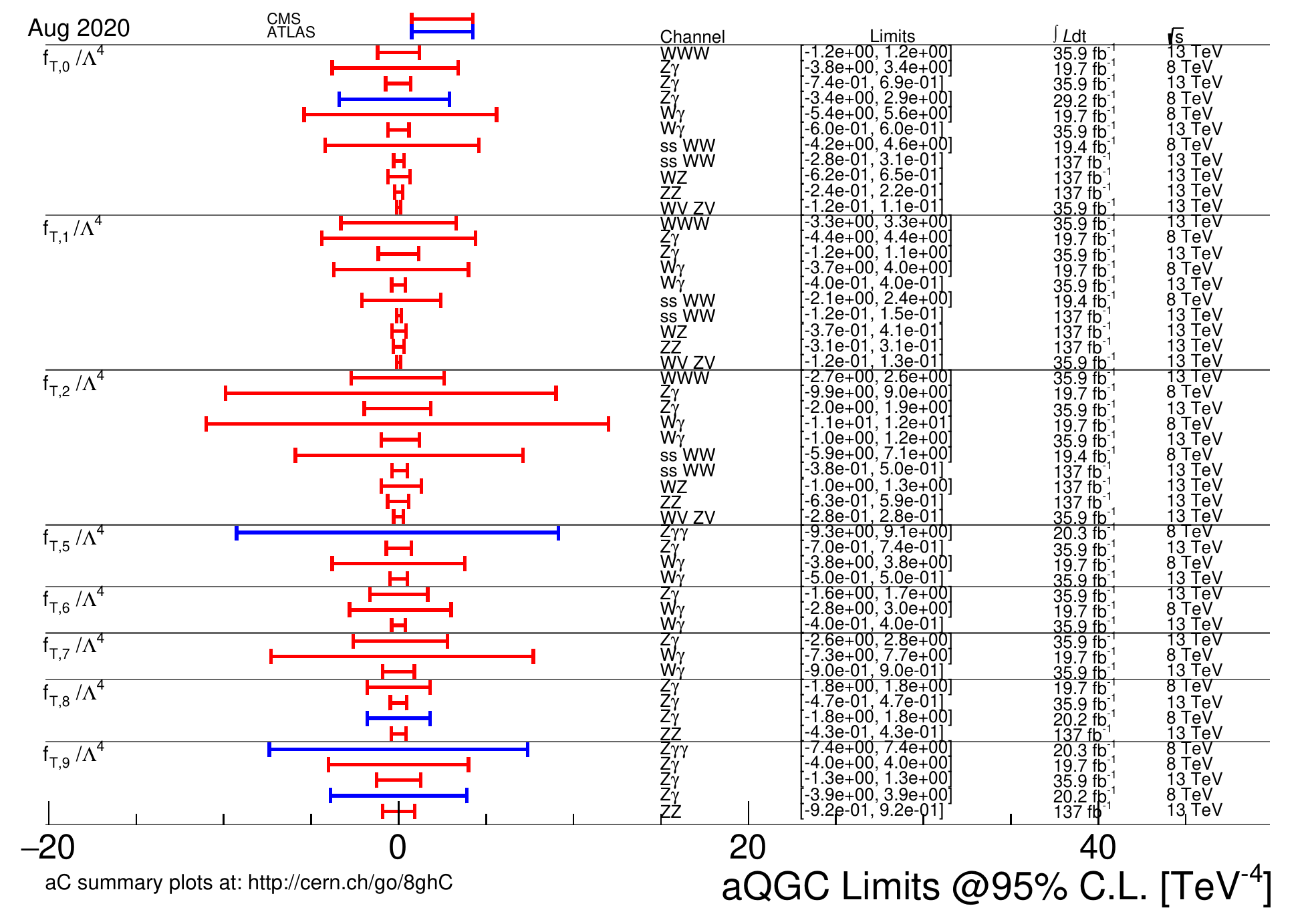}
\caption{Current constraints on transverse dimension-8 operator couplings from various ATLAS (blue) and CMS (red) analyses at $7$, $8$, and $13\TeV$, with corresponding integrated luminosities. The figure is taken from Ref.~\cite{aQGCcomp}.}
\label{fig:aqgc3}
\end{figure*}

Figures~\ref{fig:aqgc}-\ref{fig:aqgc3} show a compilation of the existing limits on
dimension-8 operator couplings. The couplings are defined as the ratio of the Wilson coefficient and the
power of the EFT new-physics scale appropriate for dimension-8 operators.
They are therefore expressed in units of $\TeV^{-4}$. 
As stated in Sec.~\ref{sec:experim}, semi-leptonic signatures are an experimental challenge, but in most cases provides the strongest handle on dimension-8 EFT operators. 
For transverse operators, different final states can be more or less sensitive to specific sets of operators, as only those with the correct combination of gauge
fields contribute to the related aQGC vertices.
Exclusive production, discussed in Sec.~\ref{sec:ggww}, gives the best results for operators sensitive to the
$\gamma\gamma\PW\PW$ interaction. It is important to notice that the presence of non-zero aQGCs would violate tree-level
unitarity at sufficiently high energy. Limits that take this effect into account can be set by cutting off the EFT
integration at the unitarity limit and just considering the expected SM contribution for generated events
with diboson invariant masses above the unitarity limit. The unitarity limits for each aQGC
parameter, typically about 1.5-2.5 $\TeV$, are usually calculated using the {\sc VBFNLO} program~\cite{Arnold:2008rz}
or taken from Ref.~\cite{Almeida:2020ylr}. These limits are typically less stringent than the naive ones, where the unitarity
violation is not taken into account.

If existing, BSM physics is unlikely to be confined to VBS processes.
The choice of selecting operators generating aQGC in the absence of aTGC is therefore not obvious, as it somehow breaks the EFT paradigm 
where operators with lower powers of $\Lambda$ should be constrained first, while data are not excluding yet all possible aTGC.
Moreover, in most VBS processes, aTGC effects enter directly, for example through specific $s$-channel diagrams.
In a dimension-6 realization of the EFT, operators affecting VBS analyses are also relevant for non-VBS $VV$ production, triple-gauge boson and Higgs boson production.
They should eventually therefore be constrained together in a larger-scope fit.
Studies of dimension-6 effects on specific VBS final states can be found in Refs.~\cite{Jager:2013iza,Gomez-Ambrosio:2018pnl,Dedes:2020xmo}.
Advancing in this direction, a very recent phenomenological 
work~\cite{Ethier:2021ydt} attempts a parameterization of many existing VBS 
results and compares the limits on several Wilson coefficients considering
just inclusive $VV$ production versus inclusion of VBS, finding $15-50\%$
weaker constraints when VBS is not included, depending on the specific operator.

In explicit BSM models, new resonances in the EW sector would also likely couple to
the vector bosons and Higgs boson such that many other production mechanism would be impacted. Therefore,
searches for new physics in diboson and Higgs events have strong implications for new
physics searches in VBS channels. Of the many possible models predicting modifications to
the EW sector, those involving additional Higgs bosons with narrow or broad natural widths are of particular
interest. In the particular cases where couplings of the new resonances to fermions are suppressed or absent,
in fact, the main production and decay modes would produce signals that are experimentally equivalent to
VBS, but with resonant $VV$ invariant masses.
As these analyses are part of a more general search program in ATLAS and CMS, also
involving other production mechanisms and final states, they will not be further reviewed here.


\section{Vector-boson scattering processes at the LHC}
\label{sec:processes}

\subsection{The $\ssWW \rightarrow \ell^\pm \nu \ell'^\pm \nu \Pj\Pj$ final state}\label{sec:ssww}
The \ssWW\ process is considered to be the \emph{golden channel} in the study of VBS.
The cross-section ratio of the EW component containing the VBS production compared to the QCD one is very large (see Sec.~\ref{sec:sob} for a precise definition of the EW and QCD contributions),
of order 4-6 in typical fiducial regions, while it is usually $\ll 1$ for other processes.
This is due to charge conservation which prevents gluon-initiated processes in the QCD background as opposed to \WZjj, \ZZjj, or \osWW.
In addition, the application of particular event selections allows to further enhance the EW component of the cross section (also see Sec.~\ref{sec:sob}).
For this reason, the \ssWW\ channel is the most sensitive to potential new-physics effects, including those affecting polarisation and anomalous quartic gauge couplings.

\subsubsection{Theoretical calculations}

From a theoretical point of view, the \ssWW\ channel is without a doubt the most accessible one, because of the reduced number of partonic channels and Feynman diagrams.
Calculations started already ten years ago with the computation of the QCD corrections to the EW process at order $\mathcal{O}\left(\alpha_{\rm s} \alpha^6 \right)$ in the VBS 
approximation~\cite{Jager:2009xx,Denner:2012dz}.
Such corrections have been then matched to parton shower in programs such as {\sc POWHEG} \cite{Jager:2011ms} or {\sc VBFNLO} \cite{Arnold:2008rz,Arnold:2011wj,Baglio:2014uba}.
The QCD background is also known since some time at NLO \cite{Melia:2010bm,Campanario:2013gea} and has been matched to PS \cite{Melia:2011gk}.
Only recently the NLO EW corrections of order $\mathcal{O}\left(\alpha^7 \right)$ have been computed \cite{Biedermann:2016yds} and found to be large.
The full tower of NLO corrections has been computed a few months later in Ref.~\cite{Biedermann:2017bss}.
Given the size of the EW corrections, these have been implemented in the computer program {\sc POWHEG} \cite{Chiesa:2019ulk} so that they can be combined with other matched predictions.

\begin{figure*}[t]
\centering
          \includegraphics[width=0.42\linewidth,height=0.35\textheight,clip=true,trim={0.cm 0.cm 0.cm 0.cm}]{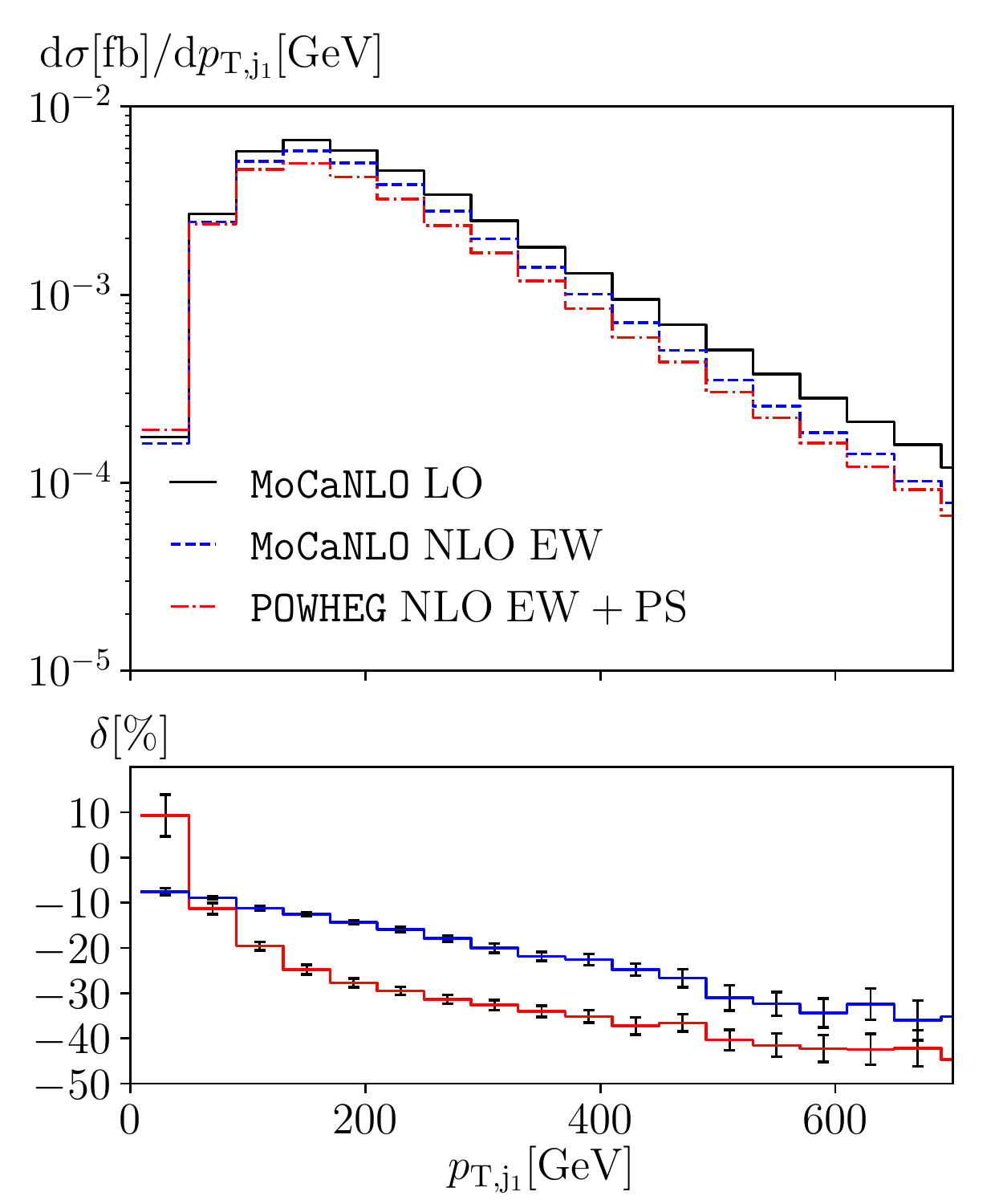}
          \includegraphics[width=0.54\linewidth,height=0.34\textheight,clip=true,trim={0.cm 0.cm 0.cm 0.cm}]{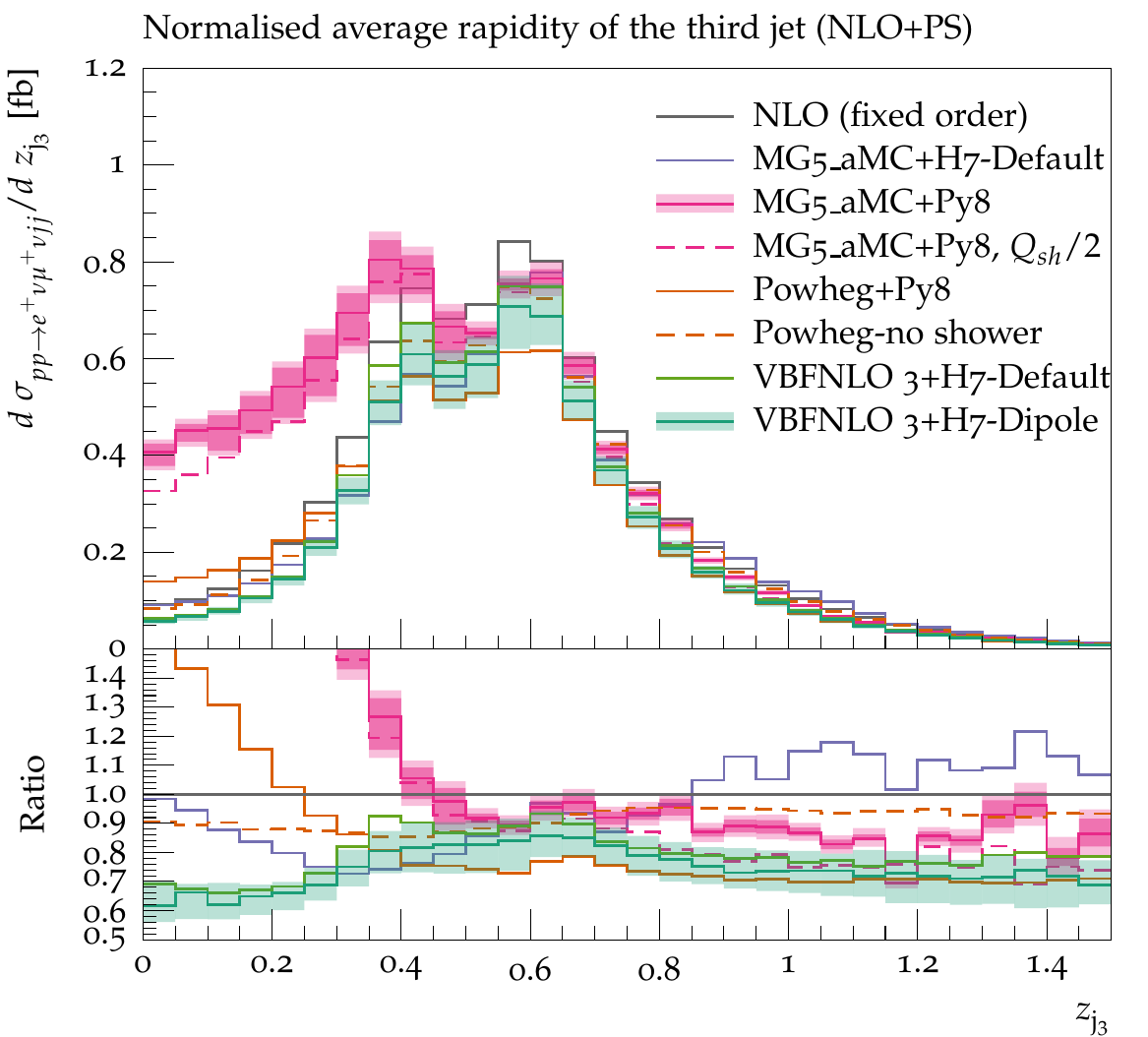}
\caption{Differential distributions for the \ssWW\ channel.
Left: transverse momentum of the hardest jet with LO, NLO EW, and NLO EW + PS accuracy.
Right: normalised average rapidity of the third jet at NLO QCD + PS accuracy for different predictions.
These figures are taken from Ref.~\cite{Chiesa:2019ulk} and Ref.~\cite{Ballestrero:2018anz}, respectively.}
\label{fig:ssWWTH}
\end{figure*}

As \ssWW\ is a representative channel for all VBS processes, in Ref.~\cite{Ballestrero:2018anz} several computer programs 
\cite{Ballestrero:2007xq,Kilian:2007gr,Moretti:2001zz,Schwan:2018nhl,Nason:2004rx,Frixione:2007vw,Alioli:2010xd,Arnold:2008rz,Arnold:2011wj,Baglio:2014uba,Alwall:2014hca}
have been used for a comparative study of fixed and matched predictions.
One of the main findings of this study is that different NLO QCD predictions matched to parton shower can vastly differ for observables involving the third jet 
(\emph{i.e.}\ a non-tagging jet).
This is particularly apparent on the right-hand side of Fig.~\ref{fig:ssWWTH}.
Nonetheless, we would like to emphasize that, upon using {\sc Pythia8} with the correct recoil scheme or {\sc Herwig}, 
reliable predictions can be obtained even for the third-jet observables.
This has been discussed in Sec.~\ref{sec:theory} and studied in detail in Ref.~\cite{Jager:2020hkz} for Higgs production via VBF.
It implies that jet veto in central regions can be used in experimental analysis provided that good care is taken in using appropriate theoretical predictions.
The second main finding of this study is that the VBS approximation is good up to few per cent for typical VBS event selections.
This implies that the VBS approximations at fixed order or used in combination with parton shower are reliable as long as selection cuts are able to suppress non-VBS contributions such as tri-boson contributions
(see further discussion in Sec.~\ref{sec:zz}).

After the publication of Ref.~\cite{Ballestrero:2018anz}, other comparative studies have appeared~\cite{ATLAS:WWMC,ATLAS:2020ryx}.
It should be made clear that, while the former study was a tuned comparison among the different event generators, where all
the parameters relevant for the partonic cross sections were identical, this is not always the case for the latter studies. In the
first case, discrepancies among predictions are unambiguously due to the different shower and hadronisation models of the Monte Carlo programs.
In the second case a certain degree of ambiguity remains in the origin of the discrepancies, which may jeopardise a proper understanding of these effects.

A summary of the available predictions is provided in Table~\ref{tab:ssWWTH}.
If it fair to say that the theoretical status is rather good, given the experimental accuracy available now and expected in the next ten years.
In particular, almost all NLO orders matched to parton shower are known.
The only exception is the order $\mathcal{O}\left(\alpha_{\rm s}^2 \alpha^5 \right)$, which has been shown in Ref.~\cite{Biedermann:2017bss} to be suppressed.
Also the order $\mathcal{O}\left(\alpha_{\rm s} \alpha^6 \right)$ matched to PS is only known in the VBS approximation.
Nonetheless, provided that typical VBS phase-space regions are used this should be a very good one.
Going beyond this approximation would require a method to match mixed-type corrections, which is currently not existing.

\begin{table}[htb]
\caption{Summary of higher-order predictions currently available for the ss-WW channel: at fixed order and matched to parton shower.
The symbols {\bf \color{green} \checkmark}, {\bf \color{green} \checkmark$^*$}, and {\bf \color{red} X}
means that the corresponding predictions are available, in the VBS approximation, or not available yet.
}
\center
{\begin{tabular}{l|cccc}
Order & $\mathcal{O}\left(\alpha^7 \right)$ & 
$\mathcal{O}\left({\alpha_{\rm s}} {\alpha}^6 \right)$ & 
$\mathcal{O}\left({\alpha_{\rm s}}^2 {\alpha}^5 \right)$ & 
$\mathcal{O}\left({\alpha_{\rm s}}^3 {\alpha}^4 \right)$\\
\hline
NLO & {\bf \color{green} \checkmark} &  {\bf \color{green} \checkmark} & {\bf \color{green} \checkmark} & {\bf \color{green} \checkmark} \\
NLO+PS & {\bf \color{green} \checkmark} & {\bf \color{green} \checkmark$^*$} & {\bf \color{red} X} & {\bf \color{green} \checkmark}
\end{tabular} \label{tab:ssWWTH}}
\end{table}

\subsubsection{Experimental approaches}

Both ATLAS and CMS have reported the observation of electroweak \ssWW\ production using a partial 13 TeV data set~\cite{ATLAS:ssWW,CMS:ssWW}. 
CMS has already published the same search on the full data set, in combination with the \WZjj\ final state~\cite{CMS:ssWWandWZ}.

\paragraph{Monte Carlo simulation}

ATLAS and CMS use Monte Carlo simulations to evaluate the signal and several background contributions to the selected
data samples. 

CMS uses the \madgraphbis generator~\cite{Alwall:2014hca}, version 2.4.2 to simulate the electroweak,
strong, and interference components separately at LO. All samples have no extra partons beyond the two quarks in the
simulated process and are hence inclusive in the number of extra jets. The interference is estimated to be about $4\%$
of the signal and is included in the signal yield. 
Since the CMS analysis is more recent, it could benefit from the complete application of
NLO QCD+EW corrections computed in~\cite{Biedermann:2017bss} (see previous section), that decrease the cross sections by
10-15\%, the correction being larger at higher values of \mjj\ and $m_{\ell\ell'}$. Similar settings are used to
simulate the \WZjj\ component, that is analyzed together in a single study. Other minor backgrounds, including
tribosons, processes with at least a top quark ($\Pt {\bar \Pt}$, $\Pt {\bar \Pt}\PW$, $\Pt {\bar \Pt}\PZ$, $\Pt\PW$, $\Pt\PZ q$ etc.)
as well as other diboson processes, are simulated with either \POWHEG~\cite{Nason:2004rx,Frixione:2007vw,Alioli:2010xd} or \madgraphbis, mostly
with inclusive NLO QCD accuracy. 

ATLAS uses the \Sherpa generator, version 2.2.2~\cite{Bothmann:2019yzt} to simulate the electroweak, strong and interference processes at LO. All samples are simulated with up to 1 extra parton beyond the two quarks and the 0- and 1-parton processes are merged using the MEPS matching scheme included in \Sherpa. The interference is estimated to be about 6\% of the signal. 
An alternative description of the VBS signal process is obtained using \POWHEG at NLO in QCD~\cite{Jager:2009xx}.
A large difference between the two theoretical fiducial cross sections is found
\footnote{In Ref.~\cite{Hoeche} it has been documented that there was an issue in \Sherpa regarding the colour flow setup when using parton shower for VBS-like processes.
To our knowledge this issue has been resolved, but the corresponding results have not yet appeared in any further publication.}. We believe that these differences should be investigated beyond the work done in Ref.~\cite{ATLAS:WWMC} 
 (see remarks in the previous section, as well as in the corresponding part of Sec.~\ref{sec:wz} for \WZjj).
Other backgrounds considered (in general the same as CMS, but with more emphasis
on $V\gamma$ and electroweak $V\gamma \Pj\Pj$, which are found to be an important contribution to the background) are generated
using different tools, perturbative accuracies, and extra-parton multiplicities.

\paragraph{Fiducial region definitions and reconstruction-level selections}

Fiducial regions considered in the ATLAS and CMS analyses are compared in Table~\ref{tab:ssWWfr}. In both analyses
leptons from $\tau$ decays are not considered in the fiducial region and lepton momenta are corrected by adding
possible final-state photon radiations in a cone of $\Delta R < 0.1$ around the lepton direction.

\begin{table}[htb]
\caption{Comparison of \ssWW\ fiducial region definitions and related EW (VBS) cross-section values in the ATLAS and CMS measurements~\cite{ATLAS:ssWW,CMS:ssWWandWZ}.
\emph{JRS} stands for generic Jet-Rapidity Separation selections. ATLAS $\sigma_\mathrm{LO}$ has the issues reported in the text.}
\center
{\begin{tabular}{@{}ccc@{}} \toprule
Variable & ATLAS  & CMS  \\
\midrule
$p_{\rm T}(\ell)$ & $> 27$ GeV & $> 20$ GeV \\
$|\eta(\ell)|$ & $< 2.5$ &  $< 2.5$ \\
$\Delta R(\ell\ell')$ & $> 0.3$ &  - \\
$m_{\ell\ell'}$ & $> 20$ GeV & $> 20$ GeV \\
$p_{\rm T,miss}$ & $> 30$ GeV & - \\
$p_{\rm T}(\Pj)$ & $> 65/35$ GeV & $> 50$ GeV \\
$|\eta(j)|$ & $< 4.5$ &  $< 4.7$ \\
\mjj & $>500$ GeV & $> 500$ GeV \\
JRS & $ \dyjj > 2$ &  $\dejj > 2.5$ \\
\midrule
$\sigma_\mathrm{LO} $ & $2.0^{+0.3}_{-0.2}\fb$ & $3.9 \pm 0.6\fb$ \\
$\sigma_\mathrm{NLO}$ & $3.1^{+0.4}_{-0.5}\fb$ (NLO QCD) & $3.3 \pm 0.5\fb$  (NLO QCD+EW) \\
\bottomrule
\end{tabular} \label{tab:ssWWfr}}
\end{table}

Reconstruction-level selections follow closely the definition of the fiducial regions. In both analyses, events
are selected at the trigger level by the presence of just one electron or muon, in order to increase efficiency.
Both experiment veto events with jets likely originated from a bottom quark, in order to reject backgrounds featuring
top-quark decays. CMS requires that the leading lepton has $p_{\rm T} > 25\GeV$, and that $p_{\rm T,miss} > 30$ GeV. In both ATLAS and CMS, background
from wrong charge reconstruction in $\Pe^\pm \Pe^\pm jj$ events is reduced by requiring $|m_{\Pe\Pe} - m_\PZ| > 15$ GeV, and in
ATLAS dielectron events with $|\eta(\Pe)| > 1.4$ are also rejected. In CMS the maximum Zeppenfeld variable $z^*_\ell$ of the
two leptons must be smaller than 0.75, where~\cite{Rainwater:1996ud}:
\begin{equation}
z^*_\ell = \left|\frac{\eta(\ell) - [\eta(\Pj_1)+\eta(\Pj_2)]/2}{\dejj}\right| .
\end{equation}

\paragraph{Analysis strategy and background estimation}

Both experiments fit the observed data after estimating backgrounds from either simulation or control
regions. In ATLAS, the contribution from~\emph{non-prompt leptons} is estimated in different control regions, depending
if they are originated from heavy-flavored meson decays or from $V \gamma$ events with photon conversions (only for
final states with electrons). Such regions are enriched in $\Pb {\bar \Pb}$ events and $\gamma$ from final-state radiation
in $\PZ$ decays, respectively. CMS uses events which pass the final selection except for one rejected lepton, which
is selected with looser requirements to enter the control region and further validates non-prompt leptons from
heavy-flavor decays by inverting the $\Pb$-jet veto. $\PW\PZ$ events are fit together in the CMS analysis, while
in ATLAS the normalization of a control region with $3\ell$ selected events is floated in the fit. Both analyses
use fully selected events with $200 < \mjj < 500\GeV$  to constrain background-component normalizations.

In the ATLAS analysis the signal-region data in four \mjj\ bins with different signal purities are fit together
with the $3\ell$ and the low-\mjj regions, in order to optimize significance. In CMS, which analyzes a
larger dataset, a similar technique is used, but using two-dimensional distributions in bins of
\mjj\ and $m_{\ell\ell'}$ in the signal region and three control regions, leaving free in the fit the normalization
of two on them, in addition to the EW and strong cross sections.

\begin{figure*}[t]
\centering
\includegraphics[width=0.4\textwidth]{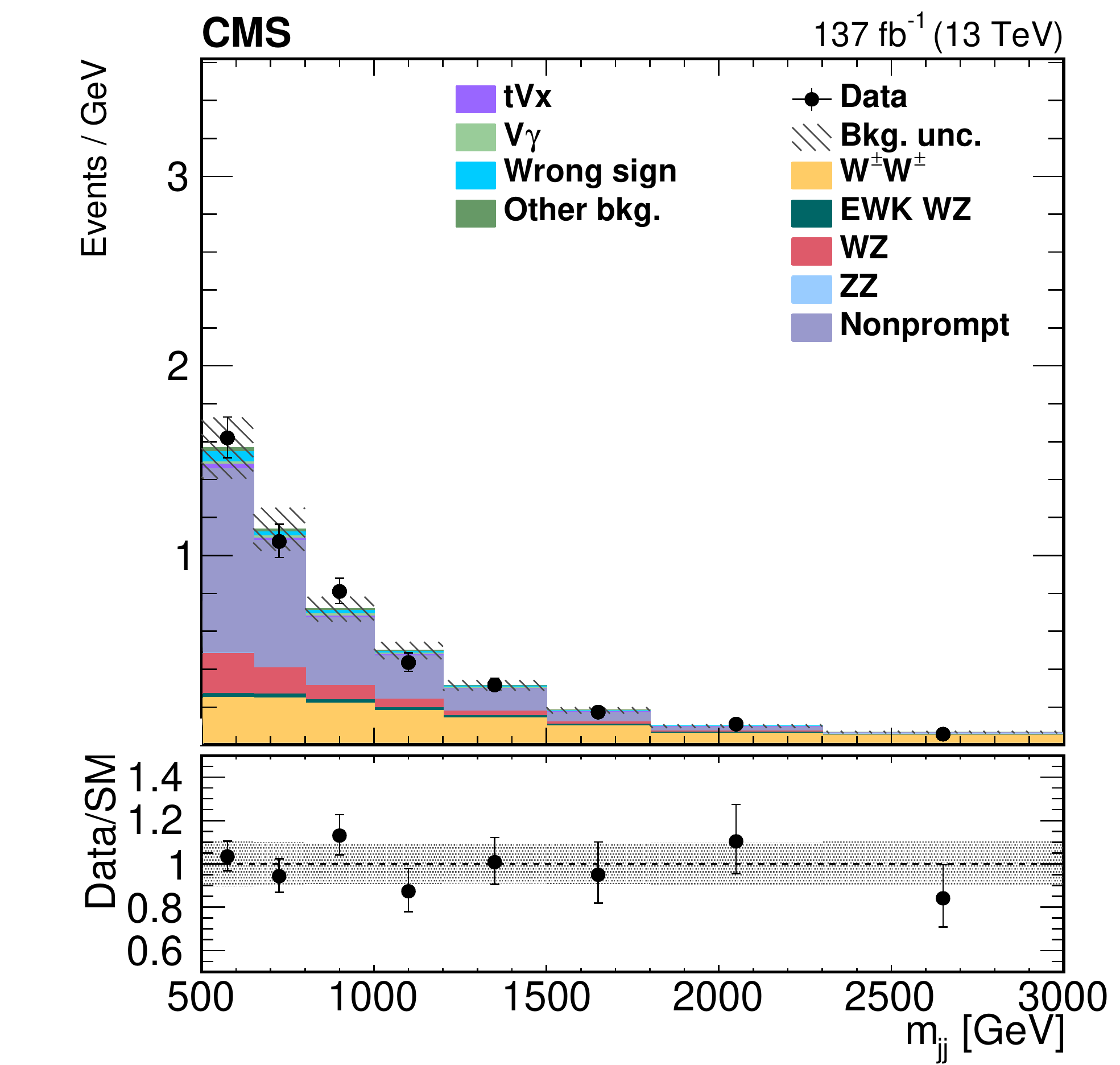} 
\includegraphics[width=0.5\textwidth]{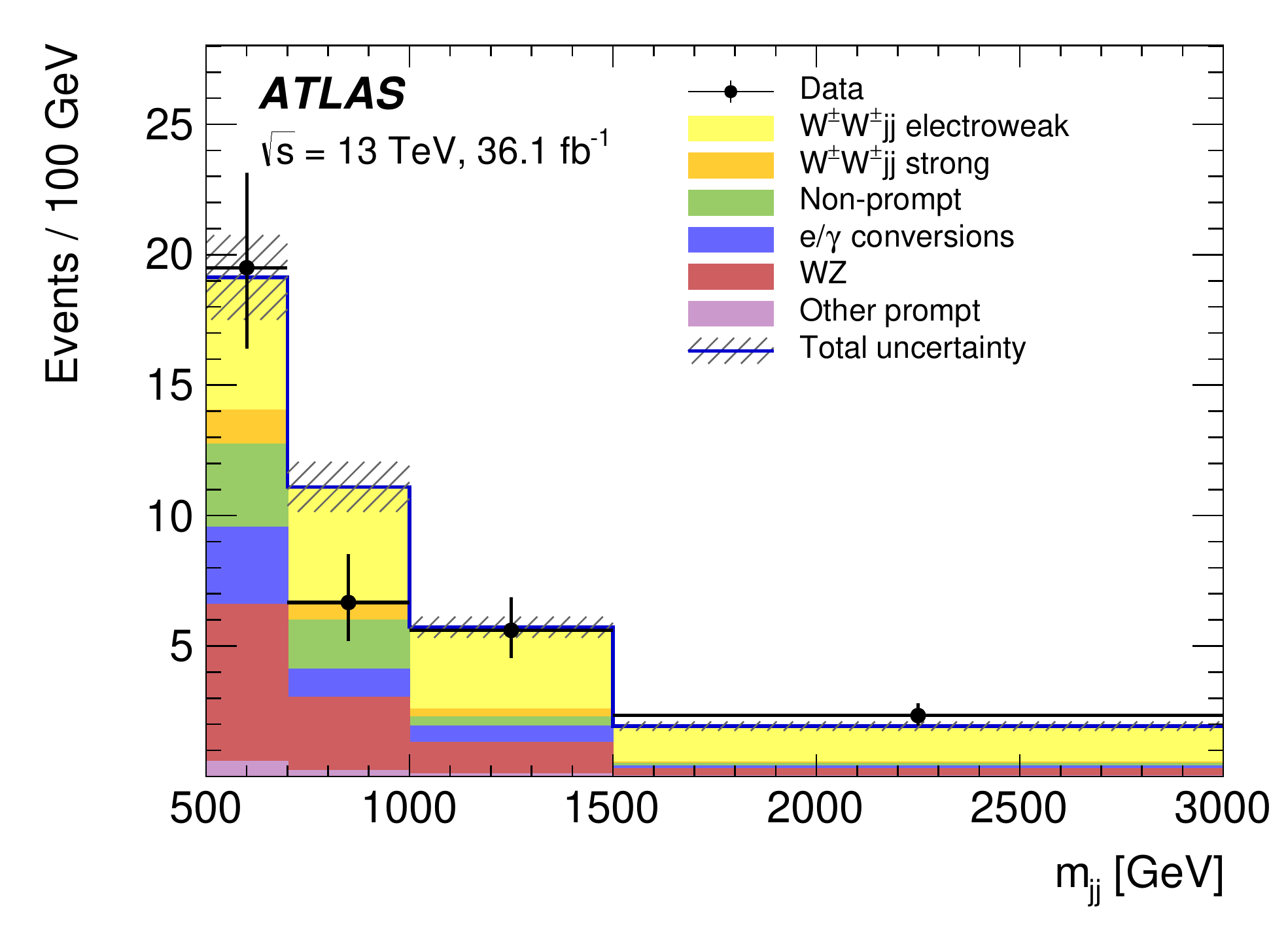}
\caption{Post-fit \mjj\ distributions in CMS (left) and ATLAS (right) in the \ssWW\ analysis. \emph{Nonprompt} in CMS
includes the photon conversion background. In ATLAS, \emph{e/$\gamma$ conversions} includes events with wrong-sign electron reconstruction.}
\label{fig:ssww}
\end{figure*}

The ATLAS and CMS data with superimposed signal and background components are shown in Figure~\ref{fig:ssww}. While
the labelling of the process composition is different, its relative amount is similar between the two analyses, 
CMS exhibiting more non-prompt background because of the softer lepton selections.

\paragraph{Systematic uncertainties}

Both ATLAS and CMS list systematic uncertainties according to their impact on the measured cross sections. In CMS the
dominant uncertainties come from the estimate of the non-prompt background, the limited size of simulated samples
in the two-dimensional distributions, and theoretical errors on the various simulated components. In ATLAS,
similar uncertainties are considered, but a larger impact from jet-energy corrections and the related $p_{\rm T,miss}$
measurement is present. No single contribution has an impact larger than $4\%$ in either analysis.

\paragraph{Results}

ATLAS reports a measured VBS fiducial cross-section of $\sigma_{\mathrm{EW}} = 2.89^{+0.59}_{-0.55}\fb$, 
where the total uncertainty is dominated by the statistical one, 
in agreement with the NLO QCD estimate of the SM cross section.
It corresponds to a background-only hypothesis rejection with a significance of 6.5$\sigma$.

CMS similarly reports $\sigma_{\mathrm{EW}} = 3.98 \pm 0.45\fb$ , also statistically dominated and in agreement with the NLO QCD+EW
estimation in the respective fiducial region. It corresponds to a background-only hypothesis rejection with a significance much larger than 5$\sigma$. The total \ssWW\ cross section
including EW and strong components is also measured to be $\sigma_{\mathrm{tot}} = 4.42 \pm 0.47\fb$.

\begin{figure*}[t]
\centering
\includegraphics[width=0.47\textwidth]{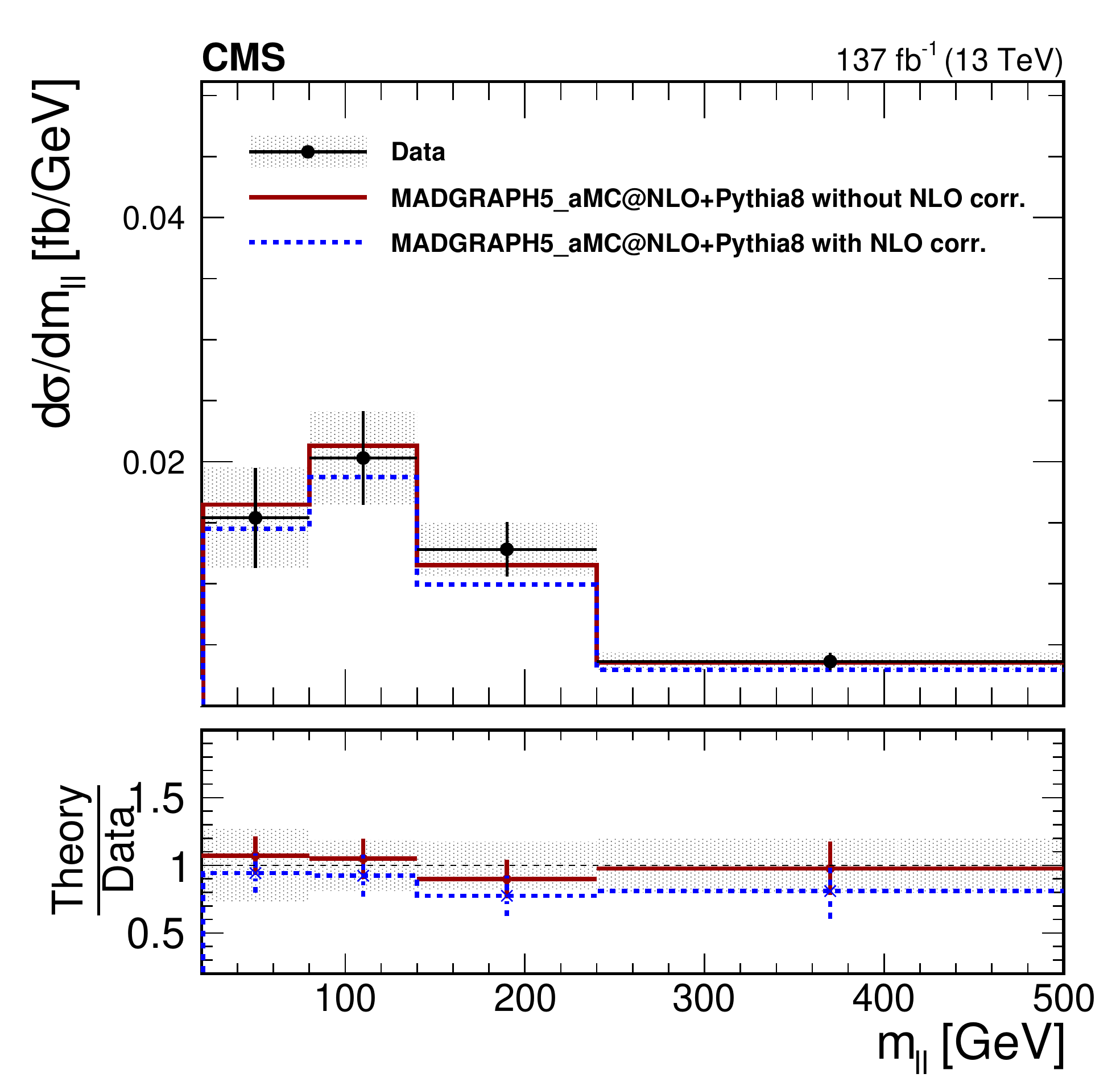} 
\includegraphics[width=0.42\textwidth]{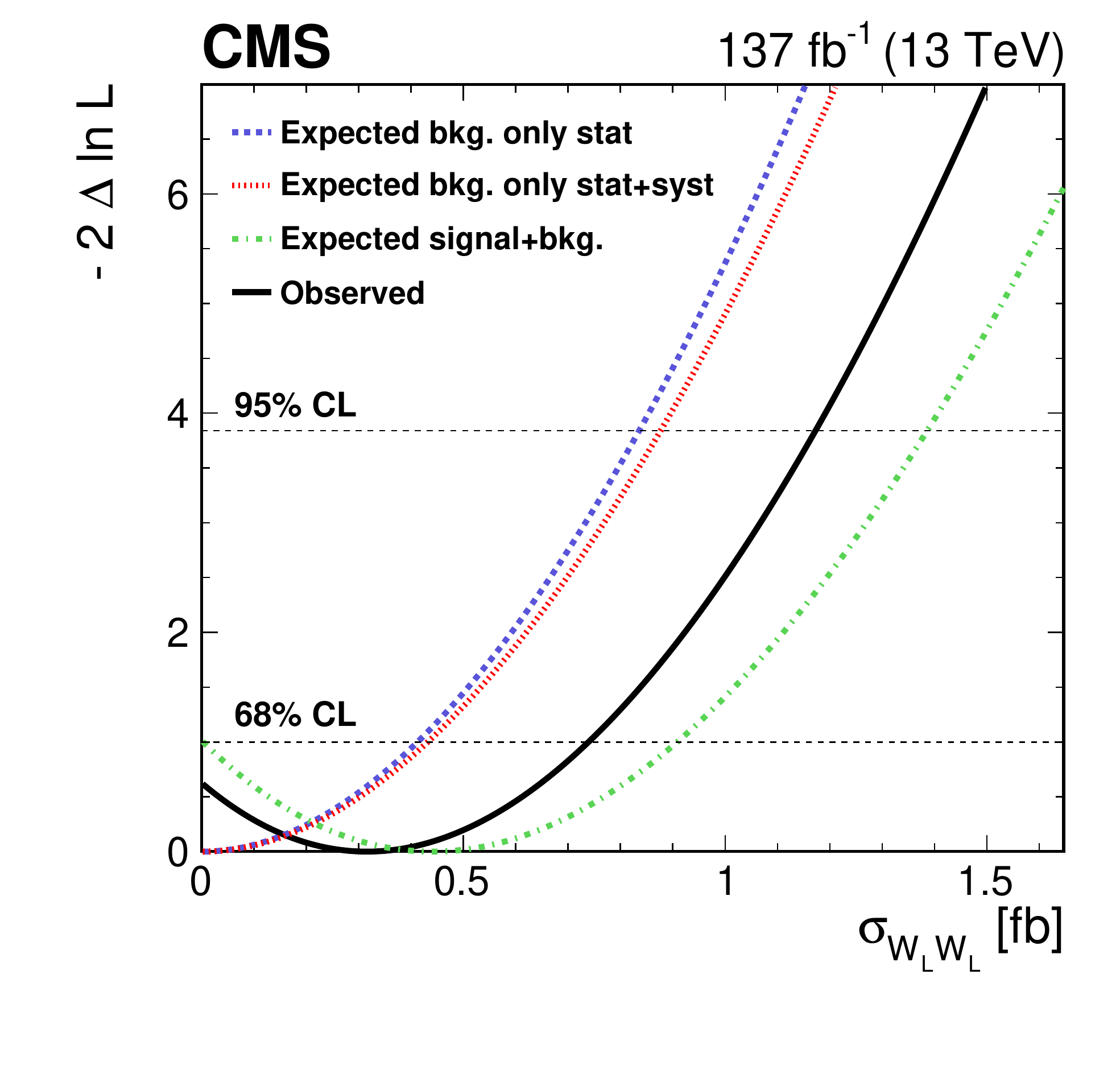}
\caption{CMS \ssWW\ analysis. Left: EW differential cross sections as a function of $m_{\ell\ell'}$ measured in data, in LO and NLO-corrected simulation.
Right: likelihood scan as a function of the cross section of $\PW_L^\pm \PW_L^\pm$ events.}
\label{fig:ssww2}
\end{figure*}

Differential cross-sections in four bins of \mjj, $m_{\ell\ell'}$ and the leading lepton $p_{\rm T}$ are obtained by fitting simultaneously
the corresponding regions of the phase space, with negligible bin-migration effects, and when needed replacing the fitted observables
with the ones under study. All the results are in agreement with SM expectations, although the
experimental uncertainties are of the order of $20\%$ because of limited statistics. Figure~\ref{fig:ssww2} (left) shows the EW differential cross sections as a function of $m_{\ell\ell'}$.

Constraints on aQGC are set by fitting the diboson transverse mass distribution\footnote{The diboson transverse mass is defined as
$m_{\rm T}(\PW\PW) = \sqrt{(\Sigma E_i)^2 - (\Sigma p_{z,i})^2}$, where $E_i$ and $p_{z,i}$ are the energies and
$z$ components of the momenta of all particles from the decay of the $\PW$ in the event. The four-momentum of
the di-neutrino system is defined using the $p_{\rm T,miss}$ vector and 
assuming that the values of the longitudinal component of the momentum and the invariant
mass are zero.} in the signal region: no BSM excess is found and 
the limits set on $f_{T,0}$, $f_{T,1}$, $f_{T,2}$, $f_{M,0}$, $f_{M,1}$,
$f_{M,7}$, and $f_{S,0}$ are the world second-best limits after the CMS semi-leptonic VBS analysis. 

\paragraph{Polarisation measurement}

In a separate analysis, CMS~\cite{CMS:WWpolar} examines the same selected dataset in order to measure
the polarisation of $\PW$ bosons in \ssWW\ events. The analysis is identical to the previously described CMS
results, for what concerns the simulation and estimation of the backgrounds, event selection,
systematic uncertainties, and fitting techniques.

Polarised signal for the three possible combinations $\PW_{\rm L}^\pm \PW_{\rm L}^\pm$, $\PW_{\rm L}^\pm \PW_{\rm T}^\pm$, and $\PW_{\rm T}^\pm \PW_{\rm T}^\pm$
are generated using a new version of \madgraphbis~\cite{BuarqueFranzosi:2019boy}. The two-dimensional fits use different
variables than in the original analysis: both are output scores of BDT algorithms, an
\emph{inclusive} one optimized to select EW \ssWW\ over backgrounds, and a \emph{signal} BDT defined in two versions, alternatively
optimized to select purely longitudinal or longitudinal-unpolarised ($\PW_{\rm L}^\pm \PW_X^\pm$) signals over
other polarisation combinations. Training variables, besides those already used in the event selection, include
the transverse mass, $p_{\rm T}$, angular or $\Delta R$ separations between leptons and jets, and the ratio of $p_{\rm T}$ products
between leptons and jets.

The resulting cross-sections of $\sigma_{\mathrm{fid}} = 1.2^{+0.6}_{-0.5}\fb$ for the $\PW_{\rm L}^\pm \PW_X^\pm$ process and
$\sigma_{\mathrm{fid}} < 1.17\fb$  at the 95\% Confidence Level (CL) for the $\PW_{\rm L}^\pm \PW_{\rm L}^\pm$ process
are in agreement with the SM. There is not yet an evidence even for a single-boson polarisation state, the significance
of the $\PW_L^\pm \PW_X^\pm$ background-hypothesis rejection being only 2.3$\sigma$.
Figure~\ref{fig:ssww2} (right) shows the likelihood scan as a function of the cross section of $\PW_{\rm L}^\pm \PW_{\rm L}^\pm$ events.
Results are also extracted by considering polarisation eigenstates not in the default helicity frame, but in the colliding-parton frame.
Results in both frames have been found to be comparable.

\subsection{The $\WZjj \rightarrow \ell^+ \ell^- \ell'^\pm \nu \Pj\Pj$ final state}
\label{sec:wz}
The \WZjj\ VBS process has a larger cross section than \ssWW, which becomes however comparable when considering the
much less abundant leptonic $\PZ$ decays. In addition, production with at least two strong vertices is a dominant
background, because $\PW^\pm \PZ$ production at the LHC is possible from a quark-antiquark initial state at
tree level, and NLO QCD corrections resulting in two-jet events are fairly large. 
Therefore searches for this VBS process are much more challenging, with
typical signal-to-background ratios of 1/10, and require advanced analysis techniques.

\subsubsection{Theoretical calculations}

With respect to the \ssWW\ channel, \WZjj\ is more complicated as it has more partonic channels and more involved Feynman diagrams.
Therefore the state of the art in the theoretical knowledge is not as advanced as for \ssWW.
Nonetheless, QCD corrections have been known for some time in the VBS approximation \cite{Bozzi:2007ur} and have also been matched to parton shower \cite{Jager:2018cyo}.
The QCD background is also known at NLO QCD accuracy since some time \cite{Campanario:2013qba} and NLO predictions matched to parton shower can nowadays be obtained from automatised tools.

\begin{figure*}[t]
\centering
          \includegraphics[width=0.48\linewidth,height=0.35\textheight,clip=true,trim={0.3cm 0.cm 0.cm 0.cm},page=14]{Figures/nlos_wz}
          \includegraphics[width=0.48\linewidth,height=0.34\textheight,clip=true,trim={0.cm 0.cm 0.cm 0.cm}]{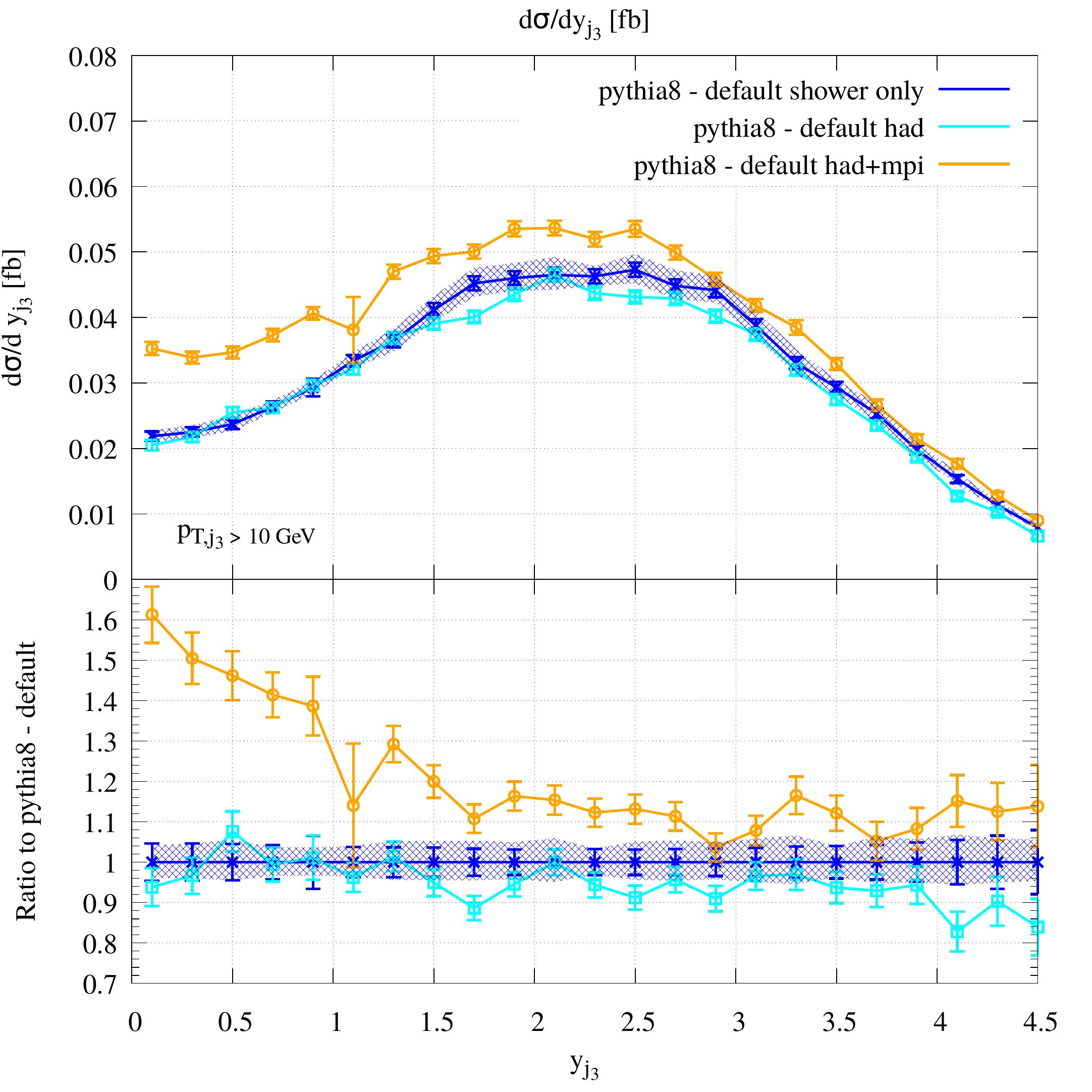}
\caption{Various differential distributions for the WZ channel.
Left: invariant mass of the two tagging jets with NLO EW + QCD corrections.
Right: Zeppenfeld variable for the third jet at NLO QCD + PS accuracy including hadronisation or multiple-parton interaction.
These figures are taken from Ref.~\cite{Denner:2019tmn} and Ref.~\cite{Jager:2018cyo}, respectively.}
\label{fig:WZTH}
\end{figure*}

Recently the full NLO computations for the orders $\mathcal{O}\left(\alpha^7 \right)$ and  $\mathcal{O}\left(\alpha_{\rm s} \alpha^6 \right)$ have been obtained \cite{Denner:2019tmn}.
It is worth noticing that it confirmed that the EW corrections are in general large for VBS processes at the LHC.
Such results can for example be seen in the left-hand side of Fig.~\ref{fig:WZTH}.

Along the lines of Ref.~\cite{Ballestrero:2018anz}, Ref.~\cite{Jager:2018cyo} showed that the details of the parton shower can be particularly relevant for phenomenological studies.
In addition, the authors showed that the effect of hadronisation and multiple-parton interaction can be significant, especially for observables involving the third jet.
This is exemplified in the right hand-side of Fig.~\ref{fig:WZTH}.

Finally, in the proceedings of Les Houches 2017 \cite{Bendavid:2018nar}, a tuned comparison of theoretical predictions up to LO+PS accuracy has been performed.
It highlights the same features as the NLO comparative study of \ssWW\ \cite{Ballestrero:2018anz}.
In particular, all differential distributions are in very good agreement at LO and LO+PS accuracy,
the only exception being, as in Ref.~\cite{Ballestrero:2018anz}, observables involving the third jet.
It is worth emphasising that the \Sherpa predictions of Ref.~\cite{Bendavid:2018nar} are LO or LO+PS predictions and 
are thus different from the \Sherpa predictions usually adopted by the ATLAS collaboration, for example in Ref.~\cite{ATLAS:2019hoc}, the latter being merged predictions with 2 and 3 jets matched to PS and largely differing from other predictions.

A summary of the available predictions is provided in Table~\ref{tab:WZTH}.
It is perfectly reasonable to expect that the accuracy obtained for the \ssWW\ channel can be also achieved in the next few years for the \WZjj\ channel.

\begin{table}[htb]
\caption{Summary of higher-order predictions currently available for the WZ channel: at fixed order and matched to parton shower.
The symbols {\bf \color{green} \checkmark}, {\bf \color{green} \checkmark$^*$}, and {\bf \color{red} X}
means that the corresponding predictions are available, in the VBS approximation, or not yet.}
\center
{\begin{tabular}{l|cccc}
Order & $\mathcal{O}\left(\alpha^7 \right)$ & 
$\mathcal{O}\left({\alpha_{\rm s}} {\alpha}^6 \right)$ & 
$\mathcal{O}\left({\alpha_{\rm s}}^2 {\alpha}^5 \right)$ & 
$\mathcal{O}\left({\alpha_{\rm s}}^3 {\alpha}^4 \right)$\\
\hline 
NLO & {\bf \color{green} \checkmark} &  {\bf \color{green} \checkmark} & {\bf \color{red} X} & {\bf \color{green} \checkmark} \\
NLO+PS & {\bf \color{red} X} & {\bf \color{green} \checkmark$^*$} & {\bf \color{red} X} & {\bf \color{green} \checkmark}
\end{tabular} \label{tab:WZTH}}
\end{table}

\subsubsection{Experimental approaches}

ATLAS has reported observation of electroweak \WZjj\ production using a partial Run-2 data set~\cite{ATLAS:WZ}.
CMS performed the same search on the full data set, in combination with the \ssWW\ final state~\cite{CMS:ssWWandWZ}, also leading to an observation of this process.

\paragraph{Monte Carlo simulation}

Since the CMS study combines the \ssWW\ and \WZjj\ final states, the simulations uses the same settings as in the previous section.
The EW-QCD interference is positive and estimated to be $1\%$ of the EW signal in the fiducial region.
The QCD-induced $\PW^\pm \PZ+{\rm jets}$ background is simulated at LO with up to 3 additional partons using \madgraphbis,
merging the jet multiplicities according to the MLM scheme~\cite{Alwall:2007fs}, and normalizing the total cross-section to diboson NNLO QCD predictions~\cite{Grazzini:2017ckn}\footnote{In the cases where QCD background is the dominant component, it is a common experimental practice to simulate all parton multiplicities starting from zero, even if only 2-jet events should in principle pass selections, provided the PS-matching scale is lower than the jet $p_{\rm T}$ thresholds. This ensures that events with non-signal (fake or pileup) jets are correctly taken into account in the simulation}.

ATLAS uses Sherpa, version 2.2.2, to simulate the electroweak process at LO, inclusive in the number of jets, while the strong process is simulated at NLO QCD with up to one extra jet.
The interference is estimated separately with \madgraphbis at LO to be about $10\%$ of the EW signal, in noticeable disagreement with the CMS estimation.
Nevertheless, its full amount is conservatively used as an uncertainty.
Alternative descriptions of the EW and strong processes are obtained using \madgraphbis at LO or \POWHEG at NLO in QCD~\cite{Melia:2011tj}.
Other backgrounds, such as $\PZ\PZ$, tribosons, $\Pt{\bar \Pt} \PW^\pm$, and  $\Pt{\bar \Pt} \PZ$ are generated using different tools, QCD accuracies, and extra-parton multiplicities.

\paragraph{Fiducial region definitions and reconstruction-level selections}

Fiducial regions considered in the ATLAS\footnote{All ATLAS cross sections are reported for a single lepton flavor and are therefore multiplied by four to compare to CMS.} and CMS analyses are compared in Table~\ref{tab:WZfr}.
The same assumptions as in \ssWW\ are used for $\tau$ decays and lepton ``dressing''.

\begin{table}[htb]
\caption{Comparison of \WZjj\ fiducial region definitions and related EW (VBS) cross-section values in the ATLAS and CMS measurements~\cite{ATLAS:WZ,CMS:ssWWandWZ}.
\emph{JRS} stands for generic Jet-Rapidity Separation selections.}
\center
{\begin{tabular}{@{}ccc@{}} \toprule
Variable & ATLAS  & CMS  \\
\midrule
$p_{\rm T}(\ell)$ & $> 20/15/15$ GeV & $> 20$ GeV \\
$|\eta(\ell)|$ & $< 2.5$ &  $< 2.5$ \\
$\Delta R(\ell\ell')$ & $> 0.3/0.2$ &  - \\
$m_{\ell\ell}$ & $[81,101]$ GeV & $[76,106]$ GeV \\
$m_{\rm T}(\PW^\pm)$ & $> 30$ GeV & - \\
$p_{\rm T}(\Pj)$ & $> 40$ GeV & $> 50$ GeV \\
$|\eta(\Pj)|$ & $< 4.5$ &  $< 4.7$ \\
\mjj & $>500$ GeV & $> 500$ GeV \\
JRS & $ \eta(\Pj_1)\cdot\eta(\Pj_2) < 0$ &  $\dejj > 2.5$ \\
\midrule
$\sigma$ LO & $1.28 \pm 0.12$ fb & $1.41 \pm 0.21$ fb \\
$\sigma$ NLO QCD+EW & - & $1.24 \pm 0.18\fb$ \\
\bottomrule
\end{tabular} \label{tab:WZfr}}
\end{table}

Reconstruction-level selections follow the fiducial regions requirements. Both analyses use
single-lepton triggers and use a $\Pb$-jet veto.
In ATLAS (CMS) the leading lepton is required to have $p_{\rm T} > 27 (25) \GeV$, and both experiments require $p_{\rm T,miss} > 30\GeV$.
In ATLAS 4$\ell$ events are explicitly vetoed, while CMS requires that $m_{\ell\ell\ell'} > 100\GeV$.
In CMS the maximum $z^*_\ell$ of the three leptons must be smaller than 1.

\paragraph{Analysis strategy and background estimation}

Both ATLAS and CMS use BDT algorithms to isolate the EW signal over the large QCD background, combining 11 to 12
variables that are related to jet kinematics, vector-boson kinematics, or to both jets and leptons kinematics
at the same time. It must be noticed that variables such as the $\PW^\pm$ rapidity or $m_{\rm T}(\PW\PZ)$ are computed
with much looser assumptions than in the \ssWW\ analysis, since just one neutrino is missing in this case
and its longitudinal momentum can be inferred by a $\PW$-mass constraint on the $\ell$-neutrino pair ($\ell$
being the lepton not associated to the $\PZ$ decay).

\begin{figure*}[t]
\centering
\includegraphics[width=0.5\linewidth,height=0.35\textheight,clip=true,trim={1.5cm 0.cm 0.cm 0.cm}]{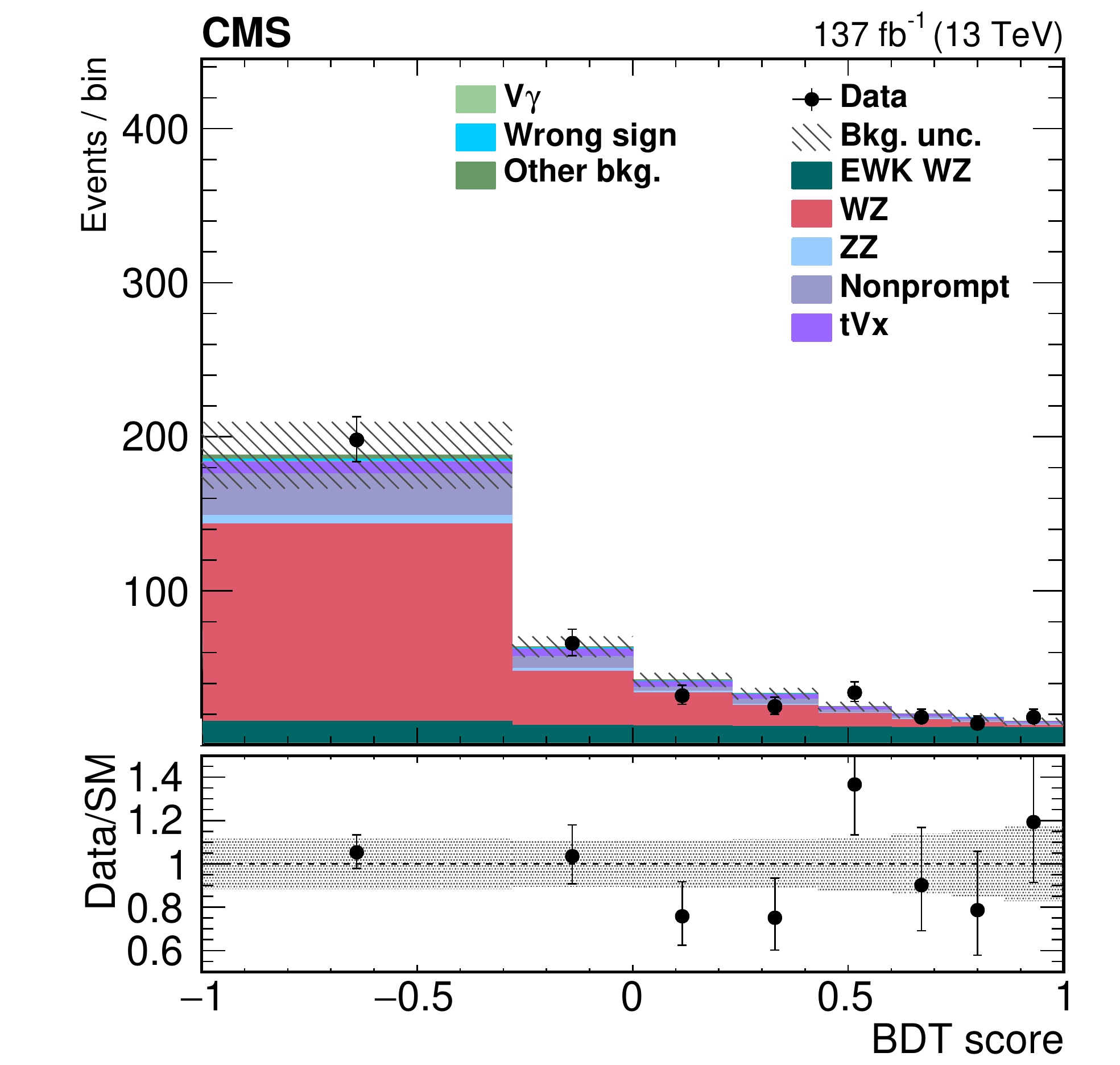} 
\includegraphics[width=0.48\linewidth,height=0.35\textheight,clip=true,trim={0.2cm 0.cm 0.cm 0.cm}]{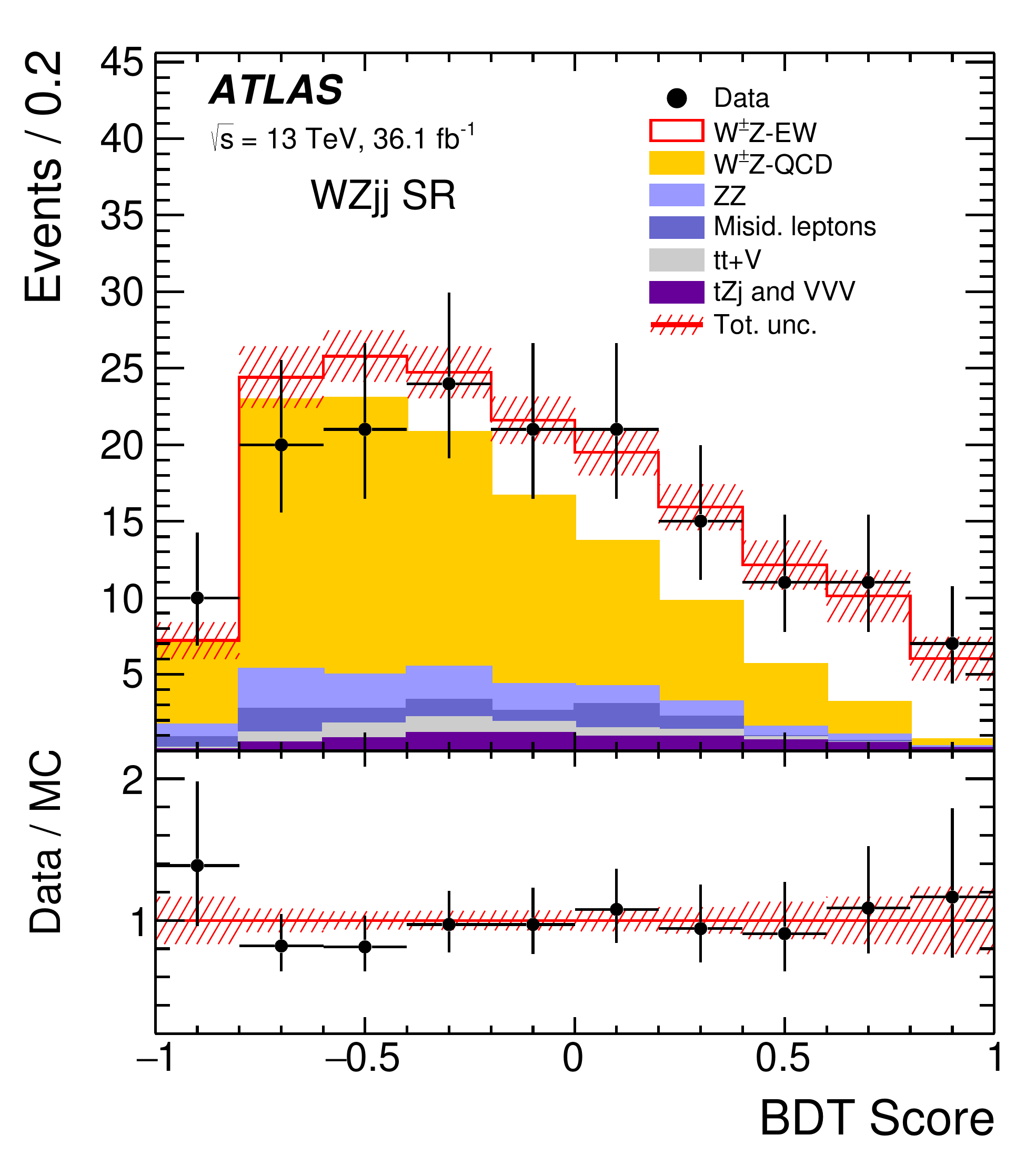}
\caption{Post-fit BDT-score distributions in CMS (left) and ATLAS (right) in the \WZjj\ analysis. Process labeling
is consistent, with the exception of events containing top quarks and vector bosons, which are merged in one
category in the CMS plot.}
\label{fig:wzjj}
\end{figure*}

Background contributions from other SM processes or from non-prompt leptons are sub-dominant
and are estimated in a similar way as for the \ssWW\ analyses.
The ATLAS and CMS data with superimposed signal and background components are shown in Figure~\ref{fig:wzjj}.

\paragraph{Systematic uncertainties}

In both ATLAS and CMS, two of the dominant uncertainties are the limited size of simulated samples and the theory errors on the \WZjj\ production
via 2 strong vertices. Among the experimental ones, lepton momentum and efficiency determinations bring the largest uncertainty in CMS, 
while in ATLAS jet energy scale and resolution uncertainties have a larger impact.
No single contribution has an impact larger than $6\%$ in either analysis, the EW search analysis being
statistically dominated in both experiments.

\paragraph{Results}

ATLAS reports a measured fiducial cross-section of $\sigma_{\mathrm{EW}} = 2.28^{+0.48}_{-0.42}\fb$, where the total
uncertainty is dominated by the statistical one, finding a fairly large measured-to-SM ratio of 1.77.
It corresponds to a background-only hypothesis rejection with a significance of 5.3$\sigma$, while only 3.2$\sigma$ is expected. CMS similarly reports $\sigma_{\mathrm{EW}} = 1.81 \pm 0.41\fb$,
also larger but more in agreement with the NLO QCD+EW estimations in the respective fiducial region. It corresponds to a background-only hypothesis rejection with a significance of 6.8$\sigma$.

The total \WZjj\ cross section
including EW and strong components is also measured to be $\sigma_{\mathrm{tot}} = 6.7 \pm 1.0$ fb in ATLAS and $\sigma_{\mathrm{tot}} = 4.97 \pm 0.46$ fb in CMS.
Differential cross-sections are reported only for the EW+strong case,
as a function of \mjj\ in CMS and as a function of many variables in the ATLAS analysis. In CMS, agreement is found
with the SM predictions.
In ATLAS the same conclusion is reached, but only after scaling up these predictions
by the quite large measured-to-SM cross-section ratio.

In the CMS study, limits on aQGC are set by fitting the diboson transverse mass distribution in the signal region and 
the results are statistically combined to those of the \ssWW\ analysis to obtain more stringent limits.

\subsection{The $\ZZjj \rightarrow \ell^+ \ell^- \ell'^+ \ell'^- \Pj\Pj$ and $\rightarrow \ell^+ \ell^- \nu {\bar \nu} \Pj\Pj$ final states}
\label{sec:zz}
The \ZZjj\ VBS process has the smallest cross section among the final states containing heavy vector bosons,
and is one of the rarest SM processes observed to date. Since strong production is also a dominant
background over the EW signal, the search for \ZZjj\ is an experimental challenge. However, its experimental
signature is very clean and, being the only final state where full reconstruction of the 6 final-state fermions
is possible, it could become important for polarization measurements with larger data sets in the future.

\subsubsection{Theoretical calculations}

The theoretical status of the \ZZjj\ channel is rather similar to the \WZjj\ one.
In particular, the QCD corrections to EW contributions are known since some time \cite{Jager:2006cp} and have been matched subsequently to PS \cite{Jager:2013iza}.
NLO QCD corrections have also been computed for the QCD background in Ref.~\cite{Campanario:2014ioa} and the matched results to parton-shower can be obtained from modern Monte Carlo generators.

\begin{figure*}[t]
\centering
\hspace{-0.4cm}
          \includegraphics[width=0.48\linewidth,height=0.34\textheight,clip=true,trim={2.1cm 0.5cm 0.cm 0.cm}]{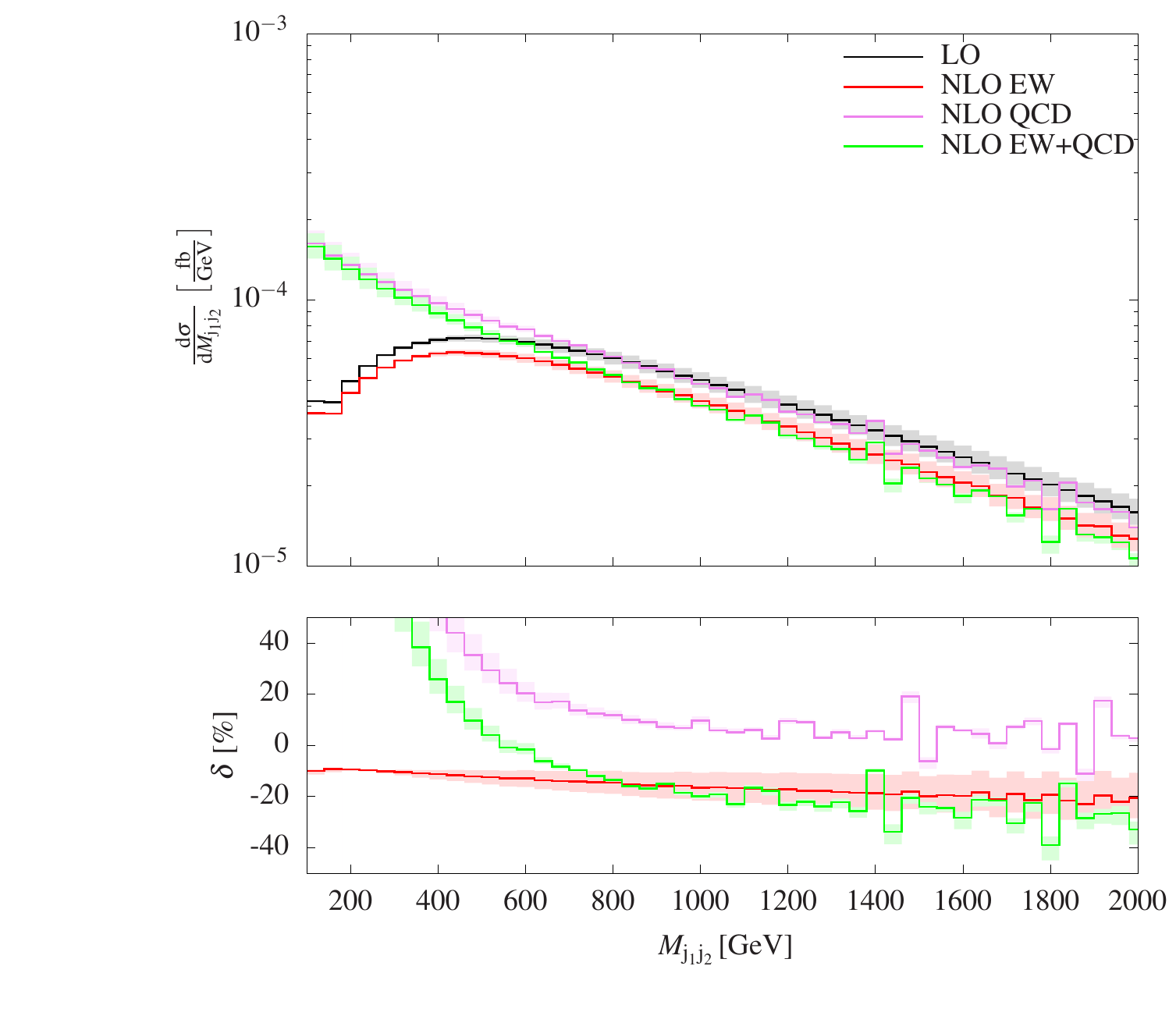}
          \includegraphics[width=0.53\linewidth,height=0.34\textheight,clip=true,trim={0.cm 0.cm 0.5cm 0.6cm}]{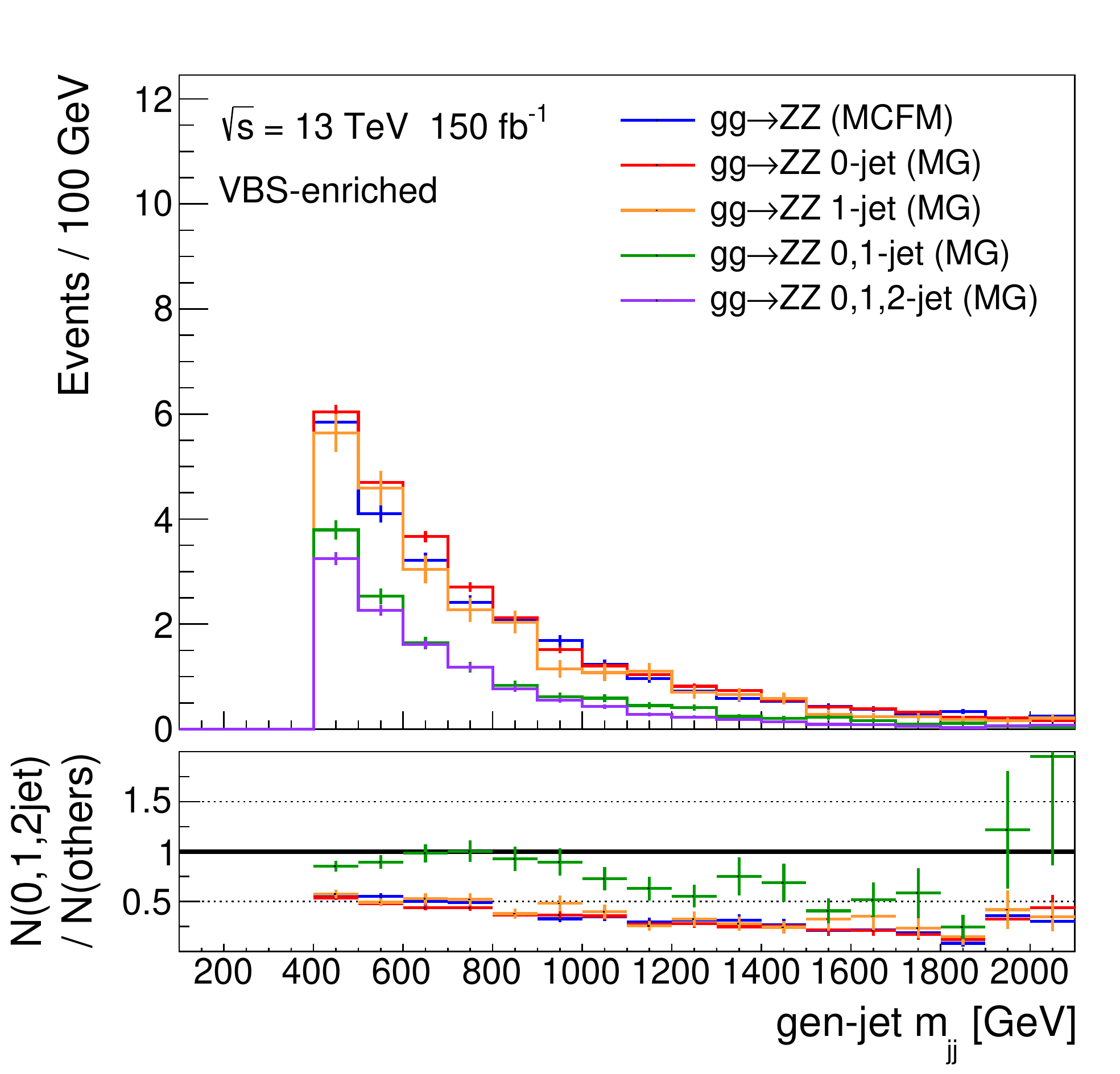}
\caption{Various differential distributions for the ZZ channel.
Left: invariant mass of the two tagging jets with NLO EW + QCD corrections.
Right: invariant mass of the two tagging jets with matching and merging of different jet multiplicities.
These figures are taken from Ref.~\cite{Denner:2020zit} and Ref.~\cite{Li:2020nmi}, respectively.}
\label{fig:ZZTH}
\end{figure*}

The full NLO contributions of orders $\mathcal{O}\left(\alpha^7 \right)$ and $\mathcal{O}\left(\alpha_{\rm s} \alpha^6 \right)$ 
and all contributing leading orders along with the loop-induced contribution of order $\mathcal{O}\left(\alpha_{\rm s}^4 \alpha^4 \right)$ have been computed in Ref.~\cite{Denner:2020zit}.
Results show the same hierarchy as for \ssWW\ and \WZjj\ and, furthermore, that loop-induced channels can contribute significantly in typical fiducial regions.
This can be seen for example in Fig.~\ref{fig:ZZTH} where, beyond a di-jet invariant mass of $500\GeV$, QCD corrections are small while the EW ones are substantial.
In Ref.~\cite{Denner:2020zit} the loop-induced contribution simply amounts to include the $\Pg\Pg\to\Pe^+\Pe^-\mu^+\mu^-\Pg\Pg$ channel.
Reference~\cite{Li:2020nmi} went beyond, by studying loop-induced ZZ diboson production with up to 2 jets merged and matched to parton showers.
\footnote{Very recently NLO QCD corrections matched to parton shower for the loop-induced process $\Pg\Pg\to\Pe^+\Pe^-\mu^+\mu^-$ were presented in Refs.~\cite{Alioli:2021wpn,Grazzini:2021iae}}
In particular, a proper description of loop-induced contributions can have a significant impact in relevant phase-space regions (see Fig.~\ref{fig:ZZTH}).

A summary of the available predictions is provided in Table~\ref{tab:ZZTH}.
Note that the loop-induced contributions are not indicated there as they are only known at LO.

\begin{table}[htb]
\caption{Summary of higher-order predictions currently available for the ZZ channel: at fixed order and matched to parton shower.
The symbols {\bf \color{green} \checkmark}, {\bf \color{green} \checkmark$^*$}, and {\bf \color{red} X}
means that the corresponding predictions are available, in the VBS approximation, or not yet.}
\center
{\begin{tabular}{l|cccc}
Order & $\mathcal{O}\left(\alpha^7 \right)$ & 
$\mathcal{O}\left({\alpha_{\rm s}} {\alpha}^6 \right)$ & 
$\mathcal{O}\left({\alpha_{\rm s}}^2 {\alpha}^5 \right)$ & 
$\mathcal{O}\left({\alpha_{\rm s}}^3 {\alpha}^4 \right)$ \\
\hline 
NLO & {\bf \color{green} \checkmark} &  {\bf \color{green} \checkmark} & {\bf \color{red} X} & {\bf \color{green} \checkmark} \\
NLO+PS & {\bf \color{red} X} & {\bf \color{green} \checkmark$^*$} & {\bf \color{red} X} & {\bf \color{green} \checkmark}
\end{tabular} \label{tab:ZZTH}}
\end{table}

In Ref.~\cite{Ballestrero:2018anz}, it has been pointed out that for \ZZjj\ the VBS approximation is less accurate at LO than at NLO.
The main reason is that performing the full computation implies including also tri-boson contributions where one of the gauge boson decays hadronically.
This means that, when including real QCD radiation, 
the two tagging jets could be the gluon radiation and a jet from the aforementioned heavy gauge boson. The gauge boson can be resonant if the two quarks originating from its decay recombine in a single jet.
While rather suppressed, such configurations can be significant when low di-jet invariant masses are used.
This is particularly explicit in the left-hand side of Fig.~\ref{fig:ZZTH} where at low invariants, the QCD corrections are large while they become small for larger invariant masses.

While such corrections are perfectly legitimate as they describe a physical effect due to the interplay between the nature of the process and the event selection, in experimental analysis, the tri-boson contributions are often subtracted using LO Monte-Carlo simulations.
First, as argued previously, this makes the measurement more dependent on theoretical input.
Second, when using low di-jet invariant masses, the physical effect described above will then never been described by the theoretical predictions not including tri-boson contributions.
This example illustrates perfectly the need to perform measurements as theory-independent as possible and have exchange between experiment and theory.

\subsubsection{Experimental approaches}

ATLAS has reported observation of electroweak \ZZjj\ production using the full Run-2 data set~\cite{ATLAS:ZZ}
and combining the two $ZZ$ decay channels. CMS only analyzed events with four charged leptons and, after
an earlier publication with limited sensitivity~\cite{CMS:ZZ}, has recently reported a strong evidence for the
EW production with the entire 13-TeV statistics~\cite{CMS:ZZ2}. 

\paragraph{Monte Carlo simulation}

CMS uses \madgraphbis at LO without additional partons to simulate the EW and interference components. The EW-QCD
interference is positive and estimated to be 3-9\% of the EW signal in the various fiducial regions considered
in the analysis. An alternative estimation at NLO QCD using \POWHEG~is also considered~\cite{Jager:2013iza}.
The QCD-induced $\PZ\PZ+{\rm jets}$ process (without loop-induced contributions) is simulated at NLO with up to one extra partons using \madgraphbis,
merging the jet multiplicities according to the FxFx scheme~\cite{Frederix:2020trv}, and normalizing the
total cross section to NNLO QCD predictions for diboson production~\cite{Grazzini:2018owa}.
While the loop-induced contribution only appears at NNLO in QCD, this contribution is significant because of the $\Pg\Pg$ initial state,
which is dominant at the LHC.
This contribution is included via a dedicated simulation at LO with up to two extra partons,
using advanced \madgraphbis settings as described in Ref.~\cite{Li:2020nmi}.
Parton Distribution Functions and scale choices follow those made in the \ssWW\ analysis.

ATLAS uses the same simulation as CMS for EW and interference, but with LO PDFs, and the full size of the
interference component is conservatively used as an uncertainty. \Sherpa 2.2.2 is used instead to simulate both
the loop-induced and tree-level strong processes, at LO (NLO) and with up to one (three) extra partons,
depending on the final state. In the $2\ell 2\nu \Pj\Pj$ channel, a dedicated Monte Carlo generator, {\sc gg2VV},
is used for the inclusive diboson loop-induced component~\cite{Kauer:2013qba}.

Other backgrounds, such as tribosons, $\Pt{\bar \Pt} \PW^\pm$, and  $\Pt{\bar \Pt} \PZ$, which are generated using different
tools, QCD accuracies, and extra-parton multiplicities in the two experiments.
These are in general a minor contribution
to the selected data samples. The only exception is the most inclusive fiducial region of the CMS analysis, 
where the triboson contribution is significant at low \mjj\ values and is subtracted using simulation. 

\paragraph{Fiducial region definitions and reconstruction-level selections}

Fiducial regions considered in the ATLAS and CMS analyses are compared in Table~\ref{tab:ZZfr}. In ATLAS
a single value for the combined 4$\ell jj$ and $2\ell2\nu jj$ phase space is given. In CMS, results are given
in three different fiducial region corresponding to increasing degrees of VBS enrichment.

\begin{table}[htb]
\caption{Comparison of \ZZjj\ fiducial region definitions and related EW (VBS) cross-section values in the ATLAS and CMS
measurements ~\cite{ATLAS:ZZ,CMS:ZZ2}. \emph{JRS} stands for generic Jet-Rapidity Separation selections.}
\center
{\begin{tabular}{@{}cccc@{}} \toprule
Variable & ATLAS $4\ell \Pj\Pj$ & ATLAS $2\ell 2\nu \Pj\Pj$ & CMS (inclusive/loose/tight)  \\
\midrule
$p_{\rm T}(\ell)$ & $> 20/20/10/7\GeV$  & $> 30/20$ GeV & $> 20/10/5/5$ GeV \\
$|\eta(\ell)|$ & $< 2.7/2.5$ &  $< 2.5$  & $< 2.5$ \\
$\Delta R(\ell\ell')$ & $> 0.2$ & - & - \\
$m_{\ell\ell}$ & $[60,120]$ GeV & $[80,100]$ GeV & $[60,120]$ GeV \\
$m_{4\ell}$ & - & - & $> 180$ GeV \\
$p_{\rm T,miss}$ & - & $> 130$ GeV & - \\
$p_{\rm T}(\Pj)$ & $> 40/30$ GeV & $> 60/40$ GeV & $> 30$ GeV \\
$|\eta(\Pj)|$ & $< 4.5$ & $<4.5$ & $< 4.7$ \\
\mjj & $>300$ GeV & $> 400$ GeV & $>100/400/1000$ GeV \\
JRS & $ \dyjj > 2$ & $ \dyjj > 2$ & $\dejj > 0/2.4/2.4$ \\
\midrule
$\sigma$ LO & \multicolumn{2}{c}{$0.61 \pm 0.03$ fb} & $0.28\pm 0.02\,/\,0.19 \pm 0.02\,/\,0.10\pm 0.01$ fb \\
$\sigma$ NLO QCD & \multicolumn{2}{c}{-} & $0.28\pm 0.02\,/\,0.20 \pm 0.02\,/\,0.11\pm 0.01$ fb  \\
\bottomrule
\end{tabular} \label{tab:ZZfr}}
\end{table}

Reconstruction-level selections follow the fiducial regions requirements very closely with minor
additions (in CMS electron $p_{\rm T}$ thresholds are raised to 7 GeV, in ATLAS the $\PZ$-mass requirement is
$[66,116]$ GeV and the $p_{\rm T,miss}$ variable is replaced by its significance). Double or single-lepton
triggers are used to select data, keeping in general a very high efficiency. 

\paragraph{Analysis strategy and background estimation}

Both ATLAS and CMS use multivariate analyses to isolate the EW signal over the large QCD background.
In the 4$\ell \Pj\Pj$ channel, ATLAS uses a BDT comprised of 12 variables: \mjj, \dyjj, jet $p_{\rm T}$ and rapidities,
$\PZ$ candidate $p_{\rm T}$ and rapidities, and combinations of full-event variables. A similar choice is done for the
$2\ell2\nu \Pj\Pj$ channel, replacing the quantities related to the undetected $\PZ$ with $p_{\rm T,miss}$, its
significance, and related variables. In CMS, the kinematic EW discriminant $K_D$ is instead built from analytical
matrix elements of the EW and strong processes at LO, obtained from the {\sc MCFM} generator and using
the {\sc MELA} event-probability calculator~\cite{Gao:2010qx,Bolognesi:2012mm,Gritsan:2020pib}.

Both ATLAS and CMS use QCD-enriched control regions to validate background estimates in the 4$\ell$ channel,
while the background from one $\PZ$ boson and non-prompt leptons is sub-dominant. In the ATLAS $2\ell2\nu \Pj\Pj$ channel,
the evaluation of the QCD contribution is instead simulation-based while control regions are used to estimate
other important background processes such as \WZjj, $\PW^+\PW^-\Pj\Pj$, and $\Pt{\bar \Pt}$. 

\begin{figure*}[t]
\centering
\includegraphics[width=0.48\textwidth]{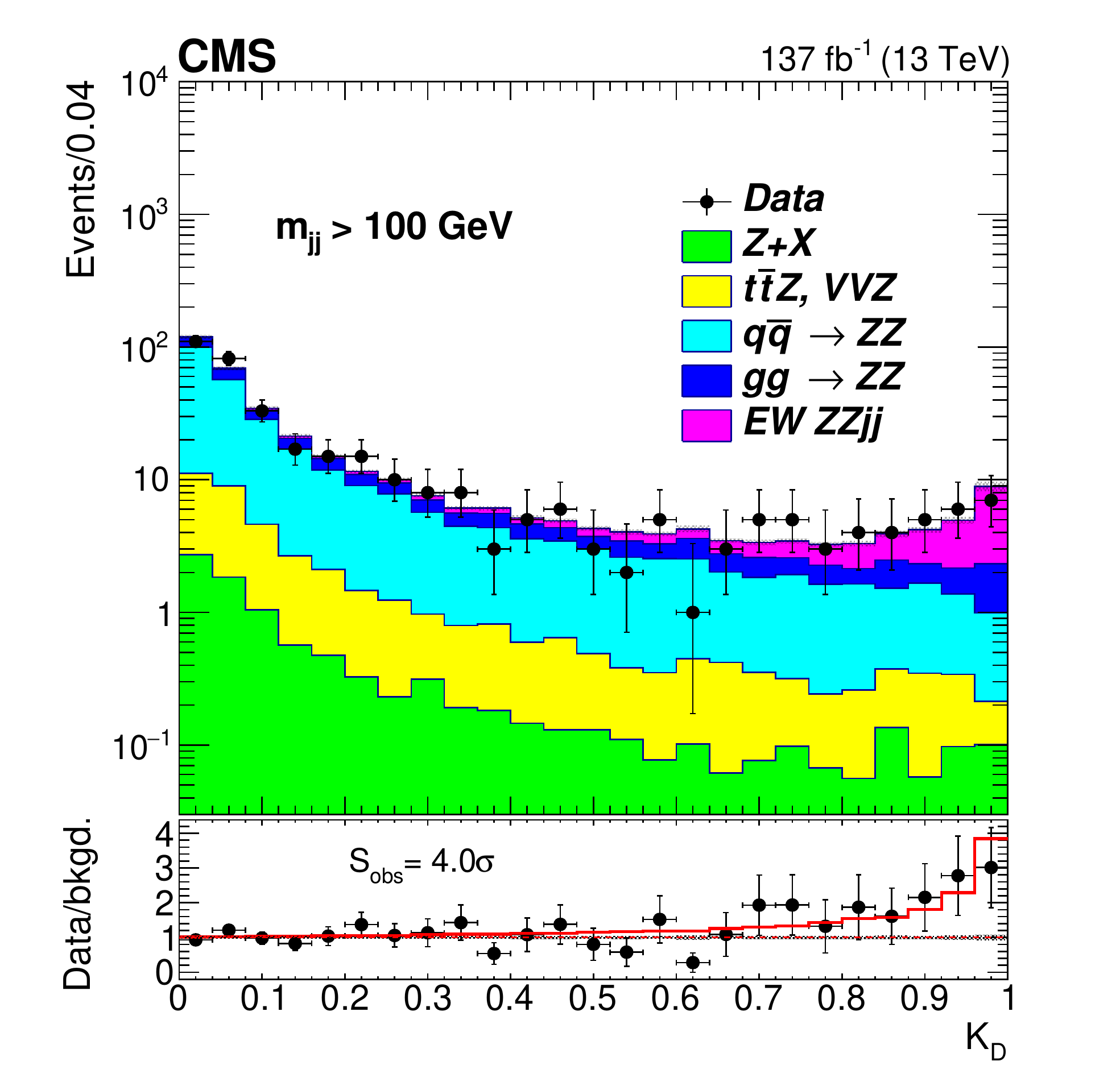} 
\includegraphics[width=0.41\textwidth]{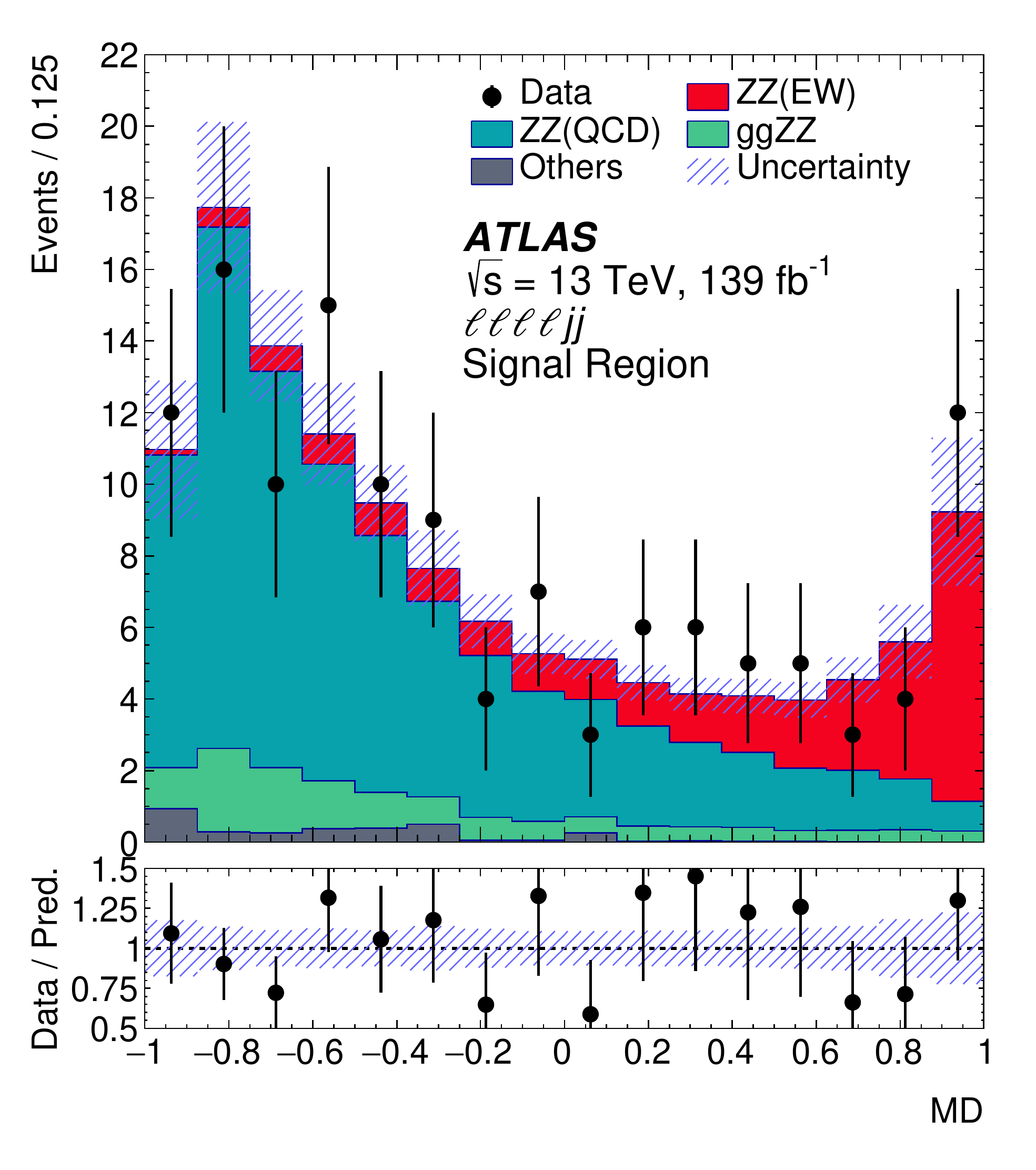}
\caption{Post-fit $K_D$ distribution in CMS (left) and BDT score distribution in ATLAS (right) in the \ZZjj\ analysis. 
The very different signal-to-background ratio is due to the different VBS-enrichment level in the two figures.}
\label{fig:zzjj}
\end{figure*}

The ATLAS and CMS $4\ell \Pj\Pj$ data with superimposed signal and background components are shown in Figure~\ref{fig:zzjj}.

\paragraph{Systematic uncertainties}

In both ATLAS and CMS uncertainties are given as fraction of predicted yields and not as impacts on the
cross-section measurements, which makes comparisons more difficult. The ATLAS uncertainties appear to be
dominated by a very large ($30\%$) theory uncertainty on the strong \ZZjj\ production simulated with \Sherpa.
In CMS this uncertainty, EW uncertainties, as well as uncertainties on lepton and jet measurements contribute
in similar amounts.

\paragraph{Results}

ATLAS reports a measured fiducial cross-section of $\sigma_{\mathrm{EW}} = 0.82 \pm 0.21\fb$, where the total
uncertainty is dominated by the statistical one.
It corresponds to a background-only hypothesis rejection with a significance of 5.5$\sigma$, while 4.3$\sigma$ is expected, with the 4$\ell jj$ channel exhibiting a much larger sensitivity. In the three fiducial regions
CMS reports $\sigma_{\mathrm{EW,incl}} = 0.33^{+0.12}_{-0.11}\fb$, $\sigma_{\mathrm{EW,loose}} = 0.18^{+0.09}_{-0.08}\fb$, and $\sigma_{\mathrm{EW,tight}} = 0.09^{+0.05}_{-0.04}\fb$, 
all in agreement with SM expectations at both LO and NLO in QCD. The background-only hypothesis is rejected with a significance of 4.0$\sigma$.

In ATLAS, the total \ZZjj\ cross section
including EW and strong components is also measured to be $\sigma_{\mathrm{tot}} = 1.27 \pm 0.14\fb$ in the
4$\ell \Pj\Pj$ channel and $\sigma_{\mathrm{tot}} = 1.2 \pm 0.3\fb$ in the 2$\ell 2\nu \Pj\Pj$ channel.
In CMS the measured value is $\sigma_{\mathrm{tot}} = 5.3 \pm 0.6\fb$, in the most inclusive fiducial region.

In the CMS analysis, limits on aQGC are set by fitting the $\PZ\PZ$ invariant mass distribution in the most inclusive region.
These are particularly constraining for the T8 and T9 operators, which only involve neutral fields and to which this channel has therefore the greatest sensitivity.

\subsection{The $\osWW \rightarrow \ell^\pm \nu \ell'^\mp \nu \Pj\Pj$ final state}
\label{sec:osww}
\subsubsection{Theoretical calculations}

Finally the \osWW\ channel raised a lot of theoretical interest in the past years.
The QCD corrections to the EW signal are known in the VBS approximation \cite{Jager:2006zc} and have been matched to parton shower \cite{Jager:2013mu}.
The QCD background is also known at the same accuracy, namely NLO QCD \cite{Melia:2011dw,Greiner:2012im} and matched to parton shower \cite{Rauch:2016upa}.
A summary of the available predictions is provided in Table~\ref{tab:osWWTH}.
It is reasonable to think that the theoretical accuracy for this final state will be comparable to the one of the other channels in the next few years.
In order not to include resonant top quarks contributions, care has to be taken.
One should either use the 4-flavour scheme or use the 5-flavour scheme and carefully remove top-quark contributions (see for example Ref.~\cite{Melia:2011dw}).

\begin{table}[htb]
\caption{Summary of higher-order predictions currently available for the os-WW channel: at fixed order and matched to parton shower.
The symbols {\bf \color{green} \checkmark}, {\bf \color{green} \checkmark$^*$}, and {\bf \color{red} X}
means that the corresponding predictions are available, in the VBS approximation, or not yet.}
\center
{\begin{tabular}{l|cccc}
Order & $\mathcal{O}\left(\alpha^7 \right)$ & 
$\mathcal{O}\left({\alpha_{\rm s}} {\alpha}^6 \right)$ & 
$\mathcal{O}\left({\alpha_{\rm s}}^2 {\alpha}^5 \right)$ & 
$\mathcal{O}\left({\alpha_{\rm s}}^3 {\alpha}^4 \right)$\\
\hline 
NLO & {\bf \color{red} X} &  {\bf \color{green} \checkmark$^*$} & {\bf \color{red} X} & {\bf \color{green} \checkmark} \\
NLO+PS & {\bf \color{red} X} & {\bf \color{green} \checkmark$^*$} & {\bf \color{red} X} & {\bf \color{green} \checkmark}
\end{tabular} \label{tab:osWWTH}}
\end{table}

\subsubsection{Experimental approaches}

There are so far no published results from ATLAS and CMS on the \osWW\ channel. With respect to other channels,
in this case the top-pair background with dileptonic $\PW^\pm$ decays is dominant and requires a very efficient
b-quark jet veto, bringing larger experimental uncertainties. The analysis method and preliminary results are
available in Ref.~\cite{tesiCardini}.

\subsection{The semileptonic final states}
\label{sec:semilep}
\subsubsection{Theoretical calculations}

As seen in the previous sections, theoretical predictions for VBS signatures with fully leptonic final states are numerous and rather complete.
Unfortunately, to date, there are no public predictions beyond LO accuracy for any of the semileptonic signatures.
Such processes include $\Pp\Pp\to\ell^+\ell^-\Pj\Pj\Pj\Pj$ and $\Pp\Pp\to\ell\nu_\ell\Pj\Pj\Pj\Pj$, meaning that they involve four jets and two leptons in the final state.
This is currently at the edge of the state of the art, as only two such computations have been performed \cite{Denner:2017kzu,Anger:2017glm} and both describe processes with
bottom quarks in the final state, meaning that in a VBS analysis they would probably be discarded because of b-jet vetoes.

While the necessary technology is already there, the two aforementioned semileptonic processes require the combination of two (\WZjj\ and \ZZjj) or three (\ssWW, \WZjj, and \osWW) VBS processes, respectively, implying a significant computing burden for such signatures.
Nonetheless, generating the on-shell gauge bosons at NLO QCD and then decaying them either leptonically or hadronically would be an option.
This would lead to neglecting some non-resonant and interference contributions which are expected to be small, or even negligible, but it would considerably reduce the computing time.
For example, for the $\ell\nu_\ell\Pj\Pj\Pj\Pj$ final state, this would imply computation at NLO QCD of the on-shell processes
$\Pp\Pp\to\PW^+\PW^-\Pj\Pj$, $\Pp\Pp\to\PW^\pm\PZ\Pj\Pj$, and $\Pp\Pp\to\PW^\pm\PW^\pm\Pj\Pj$, before adding the corresponding decays of the heavy gauge bosons.
This can be done for both the background and the signal, provided care is taken regarding the VBS approximations.

\subsubsection{Experimental approaches}

Both ATLAS and CMS have reported studies of semileptonic VBS final states in the 2016 subset of the
13-TeV LHC data, including $\PW^\pm V$ and $\PZ V$. However,
the two analyses are rather different in their principles and scopes.

The ATLAS study~\cite{ATLAS:VVsemilep} is targeting a SM VBS measurement, therefore combining a
larger set of experimental signatures to cover the largest possible phase space at the price of very high
backgrounds. The study employs a complex multivariate analysis based on BDTs in a total of nine event categories.
Events in categories differ in the number of charged leptons used to
identify the leptonic vector-boson decay (0 leptons for $\PZ \rightarrow \nu {\bar \nu}$, 1 lepton for
$\PW^\pm \rightarrow \ell^\pm \nu$, 2 leptons for $\PZ \rightarrow \ell^+ \ell^-$) and in the way the hadronic
vector-boson decay (denoted as $V$) is identified: two categories are used for \emph{merged} single-jet
reconstruction, which differ
in the purity of the selection working point, and one for the \emph{resolved} two-jet reconstruction. The
reconstruction of the merged jet employs the anti-$k_{\rm T}$ algorithm with a large distance parameter $R = 1.0$ and
its identification is based on the \emph{jet trimming} technique~\cite{Krohn:2009th}, which looks for candidate
subjets inside the larger-area jet and discards constituents not associated to those subjets. The invariant
mass of the merged jet $m_{\rm J}$ is computed after trimming and the jet is re-calibrated.

The CMS study~\cite{CMS:VVsemilep} is only
devoted to BSM searches and therefore it is just a cut-based analysis, optimized for $VV$ high-invariant-mass
regions. Only two categories are
considered, $\PW^\pm \rightarrow \ell^\pm \nu$ and $\PZ \rightarrow \ell^+ \ell^-$ plus a merged jet.
Sensitivity to SM VBS is not evaluated, while the absence of excesses at high mass are interpreted in terms
of constraints on aQGC in EFT or on a complete BSM theory, the Georgi-Machacek model~\cite{Georgi:1985nv}. The merged
jet reconstruction in CMS uses the anti-$k_{\rm T}$ algorithm with $R = 0.8$: jets are identified by the more
recent modified mass-drop algorithm~\cite{Larkoski:2014wba}, providing a cleaner estimate of the invariant mass.
The ``$\tau_N$-subjettiness'' variable, related to the compatibility of the large-area with being composed
of $N$ subjets, can be used to select jet candidates compatible with hadronic $V$ decays.

\paragraph{Monte Carlo simulation}

CMS uses \madgraphbis at LO without additional partons to simulate the EW, strong, and interference components for all the
possible final states considered in the analysis, as well as for all BSM samples.
An important feature of semileptonic searches is that
there are important contributions from the single-$\PW$ and Drell-Yan processes, where additional jets
from QCD are misreconstructed as the hadronic $V$ decay, as well as from processes with top quarks.
The processes $V$+jets with up to four outgoing partons at Born level are simulated
at QCD LO accuracy using \madgraphbis and merged using the MLM matching scheme.
The $\Pt{\bar \Pt}$, $\Pt{\bar \Pt}V$, and
single-top processes are generated at NLO accuracy using {\sc POWHEG}. The simulated samples
of background processes are normalized to the best predictions available for the total cross sections.
PDF and scale choices follow those made in other 13-TeV CMS analyses for 2016 simulations.

The ATLAS simulation choices are very similar. The $\PW$+jets and Drell-Yan processes are generated using \Sherpa 2.2.1,
and a NLO alternative description with up to two extra partons is also used. Strong $VV$ production is also simulated with \Sherpa 
and is more advanced than in CMS, up to one additional parton at NLO and up to three additional partons at LO.
Matching and merging for Sherpa samples are performed in the MEPS scheme. Interference between EW and strong
$VV$ production is neglected and therefore used as a systematic uncertainty in the measurement, variable between
5 and 10\% at different values of the BDT score (see below).

\paragraph{Fiducial region definitions and/or reconstruction-level selections}

Fiducial regions considered in the ATLAS analysis are shown in Table~\ref{tab:semilepfr}. Since CMS only targets
BSM constraints, there is no corresponding fiducial region, so offline event selection requirements are shown
instead.

\begin{table}[htb]
\caption{Fiducial region definitions and related EW (VBS) cross-section values in the ATLAS semileptonic VBS measurement~\cite{ATLAS:VVsemilep}, and
the reconstruction-level selections in the analogous CMS analysis~\cite{CMS:VVsemilep}. The symbol $J$ (capitalized) stands for a
merged jet. The subscripts $V$ and \emph{tag} stand for jets from $V$ decays or VBS-tagging jets, respectively.}
\center
{\begin{tabular}{@{}ccc@{}} \toprule
Variable & ATLAS & CMS (reconstruction level)  \\
\midrule
$p_{\rm T}(\ell)$ & $> 27$ GeV (1$\ell$), $> 28/20$ GeV (2$\ell$) & $> 50$ GeV (1$\ell$), $> 50/30$ GeV (2$\ell$) \\
$|\eta(\ell)|$ & $< 2.5$ & $< 2.4/2.5$ \\
$m_{\ell\ell}$ & $[83,99]$ GeV (2$\ell$) & $[76,106]$ GeV (2$\ell$)\\
$p_{\rm T,miss}$ & $> 200$ GeV (0$\ell$), $> 80$ GeV (1$\ell$) & $> 50/80$ GeV (1$e$/1$\mu$) \\
$p_{\rm T}({\rm J})$ & $> 200$ GeV (merged) & $> 200$ GeV \\
$|\eta({\rm J})|$ & $< 2.0$ (merged) & $< 2.4$ \\
$\tau_2/\tau_1({\rm J})$ & - & $< 0.55$ \\
$|\eta(\Pj)|$ & $< 4.5$ & $< 5.0 $\\
$p_{\rm T}(\Pj_{V})$ & $> 40/20$ GeV (resolved) & - \\
$m_{\Pj\Pj,V}$ or $m_{\rm J}$ &  $[64,106]$ GeV & $[65,105]$ GeV \\
$p_{\rm T}(\Pj\Pj_{\rm tag})$ & $> 30$ GeV & $> 30$ GeV \\
$m_{\Pj\Pj, \rm tag}$ & $> 400$ GeV & $> 800$ GeV \\
JRS & $\eta_{\rm \Pj_1,tag} \cdot \eta_{\rm \Pj_2,tag} < 0$ & $ \Delta\eta_{\rm \Pj\Pj,tag} > 4.0$ \\
\midrule
$\sigma$ LO & $43.0 \pm 2.4$ fb & - \\
\bottomrule
\end{tabular} \label{tab:semilepfr}}
\end{table}

In both ATLAS and CMS the 2$\ell$ and 1$\ell$ channel events are selected with single-electron
or single-muon triggers while events for the ATLAS 0$\ell$ channel were recorded with triggers requiring large
$p_{\rm T,miss}$. Both experiment require strictly zero $\Pb$-tagged jets in channels with reconstructed $\PW^\pm$,
to strongly suppress top-quark background.

In ATLAS, reconstruction-level selections follow the fiducial regions requirements very closely with minor
additions (tightened \PZ-mass requirement, jet-lepton and jet-jet angular separation, multijet suppression
in the 0$\ell$ channel). Enhancement of the BSM EW components in the 1-$\ell$ channel is performed in CMS using Zeppenfeld variables
previously defined and the boson \emph{centrality}, defined as: 
$\Theta = \min[\min(\eta_\PW, \eta_V) - \min(\eta_{\rm \Pj_1,tag}, \eta_{\rm \Pj_2,tag}), \max(\eta_\PW, \eta_V) - \max(\eta_{\rm \Pj_1,tag}, \eta_{\rm \Pj_2,tag})]$.

\paragraph{Analysis strategy and background estimation}

The ATLAS analyses uses different BDTs for the resolved- and merged-jet categories, comprising many variables
(16 and 23, respectively, when considering all lepton channels) to isolate the SM VBS signal over the large
backgrounds. Among the list of variables, not just kinematic and jet-identification ones are used, but also
variables sensitive to the quark or gluon origin of small-area jets such as the jet width, the number of
charged tracks inside the jet, and the number of ``track jets'' which are built as alternative to calorimeter-based
jets with only charged tracks compatible with the hardest event vertex.

CMS uses the transverse mass $m_{\rm T}(\PW V)$ or the invariant mass $m(\PZ V)$ for the final fits.
The background estimation is obtained by analytical fits with suitable empirical functions. Fit template
shapes for the various background components are obtained using the $m_{\rm J}$ sidebands, and correcting with
sideband-to-signal transfer factors obtained from simulation. In ATLAS, the estimation of the two main
backgrounds $\PW/\PZ$+jets and $\Pt{\bar \Pt}$ is performed in dedicated control regions, obtained by using the
$m_{\rm J}$ sidebands or inverting the $\Pb$-tag requirement, respectively. From these regions, simulation-to-data
correction factors are obtained as a function of $m_{\rm \Pj\Pj,tag}$ and applied to simulated BDT shapes in the
signal regions. The nine BDT output
distributions, as well as the $m_{\rm \Pj\Pj,tag}$ shapes in the control regions, are used in the statistical analysis.

\begin{figure*}[t]
\centering
\includegraphics[width=0.48\textwidth]{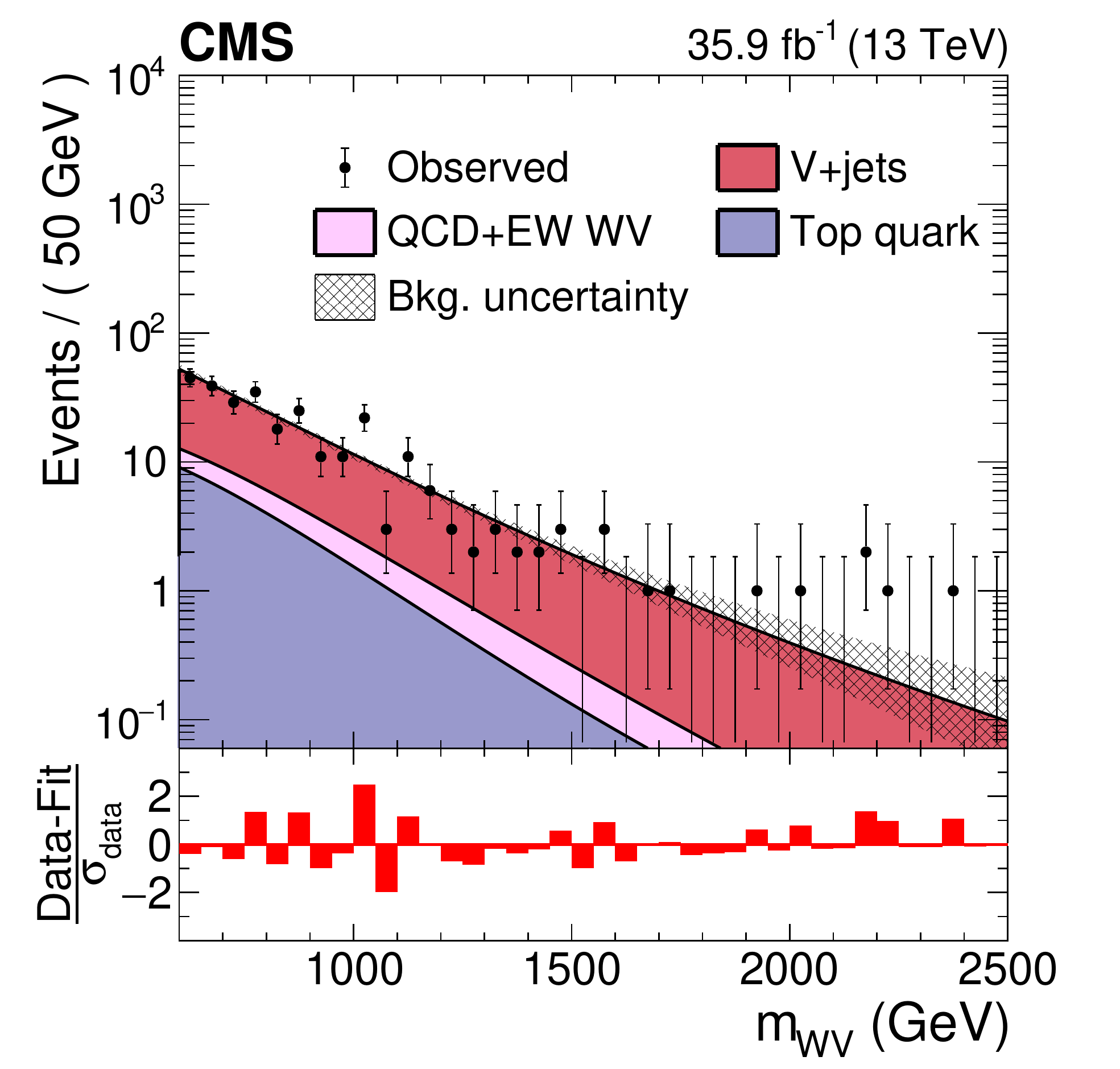} 
\includegraphics[width=0.41\textwidth]{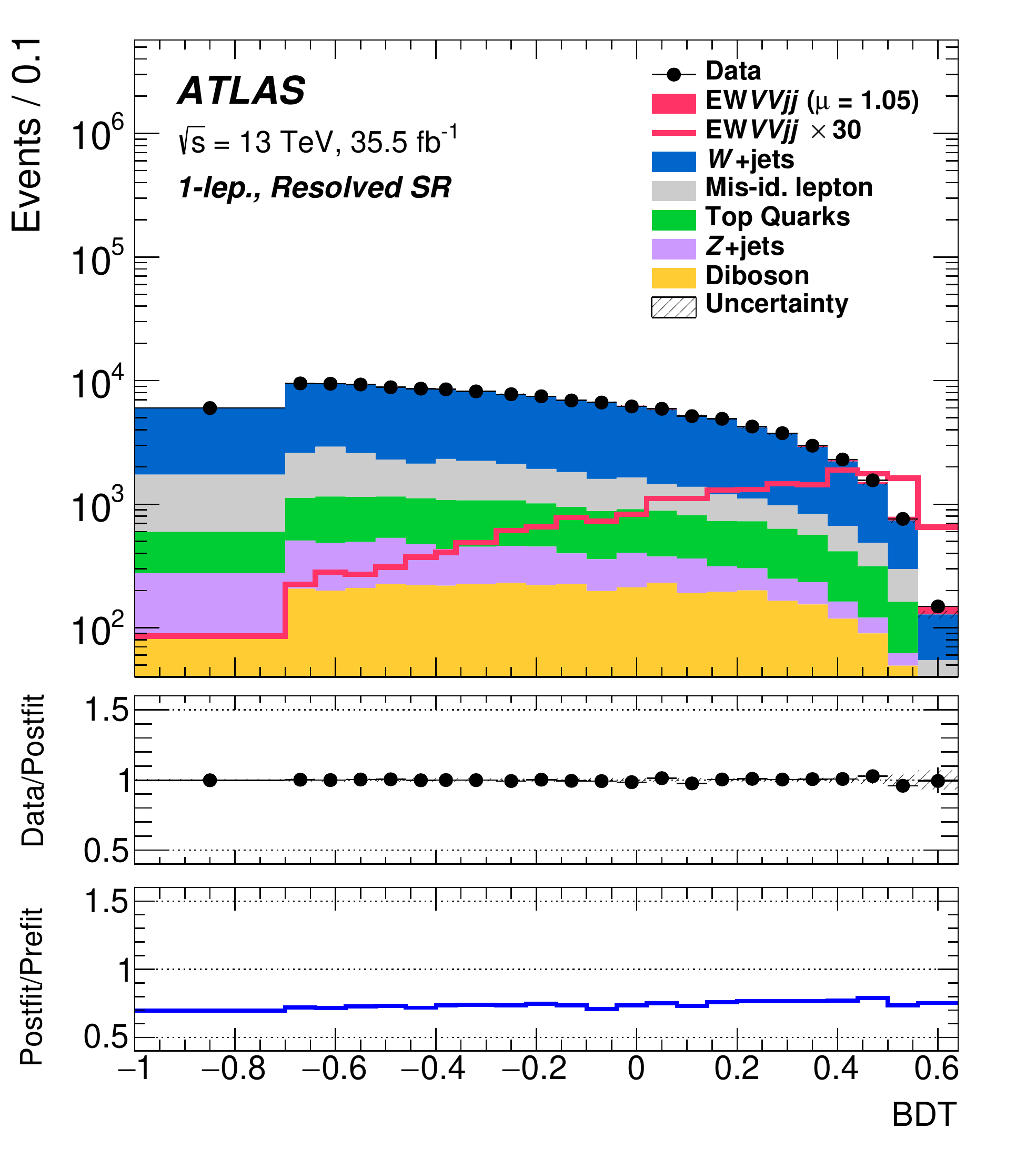}
\caption{Distribution of $m_{\rm T}(\PW\PZ)$ in the 1$\ell$ CMS signal region (left) and of the BDT score in the 1$\ell$
ATLAS signal region with resolved jets. The dominance of $\PW$+jets background is evident in both cases. In the
CMS plot, the continuous line indicates the analytical fit.}
\label{fig:semilep}
\end{figure*}

The ATLAS and CMS $\PW V$ data with superimposed signal and background components are shown in Figure~\ref{fig:semilep}.

\paragraph{Systematic uncertainties}

In ATLAS, many systematic uncertainties contribute to a total uncertainty which is larger than the statistical one.
Among the theoretical and modelling uncertainties, the main contributions are the limited size of the simulated
samples and the modelling of the $\PZ$+jet component; while among the experimental uncertainties, the largest
are the uncertainties on the $\Pb$-tag veto and the identification and energy calibration of merged jets.

CMS uncertainties are defined by variation of single components and their impact on the results is not provided.
In general, larger theory uncertainties than in ATLAS appear to contribute for both the EW, QCD, and BSM components. Other significant contributions come from limited simulated statistics and merged-jet description
in agreement with the ATLAS study.

\paragraph{Results}

ATLAS reports a measured fiducial cross-section of $\sigma_{\mathrm{EW}} = 45 \pm 18\fb$, where the
systematic uncertainty is about twice the statistical one.
It corresponds to a background-only hypothesis rejection with a significance of 2.7$\sigma$, while 2.5$\sigma$ is expected. 
The merged analysis has a much better systematic/statistical ratio with respect to the resolved one,
and gives the largest contribution to the significance.

In the CMS analysis, limits on aQGC are set by fitting the mass distributions mentioned before. The list of
constrained Wilson coefficients is the same as in the combined leptonic \ssWW\ and \WZjj\ CMS analysis. While obtained
with about one fourth of the data, constraints using the semileptonic analysis are comparable or better. In some
case the limits supersede those obtained in the leptonic analysis by a factor of 3-5, in particular for what
concerns the $f_{S,1}$ and the mixed operators.

\subsection{The $\Wgamma$ and $\Zgamma$ final states}
\label{sec:wzgamma}
While not immediately connected to the EWSB mechanism, VBS production of \Wgamma\ and \Zgamma\ is still interesting
for both verifying SM calculations and BSM searches. In particular, it shares many theoretical and experimental features with the scattering of heavy gauge bosons.
The direct detection of the high-energy photon without decays to secondary particles makes the fiducial cross sections particularly large. In both cases,
only leptonic \PW\ and \PZ\ decays are considered. With this choice, 
$V \gamma$jj production from diagrams involving QCD vertices is the dominant background, even though in
the \Wgamma\ case some contamination from SM processes with top-quarks and photons remains.

\subsubsection{Theoretical calculations}

At the moment, the theory predictions for the production of a heavy gauge boson in association with a photon via VBS are rather limited.
For \Wgamma, they consist in the NLO QCD corrections to the EW signal \cite{Campanario:2013eta} and to the QCD background \cite{Campanario:2014dpa}.
The same holds true for \Zgamma, with NLO QCD corrections available in the VBS approximation \cite{Campanario:2017ffz} and fully for the background \cite{Campanario:2014wga}.

We note that the $\gamma\gamma$ case, which has two jets and two photons in the final state, has also been studied theoretically at NLO QCD for the signal \cite{Campanario:2020xaf} 
and the background \cite{Gehrmann:2013bga,Badger:2013ava,Bern:2014vza}.
Unfortunately, as opposed to \Wgamma\ and \Zgamma\, this EW production has not been measured yet at the LHC, due to the overwhelming multijet background.

\subsubsection{Experimental approaches}

ATLAS and CMS have reported evidence of electroweak \Zgamma\ production using a partial Run-2 data set~\cite{ATLAS:Zgamma,CMS:Zgamma}.
The CMS result is used in combination with $8\TeV$ data to obtain the observed evidence. Only CMS performed the search for the \Wgamma\ final state~\cite{CMS:Wgamma} in the same data set, 
leading to the observation of this process.

Photon reconstruction and identification proceed in similar ways in ATLAS and CMS~\cite{Khachatryan:2015iwa,Aaboud:2018yqu} and have many points is common
with the corresponding procedures for electrons. The electromagnetic shower reconstruction in the calorimeter
is analogous and the
identification criteria include shower shape (in the ATLAS case information is used in each single LAr layer)
and small energy leakage in the hadronic calorimeter. Isolation requirements in trackers and calorimeters
are also similar to those used to select electrons, even if in tracker isolation no charged particle is excluded
from the cone. In ATLAS, identification is performed separately for photons
converted in $\Pe^+\Pe^-$ pairs in the detector material, while in CMS this contribution is not considered. Selection
requirements are in general tighter than for electrons and different for the barrel and endcap regions of the
detectors, with barrel photons showing higher purity because of smaller jet backgrounds~\footnote{Both ATLAS
and CMS exclude from photon acceptance a small $|\eta|$ region corresponding to the transition between
barrel and endcap calorimetry.}.

\paragraph{Monte Carlo simulation}

The \Wgamma\ and \Zgamma\ analyses adopt similar simulation strategies as other VBS analyses for signal and background.
VBS signals are simulated at LO with \madgraphbis or \Sherpa, with no additional partons besides the two tagging jets. 
The EW-QCD interferences are estimated to be 1-3\% (3-8\%) of the EW \Wgamma\ (\Zgamma) signal in the fiducial
regions.
The corresponding QCD-induced processes are simulated either at LO with a number of extra partons exceeding
two or at NLO with a number of extra partons of at least two using the \madgraphbis and \Sherpa generators. In general,
the Monte Carlo generator results are used to normalize the cross sections of simulated samples, since higher-order
QCD calculations are not matching the analysis phase spaces~\cite{Campanario:2014wga,Campanario:2014ioa}.
A unique background to these searches is the $\Pt {\bar \Pt} \gamma$ SM process, with at least one top quark
decaying leptonically, which is simulated using \madgraphbis at LO (NLO) in ATLAS (CMS).

\paragraph{Fiducial region definitions and reconstruction-level selections}

Fiducial regions considered in the ATLAS and CMS analyses are compared in Table~\ref{tab:gammafr}. The same assumptions as in \ssWW\ are used for $\tau$ decays and lepton ``dressing''.

\begin{table}[tb]
\caption{Comparison of \Wgamma\ and \Zgamma\ fiducial region definitions and related EW (VBS) cross-section values in the ATLAS and CMS measurements~\cite{ATLAS:Zgamma,CMS:Zgamma,CMS:Wgamma}.
\emph{JRS} stands for generic Jet-Rapidity Separation selections.
The large differences in the \Zgamma\ fiducial-region choice for the ATLAS and CMS searches is noticeable.}
\center
{\begin{tabular}{@{}cccc@{}} \toprule
Variable & ATLAS \Zgamma & CMS \Zgamma & CMS \Wgamma  \\
\midrule
$p_{\rm T}(\ell)$ & $> 20$ GeV & $> 25/20$ GeV & $> 30$ GeV \\
$|\eta(\ell)|$ & $< 2.5$ &  $< 2.5/2.4$ & $< 2.4$ \\
$\Delta R(\ell\gamma)$ & $> 0.4$ & $> 0.7$ & - \\
$m_{\ell\ell}/m_{\rm T}(\PW)$ & $> 40$ GeV & $[70,110]$ GeV & $> 30$ GeV \\
$p_{\rm T}(\gamma)$ & $> 15$ GeV & $> 20$ GeV & $> 25$ GeV \\
$|\eta(\gamma)|$ & $< 2.37$ &  $< 2.5$ & $< 2.5$ \\
$m(\ell\ell\gamma) + m(\ell\ell)$ & $> 182$ GeV & - & - \\
$p_{\rm T,miss}$ & - & - & $> 30$ GeV \\
$p_{\rm T}(\Pj)$ & $> 50$ GeV & $> 30$ GeV & $> 50/40$ GeV \\
$|\eta(\Pj)|$ & $< 4.5$ &  $< 4.7$ & $< 4.7$ \\
\mjj & $> 150$ GeV & $> 500$ GeV & $> 500$ GeV \\
JRS & $ \dyjj > 1$ &  $\dejj > 2.5$ & $\dejj > 2.5$ \\
centrality & $< 5$ & - & - \\
\midrule
$\sigma$ LO & $7.8 \pm 0.5\fb$ & $5.0 \pm 0.3\fb$ & $17.03\fb$ \\
\bottomrule
\end{tabular} \label{tab:gammafr}}
\end{table}

Reconstruction-level selections follow the fiducial regions requirements. All analyses use
single- or double-lepton triggers to select events online.

In the ATLAS \Zgamma\ analysis, where particle-flow categorization is not
employed, a complex procedure is used to remove potential overlaps between detector signals identified as lepton
QED radiation, photons and jets.

In CMS, a selection on $m(V\gamma) > 100$ GeV reduces the contribution from
final-state radiation in Z-boson decays. For the \PW\ case, the invariant mass is computed by imposing a \PW-mass
constraint to determine the longitudinal momentum of the missing neutrino. Further VBS enrichment is obtained
using the Zeppenfeld variable of the $V\gamma$ system and its azimuthal-angle difference with the dijet system.

\paragraph{Analysis strategy and background estimation}

The ATLAS \Zgamma\ analysis uses a BDT algorithm to isolate the EW signal over the backgrounds, where the 13 input
variables are related to the kinematic properties of the two tagging jets, the photon, and the reconstructed Z boson.

CMS uses a two-dimensional fits using the most discriminating variables, which are (\mjj, \dejj) in the \Zgamma\
case and (\mjj, $m(\ell\gamma)$) in the \Wgamma\ case. In both analyses, events are first separated by
lepton flavor and central or forward rapidity regions, and control regions with small VBS yields are fit together
with the signal regions to constrain background normalizations from data.

\begin{figure*}[t]
\centering
\includegraphics[width=0.45\textwidth]{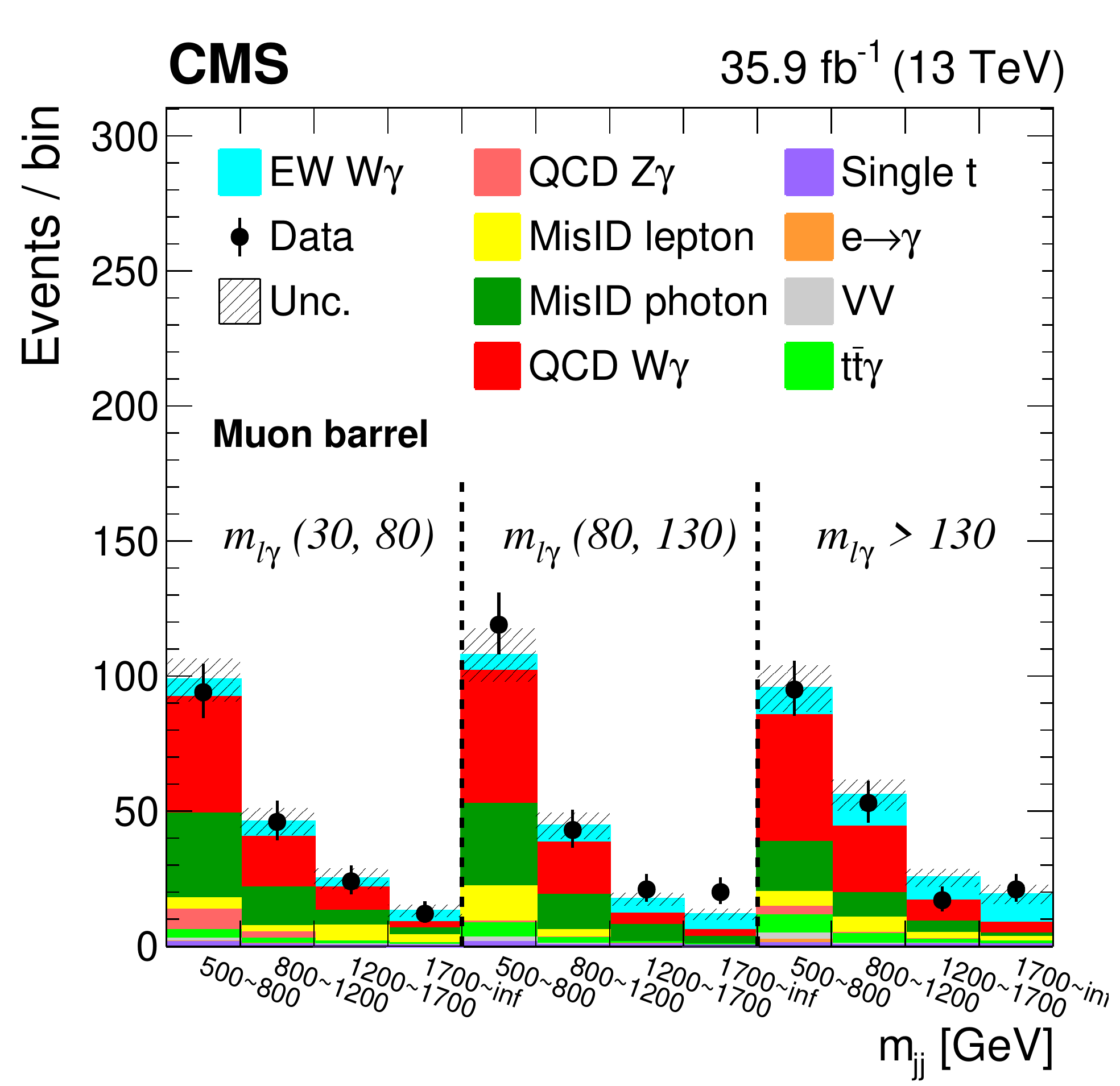} 
\includegraphics[width=0.44\textwidth]{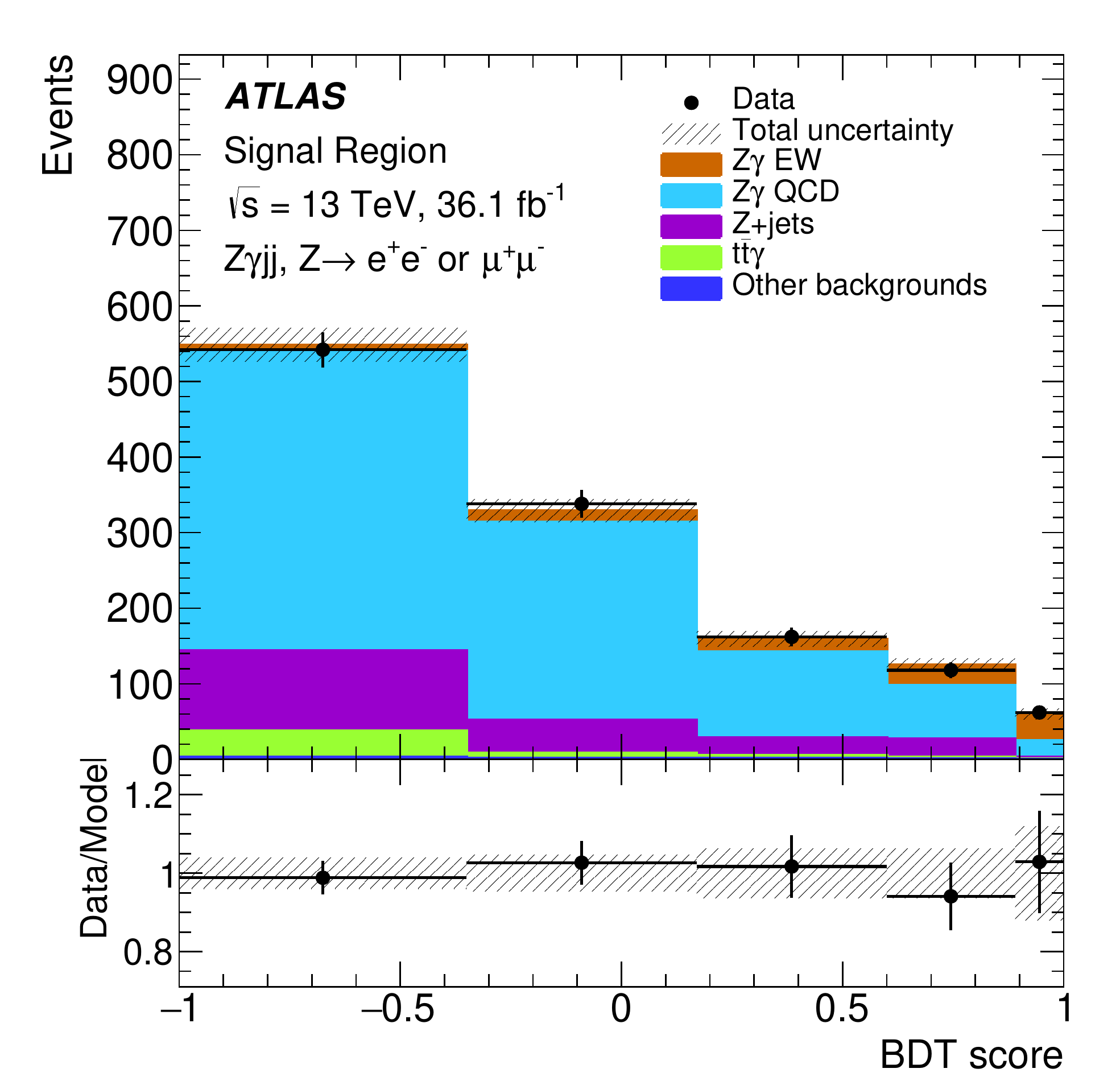}
\caption{Post-fit \mjj\ distributions in $m(\ell\gamma)$ intervals in the CMS \Wgamma\ analysis
(central-rapidity $\mu\gamma$ channel, left) and ATLAS output BDT score in the \Zgamma\ analysis (right).
The ``MisID photon'' event category in CMS is dominated by \PW+jets events.}
\label{fig:wzgamma}
\end{figure*}

In all cases the
largest background component is the QCD \Wgamma\ or \Zgamma\ production and the next-to-largest comes from $V$+jets
events where the photon is misidentified or not coming from the hard-scattering event.
The former is estimated from simulation while the second is obtained using control regions in data with relaxed photon selections.
The ATLAS and CMS data with superimposed signal and background components are shown in Figure~\ref{fig:wzgamma}.

\paragraph{Systematic uncertainties}

In ATLAS the largest systematic impacts are from the limited size of simulated samples, theory errors
on the \Zgamma\ EW production, and non-prompt-photon background modelling from data.
In CMS, the uncertainties are given as variations on a specific process and not as
total impacts on the measurements. From their absolute size, it can be inferred that a larger uncertainty
than in ATLAS is considered for the theory modelling of the QCD production, at least for \Wgamma. 
Among the experimental uncertainties, jet energy scale and resolution uncertainties have a larger impact with
respect to lepton and photon identification, in both experiments.

\paragraph{Results}

ATLAS reports a measured fiducial cross-section of $\sigma_{\mathrm{EW}} = 7.8 \pm 2.0\fb$, where the total
uncertainty is almost equally shared between statistical and systematic+modelling uncertainties.
It corresponds to a background-only hypothesis rejection with a significance of 4.1$\sigma$, in perfect agreement with expectations according to the \madgraphbis simulation, 
while \Sherpa predicts a somehow larger cross section.

For \Zgamma\ CMS similarly reports $\sigma_{\mathrm{EW}} = 3.2 \pm 1.2\fb$, where the ratio to the expectation
is $0.65$. It corresponds to a background-only hypothesis rejection with a significance of 3.9$\sigma$, while 5.2$\sigma$ was expected, 
and the compatibility with the SM is approximately at the 1.5$\sigma$ level. In the CMS case, the fiducial region is more restrictive, hence the statistical uncertainty is dominant. 
For \Wgamma\ the corresponding CMS results are $\sigma_{\mathrm{EW}} = 20 \pm 5\fb$, where the ratio to the expectation is $1.2$ and in agreement with the SM. 
It corresponds to a background-only hypothesis rejection with a significance of 5.3$\sigma$, while 4.8$\sigma$ was expected. 

The total cross sections
including EW and strong components are measured to be $\sigma_{\mathrm{\Zgamma,tot}} = 71^{+23}_{-19}\fb$ in ATLAS, 
$\sigma_{\mathrm{\Zgamma, tot}} = 14 \pm 3\fb$ and $\sigma_{\mathrm{\Wgamma, tot}} = 108 \pm 16\fb$ in CMS, 
all in good agreement with the SM.

In the CMS analyses, limits on aQGC are set by fitting the diboson-mass distributions in the signal region with 
additional requirements on the minimum transverse momentum of the photon, and/or \mjj. The range of explored
aQGC includes both mixed and transverse operators. The \Wgamma\ analysis is the most sensitive to the operators
which do not affect anomalous production of heavy vector bosons (where the semileptonic or exclusive
analyses give the most stringent limits), namely $f_{M,2}$, $f_{M,3}$, $f_{M,4}$, $f_{M,5}$, $f_{T,5}$, $f_{T,6}$, and $f_{T,7}$.

\subsection{The exclusive $\ggWW$ final state}
\label{sec:ggww}
Central exclusive production processes in proton-proton collisions are those in which the scattering
is mediated through photon-photon fusion. The studies of these processes
can be carried out in particularly clean experimental conditions
thanks to the absence of proton remnants. However, since the direction of the scattered protons is almost
parallel to the beam line, these particles escape detection in ATLAS and CMS. Considering this fact, there are
two ways to perform exclusive studies: the first is to just use information from the main apparati, where
protons are undetected, and
require very low track and neutral activity in the selected events, in addition to the hard-scattering products; the
second is to \emph{tag} the protons using dedicated detectors placed along the LHC beams at distance from the main
detectors. The Precision Proton Spectrometer (PPS) is a recent example~\cite{PPS}.

The exclusive \ggWW\ is a production process of the VBS type and, as such, is sensitive to BSM physics and in particular aQGC. It can
be performed in the leptonic final state, using events with only two reconstructed leptons and large missing momentum in the main detectors,
or in hadronic final states, where the use of proton tagging constrains the event kinematics and hence is a
very powerful background-rejection technique.

\subsubsection{Theoretical calculations}

Compared to standard LHC measurement, the corresponding theoretical predictions are not very much advanced.
In particular, for the elastic part \emph{i.e.}\ $\gamma\gamma\to\PW^+\PW^-$, these are usually LO predictions matched to parton shower.
They can be obtained from standard Monte Carlo generators and have to be combined with the corresponding photon flux (see for example Ref.~\cite{Budnev:1974de}).

The main background for such a measurement is $q\bar q \to \PW^+\PW^-$ which is very well known from a theoretical point of view with
NLO QCD \cite{Ohnemus:1991kk,Baur:1995uv,Campbell:1999ah}, 
NLO QCD + PS \cite{Frixione:2002ik,Hamilton:2010mb,Nason:2013ydw}, 
NNLO QCD \cite{Gehrmann:2014fva,Grazzini:2016ctr},
gluon--gluon loop-induced \cite{Caola:2015rqy,Grazzini:2020stb},
resummed \cite{Grazzini:2015wpa},
and NLO EW \cite{Kuhn:2011mh,Bierweiler:2012kw,Baglio:2013toa,Gieseke:2014gka,Biedermann:2016guo,Kallweit:2017khh} predictions.
Further predictions even feature combinations of the above mentioned calculations \cite{Re:2018vac,Kallweit:2019zez,Kallweit:2020gva,Brauer:2020kfv,Chiesa:2020ttl}.

\subsubsection{Experimental approaches}

ATLAS and CMS studies so far employ the technique which does not require proton tagging. ATLAS
recently reported observation of \ggWW\ in Run-2 data~\cite{ATLAS:ggWW} while CMS has issued only results
on less recent data sets at 7 and 8 TeV, finding evidence for this process and stringent
aQGC limits on the operators $f_{M,0}$ and $f_{M,1}$~\cite{CMS:ggWW}.

Since this work deals with LHC Run-2 results, we will just briefly review the ATLAS analysis in the following.
The analysis proceeds by requiring an electron and a muon of opposite signs with large dilepton transverse momentum and exactly zero additional charged particles in the event.
\footnote{The same-flavor case is not considered because of the dominant $\gamma\gamma \rightarrow \ell^+ \ell^-$ background.}

Simulation of the signal events
and of the $\gamma\gamma \rightarrow \ell^+ \ell^-$ background without intermediate \PW\ bosons proceeds through
the \HERWIG generator~\cite{Bellm:2015jjp}, interfaced to a suitable photon-flux provider and QED-dedicated
PDFs. The main background, which is $\PW^+\PW^-$ production through quark-antiquark annihilation, is simulated
using \POWHEG. \cite{Nason:2013ydw}

The fiducial region is defined by event with an electron and a muon with 
$p_{\rm T,\ell} > 27/20\GeV$  (after QED FSR recovery), $|\eta_{\ell}| < 2.5$,
$m_{e \mu} > 20\GeV$, $p_{\rm T,e \mu} > 30\GeV$ and $n_{trk} = 0$ where $trk$ is
any charged particle with $p_{\rm T} > 0.5\GeV$ and $|\eta| < 2.5$. Further offline
requirements are applied for the leptons to be compatible with a common vertex.

Background estimation is quite complex because of the multiple sources that can feed additional charged particles
in the event and therefore fail the $n_{trk} = 0$ requirement. To improve pile-up descriptions, data-driven
methods are used to derive corrections to the simulated events, targeting the
density of proton-proton interactions and the number of tracks per interaction separately. Underlying-event modelling
of non-exclusive $\PW^+\PW^-$ production is also corrected from a Drell-Yan control sample in data, while a
correction factor
for the proton-rescattering effect is derived from a $\gamma\gamma \rightarrow \ell^+ \ell^-$ control sample.

Signal systematic uncertainties are dominated by the mass-dependent transfer 
factors between \ggWW\ and the $\gamma\gamma \rightarrow \ell^+ \ell^-$ control sample, while the main systematic background uncertainties arise from 
variations from different theory predictions of non-exclusive $\PW^+\PW^-$ production. Overall, the largest impact comes from simulated statistics in 
background samples.

The fiducial cross-section is determined by fitting the signal region 
($p_{\rm T,e \mu} > 30$ GeV and $n_{trk} = 0$) after checking data/prediction 
agreement in control regions where either $p_{\rm T,e \mu} < 30$ GeV or $1 \leq n_{trk} \leq 4$. The significance of the observation is 8.4$\sigma$ and the 
corresponding result is: $\sigma_{\mathrm{fid}} = 3.1 \pm 0.4\fb$, where the
statistical and systematic uncertainties are of the same order. Comparing
with the corresponding CMS analysis, it can be inferred that constraints
on relevant dim-8 operators could be competitive or better than those obtained from deep-inelastic $\Pp\Pp$ collisions.

\section{Prospects for High-Luminosity and High-Energy LHC scenarios}
\label{sec:perspectives}
In this section we discuss prospects for VBS at the upcoming phases of the LHC and at future hadron colliders.
At the time when this review is written, not many studies on these topics are available. However, we expect the situation to improve in
the next few years when concrete realisations of plans are coming closer.
More specifically, we will focus on the High-Luminosity (HL) and High-Energy (HE) phases of the LHC, for which the main physics results are compiled in Ref.~\cite{Azzi:2019yne}.
The HL run of the LHC is planned to start in 2027 at a centre-of-mass energy $\sqrt{s} = 14\TeV$, and should provide an integrated luminosity of at least $3000\fb^{-1}$ per experiment.
On the other hand, a viable option for the HE phase is to occur at $\sqrt{s} = 27\TeV$, with expected integrated luminosities of about $15000\fb^{-1}$.

From a theoretical point of view, predictions at $\sqrt{s} = 14\TeV$ for the HL are very similar to the ones at $\sqrt{s} = 13\TeV$ that we have reviewed above.
In particular, as for the total cross sections, the QCD or EW relative corrections only differ by few per cent.
At $\sqrt{s} = 100\TeV$, some results are available in Ref.~\cite{Mangano:2016jyj} at LO which are further discussed in Ref.~\cite{Jager:2017owh}.
In Table~\ref{tab:EW_HE}, we have compiled LO [$\mathcal{O}\left(\alpha^6\right)$] and NLO EW [$\mathcal{O}\left(\alpha^7\right)$] predictions for $\PW^+\PW^+$
at several centre-of-mass energies: $14\TeV$, $27\TeV$, and $100\TeV$.
These results are based on Refs.~\cite{Biedermann:2016yds,Azzi:2019yne}.
It is interesting to see that the cross sections increase significantly with the centre-of-mass energy while the EW corrections become negatively very large.
This can simply be understood by the fact that the average typical scale that enters the Sudakov logarithms is increasing.
While we focus here on the $\PW^+\PW^+$ channel, these corrections should be rather representative of other channels.
Indeed, the Leading-Log approximation obtained following Ref.~\cite{Denner:2000jv} is identical for all scattering processes of EW bosons 
as they result from the same $\rm SU(2)_w$ coupling \cite{Biedermann:2016yds,Denner:2019tmn,Denner:2020zit}.
At $100\TeV$, the corrections reach $-25\%$ making the use of resummation techniques as well as the inclusion of heavy gauge-bosons radiations necessary.

\begin{table}[htb]
\caption{Cross sections at LO and NLO EW for VBS $\PW^+\PW^+$ at a centre-of-mass energy of $14\TeV$, $27\TeV$, and $100\TeV$.
These numbers are based on Refs.~\cite{Biedermann:2016yds,Azzi:2019yne}.}
\center
{\begin{tabular}{l|ccc}
$\sqrt{s}$ & $\sigma^{\rm LO} [\fb]$ & $\sigma^{\rm NLO}_{\rm EW} [\fb]$ & $\delta_{\rm EW} [\%]$ \\
\hline
$14\TeV$ & $1.4282(2)$ &  $1.213(5)$ & $-15.1$ \\
$27\TeV$ & $4.7848(5)$ & $3.881(7)$ & $-18.9$ \\
$100\TeV$ & $25.485(9)$ & $19.07(6)$ & $-25.2$
\end{tabular} \label{tab:EW_HE}}
\end{table}

Experimental perspectives for VBS at the HL-LHC are reviewed 
in Ref.~\cite{Azzi:2019yne} using simulations of upgraded ATLAS and CMS detectors,
a $\sqrt{s} = 14\TeV$ hypothesis and a range of integrated luminosities from 
$300\fb^{-1}$ through $8000\fb^{-1}$, where the first value corresponds to one
year of data taking, and the latter to 10 years of combined data sets 
collected by the ATLAS and CMS experiments in the most optimistic scenario.
All perspective analysis mimic or simplify event selections described in 
the previous chapters.

For the \ssWW\ final state CMS expects an overall uncertainty on the total 
fiducial VBS cross section of $4.5\%$ with a luminosity of $3000\fb^{-1}$, while 
for ATLAS this estimation is about $5-6\%$. Concerning the purely 
longitudinal component of VBS, $\PW_{\rm L}^\pm \PW_{\rm L}^\pm$, a combination of ATLAS
and CMS results, using fully simulated ATLAS events and improved electron 
efficiency, is expected to reach an expected significance of $3\sigma$ 
with $4000\fb^{-1}$ of total integrated luminosity. 

For the \WZjj\ final state, the predicted uncertainty at $3000\fb^{-1}$ in CMS
is about $5.5\%$. In ATLAS, this estimation is provided as a function of the
predicted background uncertainty, reflecting the difficulty to anticipate progress
in theory calculations, and approaches $5\%$ with the current knowledge. For the
polarized components, ATLAS estimate considers inclusive estimations for each
boson separately, \emph{i.e.}\ sensitivities for $\PW_{\rm L}^+ \PZ_X$,  $\PW_{\rm L}^- \PZ_X$, and
$\PW_X^\pm \PZ_{\rm L}$. The significance is estimated to
be between 1.5 and 2.5$\sigma$ for $\PZ_{\rm L}$ and $\PW^+_{\rm L}$ and between 0.7 and 1.5$\sigma$ for $\PW^-_{\rm L}$ depending on the final selection and systematic assumptions 
on the total background normalisation. In CMS the doubly longitudinally 
polarized component is considered and the expected significance is about 
1.3-1.4$\sigma$ at $3000\fb^{-1}$. 

For the \ZZjj\ final state, estimates in ATLAS and CMS strongly depend on the
assumption on the evolution of the theory uncertainty of the QCD component,
especially the loop-induced, gluon-initiated one, which is the most challenging from the computational point of view.
Estimates on the achievement of the evidence/observation of the process
are already superseded by recent Run-2 results. ATLAS also shows perspectives
of differential cross-section measurements, that however have uncertainties
of $60\%$ or larger after $3000\fb^{-1}$. CMS estimates to have a significance
for the detection of the purely longitudinal component,  $\PZ_{\rm L} \PZ_{\rm L}$, varying
between 1.2 and 1.4$\sigma$, depending on the detector extended muon
coverage.

\section{Conclusion}
\label{sec:conclusion}
In the present work, we have reviewed the state of the art of vector-boson scattering (VBS) at the LHC.
In particular, for the theoretical part, we have almost exclusively focused our discussion on results in the Standard Model of particle physics.
This study therefore summarizes our present understanding of the Standard Model and in particular the electroweak (EW) theory, through the magnifying glass of VBS.

The first part is a general introduction to the most important concepts in the study of VBS at the LHC.
First, we have provided a precise definition of VBS at the LHC which should hopefully help for a better use of theoretical progress in experimental analysis.
Then, the formalism of polarised VBS is discussed in detail.
After this, general theoretical aspects are examined, where emphasis is put on higher-order corrections of QCD and EW origins.
A presentation of the most relevant experimental aspects of VBS measurements follows.
These include a brief description of the ATLAS and CMS detectors, the details of VBS tagging-jet and vector-boson reconstruction and the simulations used in data analysis.
Finally, a short review of constraints on new-physics theories obtained from VBS measurements is provided, with special emphasis on Effective Field Theory interpretations.

In the second part of this report, all possible signatures of VBS at the LHC are examined.
For each of the signatures, both the theoretical and experimental aspects are discussed and specificities of each channels are highlighted and explained.
Great attention is also given to aspects related to the interface between experiment and theory and, in that respect, some directions are also given in order to improve current methodologies.

The last section is devoted to prospects for the measurements of VBS at future collider experiments. It is rather short, but should be interesting for future VBS studies.

With this work, we have gathered in one unique place, all relevant references for in-depth analysis of VBS at the LHC, which should show useful for both the experimental and theoretical communities.
In particular, during the whole review we have tried to keep the discussion at a level which should be understandable by LHC physicists irrespectively of their background.
The reason for this is that we are convinced that advance in fundamental physics can only proceed through a deep understanding of both experimental and theoretical aspects, which are two sides of the same coin and can therefore not be isolated.
In particular, we believe that this is the way to get the most accurate physics results out of VBS, at present and future high-energy experiments.

Finally, it should be clear from the reading that a lot has already been achieved in the past few years, and that the detailed study of VBS at collider experiments possesses a huge potential for new innovative investigations.
In this context, we hope that the present work will contribute to the great progresses that will be achieved in the next years regarding our understand of the EW theory of the Standard Model.

\section*{Acknowledgements}

We are grateful to the ``VBSCan'' COST action network CA16108 for offering a stimulating and dynamic atmosphere over the past few years.
Without this action, this review would probably not have seen birth. Also, we are deeply indebted to our colleagues and collaborators for the numerous discussions on the topic of VBS.
In particular, we would like to thank Pietro Govoni for his commitment in the VBSCan action as well as for feedback on the present manuscript.
RC acknowledges support from the Italian and Serbian Ministries for Foreign Affairs through the Researcher Mobility Program RS19MO06.
MP acknowledges support from the German Research Foundation (DFG) through the Research Training Group RTG2044.
MZ is supported by the ``Programma per Giovani Ricercatori Rita Levi Montalcini'' granted by the Italian Ministero dell'Universit\`a e della Ricerca (MUR). 

\bibliographystyle{utphys.bst}

\bibliography{vbs}

\providecommand{\href}[2]{#2}\begingroup\raggedright\begin{thebibliography}{100}

\bibitem{Glashow:1961tr}
S.~L. Glashow, {\em {Partial Symmetries of Weak Interactions}}.
  \href{http://dx.doi.org/10.1016/0029-5582(61)90469-2}{Nucl. Phys. {\bf 22}
  (1961)  579--588}.

\bibitem{Weinberg:1967tq}
S.~Weinberg, {\em {A Model of Leptons}}.
  \href{http://dx.doi.org/10.1103/PhysRevLett.19.1264}{Phys. Rev. Lett. {\bf
  19} (1967)  1264--1266}.

\bibitem{Salam:1968rm}
A.~Salam, {\em {Weak and Electromagnetic Interactions}}.
  \href{http://dx.doi.org/10.1142/9789812795915_0034}{Conf. Proc. C {\bf
  680519} (1968)  367--377}.

\bibitem{Higgs:1964ia}
P.~W. Higgs, {\em {Broken symmetries, massless particles and gauge fields}}.
  \href{http://dx.doi.org/10.1016/0031-9163(64)91136-9}{Phys. Lett. {\bf 12}
  (1964)  132--133}.

\bibitem{Higgs:1964pj}
P.~W. Higgs, {\em {Broken Symmetries and the Masses of Gauge Bosons}}.
  \href{http://dx.doi.org/10.1103/PhysRevLett.13.508}{Phys. Rev. Lett. {\bf 13}
  (1964)  508--509}.

\bibitem{Englert:1964et}
F.~Englert and R.~Brout, {\em {Broken Symmetry and the Mass of Gauge Vector
  Mesons}}. \href{http://dx.doi.org/10.1103/PhysRevLett.13.321}{Phys. Rev.
  Lett. {\bf 13} (1964)  321--323}.

\bibitem{Guralnik:1964eu}
G.~S. Guralnik, C.~R. Hagen, and T.~W.~B. Kibble, {\em {Global Conservation
  Laws and Massless Particles}}.
  \href{http://dx.doi.org/10.1103/PhysRevLett.13.585}{Phys. Rev. Lett. {\bf 13}
  (1964)  585--587}.

\bibitem{Higgs:1966ev}
P.~W. Higgs, {\em {Spontaneous Symmetry Breakdown without Massless Bosons}}.
  \href{http://dx.doi.org/10.1103/PhysRev.145.1156}{Phys. Rev. {\bf 145} (1966)
   1156--1163}.

\bibitem{Kibble:1967sv}
T.~W.~B. Kibble, {\em {Symmetry breaking in non-Abelian gauge theories}}.
  \href{http://dx.doi.org/10.1103/PhysRev.155.1554}{Phys. Rev. {\bf 155} (1967)
   1554--1561}.

\bibitem{Aad:2012tfa}
{\bf ATLAS} Collaboration, G.~Aad {\em et al.}, {\em {Observation of a new
  particle in the search for the Standard Model Higgs boson with the ATLAS
  detector at the LHC}}.
  \href{http://dx.doi.org/10.1016/j.physletb.2012.08.020}{Phys. Lett. B {\bf
  716} (2012)  1--29}, \href{http://arxiv.org/abs/1207.7214}{{\tt
  arXiv:1207.7214 [hep-ex]}}.

\bibitem{Chatrchyan:2012ufa}
{\bf CMS} Collaboration, S.~Chatrchyan {\em et al.}, {\em {Observation of a New
  Boson at a Mass of 125 GeV with the CMS Experiment at the LHC}}.
  \href{http://dx.doi.org/10.1016/j.physletb.2012.08.021}{Phys. Lett. B {\bf
  716} (2012)  30--61}, \href{http://arxiv.org/abs/1207.7235}{{\tt
  arXiv:1207.7235 [hep-ex]}}.

\bibitem{Gallinaro:2020cte}
M.~Gallinaro {\em et al.}, ``{Beyond the Standard Model in Vector Boson
  Scattering Signatures},''
\newblock 5, 2020.
\newblock \href{http://arxiv.org/abs/2005.09889}{{\tt arXiv:2005.09889
  [hep-ph]}}.

\bibitem{Szleper:2014xxa}
M.~Szleper, {\em {The Higgs boson and the physics of $WW$ scattering before and
  after Higgs discovery}}. \href{http://arxiv.org/abs/1412.8367}{{\tt
  arXiv:1412.8367 [hep-ph]}}.

\bibitem{Green:2016trm}
D.~R. Green, P.~Meade, and M.-A. Pleier, {\em {Multiboson interactions at the
  LHC}}. \href{http://dx.doi.org/10.1103/RevModPhys.89.035008}{Rev. Mod. Phys.
  {\bf 89} (2017) no.~3, 035008}, \href{http://arxiv.org/abs/1610.07572}{{\tt
  arXiv:1610.07572 [hep-ex]}}.

\bibitem{Rauch:2016pai}
M.~Rauch, {\em {Vector-Boson Fusion and Vector-Boson Scattering}}.
\href{http://arxiv.org/abs/1610.08420}{{\tt arXiv:1610.08420 [hep-ph]}}.

\bibitem{Tricoli:2020vgn}
A.~Tricoli, M.~Sch{\"o}nherr, and P.~Azzurri, {\em {Vector Bosons and Jets in
  Proton Collisions}}. \href{http://arxiv.org/abs/2012.13967}{{\tt
  arXiv:2012.13967 [hep-ex]}}.

\bibitem{Dawson:1984gx}
S.~Dawson, {\em {The Effective W Approximation}}.
  \href{http://dx.doi.org/10.1016/0550-3213(85)90038-0}{Nucl. Phys. B {\bf 249}
  (1985)  42--60}.

\bibitem{Duncan:1985vj}
M.~J. Duncan, G.~L. Kane, and W.~Repko, {\em {WW Physics at Future Colliders}}.
  \href{http://dx.doi.org/10.1016/0550-3213(86)90234-8}{Nucl. Phys. B {\bf 272}
  (1986)  517--559}.

\bibitem{Cahn:1983ip}
R.~Cahn and S.~Dawson, {\em {Production of Very Massive Higgs Bosons}}.
  \href{http://dx.doi.org/10.1016/0370-2693(84)91180-8}{Phys. Lett. B {\bf 136}
  (1984)  196}. [Erratum: Phys.Lett.B 138, 464 (1984)].

\bibitem{Kuss:1995yv}
I.~Kuss and H.~Spiesberger, {\em {Luminosities for vector boson - vector boson
  scattering at high-energy colliders}}.
  \href{http://dx.doi.org/10.1103/PhysRevD.53.6078}{Phys. Rev. D {\bf 53}
  (1996)  6078--6093}, \href{http://arxiv.org/abs/hep-ph/9507204}{{\tt
  arXiv:hep-ph/9507204}}.

\bibitem{Accomando:2006hq}
E.~Accomando, A.~Denner, and S.~Pozzorini, {\em {Logarithmic electroweak
  corrections to ${\rm e}^+ {\rm e}^- \to \nu_{\rm e} \bar \nu_{\rm e} {\rm
  W}^+ {\rm W}^-$}}.
  \href{http://dx.doi.org/10.1088/1126-6708/2007/03/078}{JHEP {\bf 03} (2007)
  078}, \href{http://arxiv.org/abs/hep-ph/0611289}{{\tt arXiv:hep-ph/0611289}}.

\bibitem{Figy:2003nv}
T.~Figy, C.~Oleari, and D.~Zeppenfeld, {\em {Next-to-leading order jet
  distributions for Higgs boson production via weak boson fusion}}.
  \href{http://dx.doi.org/10.1103/PhysRevD.68.073005}{Phys. Rev. D {\bf 68}
  (2003)  073005}, \href{http://arxiv.org/abs/hep-ph/0306109}{{\tt
  arXiv:hep-ph/0306109}}.

\bibitem{Ballestrero:2018anz}
A.~Ballestrero {\em et al.}, {\em {Precise predictions for same-sign W-boson
  scattering at the LHC}}.
  \href{http://dx.doi.org/10.1140/epjc/s10052-018-6136-y}{Eur. Phys. J. C {\bf
  78} (2018) no.~8, 671}, \href{http://arxiv.org/abs/1803.07943}{{\tt
  arXiv:1803.07943 [hep-ph]}}.

\bibitem{Biedermann:2017bss}
B.~Biedermann, A.~Denner, and M.~Pellen, {\em {Complete NLO corrections to
  W$^{+}$W$^{+}$ scattering and its irreducible background at the LHC}}.
  \href{http://dx.doi.org/10.1007/JHEP10(2017)124}{JHEP {\bf 10} (2017)  124},
  \href{http://arxiv.org/abs/1708.00268}{{\tt arXiv:1708.00268 [hep-ph]}}.

\bibitem{CMS:ssWWandWZ}
{\bf CMS} Collaboration, A.~M. Sirunyan {\em et al.}, {\em {Measurements of
  production cross sections of WZ and same-sign WW boson pairs in association
  with two jets in proton-proton collisions at $\sqrt{s} =$ 13 TeV}}.
  \href{http://dx.doi.org/10.1016/j.physletb.2020.135710}{Phys. Lett. B {\bf
  809} (2020)  135710}, \href{http://arxiv.org/abs/2005.01173}{{\tt
  arXiv:2005.01173 [hep-ex]}}.

\bibitem{CMS:ZZ2}
{\bf CMS} Collaboration, A.~M. Sirunyan {\em et al.}, {\em {Evidence for
  electroweak production of four charged leptons and two jets in proton-proton
  collisions at $\sqrt {s}$ = 13 TeV}}.
  \href{http://dx.doi.org/10.1016/j.physletb.2020.135992}{Phys. Lett. B {\bf
  812} (2021)  135992}, \href{http://arxiv.org/abs/2008.07013}{{\tt
  arXiv:2008.07013 [hep-ex]}}.

\bibitem{Lee:1977yc}
B.~W. Lee, C.~Quigg, and H.~Thacker, {\em {The Strength of Weak Interactions at
  Very High-Energies and the Higgs Boson Mass}}.
  \href{http://dx.doi.org/10.1103/PhysRevLett.38.883}{Phys. Rev. Lett. {\bf 38}
  (1977)  883--885}.

\bibitem{Lee:1977eg}
B.~W. Lee, C.~Quigg, and H.~Thacker, {\em {Weak Interactions at Very
  High-Energies: The Role of the Higgs Boson Mass}}.
  \href{http://dx.doi.org/10.1103/PhysRevD.16.1519}{Phys. Rev. D {\bf 16}
  (1977)  1519}.

\bibitem{Chanowitz:1984ne}
M.~S. Chanowitz and M.~K. Gaillard, {\em {Multiple Production of W and Z as a
  Signal of New Strong Interactions}}.
  \href{http://dx.doi.org/10.1016/0370-2693(84)91141-9}{Phys. Lett. B {\bf 142}
  (1984)  85--90}.

\bibitem{Chanowitz:1985hj}
M.~S. Chanowitz and M.~K. Gaillard, {\em {The TeV Physics of Strongly
  Interacting W's and Z's}}.
  \href{http://dx.doi.org/10.1016/0550-3213(85)90580-2}{Nucl. Phys. B {\bf 261}
  (1985)  379--431}.

\bibitem{Stirling:2012zt}
W.~Stirling and E.~Vryonidou, {\em {Electroweak gauge boson polarisation at the
  LHC}}. \href{http://dx.doi.org/10.1007/JHEP07(2012)124}{JHEP {\bf 07} (2012)
  124}, \href{http://arxiv.org/abs/1204.6427}{{\tt arXiv:1204.6427 [hep-ph]}}.

\bibitem{Bern:2011ie}
Z.~Bern {\em et al.}, {\em {Left-Handed W Bosons at the LHC}}.
  \href{http://dx.doi.org/10.1103/PhysRevD.84.034008}{Phys. Rev. D {\bf 84}
  (2011)  034008}, \href{http://arxiv.org/abs/1103.5445}{{\tt arXiv:1103.5445
  [hep-ph]}}.

\bibitem{AguilarSaavedra:2006fy}
J.~Aguilar-Saavedra, J.~Carvalho, N.~F. Castro, F.~Veloso, and A.~Onofre, {\em
  {Probing anomalous Wtb couplings in top pair decays}}.
  \href{http://dx.doi.org/10.1140/epjc/s10052-007-0289-4}{Eur. Phys. J. C {\bf
  50} (2007)  519--533}, \href{http://arxiv.org/abs/hep-ph/0605190}{{\tt
  arXiv:hep-ph/0605190}}.

\bibitem{Ballestrero:2017bxn}
A.~Ballestrero, E.~Maina, and G.~Pelliccioli, {\em {$W$ boson polarization in
  vector boson scattering at the LHC}}.
  \href{http://dx.doi.org/10.1007/JHEP03(2018)170}{JHEP {\bf 03} (2018)  170},
  \href{http://arxiv.org/abs/1710.09339}{{\tt arXiv:1710.09339 [hep-ph]}}.

\bibitem{Ballestrero:2007xq}
A.~Ballestrero, A.~Belhouari, G.~Bevilacqua, V.~Kashkan, and E.~Maina, {\em
  {PHANTOM: A Monte Carlo event generator for six parton final states at high
  energy colliders}}.
  \href{http://dx.doi.org/10.1016/j.cpc.2008.10.005}{Comput. Phys. Commun. {\bf
  180} (2009)  401--417}, \href{http://arxiv.org/abs/0801.3359}{{\tt
  arXiv:0801.3359 [hep-ph]}}.

\bibitem{Ballestrero:2019qoy}
A.~Ballestrero, E.~Maina, and G.~Pelliccioli, {\em {Polarized vector boson
  scattering in the fully leptonic WZ and ZZ channels at the LHC}}.
  \href{http://dx.doi.org/10.1007/JHEP09(2019)087}{JHEP {\bf 09} (2019)  087},
  \href{http://arxiv.org/abs/1907.04722}{{\tt arXiv:1907.04722 [hep-ph]}}.

\bibitem{Aeppli:1993cb}
A.~Aeppli, F.~Cuypers, and G.~J. van Oldenborgh, {\em {O($\Gamma$) corrections
  to W pair production in ${\rm e}^+ {\rm e}^-$ and $\gamma \gamma$
  collisions}}. \href{http://dx.doi.org/10.1016/0370-2693(93)91259-P}{Phys.
  Lett. B {\bf 314} (1993)  413--420},
  \href{http://arxiv.org/abs/hep-ph/9303236}{{\tt arXiv:hep-ph/9303236}}.

\bibitem{Aeppli:1993rs}
A.~Aeppli, G.~J. van Oldenborgh, and D.~Wyler, {\em {Unstable particles in one
  loop calculations}}.
  \href{http://dx.doi.org/10.1016/0550-3213(94)90195-3}{Nucl. Phys. B {\bf 428}
  (1994)  126--146}, \href{http://arxiv.org/abs/hep-ph/9312212}{{\tt
  arXiv:hep-ph/9312212}}.

\bibitem{Beenakker:1998gr}
W.~Beenakker, F.~A. Berends, and A.~Chapovsky, {\em {Radiative corrections to
  pair production of unstable particles: results for ${\rm e}^+ {\rm e}^- \to$
  four fermions}}. \href{http://dx.doi.org/10.1016/S0550-3213(99)00110-8}{Nucl.
  Phys. B {\bf 548} (1999)  3--59},
  \href{http://arxiv.org/abs/hep-ph/9811481}{{\tt arXiv:hep-ph/9811481}}.

\bibitem{Billoni:2013aba}
M.~Billoni, S.~Dittmaier, B.~J\"ager, and C.~Speckner, {\em {Next-to-leading
  order electroweak corrections to ${\rm p}{\rm p} \to {W W} \to$ 4 leptons at
  the LHC in double-pole approximation}}.
  \href{http://dx.doi.org/10.1007/JHEP12(2013)043}{JHEP {\bf 12} (2013)  043},
  \href{http://arxiv.org/abs/1310.1564}{{\tt arXiv:1310.1564 [hep-ph]}}.

\bibitem{Denner:2000bj}
A.~Denner, S.~Dittmaier, M.~Roth, and D.~Wackeroth, {\em {Electroweak radiative
  corrections to ${e}^+ {e}^- \to {W W} \to$ 4 fermions in double pole
  approximation: The RACOONWW approach}}.
  \href{http://dx.doi.org/10.1016/S0550-3213(00)00511-3}{Nucl. Phys. {\bf B587}
  (2000)  67--117},
\href{http://arxiv.org/abs/hep-ph/0006307}{{\tt arXiv:hep-ph/0006307
  [hep-ph]}}.

\bibitem{Denner:2016jyo}
A.~Denner and M.~Pellen, {\em {NLO electroweak corrections to off-shell
  top-antitop production with leptonic decays at the LHC}}.
  \href{http://dx.doi.org/10.1007/JHEP08(2016)155}{JHEP {\bf 08} (2016)  155},
  \href{http://arxiv.org/abs/1607.05571}{{\tt arXiv:1607.05571 [hep-ph]}}.

\bibitem{Denner:2020bcz}
A.~Denner and G.~Pelliccioli, {\em {Polarized electroweak bosons in ${\bf
  \text{W}^+\text{W}^-}$ production at the LHC including NLO QCD effects}}.
  \href{http://dx.doi.org/10.1007/JHEP09(2020)164}{JHEP {\bf 09} (2020)  164},
  \href{http://arxiv.org/abs/2006.14867}{{\tt arXiv:2006.14867 [hep-ph]}}.

\bibitem{Aaboud:2019gxl}
{\bf ATLAS} Collaboration, M.~Aaboud {\em et al.}, {\em {Measurement of
  $W^{\pm}Z$ production cross sections and gauge boson polarisation in $pp$
  collisions at $\sqrt{s} = 13$ TeV with the ATLAS detector}}.
  \href{http://dx.doi.org/10.1140/epjc/s10052-019-7027-6}{Eur. Phys. J. C {\bf
  79} (2019) no.~6, 535}, \href{http://arxiv.org/abs/1902.05759}{{\tt
  arXiv:1902.05759 [hep-ex]}}.

\bibitem{Grossi:2020orx}
M.~Grossi, J.~Novak, B.~Kersevan, and D.~Rebuzzi, {\em {Comparing traditional
  and deep-learning techniques of kinematic reconstruction for polarization
  discrimination in vector boson scattering}}.
  \href{http://dx.doi.org/10.1140/epjc/s10052-020-08713-1}{Eur. Phys. J. {\bf
  C80} (2020) no.~12, 1144},
\href{http://arxiv.org/abs/2008.05316}{{\tt arXiv:2008.05316 [hep-ph]}}.

\bibitem{Ballestrero:2020qgv}
A.~Ballestrero, E.~Maina, and G.~Pelliccioli, {\em {Different polarization
  definitions in same-sign $WW$ scattering at the LHC}}.
  \href{http://dx.doi.org/10.1016/j.physletb.2020.135856}{Phys. Lett. {\bf
  B811} (2020)  135856},
\href{http://arxiv.org/abs/2007.07133}{{\tt arXiv:2007.07133 [hep-ph]}}.

\bibitem{Baglio:2019nmc}
J.~Baglio and L.~D. Ninh, {\em {Polarization observables in WZ production at
  the 13 TeV LHC: Inclusive case}}.
  \href{http://dx.doi.org/10.15625/0868-3166/30/1/14461}{Commun. in Phys. {\bf
  30} (2020) no.~1, 35--47}, \href{http://arxiv.org/abs/1910.13746}{{\tt
  arXiv:1910.13746 [hep-ph]}}.

\bibitem{Denner:2020eck}
A.~Denner and G.~Pelliccioli, {\em {NLO QCD predictions for doubly-polarized WZ
  production at the LHC}}. \href{http://arxiv.org/abs/2010.07149}{{\tt
  arXiv:2010.07149 [hep-ph]}}.

\bibitem{BuarqueFranzosi:2019boy}
D.~Buarque~Franzosi, O.~Mattelaer, R.~Ruiz, and S.~Shil, {\em {Automated
  predictions from polarized matrix elements}}.
  \href{http://dx.doi.org/10.1007/JHEP04(2020)082}{JHEP {\bf 04} (2020)  082},
  \href{http://arxiv.org/abs/1912.01725}{{\tt arXiv:1912.01725 [hep-ph]}}.

\bibitem{Alwall:2014hca}
J.~Alwall, R.~Frederix, S.~Frixione, V.~Hirschi, F.~Maltoni, O.~Mattelaer,
  H.~S. Shao, T.~Stelzer, P.~Torrielli, and M.~Zaro, {\em {The automated
  computation of tree-level and next-to-leading order differential cross
  sections, and their matching to parton shower simulations}}.
  \href{http://dx.doi.org/10.1007/JHEP07(2014)079}{JHEP {\bf 07} (2014)  079},
  \href{http://arxiv.org/abs/1405.0301}{{\tt arXiv:1405.0301 [hep-ph]}}.

\bibitem{Artoisenet:2012st}
P.~Artoisenet, R.~Frederix, O.~Mattelaer, and R.~Rietkerk, {\em {Automatic
  spin-entangled decays of heavy resonances in Monte Carlo simulations}}.
  \href{http://dx.doi.org/10.1007/JHEP03(2013)015}{JHEP {\bf 03} (2013)  015},
  \href{http://arxiv.org/abs/1212.3460}{{\tt arXiv:1212.3460 [hep-ph]}}.

\bibitem{Kaplan:1983fs}
D.~B. Kaplan and H.~Georgi, {\em {SU(2) x U(1) Breaking by Vacuum
  Misalignment}}. \href{http://dx.doi.org/10.1016/0370-2693(84)91177-8}{Phys.
  Lett. B {\bf 136} (1984)  183--186}.

\bibitem{Kaplan:1983sm}
D.~B. Kaplan, H.~Georgi, and S.~Dimopoulos, {\em {Composite Higgs Scalars}}.
  \href{http://dx.doi.org/10.1016/0370-2693(84)91178-X}{Phys. Lett. B {\bf 136}
  (1984)  187--190}.

\bibitem{Georgi:1984af}
H.~Georgi and D.~B. Kaplan, {\em {Composite Higgs and Custodial SU(2)}}.
  \href{http://dx.doi.org/10.1016/0370-2693(84)90341-1}{Phys. Lett. B {\bf 145}
  (1984)  216--220}.

\bibitem{Dugan:1984hq}
M.~J. Dugan, H.~Georgi, and D.~B. Kaplan, {\em {Anatomy of a Composite Higgs
  Model}}. \href{http://dx.doi.org/10.1016/0550-3213(85)90221-4}{Nucl. Phys. B
  {\bf 254} (1985)  299--326}.

\bibitem{Contino:2003ve}
R.~Contino, Y.~Nomura, and A.~Pomarol, {\em {Higgs as a holographic
  pseudoGoldstone boson}}.
  \href{http://dx.doi.org/10.1016/j.nuclphysb.2003.08.027}{Nucl. Phys. B {\bf
  671} (2003)  148--174}, \href{http://arxiv.org/abs/hep-ph/0306259}{{\tt
  arXiv:hep-ph/0306259}}.

\bibitem{Agashe:2004rs}
K.~Agashe, R.~Contino, and A.~Pomarol, {\em {The Minimal composite Higgs
  model}}. \href{http://dx.doi.org/10.1016/j.nuclphysb.2005.04.035}{Nucl. Phys.
  B {\bf 719} (2005)  165--187},
  \href{http://arxiv.org/abs/hep-ph/0412089}{{\tt arXiv:hep-ph/0412089}}.

\bibitem{Contino:2006qr}
R.~Contino, L.~Da~Rold, and A.~Pomarol, {\em {Light custodians in natural
  composite Higgs models}}.
  \href{http://dx.doi.org/10.1103/PhysRevD.75.055014}{Phys. Rev. D {\bf 75}
  (2007)  055014}, \href{http://arxiv.org/abs/hep-ph/0612048}{{\tt
  arXiv:hep-ph/0612048}}.

\bibitem{Agashe:2006at}
K.~Agashe, R.~Contino, L.~Da~Rold, and A.~Pomarol, {\em {A Custodial symmetry
  for $Zb \bar b$}}.
  \href{http://dx.doi.org/10.1016/j.physletb.2006.08.005}{Phys. Lett. B {\bf
  641} (2006)  62--66}, \href{http://arxiv.org/abs/hep-ph/0605341}{{\tt
  arXiv:hep-ph/0605341}}.

\bibitem{Bellazzini:2014yua}
B.~Bellazzini, C.~Cs\'aki, and J.~Serra, {\em {Composite Higgses}}.
  \href{http://dx.doi.org/10.1140/epjc/s10052-014-2766-x}{Eur. Phys. J. C {\bf
  74} (2014) no.~5, 2766}, \href{http://arxiv.org/abs/1401.2457}{{\tt
  arXiv:1401.2457 [hep-ph]}}.

\bibitem{Panico:2015jxa}
G.~Panico and A.~Wulzer,
  \href{http://dx.doi.org/10.1007/978-3-319-22617-0}{{\em {The Composite
  Nambu-Goldstone Higgs}}}, vol.~913.
\newblock Springer, 2016.
\newblock \href{http://arxiv.org/abs/1506.01961}{{\tt arXiv:1506.01961
  [hep-ph]}}.

\bibitem{Han:1992hr}
T.~Han, G.~Valencia, and S.~Willenbrock, {\em {Structure function approach to
  vector boson scattering in pp collisions}}.
  \href{http://dx.doi.org/10.1103/PhysRevLett.69.3274}{Phys. Rev. Lett. {\bf
  69} (1992)  3274--3277}, \href{http://arxiv.org/abs/hep-ph/9206246}{{\tt
  arXiv:hep-ph/9206246}}.

\bibitem{Jager:2006zc}
B.~J{\"a}ger, C.~Oleari, and D.~Zeppenfeld, {\em {Next-to-leading order QCD
  corrections to $W^+W^-$ production via vector-boson fusion}}.
  \href{http://dx.doi.org/10.1088/1126-6708/2006/07/015}{JHEP {\bf 07} (2006)
  015}, \href{http://arxiv.org/abs/hep-ph/0603177}{{\tt arXiv:hep-ph/0603177}}.

\bibitem{Jager:2006cp}
B.~J{\"a}ger, C.~Oleari, and D.~Zeppenfeld, {\em {Next-to-leading order QCD
  corrections to Z boson pair production via vector-boson fusion}}.
  \href{http://dx.doi.org/10.1103/PhysRevD.73.113006}{Phys. Rev. D {\bf 73}
  (2006)  113006}, \href{http://arxiv.org/abs/hep-ph/0604200}{{\tt
  arXiv:hep-ph/0604200}}.

\bibitem{Bozzi:2007ur}
G.~Bozzi, B.~J{\"a}ger, C.~Oleari, and D.~Zeppenfeld, {\em {Next-to-leading
  order QCD corrections to $W^+Z$ and $W^-Z$ production via vector-boson
  fusion}}. \href{http://dx.doi.org/10.1103/PhysRevD.75.073004}{Phys. Rev. D
  {\bf 75} (2007)  073004}, \href{http://arxiv.org/abs/hep-ph/0701105}{{\tt
  arXiv:hep-ph/0701105}}.

\bibitem{Jager:2009xx}
B.~J{\"a}ger, C.~Oleari, and D.~Zeppenfeld, {\em {Next-to-leading order QCD
  corrections to $W^+ W^+ jj$ and $W^- W^- jj$ production via weak-boson
  fusion}}. \href{http://dx.doi.org/10.1103/PhysRevD.80.034022}{Phys. Rev. D
  {\bf 80} (2009)  034022}, \href{http://arxiv.org/abs/0907.0580}{{\tt
  arXiv:0907.0580 [hep-ph]}}.

\bibitem{Denner:2019tmn}
A.~Denner, S.~Dittmaier, P.~Maierh\"ofer, M.~Pellen, and C.~Schwan, {\em {QCD
  and electroweak corrections to WZ scattering at the LHC}}.
  \href{http://dx.doi.org/10.1007/JHEP06(2019)067}{JHEP {\bf 06} (2019)  067},
  \href{http://arxiv.org/abs/1904.00882}{{\tt arXiv:1904.00882 [hep-ph]}}.

\bibitem{Denner:2020zit}
A.~Denner, R.~Franken, M.~Pellen, and T.~Schmidt, {\em {NLO QCD and EW
  corrections to vector-boson scattering into ZZ at the LHC}}.
  \href{http://dx.doi.org/10.1007/JHEP11(2020)110}{JHEP {\bf 11} (2020)  110},
  \href{http://arxiv.org/abs/2009.00411}{{\tt arXiv:2009.00411 [hep-ph]}}.

\bibitem{Bolzoni:2010xr}
P.~Bolzoni, F.~Maltoni, S.-O. Moch, and M.~Zaro, {\em {Higgs production via
  vector-boson fusion at NNLO in QCD}}.
  \href{http://dx.doi.org/10.1103/PhysRevLett.105.011801}{Phys. Rev. Lett. {\bf
  105} (2010)  011801}, \href{http://arxiv.org/abs/1003.4451}{{\tt
  arXiv:1003.4451 [hep-ph]}}.

\bibitem{Bolzoni:2011cu}
P.~Bolzoni, F.~Maltoni, S.-O. Moch, and M.~Zaro, {\em {Vector boson fusion at
  NNLO in QCD: SM Higgs and beyond}}.
  \href{http://dx.doi.org/10.1103/PhysRevD.85.035002}{Phys. Rev. D {\bf 85}
  (2012)  035002}, \href{http://arxiv.org/abs/1109.3717}{{\tt arXiv:1109.3717
  [hep-ph]}}.

\bibitem{Cacciari:2015jma}
M.~Cacciari, F.~A. Dreyer, A.~Karlberg, G.~P. Salam, and G.~Zanderighi, {\em
  {Fully Differential Vector-Boson-Fusion Higgs Production at
  Next-to-Next-to-Leading Order}}.
  \href{http://dx.doi.org/10.1103/PhysRevLett.115.082002}{Phys. Rev. Lett. {\bf
  115} (2015) no.~8, 082002}, \href{http://arxiv.org/abs/1506.02660}{{\tt
  arXiv:1506.02660 [hep-ph]}}. [Erratum: Phys.Rev.Lett. 120, 139901 (2018)].

\bibitem{Dreyer:2016oyx}
F.~A. Dreyer and A.~Karlberg, {\em {Vector-Boson Fusion Higgs Production at
  Three Loops in QCD}}.
  \href{http://dx.doi.org/10.1103/PhysRevLett.117.072001}{Phys. Rev. Lett. {\bf
  117} (2016) no.~7, 072001}, \href{http://arxiv.org/abs/1606.00840}{{\tt
  arXiv:1606.00840 [hep-ph]}}.

\bibitem{Dreyer:2018qbw}
F.~A. Dreyer and A.~Karlberg, {\em {Vector-Boson Fusion Higgs Pair Production
  at N$^3$LO}}. \href{http://dx.doi.org/10.1103/PhysRevD.98.114016}{Phys. Rev.
  D {\bf 98} (2018) no.~11, 114016},
  \href{http://arxiv.org/abs/1811.07906}{{\tt arXiv:1811.07906 [hep-ph]}}.

\bibitem{Dreyer:2018rfu}
F.~A. Dreyer and A.~Karlberg, {\em {Fully differential Vector-Boson Fusion
  Higgs Pair Production at Next-to-Next-to-Leading Order}}.
  \href{http://dx.doi.org/10.1103/PhysRevD.99.074028}{Phys. Rev. D {\bf 99}
  (2019) no.~7, 074028}, \href{http://arxiv.org/abs/1811.07918}{{\tt
  arXiv:1811.07918 [hep-ph]}}.

\bibitem{Dreyer:2020xaj}
F.~A. Dreyer, A.~Karlberg, J.-N. Lang, and M.~Pellen, {\em {Precise predictions
  for double-Higgs production via vector-boson fusion}}.
  \href{http://arxiv.org/abs/2005.13341}{{\tt arXiv:2005.13341 [hep-ph]}}.

\bibitem{Liu:2019tuy}
T.~Liu, K.~Melnikov, and A.~A. Penin, {\em {Nonfactorizable QCD Effects in
  Higgs Boson Production via Vector Boson Fusion}}.
  \href{http://dx.doi.org/10.1103/PhysRevLett.123.122002}{Phys. Rev. Lett. {\bf
  123} (2019) no.~12, 122002}, \href{http://arxiv.org/abs/1906.10899}{{\tt
  arXiv:1906.10899 [hep-ph]}}.

\bibitem{Dreyer:2020urf}
F.~A. Dreyer, A.~Karlberg, and L.~Tancredi, {\em {On the impact of
  non-factorisable corrections in VBF single and double Higgs production}}.
  \href{http://arxiv.org/abs/2005.11334}{{\tt arXiv:2005.11334 [hep-ph]}}.

\bibitem{Jager:2020hkz}
B.~J\"ager, A.~Karlberg, S.~Pl\"atzer, J.~Scheller, and M.~Zaro, {\em
  {Parton-shower effects in Higgs production via Vector-Boson Fusion}}.
  \href{http://dx.doi.org/10.1140/epjc/s10052-020-8326-7}{Eur. Phys. J. C {\bf
  80} (2020) no.~8, 756}, \href{http://arxiv.org/abs/2003.12435}{{\tt
  arXiv:2003.12435 [hep-ph]}}.

\bibitem{Jager:2011ms}
B.~J{\"a}ger and G.~Zanderighi, {\em {NLO corrections to electroweak and QCD
  production of $W^+W^+$ plus two jets in the POWHEGBOX}}.
  \href{http://dx.doi.org/10.1007/JHEP11(2011)055}{JHEP {\bf 11} (2011)  055},
  \href{http://arxiv.org/abs/1108.0864}{{\tt arXiv:1108.0864 [hep-ph]}}.

\bibitem{Jager:2013mu}
B.~J{\"a}ger and G.~Zanderighi, {\em {Electroweak $W^+W^-jj$ prodution at NLO
  in QCD matched with parton shower in the POWHEG-BOX}}.
  \href{http://dx.doi.org/10.1007/JHEP04(2013)024}{JHEP {\bf 04} (2013)  024},
  \href{http://arxiv.org/abs/1301.1695}{{\tt arXiv:1301.1695 [hep-ph]}}.

\bibitem{Jager:2013iza}
B.~J\"ager, A.~Karlberg, and G.~Zanderighi, {\em {Electroweak $ZZjj$ production
  in the Standard Model and beyond in the POWHEG-BOX V2}}.
  \href{http://dx.doi.org/10.1007/JHEP03(2014)141}{JHEP {\bf 03} (2014)  141},
  \href{http://arxiv.org/abs/1312.3252}{{\tt arXiv:1312.3252 [hep-ph]}}.

\bibitem{Jager:2018cyo}
B.~J{\"a}ger, A.~Karlberg, and J.~Scheller, {\em {Parton-shower effects in
  electroweak $WZjj$ production at the next-to-leading order of QCD}}.
  \href{http://dx.doi.org/10.1140/epjc/s10052-019-6736-1}{Eur. Phys. J. C {\bf
  79} (2019) no.~3, 226}, \href{http://arxiv.org/abs/1812.05118}{{\tt
  arXiv:1812.05118 [hep-ph]}}.

\bibitem{Sjostrand:2007gs}
T.~Sj{\"o}strand, S.~Mrenna, and P.~Z. Skands, {\em {A Brief Introduction to
  PYTHIA 8.1}}. \href{http://dx.doi.org/10.1016/j.cpc.2008.01.036}{Comput.
  Phys. Commun. {\bf 178} (2008)  852--867},
  \href{http://arxiv.org/abs/0710.3820}{{\tt arXiv:0710.3820 [hep-ph]}}.

\bibitem{Sjostrand:2014zea}
T.~Sj\"ostrand, S.~Ask, J.~R. Christiansen, R.~Corke, N.~Desai, P.~Ilten,
  S.~Mrenna, S.~Prestel, C.~O. Rasmussen, and P.~Z. Skands, {\em {An
  introduction to PYTHIA 8.2}}.
  \href{http://dx.doi.org/10.1016/j.cpc.2015.01.024}{Comput. Phys. Commun. {\bf
  191} (2015)  159--177}, \href{http://arxiv.org/abs/1410.3012}{{\tt
  arXiv:1410.3012 [hep-ph]}}.

\bibitem{Bahr:2008pv}
M.~Bahr {\em et al.}, {\em {Herwig++ Physics and Manual}}.
  \href{http://dx.doi.org/10.1140/epjc/s10052-008-0798-9}{Eur. Phys. J. C {\bf
  58} (2008)  639--707}, \href{http://arxiv.org/abs/0803.0883}{{\tt
  arXiv:0803.0883 [hep-ph]}}.

\bibitem{Bellm:2015jjp}
J.~Bellm {\em et al.}, {\em {Herwig 7.0/Herwig++ 3.0 release note}}.
  \href{http://dx.doi.org/10.1140/epjc/s10052-016-4018-8}{Eur. Phys. J. C {\bf
  76} (2016) no.~4, 196}, \href{http://arxiv.org/abs/1512.01178}{{\tt
  arXiv:1512.01178 [hep-ph]}}.

\bibitem{Bellm:2019zci}
J.~Bellm {\em et al.}, {\em {Herwig 7.2 release note}}.
  \href{http://dx.doi.org/10.1140/epjc/s10052-020-8011-x}{Eur. Phys. J. C {\bf
  80} (2020) no.~5, 452}, \href{http://arxiv.org/abs/1912.06509}{{\tt
  arXiv:1912.06509 [hep-ph]}}.

\bibitem{Nason:2004rx}
P.~Nason, {\em {A New method for combining NLO QCD with shower Monte Carlo
  algorithms}}. \href{http://dx.doi.org/10.1088/1126-6708/2004/11/040}{JHEP
  {\bf 11} (2004)  040}, \href{http://arxiv.org/abs/hep-ph/0409146}{{\tt
  arXiv:hep-ph/0409146}}.

\bibitem{Frixione:2007vw}
S.~Frixione, P.~Nason, and C.~Oleari, {\em {Matching NLO QCD computations with
  Parton Shower simulations: the POWHEG method}}.
  \href{http://dx.doi.org/10.1088/1126-6708/2007/11/070}{JHEP {\bf 11} (2007)
  070}, \href{http://arxiv.org/abs/0709.2092}{{\tt arXiv:0709.2092 [hep-ph]}}.

\bibitem{Frixione:2002ik}
S.~Frixione and B.~R. Webber, {\em {Matching NLO QCD computations and parton
  shower simulations}}.
  \href{http://dx.doi.org/10.1088/1126-6708/2002/06/029}{JHEP {\bf 06} (2002)
  029}, \href{http://arxiv.org/abs/hep-ph/0204244}{{\tt arXiv:hep-ph/0204244}}.

\bibitem{Sirunyan:2017jej}
{\bf CMS} Collaboration, A.~M. Sirunyan {\em et al.}, {\em {Electroweak
  production of two jets in association with a Z boson in
  proton\textendash{}proton collisions at $\sqrt{s}= $ 13 $\,\text {TeV}$}}.
  \href{http://dx.doi.org/10.1140/epjc/s10052-018-6049-9}{Eur. Phys. J. C {\bf
  78} (2018) no.~7, 589}, \href{http://arxiv.org/abs/1712.09814}{{\tt
  arXiv:1712.09814 [hep-ex]}}.

\bibitem{Cabouat:2017rzi}
B.~Cabouat and T.~Sj\"ostrand, {\em {Some Dipole Shower Studies}}.
  \href{http://dx.doi.org/10.1140/epjc/s10052-018-5645-z}{Eur. Phys. J. C {\bf
  78} (2018) no.~3, 226}, \href{http://arxiv.org/abs/1710.00391}{{\tt
  arXiv:1710.00391 [hep-ph]}}.

\bibitem{Frederix:2020trv}
R.~Frederix, S.~Frixione, S.~Prestel, and P.~Torrielli, {\em {On the reduction
  of negative weights in MC@NLO-type matching procedures}}.
  \href{http://dx.doi.org/10.1007/JHEP07(2020)238}{JHEP {\bf 07} (2020)  238},
  \href{http://arxiv.org/abs/2002.12716}{{\tt arXiv:2002.12716 [hep-ph]}}.

\bibitem{Butterworth:2015oua}
J.~Butterworth {\em et al.}, {\em {PDF4LHC recommendations for LHC Run II}}.
  \href{http://dx.doi.org/10.1088/0954-3899/43/2/023001}{J. Phys. G {\bf 43}
  (2016)  023001}, \href{http://arxiv.org/abs/1510.03865}{{\tt arXiv:1510.03865
  [hep-ph]}}.

\bibitem{Bellan:2019xpr}
R.~Bellan {\em et al.}, ``{VBSCan Thessaloniki 2018 Workshop Summary},'' in
  {\em {2nd Vector Boson Scattering Coordination and Action Network Annual
  Meeting}}.
\newblock 6, 2019.
\newblock \href{http://arxiv.org/abs/1906.11332}{{\tt arXiv:1906.11332
  [hep-ph]}}.

\bibitem{Zaro:2010fc}
M.~Zaro, P.~Bolzoni, F.~Maltoni, and S.-O. Moch, {\em {Charged Higgs production
  via vector-boson fusion at NNLO in QCD}}.
  \href{http://dx.doi.org/10.22323/1.114.0028}{PoS {\bf CHARGED2010} (2010)
  028}, \href{http://arxiv.org/abs/1012.1806}{{\tt arXiv:1012.1806 [hep-ph]}}.

\bibitem{Zaro:2015ika}
M.~Zaro and H.~Logan, {\em {Recommendations for the interpretation of LHC
  searches for $H_5^0$, $H_5^{\pm}$, and $H_5^{\pm\pm}$ in vector boson fusion
  with decays to vector boson pairs, LHCHXSWG-2015-001}} tech. rep., 2015.

\bibitem{deFlorian:2016spz}
{\bf LHC Higgs Cross Section Working Group} Collaboration, D.~de~Florian {\em
  et al.}, {\em {Handbook of LHC Higgs Cross Sections: 4. Deciphering the
  Nature of the Higgs Sector}}. \href{http://arxiv.org/abs/1610.07922}{{\tt
  arXiv:1610.07922 [hep-ph]}}.

\bibitem{Denner:2019vbn}
A.~Denner and S.~Dittmaier, {\em {Electroweak Radiative Corrections for
  Collider Physics}}.
  \href{http://dx.doi.org/10.1016/j.physrep.2020.04.001}{Phys. Rept. {\bf 864}
  (2020)  1--163},
\href{http://arxiv.org/abs/1912.06823}{{\tt arXiv:1912.06823 [hep-ph]}}.

\bibitem{Biedermann:2016yds}
B.~Biedermann, A.~Denner, and M.~Pellen, {\em {Large electroweak corrections to
  vector-boson scattering at the Large Hadron Collider}}.
  \href{http://dx.doi.org/10.1103/PhysRevLett.118.261801}{Phys. Rev. Lett. {\bf
  118} (2017) no.~26, 261801},
\href{http://arxiv.org/abs/1611.02951}{{\tt arXiv:1611.02951 [hep-ph]}}.

\bibitem{Denner:1997kq}
A.~Denner and T.~Hahn, {\em {Radiative corrections to ${\rm W}^+{\rm W}^- \to
  {\rm W}^+{\rm W}^-$ in the electroweak standard model}}.
  \href{http://dx.doi.org/10.1016/S0550-3213(98)00287-9}{Nucl. Phys. B {\bf
  525} (1998)  27--50}, \href{http://arxiv.org/abs/hep-ph/9711302}{{\tt
  arXiv:hep-ph/9711302}}.

\bibitem{Denner:2000jv}
A.~Denner and S.~Pozzorini, {\em {One loop leading logarithms in electroweak
  radiative corrections. 1. Results}}.
  \href{http://dx.doi.org/10.1007/s100520100551}{Eur. Phys. J. C {\bf 18}
  (2001)  461--480}, \href{http://arxiv.org/abs/hep-ph/0010201}{{\tt
  arXiv:hep-ph/0010201}}.

\bibitem{Chiesa:2019ulk}
M.~Chiesa, A.~Denner, J.-N. Lang, and M.~Pellen, {\em {An event generator for
  same-sign W-boson scattering at the LHC including electroweak corrections}}.
  \href{http://dx.doi.org/10.1140/epjc/s10052-019-7290-6}{Eur. Phys. J. C {\bf
  79} (2019) no.~9, 788}, \href{http://arxiv.org/abs/1906.01863}{{\tt
  arXiv:1906.01863 [hep-ph]}}.

\bibitem{Manohar:2016nzj}
A.~Manohar, P.~Nason, G.~P. Salam, and G.~Zanderighi, {\em {How bright is the
  proton? A precise determination of the photon parton distribution function}}.
  \href{http://dx.doi.org/10.1103/PhysRevLett.117.242002}{Phys. Rev. Lett. {\bf
  117} (2016) no.~24, 242002}, \href{http://arxiv.org/abs/1607.04266}{{\tt
  arXiv:1607.04266 [hep-ph]}}.

\bibitem{Manohar:2017eqh}
A.~V. Manohar, P.~Nason, G.~P. Salam, and G.~Zanderighi, {\em {The Photon
  Content of the Proton}}.
  \href{http://dx.doi.org/10.1007/JHEP12(2017)046}{JHEP {\bf 12} (2017)  046},
  \href{http://arxiv.org/abs/1708.01256}{{\tt arXiv:1708.01256 [hep-ph]}}.

\bibitem{Martin:2004dh}
A.~Martin, R.~Roberts, W.~Stirling, and R.~Thorne, {\em {Parton distributions
  incorporating QED contributions}}.
  \href{http://dx.doi.org/10.1140/epjc/s2004-02088-7}{Eur. Phys. J. C {\bf 39}
  (2005)  155--161}, \href{http://arxiv.org/abs/hep-ph/0411040}{{\tt
  arXiv:hep-ph/0411040}}.

\bibitem{Schmidt:2015zda}
C.~Schmidt, J.~Pumplin, D.~Stump, and C.~Yuan, {\em {CT14QED parton
  distribution functions from isolated photon production in deep inelastic
  scattering}}. \href{http://dx.doi.org/10.1103/PhysRevD.93.114015}{Phys. Rev.
  D {\bf 93} (2016) no.~11, 114015},
  \href{http://arxiv.org/abs/1509.02905}{{\tt arXiv:1509.02905 [hep-ph]}}.

\bibitem{Ball:2013hta}
{\bf NNPDF} Collaboration, R.~D. Ball, V.~Bertone, S.~Carrazza, L.~Del~Debbio,
  S.~Forte, A.~Guffanti, N.~P. Hartland, and J.~Rojo, {\em {Parton
  distributions with QED corrections}}.
  \href{http://dx.doi.org/10.1016/j.nuclphysb.2013.10.010}{Nucl. Phys. B {\bf
  877} (2013)  290--320}, \href{http://arxiv.org/abs/1308.0598}{{\tt
  arXiv:1308.0598 [hep-ph]}}.

\bibitem{Bertone:2017bme}
{\bf NNPDF} Collaboration, V.~Bertone, S.~Carrazza, N.~P. Hartland, and
  J.~Rojo, {\em {Illuminating the photon content of the proton within a global
  PDF analysis}}. \href{http://dx.doi.org/10.21468/SciPostPhys.5.1.008}{SciPost
  Phys. {\bf 5} (2018) no.~1, 008}, \href{http://arxiv.org/abs/1712.07053}{{\tt
  arXiv:1712.07053 [hep-ph]}}.

\bibitem{Harland-Lang:2019pla}
L.~Harland-Lang, A.~Martin, R.~Nathvani, and R.~Thorne, {\em {Ad Lucem: QED
  Parton Distribution Functions in the MMHT Framework}}.
  \href{http://dx.doi.org/10.1140/epjc/s10052-019-7296-0}{Eur. Phys. J. C {\bf
  79} (2019) no.~10, 811}, \href{http://arxiv.org/abs/1907.02750}{{\tt
  arXiv:1907.02750 [hep-ph]}}.

\bibitem{Frixione:2014qaa}
S.~Frixione, V.~Hirschi, D.~Pagani, H.~Shao, and M.~Zaro, {\em {Weak
  corrections to Higgs hadroproduction in association with a top-quark pair}}.
  \href{http://dx.doi.org/10.1007/JHEP09(2014)065}{JHEP {\bf 09} (2014)  065},
  \href{http://arxiv.org/abs/1407.0823}{{\tt arXiv:1407.0823 [hep-ph]}}.

\bibitem{Frixione:2015zaa}
S.~Frixione, V.~Hirschi, D.~Pagani, H.~S. Shao, and M.~Zaro, {\em {Electroweak
  and QCD corrections to top-pair hadroproduction in association with heavy
  bosons}}. \href{http://dx.doi.org/10.1007/JHEP06(2015)184}{JHEP {\bf 06}
  (2015)  184}, \href{http://arxiv.org/abs/1504.03446}{{\tt arXiv:1504.03446
  [hep-ph]}}.

\bibitem{Czakon:2017wor}
M.~Czakon, D.~Heymes, A.~Mitov, D.~Pagani, I.~Tsinikos, and M.~Zaro, {\em
  {Top-pair production at the LHC through NNLO QCD and NLO EW}}.
  \href{http://dx.doi.org/10.1007/JHEP10(2017)186}{JHEP {\bf 10} (2017)  186},
  \href{http://arxiv.org/abs/1705.04105}{{\tt arXiv:1705.04105 [hep-ph]}}.

\bibitem{Azzi:2019yne}
P.~Azzi {\em et al.}, {\em {Report from Working Group 1}}.
  \href{http://dx.doi.org/10.23731/CYRM-2019-007.1}{CERN Yellow Rep. Monogr.
  {\bf 7} (2019)  1--220},
\href{http://arxiv.org/abs/1902.04070}{{\tt arXiv:1902.04070 [hep-ph]}}.

\bibitem{Denner:2019zfp}
A.~Denner, S.~Dittmaier, M.~Pellen, and C.~Schwan, {\em {Low-virtuality photon
  transitions $\gamma^*\to f\bar f$ and the photon-to-jet conversion
  function}}. \href{http://dx.doi.org/10.1016/j.physletb.2019.134951}{Phys.
  Lett. B {\bf 798} (2019)  134951},
  \href{http://arxiv.org/abs/1907.02366}{{\tt arXiv:1907.02366 [hep-ph]}}.

\bibitem{Aad:2008zzm}
{\bf ATLAS} Collaboration, G.~Aad {\em et al.}, {\em {The ATLAS Experiment at
  the CERN Large Hadron Collider}}.
  \href{http://dx.doi.org/10.1088/1748-0221/3/08/S08003}{JINST {\bf 3} (2008)
  S08003}.

\bibitem{Chatrchyan:2008aa}
{\bf CMS} Collaboration, S.~Chatrchyan {\em et al.}, {\em {The CMS Experiment
  at the CERN LHC}}.
  \href{http://dx.doi.org/10.1088/1748-0221/3/08/S08004}{JINST {\bf 3} (2008)
  S08004}.

\bibitem{Pellen:2019ywl}
M.~Pellen, {\em {Exploring the scattering of vector bosons at LHCb}}.
  \href{http://dx.doi.org/10.1103/PhysRevD.101.013002}{Phys. Rev. D {\bf 101}
  (2020) no.~1, 013002}, \href{http://arxiv.org/abs/1908.06805}{{\tt
  arXiv:1908.06805 [hep-ph]}}.

\bibitem{Aad:2020wji}
{\bf ATLAS} Collaboration, G.~Aad {\em et al.}, {\em {Operation of the ATLAS
  trigger system in Run 2}}.
  \href{http://dx.doi.org/10.1088/1748-0221/15/10/P10004}{JINST {\bf 15} (2020)
  no.~10, P10004}, \href{http://arxiv.org/abs/2007.12539}{{\tt arXiv:2007.12539
  [physics.ins-det]}}.

\bibitem{Khachatryan:2016bia}
{\bf CMS} Collaboration, V.~Khachatryan {\em et al.}, {\em {The CMS trigger
  system}}. \href{http://dx.doi.org/10.1088/1748-0221/12/01/P01020}{JINST {\bf
  12} (2017)  P01020},
\href{http://arxiv.org/abs/1609.02366}{{\tt arXiv:1609.02366
  [physics.ins-det]}}.

\bibitem{Aad:2016upy}
{\bf ATLAS} Collaboration, G.~Aad {\em et al.}, {\em {Topological cell
  clustering in the ATLAS calorimeters and its performance in LHC Run 1}}.
  \href{http://dx.doi.org/10.1140/epjc/s10052-017-5004-5}{Eur. Phys. J. C {\bf
  77} (2017)  490}, \href{http://arxiv.org/abs/1603.02934}{{\tt
  arXiv:1603.02934 [hep-ex]}}.

\bibitem{Sirunyan:2017ulk}
{\bf CMS} Collaboration, A.~Sirunyan {\em et al.}, {\em {Particle-flow
  reconstruction and global event description with the CMS detector}}.
  \href{http://dx.doi.org/10.1088/1748-0221/12/10/P10003}{JINST {\bf 12} (2017)
  no.~10, P10003}, \href{http://arxiv.org/abs/1706.04965}{{\tt arXiv:1706.04965
  [physics.ins-det]}}.

\bibitem{Cacciari:2008gp}
M.~Cacciari, G.~P. Salam, and G.~Soyez, {\em The anti-$k_\textrm{T}$ jet
  clustering algorithm}.
  \href{http://dx.doi.org/10.1088/1126-6708/2008/04/063}{JHEP {\bf 04} (2008)
  063}, \href{http://arxiv.org/abs/0802.1189}{{\tt arXiv:0802.1189 [hep-ex]}}.

\bibitem{Cacciari:2011ma}
M.~Cacciari, G.~P. Salam, and G.~Soyez, {\em {FastJet user manual}}.
  \href{http://dx.doi.org/10.1140/epjc/s10052-012-1896-2}{Eur. Phys. J. C {\bf
  72} (2012)  1896}, \href{http://arxiv.org/abs/1111.6097}{{\tt arXiv:1111.6097
  [hep-ph]}}.

\bibitem{Aaboud:2017jcu}
{\bf ATLAS} Collaboration, M.~Aaboud {\em et al.}, {\em {Jet energy scale
  measurements and their systematic uncertainties in proton-proton collisions
  at $\sqrt{s} = 13$ TeV with the ATLAS detector}}.
  \href{http://dx.doi.org/10.1103/PhysRevD.96.072002}{Phys. Rev. D {\bf 96}
  (2017) no.~7, 072002}, \href{http://arxiv.org/abs/1703.09665}{{\tt
  arXiv:1703.09665 [hep-ex]}}.

\bibitem{Khachatryan:2016kdb}
{\bf CMS} Collaboration, V.~Khachatryan {\em et al.}, {\em {Jet energy scale
  and resolution in the CMS experiment in pp collisions at 8 TeV}}.
  \href{http://dx.doi.org/10.1088/1748-0221/12/02/P02014}{JINST {\bf 12} (2017)
  no.~02, P02014}, \href{http://arxiv.org/abs/1607.03663}{{\tt arXiv:1607.03663
  [hep-ex]}}.

\bibitem{Cacciari:2007fd}
M.~Cacciari and G.~P. Salam, {\em Pileup subtraction using jet areas}.
  \href{http://dx.doi.org/10.1016/j.physletb.2007.09.077}{Phys. Lett. B {\bf
  659} (2008)  119}, \href{http://arxiv.org/abs/0707.1378}{{\tt arXiv:0707.1378
  [hep-ph]}}.

\bibitem{Aad:2015ina}
{\bf ATLAS} Collaboration, G.~Aad {\em et al.}, {\em {Performance of pile-up
  mitigation techniques for jets in $pp$ collisions at $\sqrt{s}=8$ TeV using
  the ATLAS detector}}.
  \href{http://dx.doi.org/10.1140/epjc/s10052-016-4395-z}{Eur. Phys. J. C {\bf
  76} (2016) no.~11, 581}, \href{http://arxiv.org/abs/1510.03823}{{\tt
  arXiv:1510.03823 [hep-ex]}}.

\bibitem{Aaboud:2017pou}
{\bf ATLAS} Collaboration, M.~Aaboud {\em et al.}, {\em {Identification and
  rejection of pile-up jets at high pseudorapidity with the ATLAS detector}}.
  \href{http://dx.doi.org/10.1140/epjc/s10052-017-5081-5}{Eur. Phys. J. C {\bf
  77} (2017) no.~9, 580}, \href{http://arxiv.org/abs/1705.02211}{{\tt
  arXiv:1705.02211 [hep-ex]}}. [Erratum: Eur.Phys.J.C 77, 712 (2017)].

\bibitem{Sirunyan:2020foa}
{\bf CMS} Collaboration, A.~M. Sirunyan {\em et al.}, {\em {Pileup mitigation
  at CMS in 13 TeV data}}.
  \href{http://dx.doi.org/10.1088/1748-0221/15/09/P09018}{JINST {\bf 15} (2020)
  no.~09, P09018}, \href{http://arxiv.org/abs/2003.00503}{{\tt arXiv:2003.00503
  [hep-ex]}}.

\bibitem{ATLAS:2016wzt}
{\bf ATLAS} Collaboration, {\em {Discrimination of Light Quark and Gluon Jets
  in $pp$ collisions at $\sqrt{s} = 8$ TeV with the ATLAS Detector,
  ATLAS-CONF-2016-034}} tech. rep., 2016.

\bibitem{CMS:2013kfa}
{\bf CMS} Collaboration, {\em {Performance of quark/gluon discrimination in 8
  TeV pp data, CMS-PAS-JME-13-002}} tech. rep., 2013.

\bibitem{jetsatLHC}
R.~Kogler {\em et al.}, {\em {Jet Substructure at the Large Hadron Collider:
  Experimental Review}}.
  \href{http://dx.doi.org/10.1103/RevModPhys.91.045003}{Rev. Mod. Phys. {\bf
  91} (2019) no.~4, 045003}, \href{http://arxiv.org/abs/1803.06991}{{\tt
  arXiv:1803.06991 [hep-ex]}}.

\bibitem{Aad:2016jkr}
{\bf ATLAS} Collaboration, G.~Aad {\em et al.}, {\em {Muon reconstruction
  performance of the ATLAS detector in proton\textendash{}proton collision data
  at $\sqrt{s}$ =13 TeV}}.
  \href{http://dx.doi.org/10.1140/epjc/s10052-016-4120-y}{Eur. Phys. J. C {\bf
  76} (2016) no.~5, 292}, \href{http://arxiv.org/abs/1603.05598}{{\tt
  arXiv:1603.05598 [hep-ex]}}.

\bibitem{Sirunyan:2018fpa}
{\bf CMS} Collaboration, A.~Sirunyan {\em et al.}, {\em {Performance of the CMS
  muon detector and muon reconstruction with proton-proton collisions at
  $\sqrt{s}=$ 13 TeV}}.
  \href{http://dx.doi.org/10.1088/1748-0221/13/06/P06015}{JINST {\bf 13} (2018)
  no.~06, P06015}, \href{http://arxiv.org/abs/1804.04528}{{\tt arXiv:1804.04528
  [physics.ins-det]}}.

\bibitem{Aaboud:2019ynx}
{\bf ATLAS} Collaboration, M.~Aaboud {\em et al.}, {\em {Electron
  reconstruction and identification in the ATLAS experiment using the 2015 and
  2016 LHC proton-proton collision data at $\sqrt{s}$ = 13 TeV}}.
  \href{http://dx.doi.org/10.1140/epjc/s10052-019-7140-6}{Eur. Phys. J. C {\bf
  79} (2019) no.~8, 639}, \href{http://arxiv.org/abs/1902.04655}{{\tt
  arXiv:1902.04655 [physics.ins-det]}}.

\bibitem{Aad:2019tso}
{\bf ATLAS} Collaboration, G.~Aad {\em et al.}, {\em {Electron and photon
  performance measurements with the ATLAS detector using the
  2015\textendash{}2017 LHC proton-proton collision data}}.
  \href{http://dx.doi.org/10.1088/1748-0221/14/12/P12006}{JINST {\bf 14} (2019)
  no.~12, P12006}, \href{http://arxiv.org/abs/1908.00005}{{\tt arXiv:1908.00005
  [hep-ex]}}.

\bibitem{Khachatryan:2015hwa}
{\bf CMS} Collaboration, V.~Khachatryan {\em et al.}, {\em {Performance of
  Electron Reconstruction and Selection with the CMS Detector in Proton-Proton
  Collisions at \ensuremath{\sqrt{}}s = 8 TeV}}.
  \href{http://dx.doi.org/10.1088/1748-0221/10/06/P06005}{JINST {\bf 10} (2015)
  no.~06, P06005}, \href{http://arxiv.org/abs/1502.02701}{{\tt arXiv:1502.02701
  [physics.ins-det]}}.

\bibitem{Speckmayer:2010zz}
P.~Speckmayer, A.~H{\"o}cker, J.~Stelzer, and H.~Voss, {\em {The toolkit for
  multivariate data analysis, TMVA 4}}.
  \href{http://dx.doi.org/10.1088/1742-6596/219/3/032057}{J. Phys. Conf. Ser.
  {\bf 219} (2010)  032057}.

\bibitem{Aaboud:2018tkc}
{\bf ATLAS} Collaboration, M.~Aaboud {\em et al.}, {\em {Performance of missing
  transverse momentum reconstruction with the ATLAS detector using
  proton-proton collisions at $\sqrt{s}$ = 13 TeV}}.
  \href{http://dx.doi.org/10.1140/epjc/s10052-018-6288-9}{Eur. Phys. J. C {\bf
  78} (2018) no.~11, 903}, \href{http://arxiv.org/abs/1802.08168}{{\tt
  arXiv:1802.08168 [hep-ex]}}.

\bibitem{Sirunyan:2019kia}
{\bf CMS} Collaboration, A.~M. Sirunyan {\em et al.}, {\em {Performance of
  missing transverse momentum reconstruction in proton-proton collisions at
  $\sqrt{s} =$ 13 TeV using the CMS detector}}.
  \href{http://dx.doi.org/10.1088/1748-0221/14/07/P07004}{JINST {\bf 14} (2019)
  no.~07, P07004}, \href{http://arxiv.org/abs/1903.06078}{{\tt arXiv:1903.06078
  [hep-ex]}}.

\bibitem{Alioli:2010xd}
S.~Alioli, P.~Nason, C.~Oleari, and E.~Re, {\em {A general framework for
  implementing NLO calculations in shower Monte Carlo programs: the POWHEG
  BOX}}. \href{http://dx.doi.org/10.1007/JHEP06(2010)043}{JHEP {\bf 06} (2010)
  043}, \href{http://arxiv.org/abs/1002.2581}{{\tt arXiv:1002.2581 [hep-ph]}}.

\bibitem{Bothmann:2019yzt}
{\bf Sherpa} Collaboration, E.~Bothmann {\em et al.}, {\em {Event Generation
  with Sherpa 2.2}}.
  \href{http://dx.doi.org/10.21468/SciPostPhys.7.3.034}{SciPost Phys. {\bf 7}
  (2019) no.~3, 034}, \href{http://arxiv.org/abs/1905.09127}{{\tt
  arXiv:1905.09127 [hep-ph]}}.

\bibitem{Ball:2014uwa}
{\bf NNPDF} Collaboration, R.~D. Ball {\em et al.}, {\em {Parton distributions
  for the LHC Run II}}. \href{http://dx.doi.org/10.1007/JHEP04(2015)040}{JHEP
  {\bf 04} (2015)  040}, \href{http://arxiv.org/abs/1410.8849}{{\tt
  arXiv:1410.8849 [hep-ph]}}.

\bibitem{Khachatryan:2015pea}
{\bf CMS} Collaboration, V.~Khachatryan {\em et al.}, {\em {Event generator
  tunes obtained from underlying event and multiparton scattering
  measurements}}. \href{http://dx.doi.org/10.1140/epjc/s10052-016-3988-x}{Eur.
  Phys. J. C {\bf 76} (2016) no.~3, 155},
  \href{http://arxiv.org/abs/1512.00815}{{\tt arXiv:1512.00815 [hep-ex]}}.

\bibitem{Sirunyan:2019dfx}
{\bf CMS} Collaboration, A.~M. Sirunyan {\em et al.}, {\em {Extraction and
  validation of a new set of CMS PYTHIA8 tunes from underlying-event
  measurements}}. \href{http://dx.doi.org/10.1140/epjc/s10052-019-7499-4}{Eur.
  Phys. J. C {\bf 80} (2020) no.~1, 4},
  \href{http://arxiv.org/abs/1903.12179}{{\tt arXiv:1903.12179 [hep-ex]}}.

\bibitem{GEANT}
{\bf GEANT4} Collaboration, S.~Agostinelli {\em et al.}, {\em {GEANT4}---a
  simulation toolkit}.
\href{http://dx.doi.org/10.1016/S0168-9002(03)01368-8}{Nucl. Instrum. Meth. A
  {\bf 506} (2003)  250}.

\bibitem{GEANT2}
J.~Allison {\em et al.}, {\em {GEANT4 developments and applications}}.
  \href{http://dx.doi.org/10.1109/TNS.2006.869826}{IEEE Trans. Nucl. Sci. {\bf
  53} (2006)  270}.

\bibitem{Degrande:2012wf}
C.~Degrande, N.~Greiner, W.~Kilian, O.~Mattelaer, H.~Mebane, T.~Stelzer,
  S.~Willenbrock, and C.~Zhang, {\em {Effective Field Theory: A Modern Approach
  to Anomalous Couplings}}.
  \href{http://dx.doi.org/10.1016/j.aop.2013.04.016}{Annals Phys. {\bf 335}
  (2013)  21--32}, \href{http://arxiv.org/abs/1205.4231}{{\tt arXiv:1205.4231
  [hep-ph]}}.

\bibitem{Eboli:2006wa}
O.~J.~P. \'Eboli, M.~Gonzalez\textendash{}Garcia, and J.~Mizukoshi, {\em {$p p
  \rightarrow j j e^\pm mu^\pm \nu \nu$ and $j j e^\pm mu^\mp \nu \nu$ at
  O($\alpha^6_{em}$) and O($\alpha^4_{em}\alpha^2_{s}$) for the study of the
  quartic electroweak gauge boson vertex at CERN LHC}}.
  \href{http://dx.doi.org/10.1103/PhysRevD.74.073005}{Phys. Rev. D {\bf 74}
  (2006)  073005}, \href{http://arxiv.org/abs/hep-ph/0606118}{{\tt
  arXiv:hep-ph/0606118}}.

\bibitem{Almeida:2020ylr}
E.~d.~S. Almeida, O.~J.~P. \'Eboli, and M.~Gonzalez\textendash{}Garcia, {\em
  {Unitarity constraints on anomalous quartic couplings}}.
  \href{http://dx.doi.org/10.1103/PhysRevD.101.113003}{Phys. Rev. D {\bf 101}
  (2020) no.~11, 113003}, \href{http://arxiv.org/abs/2004.05174}{{\tt
  arXiv:2004.05174 [hep-ph]}}.

\bibitem{Arnold:2008rz}
K.~Arnold {\em et al.}, {\em {VBFNLO: A Parton level Monte Carlo for processes
  with electroweak bosons}}.
  \href{http://dx.doi.org/10.1016/j.cpc.2009.03.006}{Comput. Phys. Commun. {\bf
  180} (2009)  1661--1670}, \href{http://arxiv.org/abs/0811.4559}{{\tt
  arXiv:0811.4559 [hep-ph]}}.

\bibitem{aQGCcomp}
{\bf CMS} Collaboration, {\em {Limits on anomalous triple and quartic gauge
  couplings}},
  \texttt{https://twiki.cern.ch/twiki/bin/view/CMSPublic/PhysicsResultsSMPaTGC},
  2020.
\newblock [Online].

\bibitem{Perez:2018kav}
G.~Perez, M.~Sekulla, and D.~Zeppenfeld, {\em {Anomalous quartic gauge
  couplings and unitarization for the vector boson scattering process
  $pp\rightarrow W^+W^+jjX\rightarrow \ell ^+\nu _\ell \ell ^+\nu _\ell jjX$}}.
  \href{http://dx.doi.org/10.1140/epjc/s10052-018-6230-1}{Eur. Phys. J. C {\bf
  78} (2018) no.~9, 759}, \href{http://arxiv.org/abs/1807.02707}{{\tt
  arXiv:1807.02707 [hep-ph]}}.

\bibitem{Gomez-Ambrosio:2018pnl}
R.~Gomez-Ambrosio, {\em {Studies of Dimension-Six EFT effects in Vector Boson
  Scattering}}. \href{http://dx.doi.org/10.1140/epjc/s10052-019-6893-2}{Eur.
  Phys. J. C {\bf 79} (2019) no.~5, 389},
  \href{http://arxiv.org/abs/1809.04189}{{\tt arXiv:1809.04189 [hep-ph]}}.

\bibitem{Dedes:2020xmo}
A.~Dedes, P.~Koz\'ow, and M.~Szleper, {\em {SM EFT effects in Vector-Boson
  Scattering at the LHC}}. \href{http://arxiv.org/abs/2011.07367}{{\tt
  arXiv:2011.07367 [hep-ph]}}.

\bibitem{Ethier:2021ydt}
J.~J. Ethier, R.~Gomez-Ambrosio, G.~Magni, and J.~Rojo, {\em {SMEFT analysis of
  vector boson scattering and diboson data from the LHC Run II}}.
  \href{http://arxiv.org/abs/2101.03180}{{\tt arXiv:2101.03180 [hep-ph]}}.

\bibitem{Denner:2012dz}
A.~Denner, L.~Ho{\v{s}}ekov{\'{a}}, and S.~Kallweit, {\em {NLO QCD corrections
  to ${\rm W}^+{\rm W}^+$jj production in vector-boson fusion at the LHC}}.
  \href{http://dx.doi.org/10.1103/PhysRevD.86.114014}{Phys. Rev. D {\bf 86}
  (2012)  114014}, \href{http://arxiv.org/abs/1209.2389}{{\tt arXiv:1209.2389
  [hep-ph]}}.

\bibitem{Arnold:2011wj}
J.~Baglio {\em et al.}, {\em {VBFNLO: A Parton Level Monte Carlo for Processes
  with Electroweak Bosons -- Manual for Version 2.7.0}}.
  \href{http://arxiv.org/abs/1107.4038}{{\tt arXiv:1107.4038 [hep-ph]}}.

\bibitem{Baglio:2014uba}
J.~Baglio {\em et al.}, {\em {Release Note - VBFNLO 2.7.0}}.
  \href{http://arxiv.org/abs/1404.3940}{{\tt arXiv:1404.3940 [hep-ph]}}.

\bibitem{Melia:2010bm}
T.~Melia, K.~Melnikov, R.~R{\"o}ntsch, and G.~Zanderighi, {\em {Next-to-leading
  order QCD predictions for $W^+W^+jj$ production at the LHC}}.
  \href{http://dx.doi.org/10.1007/JHEP12(2010)053}{JHEP {\bf 12} (2010)  053},
  \href{http://arxiv.org/abs/1007.5313}{{\tt arXiv:1007.5313 [hep-ph]}}.

\bibitem{Campanario:2013gea}
F.~Campanario, M.~Kerner, L.~D. Ninh, and D.~Zeppenfeld, {\em {Next-to-leading
  order QCD corrections to $W^+W^+$ and $W^-W^-$ production in association with
  two jets}}. \href{http://dx.doi.org/10.1103/PhysRevD.89.054009}{Phys. Rev. D
  {\bf 89} (2014) no.~5, 054009}, \href{http://arxiv.org/abs/1311.6738}{{\tt
  arXiv:1311.6738 [hep-ph]}}.

\bibitem{Melia:2011gk}
T.~Melia, P.~Nason, R.~R{\"o}ntsch, and G.~Zanderighi, {\em {$W^+W^+$ plus
  dijet production in the POWHEGBOX}}.
  \href{http://dx.doi.org/10.1140/epjc/s10052-011-1670-x}{Eur. Phys. J. C {\bf
  71} (2011)  1670}, \href{http://arxiv.org/abs/1102.4846}{{\tt arXiv:1102.4846
  [hep-ph]}}.

\bibitem{Kilian:2007gr}
W.~Kilian, T.~Ohl, and J.~Reuter, {\em {WHIZARD: Simulating Multi-Particle
  Processes at LHC and ILC}}.
  \href{http://dx.doi.org/10.1140/epjc/s10052-011-1742-y}{Eur. Phys. J. C {\bf
  71} (2011)  1742}, \href{http://arxiv.org/abs/0708.4233}{{\tt arXiv:0708.4233
  [hep-ph]}}.

\bibitem{Moretti:2001zz}
M.~Moretti, T.~Ohl, and J.~Reuter, {\em {O'Mega: An Optimizing matrix element
  generator}}. \href{http://arxiv.org/abs/hep-ph/0102195}{{\tt
  arXiv:hep-ph/0102195}}.

\bibitem{Schwan:2018nhl}
C.~Schwan, {\em {Vector-boson scattering at the LHC}}.
  \href{http://dx.doi.org/10.22323/1.290.0081}{PoS {\bf RADCOR2017} (2018)
  081}.

\bibitem{ATLAS:WWMC}
{\bf ATLAS} Collaboration, {\em {Modelling of the vector boson scattering
  process $pp\rightarrow W^\pm W^\pm jj$ in Monte Carlo generators in ATLAS,
  ATL-PHYS-PUB-2019-004}} tech. rep., 2019.

\bibitem{ATLAS:2020ryx}
{\bf ATLAS, CMS, LHC EWWG MB Working Group} Collaboration, {\em {Comparison of
  ATLAS and CMS VBS Monte Carlo simulation, ATL-PHYS-PUB-2020-026}} tech. rep.,
  2020.

\bibitem{ATLAS:ssWW}
{\bf ATLAS} Collaboration, M.~Aaboud {\em et al.}, {\em {Observation of
  electroweak production of a same-sign $W$ boson pair in association with two
  jets in pp collisions at $\sqrt{s}=13$ TeV with the ATLAS detector}}.
  \href{http://dx.doi.org/10.1103/PhysRevLett.123.161801}{Phys. Rev. Lett. {\bf
  123} (2019)  161801}, \href{http://arxiv.org/abs/1906.03203}{{\tt
  arXiv:1906.03203 [hep-ex]}}.

\bibitem{CMS:ssWW}
{\bf CMS} Collaboration, A.~M. Sirunyan {\em et al.}, {\em {Observation of
  electroweak production of same-sign W boson pairs in the two jet and two
  same-sign lepton final state in proton-proton collisions at $\sqrt{s} = $ 13
  TeV}}. \href{http://dx.doi.org/10.1103/PhysRevLett.120.081801}{Phys. Rev.
  Lett. {\bf 120} (2018)  081801},
\href{http://arxiv.org/abs/1709.05822}{{\tt arXiv:1709.05822 [hep-ex]}}.

\bibitem{Hoeche}
S.~H{\"o}che, ``{Status of Sherpa event generator},'' in {\em Proceedings of
  the Multi-Boson Interactions Workshop (MBI), University of Michigan}.
\newblock 8, 2018.

\bibitem{Rainwater:1996ud}
D.~L. Rainwater, R.~Szalapski, and D.~Zeppenfeld, {\em {Probing color singlet
  exchange in $Z$ + two jet events at the CERN LHC}}.
  \href{http://dx.doi.org/10.1103/PhysRevD.54.6680}{Phys. Rev. D {\bf 54}
  (1996)  6680}, \href{http://arxiv.org/abs/hep-ph/9605444}{{\tt
  arXiv:hep-ph/9605444}}.

\bibitem{CMS:WWpolar}
{\bf CMS} Collaboration, A.~M. Sirunyan {\em et al.}, {\em {Measurements of
  production cross sections of polarized same-sign W boson pairs in association
  with two jets in proton-proton collisions at $\sqrt{s} =$ 13 TeV}}.
  \href{http://dx.doi.org/10.1016/j.physletb.2020.136018}{Phys. Lett. B {\bf
  812} (2021)  136018}, \href{http://arxiv.org/abs/2009.09429}{{\tt
  arXiv:2009.09429 [hep-ex]}}.

\bibitem{Campanario:2013qba}
F.~Campanario, M.~Kerner, L.~D. Ninh, and D.~Zeppenfeld, {\em {WZ Production in
  Association with Two Jets at Next-to-Leading Order in QCD}}.
  \href{http://dx.doi.org/10.1103/PhysRevLett.111.052003}{Phys. Rev. Lett. {\bf
  111} (2013) no.~5, 052003}, \href{http://arxiv.org/abs/1305.1623}{{\tt
  arXiv:1305.1623 [hep-ph]}}.

\bibitem{Bendavid:2018nar}
{\em {Les Houches 2017: Physics at TeV Colliders Standard Model Working Group
  Report}}.
\newblock 3, 2018.
\newblock \href{http://arxiv.org/abs/1803.07977}{{\tt arXiv:1803.07977
  [hep-ph]}}.

\bibitem{ATLAS:2019hoc}
{\bf ATLAS} Collaboration, {\em {Modelling of the vector boson scattering
  process $pp\rightarrow W^\pm W^\pm jj$ in Monte Carlo generators in ATLAS}}
  tech. rep., 2019.

\bibitem{ATLAS:WZ}
{\bf ATLAS} Collaboration, M.~Aaboud {\em et al.}, {\em {Observation of
  electroweak W$^{\pm}$Z boson pair production in association with two jets in
  pp collisions at $\sqrt{s} =$ 13 TeV with the ATLAS detector}}.
  \href{http://dx.doi.org/10.1016/j.physletb.2019.05.012}{Phys. Lett. B {\bf
  793} (2019)  469}, \href{http://arxiv.org/abs/1812.09740}{{\tt
  arXiv:1812.09740 [hep-ex]}}.

\bibitem{Alwall:2007fs}
J.~Alwall {\em et al.}, {\em {Comparative study of various algorithms for the
  merging of parton showers and matrix elements in hadronic collisions}}.
  \href{http://dx.doi.org/10.1140/epjc/s10052-007-0490-5}{Eur. Phys. J. C {\bf
  53} (2008)  473}, \href{http://arxiv.org/abs/0706.2569}{{\tt arXiv:0706.2569
  [hep-ph]}}.

\bibitem{Grazzini:2017ckn}
M.~Grazzini, S.~Kallweit, D.~Rathlev, and M.~Wiesemann, {\em {$W^\pm Z$
  production at the LHC: fiducial cross sections and distributions in NNLO
  QCD}}. \href{http://dx.doi.org/10.1007/JHEP05(2017)139}{JHEP {\bf 05} (2017)
  139}, \href{http://arxiv.org/abs/1703.09065}{{\tt arXiv:1703.09065
  [hep-ph]}}.

\bibitem{Melia:2011tj}
T.~Melia, P.~Nason, R.~R{\"o}ntsch, and G.~Zanderighi, {\em {$W^+W^-$, WZ and
  ZZ production in the POWHEG BOX}}.
  \href{http://dx.doi.org/10.1007/JHEP11(2011)078}{JHEP {\bf 11} (2011)  078},
  \href{http://arxiv.org/abs/1107.5051}{{\tt arXiv:1107.5051 [hep-ph]}}.

\bibitem{Campanario:2014ioa}
F.~Campanario, M.~Kerner, L.~D. Ninh, and D.~Zeppenfeld, {\em {Next-to-leading
  order QCD corrections to ZZ production in association with two jets}}.
  \href{http://dx.doi.org/10.1007/JHEP07(2014)148}{JHEP {\bf 07} (2014)  148},
  \href{http://arxiv.org/abs/1405.3972}{{\tt arXiv:1405.3972 [hep-ph]}}.

\bibitem{Li:2020nmi}
C.~Li, Y.~An, C.~Charlot, R.~Covarelli, Z.~Guan, and Q.~Li, {\em {Loop-induced
  $ZZ$ production at the LHC: An improved description by matrix-element
  matching}}. \href{http://dx.doi.org/10.1103/PhysRevD.102.116003}{Phys. Rev. D
  {\bf 102} (2020) no.~11, 116003}, \href{http://arxiv.org/abs/2006.12860}{{\tt
  arXiv:2006.12860 [hep-ph]}}.

\bibitem{Alioli:2021wpn}
S.~Alioli, S.~Ferrario~Ravasio, J.~M. Lindert, and R.~R\"ontsch, {\em
  {Four-lepton production in gluon fusion at NLO matched to parton showers}}.
  \href{http://arxiv.org/abs/2102.07783}{{\tt arXiv:2102.07783 [hep-ph]}}.

\bibitem{Grazzini:2021iae}
M.~Grazzini, S.~Kallweit, M.~Wiesemann, and J.~Y. Yook, {\em {Four lepton
  production in gluon fusion: off-shell Higgs effects in NLO QCD}}.
  \href{http://arxiv.org/abs/2102.08344}{{\tt arXiv:2102.08344 [hep-ph]}}.

\bibitem{ATLAS:ZZ}
{\bf ATLAS} Collaboration, G.~Aad {\em et al.}, {\em {Observation of
  electroweak production of two jets and a $Z$-boson pair with the ATLAS
  detector at the LHC}}. \href{http://arxiv.org/abs/2004.10612}{{\tt
  arXiv:2004.10612 [hep-ex]}}. Submitted to Nature Phys.

\bibitem{CMS:ZZ}
{\bf CMS} Collaboration, A.~M. Sirunyan {\em et al.}, {\em {Measurement of
  vector boson scattering and constraints on anomalous quartic couplings from
  events with four leptons and two jets in proton-proton collisions at
  $\sqrt{s}= 13$ TeV}}.
  \href{http://dx.doi.org/10.1016/j.physletb.2017.10.020}{Phys. Lett. B {\bf
  774} (2017)  682},
\href{http://arxiv.org/abs/1708.02812}{{\tt arXiv:1708.02812 [hep-ex]}}.

\bibitem{Grazzini:2018owa}
M.~Grazzini, S.~Kallweit, M.~Wiesemann, and J.~Y. Yook, {\em {$ZZ$ production
  at the LHC: NLO QCD corrections to the loop-induced gluon fusion channel}}.
  \href{http://dx.doi.org/10.1007/JHEP03(2019)070}{JHEP {\bf 03} (2019)  070},
  \href{http://arxiv.org/abs/1811.09593}{{\tt arXiv:1811.09593 [hep-ph]}}.

\bibitem{Kauer:2013qba}
N.~Kauer, {\em {Interference effects for H $\to$ WW/ZZ $\to
  \ell\bar{\nu}_\ell\bar{\ell}\nu_\ell$ searches in gluon fusion at the LHC}}.
  \href{http://dx.doi.org/10.1007/JHEP12(2013)082}{JHEP {\bf 12} (2013)  082},
  \href{http://arxiv.org/abs/1310.7011}{{\tt arXiv:1310.7011 [hep-ph]}}.

\bibitem{Gao:2010qx}
Y.~Gao, A.~V. Gritsan, Z.~Guo, K.~Melnikov, M.~Schulze, and N.~V. Tran, {\em
  {Spin determination of single-produced resonances at hadron colliders}}.
  \href{http://dx.doi.org/10.1103/PhysRevD.81.075022}{Phys. Rev. D {\bf 81}
  (2010)  075022}, \href{http://arxiv.org/abs/1001.3396}{{\tt arXiv:1001.3396
  [hep-ph]}}.
[Erratum: 10.1103/PhysRevD.81.079905].

\bibitem{Bolognesi:2012mm}
S.~Bolognesi, Y.~Gao, A.~V. Gritsan, K.~Melnikov, M.~Schulze, N.~V. Tran, and
  A.~Whitbeck, {\em {On the spin and parity of a single-produced resonance at
  the LHC}}. \href{http://dx.doi.org/10.1103/PhysRevD.86.095031}{Phys. Rev. D
  {\bf 86} (2012)  095031},
\href{http://arxiv.org/abs/1208.4018}{{\tt arXiv:1208.4018 [hep-ph]}}.

\bibitem{Gritsan:2020pib}
A.~V. Gritsan, J.~Roskes, U.~Sarica, M.~Schulze, M.~Xiao, and Y.~Zhou, {\em
  {New features in the JHU generator framework: constraining Higgs boson
  properties from on-shell and off-shell production}}.
  \href{http://dx.doi.org/10.1103/PhysRevD.102.056022}{Phys. Rev. D {\bf 102}
  (2020) no.~5, 056022}, \href{http://arxiv.org/abs/2002.09888}{{\tt
  arXiv:2002.09888 [hep-ph]}}.

\bibitem{Melia:2011dw}
T.~Melia, K.~Melnikov, R.~R{\"o}ntsch, and G.~Zanderighi, {\em {NLO QCD
  corrections for $W^+W^-$ pair production in association with two jets at
  hadron colliders}}. \href{http://dx.doi.org/10.1103/PhysRevD.83.114043}{Phys.
  Rev. D {\bf 83} (2011)  114043}, \href{http://arxiv.org/abs/1104.2327}{{\tt
  arXiv:1104.2327 [hep-ph]}}.

\bibitem{Greiner:2012im}
N.~Greiner, G.~Heinrich, P.~Mastrolia, G.~Ossola, T.~Reiter, and F.~Tramontano,
  {\em {NLO QCD corrections to the production of W+ W- plus two jets at the
  LHC}}. \href{http://dx.doi.org/10.1016/j.physletb.2012.06.027}{Phys. Lett. B
  {\bf 713} (2012)  277--283}, \href{http://arxiv.org/abs/1202.6004}{{\tt
  arXiv:1202.6004 [hep-ph]}}.

\bibitem{Rauch:2016upa}
M.~Rauch and S.~Pl{\"a}tzer, {\em {Parton Shower Matching Systematics in
  Vector-Boson-Fusion WW Production}}.
  \href{http://dx.doi.org/10.1140/epjc/s10052-017-4860-3}{Eur. Phys. J. C {\bf
  77} (2017) no.~5, 293}, \href{http://arxiv.org/abs/1605.07851}{{\tt
  arXiv:1605.07851 [hep-ph]}}.

\bibitem{tesiCardini}
A.~Cardini, ``{Study of $\PW^+\PW^-$ pair production via vector boson
  scattering in proton-proton collisions at 13 TeV with the CMS experiment}.''
  Master thesis, University of Florence, 2017.

\bibitem{Denner:2017kzu}
A.~Denner and M.~Pellen, {\em {Off-shell production of top-antitop pairs in the
  lepton+jets channel at NLO QCD}}.
  \href{http://dx.doi.org/10.1007/JHEP02(2018)013}{JHEP {\bf 02} (2018)  013},
  \href{http://arxiv.org/abs/1711.10359}{{\tt arXiv:1711.10359 [hep-ph]}}.

\bibitem{Anger:2017glm}
F.~Anger, F.~Febres~Cordero, H.~Ita, and V.~Sotnikov, {\em {NLO QCD predictions
  for $Wb\bar b$ production in association with up to three light jets at the
  LHC}}. \href{http://dx.doi.org/10.1103/PhysRevD.97.036018}{Phys. Rev. D {\bf
  97} (2018) no.~3, 036018}, \href{http://arxiv.org/abs/1712.05721}{{\tt
  arXiv:1712.05721 [hep-ph]}}.

\bibitem{ATLAS:VVsemilep}
{\bf ATLAS} Collaboration, G.~Aad {\em et al.}, {\em {Search for the
  electroweak diboson production in association with a high-mass dijet system
  in semileptonic final states in pp collisions at $\sqrt{s}=13$ TeV with the
  ATLAS detector}}. \href{http://dx.doi.org/10.1103/PhysRevD.100.032007}{Phys.
  Rev. D {\bf 100} (2019)  032007},
\href{http://arxiv.org/abs/1905.07714}{{\tt arXiv:1905.07714 [hep-ex]}}.

\bibitem{Krohn:2009th}
D.~Krohn, J.~Thaler, and L.-T. Wang, {\em {Jet Trimming}}.
  \href{http://dx.doi.org/10.1007/JHEP02(2010)084}{JHEP {\bf 02} (2010)  084},
  \href{http://arxiv.org/abs/0912.1342}{{\tt arXiv:0912.1342 [hep-ph]}}.

\bibitem{CMS:VVsemilep}
{\bf CMS} Collaboration, A.~M. Sirunyan {\em et al.}, {\em {Search for
  anomalous electroweak production of vector boson pairs in association with
  two jets in proton-proton collisions at 13 TeV}}.
  \href{http://dx.doi.org/10.1016/j.physletb.2019.134985}{Phys. Lett. B {\bf
  798} (2019)  134985},
\href{http://arxiv.org/abs/1905.07445}{{\tt arXiv:1905.07445 [hep-ex]}}.

\bibitem{Georgi:1985nv}
H.~Georgi and M.~Machacek, {\em {Doubly charged Higgs bosons}}.
  \href{http://dx.doi.org/10.1016/0550-3213(85)90325-6}{Nucl. Phys. B {\bf 262}
  (1985)  463}.

\bibitem{Larkoski:2014wba}
A.~J. Larkoski, S.~Marzani, G.~Soyez, and J.~Thaler, {\em {Soft Drop}}.
  \href{http://dx.doi.org/10.1007/JHEP05(2014)146}{JHEP {\bf 05} (2014)  146},
  \href{http://arxiv.org/abs/1402.2657}{{\tt arXiv:1402.2657 [hep-ph]}}.

\bibitem{Campanario:2013eta}
F.~Campanario, N.~Kaiser, and D.~Zeppenfeld, {\em {W$\gamma$ production in
  vector boson fusion at NLO in QCD}}.
  \href{http://dx.doi.org/10.1103/PhysRevD.89.014009}{Phys. Rev. D {\bf 89}
  (2014) no.~1, 014009}, \href{http://arxiv.org/abs/1309.7259}{{\tt
  arXiv:1309.7259 [hep-ph]}}.

\bibitem{Campanario:2014dpa}
F.~Campanario, M.~Kerner, L.~D. Ninh, and D.~Zeppenfeld, {\em {Next-to-leading
  order QCD corrections to $W \gamma$ production in association with two
  jets}}. \href{http://dx.doi.org/10.1140/epjc/s10052-014-2882-7}{Eur. Phys. J.
  C {\bf 74} (2014) no.~5, 2882}, \href{http://arxiv.org/abs/1402.0505}{{\tt
  arXiv:1402.0505 [hep-ph]}}.

\bibitem{Campanario:2017ffz}
F.~Campanario, M.~Kerner, and D.~Zeppenfeld, {\em {$Z_\gamma$ production in
  vector-boson scattering at next-to-leading order QCD}}.
  \href{http://dx.doi.org/10.1007/JHEP01(2018)160}{JHEP {\bf 01} (2018)  160},
  \href{http://arxiv.org/abs/1704.01921}{{\tt arXiv:1704.01921 [hep-ph]}}.

\bibitem{Campanario:2014wga}
F.~Campanario, M.~Kerner, L.~D. Ninh, and D.~Zeppenfeld, {\em {$Z\gamma$
  production in association with two jets at next-to-leading order QCD}}.
  \href{http://dx.doi.org/10.1140/epjc/s10052-014-3085-y}{Eur. Phys. J. C {\bf
  74} (2014) no.~9, 3085}, \href{http://arxiv.org/abs/1407.7857}{{\tt
  arXiv:1407.7857 [hep-ph]}}.

\bibitem{Campanario:2020xaf}
F.~Campanario, M.~Kerner, D.~Ninh, and I.~Rosario, {\em {Diphoton production in
  vector-boson scattering at the LHC at next-to-leading order QCD}}.
  \href{http://dx.doi.org/10.1007/JHEP06(2020)072}{JHEP {\bf 06} (2020)  072},
  \href{http://arxiv.org/abs/2002.12109}{{\tt arXiv:2002.12109 [hep-ph]}}.

\bibitem{Gehrmann:2013bga}
T.~Gehrmann, N.~Greiner, and G.~Heinrich, {\em {Precise QCD predictions for the
  production of a photon pair in association with two jets}}.
  \href{http://dx.doi.org/10.1103/PhysRevLett.111.222002}{Phys. Rev. Lett. {\bf
  111} (2013)  222002}, \href{http://arxiv.org/abs/1308.3660}{{\tt
  arXiv:1308.3660 [hep-ph]}}.

\bibitem{Badger:2013ava}
S.~Badger, A.~Guffanti, and V.~Yundin, {\em {Next-to-leading order QCD
  corrections to di-photon production in association with up to three jets at
  the Large Hadron Collider}}.
  \href{http://dx.doi.org/10.1007/JHEP03(2014)122}{JHEP {\bf 03} (2014)  122},
  \href{http://arxiv.org/abs/1312.5927}{{\tt arXiv:1312.5927 [hep-ph]}}.

\bibitem{Bern:2014vza}
Z.~Bern, L.~Dixon, F.~Febres~Cordero, S.~H{\"o}che, H.~Ita, D.~Kosower, N.~A.
  Lo~Presti, and D.~Maitre, {\em {Next-to-leading order $\gamma \gamma+2$-jet
  production at the LHC}}.
  \href{http://dx.doi.org/10.1103/PhysRevD.90.054004}{Phys. Rev. D {\bf 90}
  (2014) no.~5, 054004}, \href{http://arxiv.org/abs/1402.4127}{{\tt
  arXiv:1402.4127 [hep-ph]}}.

\bibitem{ATLAS:Zgamma}
{\bf ATLAS} Collaboration, G.~Aad {\em et al.}, {\em {Evidence for electroweak
  production of two jets in association with a $Z\gamma$ pair in $pp$
  collisions at $\sqrt{s} = 13$ TeV with the ATLAS detector}}.
  \href{http://dx.doi.org/10.1016/j.physletb.2020.135341}{Phys. Lett. B {\bf
  803} (2020)  135341}, \href{http://arxiv.org/abs/1910.09503}{{\tt
  arXiv:1910.09503 [hep-ex]}}.

\bibitem{CMS:Zgamma}
{\bf CMS} Collaboration, A.~M. Sirunyan {\em et al.}, {\em {Measurement of the
  cross section for electroweak production of a Z boson, a photon and two jets
  in proton-proton collisions at $\sqrt{s} =$ 13 TeV and constraints on
  anomalous quartic couplings}}.
  \href{http://dx.doi.org/10.1007/JHEP06(2020)076}{JHEP {\bf 06} (2020)  076},
  \href{http://arxiv.org/abs/2002.09902}{{\tt arXiv:2002.09902 [hep-ex]}}.

\bibitem{CMS:Wgamma}
{\bf CMS} Collaboration, A.~M. Sirunyan {\em et al.}, {\em {Observation of
  electroweak production of W$\gamma$ with two jets in proton-proton collisions
  at $\sqrt{s} = $ 13 TeV}}.
  \href{http://dx.doi.org/10.1016/j.physletb.2020.135988}{Phys. Lett. B {\bf
  811} (2020)  135988}, \href{http://arxiv.org/abs/2008.10521}{{\tt
  arXiv:2008.10521 [hep-ex]}}.

\bibitem{Khachatryan:2015iwa}
{\bf CMS} Collaboration, V.~Khachatryan {\em et al.}, {\em {Performance of
  Photon Reconstruction and Identification with the CMS Detector in
  Proton-Proton Collisions at sqrt(s) = 8 TeV}}.
  \href{http://dx.doi.org/10.1088/1748-0221/10/08/P08010}{JINST {\bf 10} (2015)
  no.~08, P08010}, \href{http://arxiv.org/abs/1502.02702}{{\tt arXiv:1502.02702
  [physics.ins-det]}}.

\bibitem{Aaboud:2018yqu}
{\bf ATLAS} Collaboration, M.~Aaboud {\em et al.}, {\em {Measurement of the
  photon identification efficiencies with the ATLAS detector using LHC Run 2
  data collected in 2015 and 2016}}.
  \href{http://dx.doi.org/10.1140/epjc/s10052-019-6650-6}{Eur. Phys. J. C {\bf
  79} (2019) no.~3, 205}, \href{http://arxiv.org/abs/1810.05087}{{\tt
  arXiv:1810.05087 [hep-ex]}}.

\bibitem{PPS}
{\bf CMS, TOTEM} Collaboration, {\em {CMS-TOTEM Precision Proton Spectrometer,
  CERN-LHCC-2014-021}} tech. rep., 2014.

\bibitem{Budnev:1974de}
V.~Budnev, I.~Ginzburg, G.~Meledin, and V.~Serbo, {\em {The Two photon particle
  production mechanism. Physical problems. Applications. Equivalent photon
  approximation}}. \href{http://dx.doi.org/10.1016/0370-1573(75)90009-5}{Phys.
  Rept. {\bf 15} (1975)  181--281}.

\bibitem{Ohnemus:1991kk}
J.~Ohnemus, {\em {An Order $\alpha_s$ calculation of hadronic $W^{-} W^{+}$
  production}}. \href{http://dx.doi.org/10.1103/PhysRevD.44.1403}{Phys. Rev. D
  {\bf 44} (1991)  1403--1414}.

\bibitem{Baur:1995uv}
U.~Baur, T.~Han, and J.~Ohnemus, {\em {QCD corrections and nonstandard three
  vector boson couplings in $W^{+} W^{-}$ production at hadron colliders}}.
  \href{http://dx.doi.org/10.1103/PhysRevD.53.1098}{Phys. Rev. D {\bf 53}
  (1996)  1098--1123}, \href{http://arxiv.org/abs/hep-ph/9507336}{{\tt
  arXiv:hep-ph/9507336}}.

\bibitem{Campbell:1999ah}
J.~M. Campbell and R.~Ellis, {\em {An Update on vector boson pair production at
  hadron colliders}}. \href{http://dx.doi.org/10.1103/PhysRevD.60.113006}{Phys.
  Rev. D {\bf 60} (1999)  113006},
  \href{http://arxiv.org/abs/hep-ph/9905386}{{\tt arXiv:hep-ph/9905386}}.

\bibitem{Hamilton:2010mb}
K.~Hamilton, {\em {A positive-weight next-to-leading order simulation of weak
  boson pair production}}.
  \href{http://dx.doi.org/10.1007/JHEP01(2011)009}{JHEP {\bf 01} (2011)  009},
  \href{http://arxiv.org/abs/1009.5391}{{\tt arXiv:1009.5391 [hep-ph]}}.

\bibitem{Nason:2013ydw}
P.~Nason and G.~Zanderighi, {\em {$W^+ W^-$ , $W Z$ and $Z Z$ production in the
  POWHEG-BOX-V2}}. \href{http://dx.doi.org/10.1140/epjc/s10052-013-2702-5}{Eur.
  Phys. J. C {\bf 74} (2014) no.~1, 2702},
  \href{http://arxiv.org/abs/1311.1365}{{\tt arXiv:1311.1365 [hep-ph]}}.

\bibitem{Gehrmann:2014fva}
T.~Gehrmann, M.~Grazzini, S.~Kallweit, P.~Maierh{\"o}fer, A.~von Manteuffel,
  S.~Pozzorini, D.~Rathlev, and L.~Tancredi, {\em {$W^+W^-$ Production at
  Hadron Colliders in Next to Next to Leading Order QCD}}.
  \href{http://dx.doi.org/10.1103/PhysRevLett.113.212001}{Phys. Rev. Lett. {\bf
  113} (2014) no.~21, 212001}, \href{http://arxiv.org/abs/1408.5243}{{\tt
  arXiv:1408.5243 [hep-ph]}}.

\bibitem{Grazzini:2016ctr}
M.~Grazzini, S.~Kallweit, S.~Pozzorini, D.~Rathlev, and M.~Wiesemann, {\em
  {$W^+W^-$ production at the LHC: fiducial cross sections and distributions in
  NNLO QCD}}. \href{http://dx.doi.org/10.1007/JHEP08(2016)140}{JHEP {\bf 08}
  (2016)  140}, \href{http://arxiv.org/abs/1605.02716}{{\tt arXiv:1605.02716
  [hep-ph]}}.

\bibitem{Caola:2015rqy}
F.~Caola, K.~Melnikov, R.~R{\"o}ntsch, and L.~Tancredi, {\em {QCD corrections
  to $W^+W^-$ production through gluon fusion}}.
  \href{http://dx.doi.org/10.1016/j.physletb.2016.01.046}{Phys. Lett. B {\bf
  754} (2016)  275--280}, \href{http://arxiv.org/abs/1511.08617}{{\tt
  arXiv:1511.08617 [hep-ph]}}.

\bibitem{Grazzini:2020stb}
M.~Grazzini, S.~Kallweit, M.~Wiesemann, and J.~Y. Yook, {\em {$W^+W^-$
  production at the LHC: NLO QCD corrections to the loop-induced gluon fusion
  channel}}. \href{http://dx.doi.org/10.1016/j.physletb.2020.135399}{Phys.
  Lett. B {\bf 804} (2020)  135399},
  \href{http://arxiv.org/abs/2002.01877}{{\tt arXiv:2002.01877 [hep-ph]}}.

\bibitem{Grazzini:2015wpa}
M.~Grazzini, S.~Kallweit, D.~Rathlev, and M.~Wiesemann, {\em
  {Transverse-momentum resummation for vector-boson pair production at
  NNLL+NNLO}}. \href{http://dx.doi.org/10.1007/JHEP08(2015)154}{JHEP {\bf 08}
  (2015)  154}, \href{http://arxiv.org/abs/1507.02565}{{\tt arXiv:1507.02565
  [hep-ph]}}.

\bibitem{Kuhn:2011mh}
J.~K{\"u}hn, F.~Metzler, A.~Penin, and S.~Uccirati, {\em
  {Next-to-Next-to-Leading Electroweak Logarithms for W-Pair Production at
  LHC}}. \href{http://dx.doi.org/10.1007/JHEP06(2011)143}{JHEP {\bf 06} (2011)
  143}, \href{http://arxiv.org/abs/1101.2563}{{\tt arXiv:1101.2563 [hep-ph]}}.

\bibitem{Bierweiler:2012kw}
A.~Bierweiler, T.~Kasprzik, J.~H. K\"uhn, and S.~Uccirati, {\em {Electroweak
  corrections to W-boson pair production at the LHC}}.
  \href{http://dx.doi.org/10.1007/JHEP11(2012)093}{JHEP {\bf 11} (2012)  093},
  \href{http://arxiv.org/abs/1208.3147}{{\tt arXiv:1208.3147 [hep-ph]}}.

\bibitem{Baglio:2013toa}
J.~Baglio, L.~D. Ninh, and M.~M. Weber, {\em {Massive gauge boson pair
  production at the LHC: a next-to-leading order story}}.
  \href{http://dx.doi.org/10.1103/PhysRevD.94.099902}{Phys. Rev. D {\bf 88}
  (2013)  113005}, \href{http://arxiv.org/abs/1307.4331}{{\tt arXiv:1307.4331
  [hep-ph]}}. [Erratum: Phys.Rev.D 94, 099902 (2016)].

\bibitem{Gieseke:2014gka}
S.~Gieseke, T.~Kasprzik, and J.~H. K\"uhn, {\em {Vector-boson pair production
  and electroweak corrections in HERWIG++}}.
  \href{http://dx.doi.org/10.1140/epjc/s10052-014-2988-y}{Eur. Phys. J. C {\bf
  74} (2014) no.~8, 2988}, \href{http://arxiv.org/abs/1401.3964}{{\tt
  arXiv:1401.3964 [hep-ph]}}.

\bibitem{Biedermann:2016guo}
B.~Biedermann, M.~Billoni, A.~Denner, S.~Dittmaier, L.~Hofer, B.~J\"ager, and
  L.~Salfelder, {\em {Next-to-leading-order electroweak corrections to $pp \to
  W^+W^-\to$ 4 leptons at the LHC}}.
  \href{http://dx.doi.org/10.1007/JHEP06(2016)065}{JHEP {\bf 06} (2016)  065},
  \href{http://arxiv.org/abs/1605.03419}{{\tt arXiv:1605.03419 [hep-ph]}}.

\bibitem{Kallweit:2017khh}
S.~Kallweit, J.~Lindert, S.~Pozzorini, and M.~Sch\"onherr, {\em {NLO QCD+EW
  predictions for $2\ell2\nu$ diboson signatures at the LHC}}.
  \href{http://dx.doi.org/10.1007/JHEP11(2017)120}{JHEP {\bf 11} (2017)  120},
  \href{http://arxiv.org/abs/1705.00598}{{\tt arXiv:1705.00598 [hep-ph]}}.

\bibitem{Re:2018vac}
E.~Re, M.~Wiesemann, and G.~Zanderighi, {\em {NNLOPS accurate predictions for
  $W^+W^-$ production}}. \href{http://dx.doi.org/10.1007/JHEP12(2018)121}{JHEP
  {\bf 12} (2018)  121}, \href{http://arxiv.org/abs/1805.09857}{{\tt
  arXiv:1805.09857 [hep-ph]}}.

\bibitem{Kallweit:2019zez}
M.~Grazzini, S.~Kallweit, J.~M. Lindert, S.~Pozzorini, and M.~Wiesemann, {\em
  {NNLO QCD + NLO EW with Matrix+OpenLoops: precise predictions for
  vector-boson pair production}}.
  \href{http://dx.doi.org/10.1007/JHEP02(2020)087}{JHEP {\bf 02} (2020)  087},
  \href{http://arxiv.org/abs/1912.00068}{{\tt arXiv:1912.00068 [hep-ph]}}.

\bibitem{Kallweit:2020gva}
S.~Kallweit, E.~Re, L.~Rottoli, and M.~Wiesemann, {\em {Accurate single- and
  double-differential resummation of colour-singlet processes with
  MATRIX+RADISH: $W^+W^-$ production at the LHC}}.
  \href{http://dx.doi.org/10.1007/JHEP12(2020)147}{JHEP {\bf 12} (2020)  147},
  \href{http://arxiv.org/abs/2004.07720}{{\tt arXiv:2004.07720 [hep-ph]}}.

\bibitem{Brauer:2020kfv}
S.~Br{\"a}uer, A.~Denner, M.~Pellen, M.~Sch{\"o}nherr, and S.~Schumann, {\em
  {Fixed-order and merged parton-shower predictions for WW and WWj production
  at the LHC including NLO QCD and EW corrections}}.
  \href{http://dx.doi.org/10.1007/JHEP10(2020)159}{JHEP {\bf 10} (2020)  159},
\href{http://arxiv.org/abs/2005.12128}{{\tt arXiv:2005.12128 [hep-ph]}}.

\bibitem{Chiesa:2020ttl}
M.~Chiesa, C.~Oleari, and E.~Re, {\em {NLO QCD+NLO EW corrections to diboson
  production matched to parton shower}}.
  \href{http://dx.doi.org/10.1140/epjc/s10052-020-8419-3}{Eur. Phys. J. C {\bf
  80} (2020) no.~9, 849}, \href{http://arxiv.org/abs/2005.12146}{{\tt
  arXiv:2005.12146 [hep-ph]}}.

\bibitem{ATLAS:ggWW}
{\bf ATLAS} Collaboration, G.~Aad {\em et al.}, {\em {Observation of
  photon-induced $W^+W^-$ production in $pp$ collisions at $\sqrt{s}=13$ TeV
  using the ATLAS detector}}. \href{http://arxiv.org/abs/2010.04019}{{\tt
  arXiv:2010.04019 [hep-ex]}}.

\bibitem{CMS:ggWW}
{\bf CMS} Collaboration, V.~Khachatryan {\em et al.}, {\em {Evidence for
  exclusive $\gamma\gamma \to W^+ W^-$ production and constraints on anomalous
  quartic gauge couplings in $pp$ collisions at $ \sqrt{s}=7 $ and 8 TeV}}.
  \href{http://dx.doi.org/10.1007/JHEP08(2016)119}{JHEP {\bf 08} (2016)  119},
  \href{http://arxiv.org/abs/1604.04464}{{\tt arXiv:1604.04464 [hep-ex]}}.

\bibitem{Mangano:2016jyj}
M.~L. Mangano {\em et al.}, {\em {Physics at a 100 TeV pp Collider: Standard
  Model Processes}}. \href{http://dx.doi.org/10.23731/CYRM-2017-003.1}{CERN
  Yellow Rep. (2017) no.~3, 1--254},
\href{http://arxiv.org/abs/1607.01831}{{\tt arXiv:1607.01831 [hep-ph]}}.

\bibitem{Jager:2017owh}
B.~J{\"a}ger, L.~Salfelder, M.~Worek, and D.~Zeppenfeld, {\em {Physics
  opportunities for vector-boson scattering at a future 100 TeV hadron
  collider}}. \href{http://dx.doi.org/10.1103/PhysRevD.96.073008}{Phys. Rev.
  {\bf D96} (2017) no.~7, 073008},
\href{http://arxiv.org/abs/1704.04911}{{\tt arXiv:1704.04911 [hep-ph]}}.

\end{thebibliography}\endgroup
\end{document}